\documentclass[aps,prx,10pt,superscriptaddress,twocolumn,floatfix,footinbib,longbibliography]{revtex4-2}

\usepackage{array}[=2016-10-06]
\usepackage[T1]{fontenc}
\usepackage{lmodern,microtype,amsmath,amssymb,amsthm,mathtools,bm,braket}
\usepackage{graphicx,booktabs,tabularx,xcolor,tikz,enumitem,needspace}
\usepackage[colorlinks=true,linkcolor=blue!45!black,citecolor=blue!45!black,urlcolor=blue!45!black]{hyperref}
\usetikzlibrary{arrows.meta,positioning,calc,fit,backgrounds,decorations.pathreplacing,shapes.geometric}

\hypersetup{pdftitle={Compatible Check Measurements for Fault-Tolerant Quantum Code Surgery},pdfauthor={Zongyuan Wang}}

\newtheorem{theorem}{Theorem}
\newtheorem{proposition}[theorem]{Proposition}
\newtheorem{lemma}[theorem]{Lemma}
\newtheorem{corollary}[theorem]{Corollary}

\makeatletter
\DeclareRobustCommand{\citelist}[1]{{\let\NAT@cmprs\z@\cite{#1}}}
\makeatother

\newcounter{protocol}
\newcommand{\protocoltitle}[1]{\par\medskip\noindent\refstepcounter{protocol}\textbf{Protocol \theprotocol. #1.}\par\nobreak}
\newenvironment{protocolsteps}{\begin{enumerate}[label=\arabic*.,leftmargin=1.6em,labelsep=.4em,topsep=3pt,itemsep=2pt,parsep=0pt]}{\end{enumerate}}

\allowdisplaybreaks[1]

\definecolor{dFourRed}{HTML}{DB3636}
\definecolor{dFourGreen}{HTML}{159447}
\definecolor{dFourBlue}{HTML}{2467DE}

\begin{document}

\title{Compatible Check Measurements for Fault-Tolerant Quantum Code Surgery}
\author{Zongyuan Wang}
\email{zongyuan.wang15@gmail.com}
\affiliation{International Center for Quantum Materials, School of Physics, Peking University, Beijing 100871, China}

\date{\today}

\begin{abstract}
Gauging turns a logical operator into a product of local checks, providing a route to logical measurement and lattice surgery. These checks can fail to commute outside the code space, so measuring one can randomize the outcome of another. We introduce completion qubits or qudits that make the checks commute on the full Hilbert space, enabling compatible measurements. For check algebras with nontrivial central commutators, completion preserves the logical measurement and the distance of the raw gauged code. For qubits with Pauli central checks, completion maps Clifford raw checks to Clifford completed checks. We give explicit circuits for bilayer toric-code surgery through the non-Abelian $D(D_4)$ quantum double, providing a logical Toffoli-state source. On an even $L\times L$ torus with $R$ gauged rounds, the error-detecting circuit has exact fault distance $\min\{L,R\}$ under arbitrary local circuit faults, including faults at both code switches. Completion qubits enable coherent recovery of the intended logical measurement before ungauging. We extend completion to a class of check algebras with noncentral commutators. We also construct compatible checks without completion qubits or qudits, including an example for $D(S_3)$. Without completion qubits or qudits, compatible measurements can involve higher check hierarchy levels and additional measurements for recovery. For $D(S_3)$ in its fixed qubit--qutrit encoding, we show that no product-state completion gives commuting checks at finite Clifford-hierarchy levels for the same code. Our results provide a framework for fault-tolerant logical measurements using commuting checks, with applications to magic-state preparation by lattice surgery.
\end{abstract}

\maketitle

\section{Introduction}\label{sec:prx-introduction}

Lattice surgery measures joint logical operators by merging and splitting code patches. For example, in the surface code, measuring checks along their shared boundary reveals the parity $\overline{Z}_1\overline{Z}_2$ while preserving superpositions within each parity sector~\cite{Horsman2012Surgery}. Recently, lattice surgery has expanded beyond surface and color codes~\cite{LandahlRyanAnderson2014} to more general quantum low-density parity-check (LDPC) codes~\cite{Cohen2022Lattice,CowtanBurton2024CSS,Ide2025Homological}. Generalizations connect distinct code families~\cite{Cross2024QLDPC}, measure several logical Pauli operators in parallel~\cite{Cowtan2026Parallel}, and reduce the number of auxiliary qubits needed for a measurement~\cite{Yuan2026Parsimonious}.

Gauging implements a logical measurement by treating the logical operator as a symmetry and expressing it as a product of local checks. Measuring these checks and then ungauging returns the state to the parent code with the symmetry measured~\cite{WilliamsonYoder2026,Chang2026Magic}. This extends to Clifford symmetries, whose measurement can prepare magic states~\cite{Davydova2026,Williamson2026HigherForm,Hetenyi2026Cultivation,Chang2026Magic}, and connects code surgery to non-Abelian topological order~\cite{HuangHybrid2025,Sajith2025Hybrid,Christos2026NonAbelian,Zhu2026NonAbelianQLDPC}.

\label{par:prx277-prior-work}
Gauging-based logical measurements already have distance guarantees~\cite{WilliamsonYoder2026,Chang2026Magic}, and non-Abelian code protocols support universal computation with a threshold under local stochastic circuit noise~\cite{Davydova2026}. Our starting point is a further question about the measurements themselves. We call the checks obtained directly from gauging \emph{raw checks}. They commute on the code space, but need not commute on states produced by errors. An otherwise ideal measurement can then change the outcome of another, including the reported logical symmetry. The protected code space alone does not determine this behavior during repeated measurements.

We construct compatible local measurements for gauged non-Pauli codes and give simple, explicit circuits for measuring them. We analyze faults in the specified gates and code-switching operations to establish circuit-distance guarantees for logical measurements and state preparation. We use one readout ancilla per binary check. We add completion qubits or qudits that make the checks commute on the full Hilbert space while preserving the intended logical measurement. For raw checks with nontrivial central commutators, completion appends a fixed product state to the raw code space and preserves its distance exactly. For qubits with Pauli central checks, the completed checks are Clifford whenever the raw checks are Clifford.  Completion therefore changes the measurements available on error states while keeping the protected encoded physics fixed.

Our main circuit demonstration is surgery between two toric codes and the $D(D_4)$ quantum double. For a specified gate order on an even $L\times L$ torus with $R$ gauged rounds, we prove the exact circuit fault distance $d_{\mathrm{surg}}=\min\{L,R\}$. The analysis includes arbitrary local faults at individual gates, preparations, readouts and idles, including correlated errors from entangling gates and faults at both surgery interfaces. The same circuit prepares a logical Toffoli state. The result combines compatible measurements with a circuit-specific propagation argument that preserves the full distance.

Completion qubits serve two roles in this surgery. They make the measurements compatible and retain quantum information for recovery. The first makes the joint check outcomes repeatable in later ideal rounds. The second supports a coherent repair of certain detected errors at the interface into gauging, before ungauging, recovering the intended logical measurement on the original encoded input when subsequent operations are ideal. The same completion qubits can also signal certain faults in the check-measurement circuits.

We establish the scope of the construction through several extensions. Completion applies to finite-order qudit checks with central commutators and to noncentral relations arising from semidirect products of finite abelian groups. A color-code logical $T$ protocol and a dihedral family have linear circuit-distance guarantees. Explicit circuits for the gross code illustrate the construction beyond geometrically local parent codes~\cite{Bravyi2024Memory}. We also construct commuting checks without completion qubits and compare the measured observables, circuit costs and recovery operations. For $D(S_3)$ in its fixed qubit--qutrit encoding, we identify an obstruction to completion with commuting checks at finite Clifford-hierarchy levels. Finally, the physical gauging maps give symmetry-protected topological (SPT) Hamiltonians, including higher-form examples, whose ground states provide resources for these maps.

Our main example gauges two toric codes to the $D(D_4)$ quantum double and back. To see the problem motivating completion, start with its raw-check conflict in Eq.~\eqref{eq:prx332-raw-conflict} of Section~\ref{sec:prx-preliminaries}. For the central physical idea, Section~\ref{sec:prx-operational} compares a simple pair of measurements before and after completion and explains what changes when errors occur. Section~\ref{sec:prx-d4} gives the surgery circuit between two toric codes and its exact distance. Section~\ref{sec:prx-d4-recovery} explains how the retained completion qubits support coherent recovery before ungauging. Appendix~\ref{sec:sm297-coherent-recovery} gives the detailed recovery procedure for errors present when gauging begins. The general construction and circuit argument are developed in Section~\ref{sec:prx-central}. Section~\ref{sec:prx-semidirect} gives the noncentral extension, with a detailed $D_8$ example in Appendix~\ref{sec:prx278-d8}. Section~\ref{sec:sm-without-completion} compares alternatives without completion. Appendix~\ref{sec:prx314-spt-details} gives the physical gauging map, state-preparation protocol and SPT resource Hamiltonians.

\section{Preliminaries}
\label{sec:prx-preliminaries}
In this section, we review how gauging turns a logical symmetry measurement into measurements of local checks~\cite{WilliamsonYoder2026,Chang2026Magic}. We first explain the construction for a Pauli stabilizer code and a Clifford symmetry built from local operations. We then show why the resulting checks can become incompatible after errors.

\subsection{Raw checks for \texorpdfstring{$\mathbb{Z}_2$}{Z2} logical gauging}
\label{sec:prx-raw-gauging}

Let us consider a Pauli stabilizer code with generators $S_i$, stabilizer group $\mathcal{S}=\langle S_i\rangle$ and code projector $P$. The original data qubits are called \emph{matter qubits} in the language of gauging in condensed matter physics~\cite{VijayHaahFu2016,Wegner1971,Tantivasadakarn2024LongRange}. Group the data qubits into disjoint blocks of bounded size, labeled by $v$. Let $\mathcal U$ be a Clifford symmetry of the code ($\mathcal U P\mathcal U^\dagger=P$) of the form
\begin{equation}
 \mathcal{U}=\prod_v U_v,\qquad U_v^2=I,
 \label{eq:prx273-block-symmetry}
\end{equation}
where $U_v$ is a Clifford operator acting only on block $v$. These operators need not preserve the code space individually. Since each $U_v$ squares to identity, the logical observable $\mathcal U$ has eigenvalues $\pm1$. For example, consider a transversal CNOT between two copies of a CSS code. Each block $v$ contains a control qubit and its corresponding target qubit, and $U_v$ acts as CNOT on this pair.

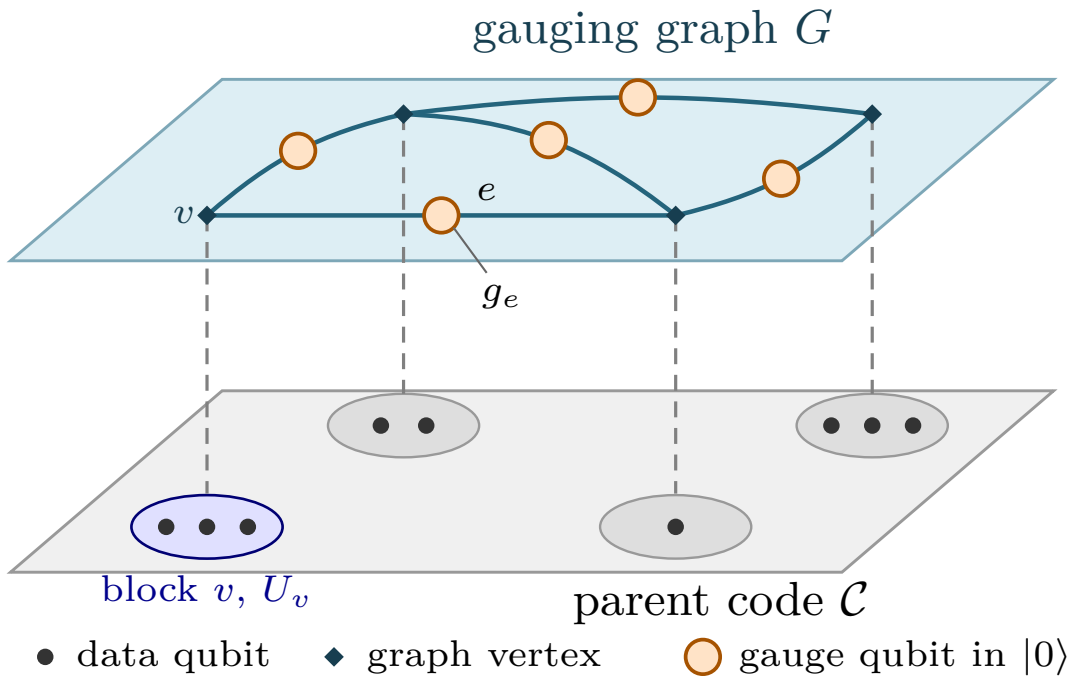
\begin{figure}[tb]
\centering
\definecolor{gaugingPlane}{HTML}{DDEEF4}
\definecolor{gaugingBorder}{HTML}{7DA7B7}
\definecolor{gaugingEdge}{HTML}{24647C}
\definecolor{gaugingVertex}{HTML}{173E50}
\begin{tikzpicture}[
  x=1cm,y=1cm,font=\scriptsize,text=black,
  edge/.style={draw=gaugingEdge,line width=.85pt},
  association/.style={draw=black!50,densely dashed,line width=.6pt},
  vertex/.style={diamond,fill=gaugingVertex,draw=none,
    minimum size=3.7pt,inner sep=0pt,outer sep=0pt},
  data/.style={circle,fill=black!80,draw=none,
    minimum size=3.1pt,inner sep=0pt,outer sep=0pt},
  gauge/.style={circle,draw=orange!65!black,fill=orange!22,
    line width=.7pt,minimum size=6.5pt,inner sep=0pt,outer sep=0pt}
]
  \path[use as bounding box] (-.10,-.76) rectangle (7.70,3.80);

  \filldraw[fill=black!6,draw=black!40,line width=.55pt]
    (.15,0)--(5.65,0)--(7.05,1.20)--(1.55,1.20)--cycle;
  \filldraw[fill=blue!12,draw=blue!45!black,line width=.5pt]
    (1.45,.30) ellipse[x radius=.50,y radius=.21];
  \foreach \x/\y in {4.55/.30,2.75/.97,5.85/.97}
    \filldraw[fill=black!13,draw=black!40,line width=.45pt]
      (\x,\y) ellipse[x radius=.50,y radius=.21];
  \foreach \x/\y in {1.18/.30,1.45/.30,1.72/.30,
                      4.55/.30,
                      2.60/.97,2.90/.97,
                      5.58/.97,5.85/.97,6.12/.97}
    \node[data] at (\x,\y) {};
  \node[anchor=north,inner sep=1pt,text=blue!55!black]
    at (1.45,.02) {block $v$, $U_v$};
  \node[anchor=north,inner sep=1pt,font=\small] at (4.85,-.02)
    {parent code $\mathcal C$};

  \filldraw[fill=gaugingPlane,draw=gaugingBorder,line width=.55pt]
    (.15,2.06)--(5.65,2.06)--(7.05,3.26)--(1.55,3.26)--cycle;
  \coordinate (v) at (1.45,2.36);
  \coordinate (u) at (4.55,2.36);
  \coordinate (w) at (2.75,3.03);
  \coordinate (r) at (5.85,3.03);
  \draw[association] (v)--(1.45,.52);
  \draw[association] (u)--(4.55,.52);
  \draw[association] (w)--(2.75,1.18);
  \draw[association] (r)--(5.85,1.18);
  \draw[edge] (v)--node[midway,gauge] (gvu) {} (u);
  \foreach \a/\b/\bend in {v/w/14,w/r/7,r/u/14,w/u/18}{
    \draw[edge] (\a) to[bend left=\bend]
      node[midway,gauge] (g\a\b) {} (\b);
  }
  \foreach \a in {v,u,w,r} \node[vertex] at (\a) {};
  \node[anchor=east,inner sep=2pt,text=gaugingVertex] at (v) {$v$};
  \node[anchor=south,inner sep=1pt] at (3.30,2.42) {$e$};
  \draw[draw=black!60,line width=.4pt] (gvu.south east)--(3.30,1.98);
  \node[anchor=north,inner sep=1pt] at (3.40,1.94) {$g_e$};
  \node[anchor=south,inner sep=1pt,text=gaugingVertex,font=\small]
    at (4.40,3.40) {gauging graph $G$};

  \node[data] at (.38,-.56) {};
  \node[anchor=west,inner sep=2pt] at (.51,-.56) {data qubit};
  \node[vertex] at (2.29,-.56) {};
  \node[anchor=west,inner sep=2pt] at (2.43,-.56) {graph vertex};
  \node[gauge] at (4.73,-.56) {};
  \node[anchor=west,inner sep=2pt] at (4.89,-.56)
    {gauge qubit in $|0\rangle$};
\end{tikzpicture}

\caption{Graph for logical gauging, adapted from~\cite{WilliamsonYoder2026,Chang2026Magic}. The parent code is shown below the graph $G$. Dashed lines identify graph vertices with disjoint blocks of data qubits, on which the operators $U_v$ act. Each graph edge carries a gauge qubit initialized in $|0\rangle$. The vertical separation is schematic.}
\label{fig:prx326-gauging-graph}
\end{figure}

To connect these blocks, choose a connected graph $G=(V,E)$ as illustrated in Fig.~\ref{fig:prx326-gauging-graph}. Each vertex represents a block on which $U_v\ne I$, and each edge $e$ carries a gauge qubit $g_e$, initially in $|0\rangle$. The raw check family has three parts. Gauss checks measure the symmetry, flux checks constrain the gauge qubits, and modified parent checks track the original stabilizers.

At each vertex, the \emph{Gauss check} $A_v$ applies $U_v$ to the matter block and flips the gauge qubits on the incident edges. A \emph{flux check} $B_p$ measures the parity of the gauge bits around a chosen closed path $\partial p$. Thus,
\begin{equation}
 A_v=U_v\prod_{e\ni v}X_{g_e},\qquad
 B_p=\prod_{e\in\partial p}Z_{g_e}.
 \label{eq:prx273-gauss-flux}
\end{equation}
Every edge flip occurs twice in the product of all Gauss checks and cancels, leaving $\prod_v A_v=\mathcal U$. Thus, if the Gauss checks give eigenvalues $a_v=\pm1$, their product $\prod_v a_v$ gives the desired logical symmetry eigenvalue.

\paragraph{Gauging circuit.}
The circuit uses one readout ancilla $\mu_v$ at each vertex, initialized in $|+\rangle$. It first couples the ancilla to the matter block, then to the incident gauge qubits, and finally measures it in $X$. The matter and gauge qubits remain in the gauged code. The first step, called \emph{symmetry enrichment}, applies
\begin{equation}
 \bm{E}_{\mathrm{enr}}=\prod_v\operatorname{ctrl}_{\mu_v}(U_v).
 \label{eq:prx273-enrichment-circuit}
\end{equation}
Each controlled-$U_v$ couples the ancilla's initial $X$ constraint to the local matter symmetry~\cite{Verresen2021Enrichment,Tantivasadakarn2024LongRange},
\begin{equation}
 \bm{E}_{\mathrm{enr}}X_{\mu_v}\bm{E}_{\mathrm{enr}}^\dagger
 =X_{\mu_v}U_v.
 \label{eq:prx335-enriched-constraint}
\end{equation}
Thus, the enriched subspace satisfies $X_{\mu_v}U_v=+1$. Within this subspace, the product $\prod_vX_{\mu_v}$ represents the original global symmetry $\mathcal U$. Enrichment changes the local representation of the symmetry while preserving its action on both eigenspaces.

The second step transfers the ancilla parity information to the edges. For each edge $e=(uv)$, apply CNOTs from $\mu_u$ and $\mu_v$ to $g_e$, then measure the ancillas in $X$. This gauges the auxiliary Pauli symmetry $\prod_v X_{\mu_v}$~\cite{WilliamsonYoder2026} and generalizes Kramers--Wannier duality to the graph $G$~\cite{KramersWannier1941,Tantivasadakarn2024LongRange}. For positive readout outcomes, the resulting nonunitary map has the operator correspondences
\begin{equation}
 \begin{aligned}
 X_{\mu_v}&\longmapsto\prod_{e\ni v}X_{g_e},\\
 Z_{\mu_u}Z_{\mu_v}&\longmapsto Z_{g_e},
 \end{aligned}
 \label{eq:prx336-kw-map}
\end{equation}
Before readout, the CNOTs send $X_{\mu_v}$ to $X_{\mu_v}\prod_{e\ni v}X_{g_e}$ and the initial edge constraint to $Z_{\mu_u}Z_{g_e}Z_{\mu_v}=+1$. These give the two correspondences above. Since the CNOTs act trivially on matter, the enriched constraint becomes $X_{\mu_v}A_v=+1$. A positive ancilla readout therefore gives
\begin{equation}
 X_{\mu_v}U_v\longmapsto
 U_v\prod_{e\ni v}X_{g_e}=A_v.
 \label{eq:prx336-enriched-duality}
\end{equation}
For general readout outcomes $a_v$, the first correspondence is $X_{\mu_v}\mapsto a_v\prod_{e\ni v}X_{g_e}$, giving $A_v=a_v$. The outcome product selects the original symmetry sector $\mathcal U=\prod_v a_v$. Together, the enrichment and CNOT gates form a controlled-$A_v$ circuit with readout ancilla $\mu_v$. This implements the initial raw-Gauss measurement. The completed protocol below adds completion factors to this circuit. Appendix~\ref{sec:sm269-physical-gauging} gives the signed derivation.

The second correspondence also explains the flux constraint $B_p=+1$. Multiplying the endpoint parities around a closed path cancels every vertex $Z$ operator, leaving the identity.

\paragraph{Modified parent checks.}
We next modify each parent generator $S_i$ to account for the local symmetry operations accompanying gauging~\cite{Chang2026Magic}. The gauge qubits record the relative symmetry choices between neighboring blocks. In an ideal gauge configuration, an edge has value $z_{uv}=x_u+x_v\bmod2$, where $x_v$ specifies whether $U_v$ acts on block $v$. Summing these edge values along a path cancels the interior bits and gives the relative choice at its endpoints.

For a given $S_i$, call a block \emph{active} when both $S_i$ and $U_v$ act nontrivially there. Choose one active block $r_i$ as a reference. For each other active block $v$, choose a path $p_i(v)$ from $r_i$ to $v$. Its parity tells us whether to apply $U_v$ to align that block's symmetry choice with the reference.

Write $Z_g(p)=\prod_{e\in p}Z_{g_e}$. For $p=p_i(v)$, its eigenvalue on the ideal configuration is $(-1)^{x_v+x_{r_i}}$. We therefore apply $U_v$ when this parity is negative and the identity when it is positive. Keeping these choices coherent gives the control unitary $\bm W_i$ and the modified check $\widetilde S_i$ below. We use $p_i(r_i)=0$ and $p_i(v)=0$ outside the active blocks, with $Z_g(0)=I$, so these blocks contribute no control factors.
\begin{align}
 \bm{W}_i
 &=\prod_{v\ne r_i}\left[
   \frac{I+Z_g(p_i(v))}{2}
   +\frac{I-Z_g(p_i(v))}{2}U_v
 \right],\nonumber\\
 \widetilde S_i&=\bm{W}_i S_i\bm{W}_i^\dagger.
 \label{eq:prx273-rooted-check}
\end{align}
\begin{figure*}[tbp]
\centering
\begingroup
\definecolor{mcRed}{HTML}{DB3636}
\definecolor{mcGreen}{HTML}{159447}
\definecolor{mcBlue}{HTML}{2467DE}
\definecolor{mcTriangle}{HTML}{C18A38}
\begin{minipage}{.80\textwidth}\centering
\begin{minipage}[t]{.32\linewidth}\centering
\textbf{(a) Parent star}\par\smallskip
\begin{tikzpicture}[x=1.20cm,y=1.20cm,font=\scriptsize,text=black]
  \path[use as bounding box] (-1.28,-1.13) rectangle (1.55,1.17);
  \draw[step=1,black!25,line width=.35pt] (-1,-1) grid (1,1);
  \draw[mcGreen,line width=1.35pt] (-1,0)--(1,0) (0,-1)--(0,1);
  \foreach \x/\y in {-.5/0,.5/0,0/-.5,0/.5}
    \node[fill=white,inner sep=.5pt,text=mcGreen!85!black,font=\small] at (\x,\y) {$X$};
  \fill[mcBlue] (0,.72) circle (2pt);
  \fill[mcBlue] (.72,0) circle (2pt);
  \fill[black!70] (0,0) circle (1pt);
  \node[anchor=north west,inner sep=1.5pt] at (.04,-.04) {$v$};
  \node[anchor=south,text=mcGreen!85!black] at (0,1.05) {$S_{g,v}$};
  \draw[-{Latex[length=2mm]},black!55,line width=.8pt] (1.12,0)--(1.48,0);
\end{tikzpicture}\par
{\footnotesize Four green $X$ operators.}
\end{minipage}\hfill
\begin{minipage}[t]{.32\linewidth}\centering
\textbf{(b) Gauge controls}\par\smallskip
\begin{tikzpicture}[x=1.20cm,y=1.20cm,font=\scriptsize,text=black]
  \path[use as bounding box] (-1.28,-1.13) rectangle (1.55,1.17);
  \draw[step=1,black!25,line width=.35pt] (-1,-1) grid (1,1);
  \filldraw[fill=mcTriangle!15,draw=mcTriangle!85!black,line width=.55pt]
    (-.72,0)--(-.5,0)--(0,.72)--cycle;
  \filldraw[fill=mcTriangle!15,draw=mcTriangle!85!black,line width=.55pt]
    (0,-.72)--(0,-.5)--(.72,0)--cycle;
  \draw[mcRed,line width=1.2pt] (-1,0)--(0,0)--(0,-1);
  \draw[mcGreen!60,line width=.8pt] (0,0)--(1,0) (0,0)--(0,1);
  \foreach \x/\y in {-.5/0,.5/0,0/-.5,0/.5}
    \node[fill=white,inner sep=.5pt,text=mcGreen!85!black,font=\small] at (\x,\y) {$X$};
  \fill[mcRed] (-.72,0) circle (2.4pt);
  \fill[mcRed] (0,-.72) circle (2.4pt);
  \fill[mcBlue] (0,.72) circle (2.4pt);
  \fill[mcBlue] (.72,0) circle (2.4pt);
  \fill[black!70] (0,0) circle (1.3pt);
  \fill[black!60] (-1,0) circle (1.1pt);
  \fill[black!60] (0,-1) circle (1.1pt);
  \node[anchor=north west,inner sep=1.5pt] at (.04,-.04) {$v$};
  \node[anchor=south,inner sep=2pt] at (-1,.03) {$v-\hat x$};
  \node[anchor=west,inner sep=2pt] at (.03,-1) {$v-\hat y$};
  \node[anchor=south] at (0,1.05) {reference $v$};
  \draw[-{Latex[length=2mm]},black!55,line width=.8pt] (1.12,0)--(1.48,0);
\end{tikzpicture}\par
{\footnotesize Red controls CZ on green--blue.}
\end{minipage}\hfill
\begin{minipage}[t]{.32\linewidth}\centering
\textbf{(c) Modified star}\par\smallskip
\begin{tikzpicture}[x=1.20cm,y=1.20cm,font=\scriptsize,text=black]
  \path[use as bounding box] (-1.28,-1.13) rectangle (1.55,1.17);
  \draw[step=1,black!25,line width=.35pt] (-1,-1) grid (1,1);
  \draw[mcGreen,line width=1.35pt] (-1,0)--(1,0) (0,-1)--(0,1);
  \draw[black!85,line width=.9pt] (-.72,0) .. controls (-.79,.43) and (-.43,.79) .. (0,.72);
  \draw[black!85,line width=.9pt] (0,-.72) .. controls (.43,-.79) and (.79,-.43) .. (.72,0);
  \foreach \x/\y in {-.5/0,.5/0,0/-.5,0/.5}
    \node[fill=white,inner sep=.5pt,text=mcGreen!85!black,font=\small] at (\x,\y) {$X$};
  \fill[mcRed] (-.72,0) circle (2.4pt);
  \fill[mcRed] (0,-.72) circle (2.4pt);
  \fill[mcBlue] (0,.72) circle (2.4pt);
  \fill[mcBlue] (.72,0) circle (2.4pt);
  \fill[black!70] (0,0) circle (1pt);
  \node[anchor=north west,inner sep=1.5pt] at (.04,-.04) {$v$};
  \node[anchor=south] at (0,1.05) {$A_{g,v}$};
\end{tikzpicture}\par
{\footnotesize Four $X$s plus two red--blue CZs.}
\end{minipage}
\end{minipage}
\endgroup
\caption{Lattice construction of a modified check using Eq.~\eqref{eq:prx273-rooted-check}. (a) The green parent star. (b) With reference $v$, the west and south red gauge qubits control CZ factors on green--blue pairs. Each shaded triangle marks the three participants of one CCZ. The green $X$ labels are check operators, not circuit-target symbols. (c) Conjugation leaves the four $X$s and adds the two black red--blue CZ curves. The other factors commute with the star and cancel. Completion is added later. See Appendix~\ref{sec:smd4-bilayer-gauging}.}
\label{fig:v435-modified-parent-check}
\end{figure*}

For a single-edge path and $U_v=\operatorname{CZ}_{q,p}$, the corresponding factor in $\bm W_i$ is $\operatorname{CCZ}_{g_e,q,p}$. Conjugating $X_q$ gives $X_q\operatorname{CZ}_{g_e,p}$. Figure~\ref{fig:v435-modified-parent-check} shows how this produces the CZ factors of a modified toric-code star.

An ideal measurement of $\widetilde S_i$ applies $\bm W_i^\dagger$, measures $S_i$, then applies $\bm W_i$. Each path factor is a product of $\operatorname{ctrl}_{g_e}(U_v)$ over the edges of the path, since $U_v^2=I$. The gauge qubits remain quantum controls throughout this procedure. The protocol below measures completed versions using the specified native circuits and gate order.

When all gauge qubits have $Z_{g_e}=+1$, the modified check reduces to $S_i$. A generator with no active blocks is unchanged. Unitary conjugation makes $\widetilde S_i$ Hermitian and an involution on the full Hilbert space.

A Gauss operation at $v\ne r_i$ changes both the local matter choice and its path parity, so the comparison is unchanged and $[A_v,\widetilde S_i]=0$. A Gauss operation at the reference block changes all relative choices together, giving
\begin{equation}
 A_{r_i}\widetilde S_i A_{r_i}
 =\bm W_i(\mathcal U S_i\mathcal U)\bm W_i^\dagger.
 \label{eq:prx344-parent-gauss}
\end{equation}
If $S_i$ is invariant under $\mathcal U$, the right-hand side equals $\widetilde S_i$, so the modified check commutes with every Gauss check. Here, $\mathcal U$ need only preserve the parent code and can map $S_i$ to a different stabilizer. Therefore, the modified parent check $\widetilde S_i$ need not commute with the reference Gauss check on the full Hilbert space. The Gauss checks $A_v$, modified parent checks $\widetilde S_i$ and flux checks $B_p$ form the raw check family.

\paragraph{Bilayer toric-code example.}
Gauging two toric codes gives the following raw $D(D_4)$ checks, where $D_4$ is the dihedral group of order eight~\cite{Tantivasadakarn2023Shortest,Tantivasadakarn2023Hierarchy,Christos2026NonAbelian}. Each edge of the square lattice carries a red gauge qubit and green and blue matter qubits. Appendix~\ref{sec:smd4-bilayer-gauging} derives the red Gauss check and applies Eq.~\eqref{eq:prx273-rooted-check} to obtain the green and blue stars. The checks below use the lattice convention of Fig.~\ref{fig:d4-completed-checks}, before adding completion qubits. Each black curve denotes a CZ operator between its colored endpoints, and each $X$ acts on the indicated color of edge qubit.

\begingroup
\definecolor{rawDfourRed}{HTML}{DB3636}
\definecolor{rawDfourGreen}{HTML}{159447}
\definecolor{rawDfourBlue}{HTML}{2467DE}
\tikzset{
  rawDfour/.style={x=1.12cm,y=1.12cm,font=\scriptsize,
    baseline={([yshift=-.5ex]current bounding box.center)}},
  rawDfourGrid/.style={draw=black!35,line width=.3pt},
  rawDfourCZ/.style={draw=black!85,line width=.65pt},
  rawDfourX/.style={fill=white,inner sep=.35pt,font=\small},
  rawDfourVertex/.style={anchor=north west,inner sep=.4pt,
    font=\scriptsize,text=black!80}
}
\begin{align}
 A_{r,v}&=
 \begin{tikzpicture}[rawDfour]
  \path[use as bounding box] (-1.08,-1.08) rectangle (1.08,1.08);
  \draw[rawDfourGrid,step=1] (-1,-1) grid (1,1);
  \draw[rawDfourRed,line width=1.35pt] (-1,0)--(1,0) (0,-1)--(0,1);
  \draw[rawDfourCZ] (.5,0) .. controls (.76,.03) and (.98,.12) .. (1,.72);
  \draw[rawDfourCZ] (0,.5) .. controls (.03,.76) and (.12,.98) .. (.72,1);
  \foreach \x/\y in {.5/0,0/.5}
    \fill[rawDfourGreen] (\x,\y) circle[radius=1.9pt];
  \foreach \x/\y in {1/.72,.72/1}
    \fill[rawDfourBlue] (\x,\y) circle[radius=1.9pt];
  \foreach \x/\y in {.28/0,-.72/0,0/.28,0/-.72}
    \node[rawDfourX,text=rawDfourRed!82!black] at (\x,\y) {$X$};
  \fill[black!65] (0,0) circle[radius=.65pt];
  \node[rawDfourVertex] at (.09,-.09) {$v$};
 \end{tikzpicture},
 \label{eq:prx332-raw-r}\\[2pt]
 A_{g,v}&=
 \begin{tikzpicture}[rawDfour]
  \path[use as bounding box] (-1.08,-1.08) rectangle (1.08,1.08);
  \draw[rawDfourGrid,step=1] (-1,-1) grid (1,1);
  \draw[rawDfourGreen,line width=1.35pt] (-1,0)--(1,0) (0,-1)--(0,1);
  \draw[rawDfourCZ] (-.72,0) .. controls (-.79,.43) and (-.43,.79) .. (0,.72);
  \draw[rawDfourCZ] (0,-.72) .. controls (.43,-.79) and (.79,-.43) .. (.72,0);
  \foreach \x/\y in {-.72/0,0/-.72}
    \fill[rawDfourRed] (\x,\y) circle[radius=1.9pt];
  \foreach \x/\y in {0/.72,.72/0}
    \fill[rawDfourBlue] (\x,\y) circle[radius=1.9pt];
  \foreach \x/\y in {.5/0,-.5/0,0/.5,0/-.5}
    \node[rawDfourX,text=rawDfourGreen!82!black] at (\x,\y) {$X$};
  \fill[black!65] (0,0) circle[radius=.65pt];
  \node[rawDfourVertex] at (.09,-.09) {$v$};
 \end{tikzpicture},
 \label{eq:prx332-raw-g}\\[2pt]
 A_{b,v}&=
 \begin{tikzpicture}[rawDfour]
  \path[use as bounding box] (-1.08,-1.08) rectangle (1.08,1.08);
  \draw[rawDfourGrid,step=1] (-1,-1) grid (1,1);
  \draw[rawDfourBlue,line width=1.35pt] (-1,0)--(1,0) (0,-1)--(0,1);
  \draw[rawDfourCZ] (-.72,-1) .. controls (-.12,-.98) and (-.03,-.76) .. (0,-.5);
  \draw[rawDfourCZ] (-1,-.72) .. controls (-.98,-.12) and (-.76,-.03) .. (-.5,0);
  \foreach \x/\y in {-.72/-1,-1/-.72}
    \fill[rawDfourRed] (\x,\y) circle[radius=1.9pt];
  \foreach \x/\y in {0/-.5,-.5/0}
    \fill[rawDfourGreen] (\x,\y) circle[radius=1.9pt];
  \foreach \x/\y in {.72/0,-.28/0,0/.72,0/-.28}
    \node[rawDfourX,text=rawDfourBlue!82!black] at (\x,\y) {$X$};
  \fill[black!65] (0,0) circle[radius=.65pt];
  \node[rawDfourVertex] at (.09,-.09) {$v$};
 \end{tikzpicture},
 \label{eq:prx332-raw-b}\\[2pt]
 B_{c,f}&=
 \begin{tikzpicture}[x=1.35cm,y=1.35cm,font=\scriptsize,
     baseline={([yshift=-.5ex]current bounding box.center)}]
  \path[use as bounding box] (-.18,-.18) rectangle (1.18,1.18);
  \draw[rawDfourGreen,line width=1.35pt] (0,0) rectangle (1,1);
  \foreach \x/\y in {.5/0,.5/1,0/.5,1/.5}
    \node[fill=white,inner sep=1pt,font=\small,text=rawDfourGreen!82!black]
      at (\x,\y) {$Z$};
  \node[text=black!80] at (.5,.5) {$f$};
 \end{tikzpicture}
 =\prod_{e\in\partial f}Z_{c,e},
 \label{eq:prx332-raw-flux}
\end{align}
\endgroup
where the index $c=r,g,b$ labels the red, green and blue qubits. $A_{r,v}$ is the Gauss check. The checks $A_{g,v}$ and $A_{b,v}$ are the green and blue toric-code stars multiplied by CZ operators coupling the red gauge qubits to blue and green matter qubits, respectively. The operator $B_{r,f}$ is the gauge flux check. Their common $+1$ eigenspace realizes $D(D_4)$ topological order~\cite{Tantivasadakarn2023Shortest,Christos2026NonAbelian,KobayashiZhuHsin2026}. These raw checks need not commute when a flux check has eigenvalue $-1$. For example,
\begin{equation}
 A_{r,v}A_{g,v}=B_{b,v}A_{g,v}A_{r,v}.
 \label{eq:prx332-raw-conflict}
\end{equation}
If $B_{b,v}=-1$, these two checks anticommute. This is the incompatibility we will remove in later sections.

\subsection{Incompatibility of raw measurements}
\label{sec:prx-raw-algebra}
The $D(D_4)$ example illustrates why preserving the parent code does not guarantee compatible measurements after gauging. The symmetry $\mathcal U$ preserves the parent stabilizer group, but can send an individual generator $S_i$ to a different stabilizer $\mathcal U S_i\mathcal U$. Both act as $+1$ on the parent code, yet their eigenvalues can differ outside that code space. Gauging turns this distinction into a possible failure of a modified parent check to commute with its reference Gauss check. The dressing paths also matter. Two modified parent checks can fail to commute when the gauge flux around a closed cycle determined by their paths is nontrivial. The second relation in Eq.~\eqref{eq:prx273-raw-commutators} below makes this dependence explicit.

For every Clifford symmetry satisfying Eq.~\eqref{eq:prx273-block-symmetry}, both effects are controlled by Pauli checks that commute with the entire raw family. Their eigenvalues determine which raw measurements are compatible.

To track the change of a parent generator, define $N_i=S_i(\mathcal U S_i\mathcal U)$. This product is itself a parent stabilizer and obeys $\mathcal U S_i\mathcal U=S_iN_i$. We call it a \emph{parent invariant} because $\mathcal U N_i\mathcal U=N_i$. It equals $I$ when the symmetry leaves $S_i$ unchanged. Its corresponding check after gauging is $\mathcal N_i=\bm W_iN_i\bm W_i^\dagger$, using the same paths as $\widetilde S_i$ in Eq.~\eqref{eq:prx273-rooted-check}. For the blockwise Clifford involutions in Eq.~\eqref{eq:prx273-block-symmetry}, the modified checks $\widetilde S_i$ are Clifford and the invariant checks $\mathcal N_i$ are Pauli.

The only pairs of raw checks that can fail to commute obey~\cite{Chang2026Magic}
\begin{align}
 \widetilde S_i A_{r_i}
   &=\mathcal N_i A_{r_i}\widetilde S_i,\nonumber\\
 \widetilde S_i\widetilde S_j
   &=Z_g(\gamma_{ij})\widetilde S_j\widetilde S_i.
 \label{eq:prx273-raw-commutators}
\end{align}
Here, $\gamma_{ij}$ is a closed edge cycle determined by the chosen paths. Choose flux faces whose boundaries span these cycles, so that $Z_g(\gamma_{ij})$ is a product of retained flux checks. Both the invariant checks and the flux checks commute with every raw check. An ideal measurement of any raw check therefore leaves their eigenvalues unchanged. Appendix~\ref{sec:sm-clifford-derivation} derives these relations, including the phase factors needed in the Clifford expressions.

The noncommutation in Eq.~\eqref{eq:prx273-raw-commutators} is an intrinsic property of the raw checks. If the factor $\mathcal N_i$ or $Z_g(\gamma_{ij})$ has eigenvalue $+1$, the corresponding pair commutes on that eigenspace. If it has eigenvalue $-1$, the pair anticommutes there. Even an ideal measurement of one check can then disturb the outcome of the other.

Ideal gauging of a parent-code state gives $+1$ for all flux and invariant checks, so the raw measurements are compatible. Ideal raw-check measurements cannot move the state out of this common $+1$ eigenspace. Errors can introduce components with eigenvalue $-1$ for a factor in Eq.~\eqref{eq:prx273-raw-commutators}, on which even ideal measurements of the corresponding raw pair can disturb one another's outcomes.

In Section~\ref{sec:prx-central}, we couple each conflicting pair to a completion qubit. The added factors cancel the sign in the raw commutation relation. The resulting checks commute on the full Hilbert space while preserving the intended measurements on the encoded input.

\section{Completion with central commutators}
\label{sec:prx-central}
We call measured unitary checks compatible when they commute on the full Hilbert space and each extraction circuit measures their spectral projectors without resolving additional observables. The resulting joint projectors specify both the outcome probabilities and the conditional quantum state. Starting from the commutation relations above, we add completion qubits so that the measured checks commute on the full Hilbert space.

\subsection{Construction and code distance}
\label{sec:prx-central-completion}
We can now state the algebraic input independently of its gauging origin. Let $C_i$ be Hermitian unitary raw checks and $\mathsf{K}_a$ mutually commuting Hermitian unitary checks. We call the $\mathsf{K}_a$ \emph{central checks} because they commute with every raw check. We require
\begin{equation}
 \begin{gathered}
 C_i^2=\mathsf{K}_a^2=I,\qquad
 [C_i,\mathsf{K}_a]=[\mathsf{K}_a,\mathsf{K}_b]=0,\\
 C_iC_j=\mathsf{K}_{ij}C_jC_i,\qquad
 \mathsf{K}_{ij}\in\langle\mathsf{K}_a\rangle.
 \end{gathered}
 \label{eq:main-central}
\end{equation}
For the preceding gauging construction, $C_i$ enumerates the Gauss and modified parent checks, and the central family contains all retained flux and invariant checks. Equation~\eqref{eq:main-central} includes the gauging relations in Eq.~\eqref{eq:prx273-raw-commutators} and the $D(D_4)$ example in Eq.~\eqref{eq:prx332-raw-conflict}. The completion construction follows directly from Eq.~\eqref{eq:main-central}, extending beyond Pauli parent codes and Clifford symmetries.

We define the conflict graph $\mathcal{H}$ with one vertex per $C_i$ and one edge per pair with $\mathsf{K}_{ij}\ne I$. Orient each edge once, as $e:i\to j$, and assign it a completion qubit $t_e$. Distinct edges receive distinct qubits even when their commutator labels coincide. For any label and qubit, write $V(\mathsf{K},t)=\operatorname{ctrl}_t(\mathsf{K})$. On edge $e$, set $\mathsf{K}_e=\mathsf{K}_{ij}$ and
\begin{equation}
 V_e=\operatorname{ctrl}_{t_e}(\mathsf{K}_e)
   =\frac{I+\mathsf{K}_e}{2}
       +\frac{I-\mathsf{K}_e}{2}Z_{t_e}.
 \label{eq:prx273-completion-factor}
\end{equation}
The completed checks are
\begin{equation}
 \widehat{C}_i=C_i\prod_{e:i\to j}X_{t_e}
                       \prod_{e:j\to i}V_e.
 \label{eq:main-completion}
\end{equation}
We denote the completed Gauss operator by $\widehat{A}_v$. To fix the added degrees of freedom on the code space, also measure the pinning checks
\begin{align}
 T_e&=T(\mathsf{K}_e,t_e)\nonumber\\
 &=H_{t_e}\operatorname{ctrl}_{t_e}(-\mathsf{K}_e)H_{t_e}
 \nonumber\\
 &=\frac{I-\mathsf{K}_e}{2}
        +\frac{I+\mathsf{K}_e}{2}X_{t_e}.
 \label{eq:main-pinning}
\end{align}
The measured family is $\mathcal{F}=\{\widehat{C}_i,\mathsf{K}_a,T_e\}$.

The following theorem shows that completion preserves the raw code distance exactly by adjoining a fixed product state to its code space. Section~\ref{sec:prx-ft} develops the circuit fault-distance guarantees by analyzing faults during extraction, gauging and ungauging.

\begin{theorem}[Completion preserves code distance]
\label{thm:prx273-central-completion}
Assume Eq.~\eqref{eq:main-central} and a nonzero projector $P_g$ onto the common $+1$ eigenspace of the raw and central checks. The family $\mathcal{F}$ consists of globally commuting Hermitian unitaries, and the code it defines has exactly the same distance as the raw code. With $c=|E(\mathcal{H})|$ and $|\Omega\rangle=|+\rangle^{\otimes c}$, its common $+1$ projector is
\begin{equation}
 \widehat{P}_g=P_g\otimes|\Omega\rangle\langle\Omega|,
 \qquad d(\widehat{P}_g)=d(P_g).
 \label{eq:main-code-distance}
\end{equation}
Here, $d(P)=\min\{|\operatorname{supp}O|:POP\notin\mathbb{C}P\}$, with the minimum of the empty set equal to infinity. On the common $+1$ eigenspace of the $\mathsf{K}_a$, with the completion register in $|\Omega\rangle$, the completed measurements preserve every compatible raw-check outcome and its conditional state.
\end{theorem}

\begin{proof}
For one conflict edge, abbreviate its factors by $\mathsf{K},X,V,T$. The identity $XV=\mathsf{K}VX$ cancels the raw endpoint commutator. Other completion factors commute by centrality and the use of separate qubits, and all factors within a completed check commute and square to identity. In the $\mathsf{K}=+1$ sector, $V=I$ and $T=X$. In the $\mathsf{K}=-1$ sector, $V=Z$ and $T=I$. Thus, the pins commute with both endpoint factors in either sector.

On the common $+1$ eigenspace of the $\mathsf{K}_a$, every $V_e=I$ and every pin fixes $t_e$ to $|+\rangle$. Restricting $\widehat{C}_i$ to this subspace gives $C_i$, proving the projector equality. Restricting $(I+b_i\widehat{C}_i)/2$ gives the corresponding raw outcome projector, proving the measurement statement.

For an arbitrary error operator $O$ on the enlarged system, compress the completion register to define
\begin{equation}
 O_{\mathrm{eff}}=(I\otimes\langle\Omega|)
                     O(I\otimes|\Omega\rangle).
 \label{eq:prx273-compressed-error}
\end{equation}
This operator acts only on the original qudits touched by $O$, and
\begin{equation}
 \widehat{P}_gO\widehat{P}_g
  =(P_gO_{\mathrm{eff}}P_g)\otimes|\Omega\rangle\langle\Omega|.
 \label{eq:prx273-distance-compression}
\end{equation}
Every error of weight below $d(P_g)$ is therefore detectable after completion, including errors on completion qubits. Conversely, an undetectable original error tensored with identity on the new register has unchanged weight. The two inequalities give the exact distance equality.
\end{proof}

\subsection{Spatial and hierarchy costs}
\label{sec:prx-spatial-costs}
Pauli and Clifford operators form the first and second levels of the Clifford hierarchy~\cite{GottesmanChuang1999}. A unitary lies in level $h\ge2$ if conjugating any Pauli operator by it gives an operator in level $h-1$.

For qubit raw checks and Pauli central checks, $V_e$ is a controlled Pauli and $T_e$ is a controlled Pauli conjugated by a Hadamard, so both are Clifford. Thus, the completed checks $\widehat{C}_i$ are Clifford whenever the raw checks $C_i$ are Clifford. More generally, multiplication on either side by a Clifford preserves every hierarchy level $h\ge2$. If $C_i$ belongs to level $h$, so does $\widehat{C}_i$, because all its added factors together form a Clifford operator. The pinning checks have level at most two, independently of the number of conflicts.

Each Hermitian unitary check $Q$ is measured by preparing one readout ancilla in $|+\rangle$, applying $\operatorname{ctrl}(Q)$, and measuring the ancilla in $X$. If $Q$ is Clifford, this controlled operation has hierarchy level at most three. Under its conjugation, the control $Z$ is fixed, the control $X$ becomes $X\otimes Q$, and a Pauli on the check qubits becomes that Pauli times a controlled Pauli. These images are all Clifford, including when the controlled Pauli carries a phase $\pm i$. Thus the completed Clifford checks and their controlled measurement operations have levels at most two and three, respectively. Theorem~\ref{thm:qudit-hierarchy-control} extends a given bound on the raw measurement circuits to the completed circuits, including for prime-dimensional qudits.

For $n$ matter qubits, $M_{\mathrm{raw}}$ raw checks including the central family, and $c$ conflicts, the qubit counts are
\begin{equation}
 \begin{aligned}
 n_{\mathrm{gauged}}&=n+|E(G)|+c,\\
 n_{\mathrm{meas}}&=M_{\mathrm{raw}}+c.
 \end{aligned}
 \label{eq:main-resource-ledger}
\end{equation}
Here, $|E(G)|$ is the number of edges in the gauging graph, with one gauge qubit per edge. The first count includes matter, gauge and completion qubits. The second uses our allocation of one readout ancilla per measured check. Each conflict adds one completion qubit and one readout ancilla for its pin.

Let $w$ bound the raw check weight and let $q$ bound the number of raw checks acting on any qubit. Both bounds include the central checks. A check can conflict only with checks that share a qubit with it, so it has at most $\Delta=w(q-1)$ conflicts. For $M$ noncentral raw checks, counting each conflict at both ends gives $c\le M\Delta/2$. Since $\mathsf{K}_e=C_iC_jC_iC_j$ acts only on qubits touched by $C_i$ or $C_j$, the completed and pinning weights satisfy
\begin{equation}
 \operatorname{wt}(\widehat{C}_i)\le w+(2w+1)\Delta,
 \qquad \operatorname{wt}(T_e)\le2w+1.
 \label{eq:prx273-completed-weights}
\end{equation}
A matter or gauge qubit occurs in at most $q\Delta$ operators $\mathsf K_e$, counting distinct conflict edges separately. Each can add that qubit to one completed check and one pin, so the qubit belongs to at most $q+2q\Delta$ measured checks. A completion qubit participates in only its two endpoint checks and its pin. If $w$ and $q$ are bounded independently of code size, both the check weights and the number of checks per qubit remain bounded. Completion therefore preserves the LDPC property, with a number of additional qubits linear in the raw check count. Geometric locality is preserved when the raw checks have bounded range and each completion qubit is placed near its two conflicting checks.

The raw family must itself have bounded check weight and a bounded number of checks per qubit. For a Pauli LDPC parent, it suffices to bound the matter-block size, the number of edges at each graph vertex, the length of each dressing path, and the number of paths sharing an edge. The number of edges in each flux check and the number of flux checks using an edge must also remain bounded. All these bounds are independent of code size.

For the bounded-weight qubit Clifford checks, use exact Clifford synthesis into $H$, $S$, and CNOT gates and a scalar phase. Control each gate with the assigned readout ancilla and apply the scalar phase as a phase gate on that ancilla. Taking these controlled gates and readout phase gates as native gives bounded-depth extraction on the check support and readout. Bounded check weights and a bounded number of checks per qubit then ensure that each circuit shares qubits with only a bounded number of others. We can divide the circuits into a constant number of groups with disjoint supports within each group. Running each group in parallel and the groups in sequence gives a constant-depth measurement round. These locality and circuit bounds enter the fault-distance analysis in \mbox{Section~\ref{sec:prx-ft}}.

\subsection{Logical measurement by gauging}
\label{sec:prx273-physical}
We now show that, for a parent-code input and ideal operations, gauging followed by ungauging measures the logical symmetry $\mathcal U$ and preserves superpositions within each eigenspace. We add the completion factors to the circuit in Appendix~\ref{sec:sm269-physical-gauging} and follow the state through measurement and correction~\cite{Verresen2021Enrichment,Tantivasadakarn2024LongRange,Chang2026Magic}.

\subsubsection{Completed Gauss measurement}
\label{sec:prx-instrument}
Prepare each vertex ancilla in $|+\rangle$ and each gauge qubit in $|0\rangle$. The enrichment circuit in Eq.~\eqref{eq:prx273-enrichment-circuit} acts as
\begin{equation}
 V_{\mathrm{enr}}|\psi\rangle
 =\bm{E}_{\mathrm{enr}}\bigl(|+\rangle_\mu\otimes|\psi\rangle_m\bigr).
 \label{eq:prx273-enrichment-isometry}
\end{equation}
The constraints $X_{\mu_v}U_v=1$ imply
\begin{equation}
 \left(\prod_vX_{\mu_v}\right)V_{\mathrm{enr}}
 =V_{\mathrm{enr}}\mathcal{U}.
 \label{eq:prx273-enrichment-intertwining}
\end{equation}
Thus, applying the product of vertex-ancilla $X$ operators after enrichment is equivalent to applying $\mathcal U$ to the input. For vertex bits $x_v\in\{0,1\}$, the link circuit in Eq.~\eqref{eq:prx273-link-circuit} writes $z_{uv}=(\delta x)_{uv}=x_u+x_v\bmod2$ onto the gauge qubit on edge $(uv)$.

Write $\widehat A_v=A_vD_v$, where $D_v$ is the product of the completion factors assigned to $A_v$ in Eq.~\eqref{eq:main-completion}. Add the controlled operations
\begin{equation}
 \bm{E}_{\mathrm{comp}}=\prod_v\operatorname{ctrl}_{\mu_v}(D_v).
 \label{eq:prx273-completion-circuit}
\end{equation}
Each central check commutes with every raw check, so $[D_v,A_u]=0$ for all $u,v$. The raw Gauss checks commute with one another, so no conflict edge joins two of them. Their completion factors use distinct completion qubits and commuting central checks, so the $D_v$ also commute. Consequently,
\begin{equation}
 \bm{E}_{\mathrm{comp}}\bm{E}_{\mathrm{link}}\bm{E}_{\mathrm{enr}}
 =\prod_v\operatorname{ctrl}_{\mu_v}(\widehat{A}_v).
 \label{eq:prx273-physical-factorization}
\end{equation}
Measuring $\mu_v$ in $X$ with outcome $a_v=\pm1$ therefore measures $\widehat A_v$. If $|a^X\rangle_\mu$ denotes the product of the recorded $X$ eigenstates, the corresponding operation, retaining the measured ancillas, is
\begin{equation}
 \mathcal{G}_a=
 |a^X\rangle\langle a^X|_\mu\,
 \bm{E}_{\mathrm{comp}}\bm{E}_{\mathrm{link}}\bm{E}_{\mathrm{enr}}.
 \label{eq:main-circuit-gauging}
\end{equation}
This equality describes the ideal measurement. The circuit-distance analysis uses the prescribed gate order, with full measurement circuits that share qubits run one after another. The vertex ancillas serve as the Gauss readout qubits. The completion qubits are kept throughout the measurement rounds.

\subsubsection{Outcomes and ungauging}
\label{sec:main-surgery}
Let $J$ append $|0\rangle_g$ and $|+\rangle_t$ to the matter input, and write $n_V=|V|$, $U(x)=\prod_vU_v^{x_v}$ and $a^x=\prod_va_v^{x_v}$. For a parent-code state $P|\psi\rangle=|\psi\rangle$, the initial gauge qubits give $+1$ for every flux check, and each invariant check reduces to a parent stabilizer. Gauss measurements preserve these $+1$ eigenvalues. With the completion qubits in $|+\rangle_t$, all completion factors therefore act as identity. Expanding the Gauss projectors gives
\begin{equation}
 \begin{aligned}
 &{}_\mu\langle a^X|\mathcal{G}_a|+\rangle_\mu J|\psi\rangle\\
 &\quad=2^{-n_V}\sum_{x\in\mathbb{F}_2^V}
 a^x|\delta x\rangle_g U(x)|\psi\rangle_m|+\rangle_t.
 \end{aligned}
 \label{eq:prx273-gauging-sum}
\end{equation}
Because the graph is connected, exactly two assignments of vertex bits give the same gauge bits. They are $x$ and $x+\mathbf1$, related by flipping every vertex bit. Their contributions combine as $a^xU(x)(I+s\mathcal U)$, where $s=\prod_va_v$. Fix a reference vertex $v_0$ and set $c_V=2^{-(n_V-1)/2}$. The gauging map
\begin{equation}
 \Gamma_a|\phi\rangle=
 c_V\sum_{x:x_{v_0}=0}
 a^x|\delta x\rangle_g U(x)|\phi\rangle_m|+\rangle_t
 \label{eq:prx273-rooted-isometry}
\end{equation}
preserves inner products. Distinct assignments with $x_{v_0}=0$ give orthogonal gauge states, and each $U(x)$ is unitary. With $\Pi_s=(I+s\mathcal U)/2$, pairing the two assignments in Eq.~\eqref{eq:prx273-gauging-sum} gives
\begin{equation}
 \mathcal{G}_a\bigl(|+\rangle_\mu J|\psi\rangle\bigr)
 =c_V|a^X\rangle_\mu\otimes\Gamma_a\Pi_s|\psi\rangle.
 \label{eq:main-flat-certificate}
\end{equation}
For a parent-code density operator $\rho$, a particular vertex record has probability
\begin{equation}
 p(a)=2^{1-n_V}\operatorname{Tr}(\Pi_s\rho).
 \label{eq:prx273-record-probability}
\end{equation}
For this fixed record $a$, the remaining state is proportional to $\Gamma_a\Pi_s\rho\Pi_s\Gamma_a^\dagger$. Summing the probabilities of the $2^{n_V-1}$ records with each product $s$ gives $\operatorname{Tr}(\Pi_s\rho)$, the probability of outcome $s$ when measuring $\mathcal U$.

To ungauge, measure the gauge qubits in $Z$, obtaining bits $z_{uv}$. Fix $x_{v_0}=0$ and find vertex bits satisfying $z_{uv}=x_u+x_v\bmod2$ on every edge, as in Refs.~\cite{WilliamsonYoder2026,Chang2026Magic}. The matter state is proportional to $a^xU(x)\Pi_s|\psi\rangle$. Applying $U(x)^\dagger$ leaves only the overall sign $a^x$, which does not affect the state. Discarding the individual records while retaining $s$ implements
\begin{equation}
 \mathcal{I}_s(\rho)=\Pi_s\rho\Pi_s.
 \label{eq:prx273-logical-instrument}
\end{equation}
Thus, the procedure projects onto the measured eigenspace without disturbing superpositions within it, even when the input is entangled with another system.

\Needspace{4\baselineskip}
\protocoltitle{Logical measurement by commuting-check $\mathbb{Z}_2$ gauging}
\label{alg:commuting-gauging}
\begin{protocolsteps}
\item \emph{Complete and initialize.} Start with a state in the parent code. Construct the checks in Eq.~\eqref{eq:main-completion}, including all central and pinning checks. Prepare the gauge qubits in $|0\rangle$ and the completion qubits in $|+\rangle$.
\item \emph{Gauge.} Measure each completed Gauss check and all remaining checks in the chosen fixed schedule. The readout-controlled operation for each completed Gauss check contains the enrichment, link and completion factors of Eq.~\eqref{eq:prx273-physical-factorization}. Use the prescribed gate order and assigned readout ancillas. Record the Gauss outcomes $a_v$ and $s=\prod_va_v$.
\item \emph{Repeat measurements if required.} Reset the readout ancillas. Measure the same full check family in a fixed schedule. Keep the gauge and completion qubits throughout. In ideal operation, all outcomes other than the Gauss outcomes are $+1$, and the Gauss outcomes repeat. Section~\ref{sec:prx-ft} specifies the rounds and tests on the records needed for the circuit fault-distance bound.
\item \emph{Ungauge and correct.} Measure the gauge qubits in $Z$, find vertex bits satisfying $z=\delta x$ with $x_{v_0}=0$, and check this equation on every edge. Apply $U(x)^\dagger$, discard gauge and completion qubits, and retain the parent postmeasurement state and reported sign $s$.
\end{protocolsteps}

\subsection{Compatible measurements after errors}
\label{sec:prx-operational}

Section~\ref{sec:prx273-physical} establishes the intended logical measurement on a parent-code input. We now explain how completion preserves compatible measurements after errors and retains quantum information for recovery.

An error can change the eigenvalue of the central operator $\mathsf K_e$ associated with a conflicting pair. On a subspace with eigenvalue $k_e=\pm1$, the raw checks obey $C_iC_j=k_eC_jC_i$. When $k_e=-1$, their ideal measurements can disturb one another even if no further error occurs. For the conflict edge $e:i\to j$, completion adds the following factors:
\begin{equation}
 \begin{array}{c|cc|c}
 k_e & \text{factor on }C_i & \text{factor on }C_j & T_e\\ \hline
 +1 & X_{t_e} & I & X_{t_e}\\
 -1 & X_{t_e} & Z_{t_e} & I
 \end{array}
 \label{eq:prx273-sector-table}
\end{equation}

\begin{table}[tbp]
\caption{Three measurement choices for one conflicting pair on the subspace $\mathsf K_e=-1$. Here, $q$ is a data qubit and $t$ is a completion qubit. The controlled choice uses no completion qubit and measures only $Z_q$, while completion makes both joint parities repeatable.}
\label{tab:prx285-measurement-comparison}
\centering
\setlength{\tabcolsep}{3pt}
\begin{tabular}{@{}p{.19\linewidth}p{.29\linewidth}p{.46\linewidth}@{}}
\toprule
Choice & Measured checks & Measurement record\\
\midrule
Raw & $X_q,\ Z_q$ & One check randomizes the other.\\
Controlled & $I,\ Z_q$ & One check is trivial, the other repeats.\\
Completed & $X_qX_t,\ Z_qZ_t$ & Both joint parities are repeatable.\\
\bottomrule
\end{tabular}
\end{table}

For $k_e=+1$, imposing the pin $T_e=+1$ fixes $t_e$ to $|+\rangle$ and recovers the raw measurements. For $k_e=-1$, the added $X$ and $Z$ factors cancel the raw anticommutation. The pin becomes identity and no longer fixes the completion qubit, so the completed checks measure joint observables of the original qubits and this added qubit.

Table~\ref{tab:prx285-measurement-comparison} compares two ways to remove this disturbance. In the controlled construction, the central eigenvalue determines whether the first check acts. It equals the raw check for $k_e=+1$ and the identity for $k_e=-1$. Thus, on the subspace shown, the raw pair $X_q,Z_q$ is replaced by $I,Z_q$, leaving only the $Z_q$ measurement. Completion measures the commuting pair $X_qX_t,Z_qZ_t$, retaining two nontrivial joint observables.

Both constructions give repeatable ideal outcomes. For recorded outcomes $b$ of the completed Hermitian checks, a measurement round acts through the operator
\begin{equation}
 M_b=\prod_{Q\in\mathcal{F}}\frac{I+b_QQ}{2},\qquad
 M_bM_{b'}=\delta_{b,b'}M_b.
 \label{eq:prx273-joint-instrument}
\end{equation}
The first round can have random outcomes. Once recorded, orthogonality ensures that every subsequent ideal round repeats them, provided no further error occurs. The same argument applies to controlled checks and to commuting finite-order qudit checks using their spectral projectors. Section~\ref{sec:prx-ft} uses this repeatability to track faults through later ideal rounds.

Raw measurement disturbance can also affect the reported logical sign. Since $\mathcal{U}=\prod_vA_v$ and only the Gauss check at the reference block conflicts with $\widetilde{S}_i$, Eq.~\eqref{eq:prx273-raw-commutators} gives
\begin{equation}
 \widetilde{S}_i\mathcal{U}
 =\mathcal{N}_i\mathcal{U}\widetilde{S}_i.
 \label{eq:prx277-logical-backaction}
\end{equation}
Denote the completed symmetry by $\widehat{\mathcal U}=\prod_v\widehat A_v$. It commutes with every measured check $Q\in\mathcal F$. When $\mathcal N_i=-1$, an ideal raw measurement can randomize a previously definite symmetry sign. Completed measurements preserve the recorded value of $\widehat{\mathcal U}$ throughout later ideal rounds. Outside the code space, this operator can act jointly on matter, gauge and completion qubits. The bilayer example below uses the retained completion qubits to recover the intended logical measurement.

In the single-edge bilayer example of Section~\ref{sec:prx-d4-recovery}, a known Pauli error before gauging can make the initial Gauss-sign product random even for an input with a definite logical symmetry. With subsequent operations ideal, the coherent repair there restores the intended logical measurement and updates the reported sign. For a Pauli logical observable, a Pauli error can only preserve or reverse a definite sign, since the two operators commute or anticommute. The circuit analysis in Section~\ref{sec:prx-ft} treats arbitrary local Kraus faults at both surgery interfaces.

Completion thus removes mutual measurement disturbance and retains quantum information for recovery. Section~\ref{sec:prx-d4-recovery} shows explicitly how a retained completion qubit controls a coherent repair before ungauging in the two-toric-code surgery. Appendix~\ref{sec:sm297-coherent-recovery} develops this recovery procedure in detail.

\subsection{Circuit fault tolerance}\label{sec:prx-ft}
We now ask when faults in the complete surgery circuit can change the intended logical measurement while passing its detection tests. First we specify the extraction circuits and fault model, including faults at individual gates. With the stated locality and reset assumptions, commutation bounds error spread through later ideal measurement rounds. The circuit-distance bound additionally depends on the specified gates and measurement order. Finally, we analyze both surgery interfaces, where gauge errors become partial symmetry operations on the parent code. This step is essential because a small gauge-error pattern can correspond to a large matter region. Combining these steps gives a general circuit fault-distance bound. Our specified $D_4$ gate order and square-lattice geometry give the sharper result in Section~\ref{sec:prx-d4}.

\paragraph{Measurement circuits and the fault model.}
The physical implementation enters through the circuit for each check. For a Hermitian unitary $Q$, prepare one measurement ancilla $a$ in $|+\rangle$, apply a controlled $Q$, and measure $a$ in $X$. With outcome $q=\pm1$, the corresponding measurement operator on the qubits checked by $Q$ is
\begin{equation}
 {}_a\!\langle q_X|\operatorname{ctrl}_a(Q)|+\rangle_a
 =\frac{I+qQ}{2}.
 \label{eq:prx273-native-projector}
\end{equation}
The controlled operation is decomposed into specified gates acting on the check qubits and readout ancilla. We call these \emph{native gates}. Each is treated as one physical operation and one possible fault location in the circuit model. The controlled operation used to measure a completed Clifford check has a decomposition using gates of hierarchy level at most three. A single ancilla fault can propagate through these gates to several code qubits, so check weight alone does not specify the circuit distance. After each readout, we reprepare the assigned ancilla $a$ for the next round, in $|+\rangle$ for the circuit above. For two successive measurements of the same check, with $D$ denoting the matter, gauge and completion qubits that remain between rounds,
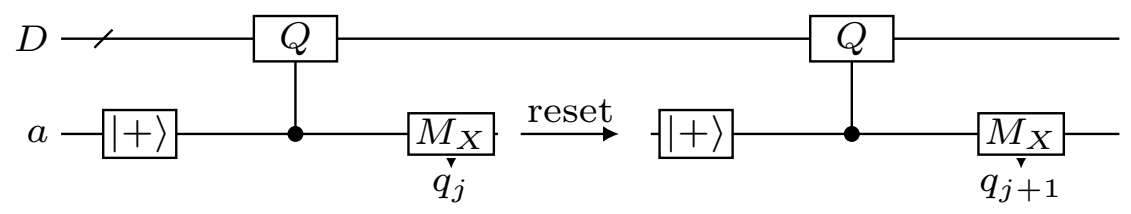
\[
\begin{tikzpicture}[x=1cm,y=1cm,line width=.45pt,font=\scriptsize,>={Latex[length=3pt,width=3pt]},
 baseline={([yshift=-.5ex]current bounding box.center)}]
\path[use as bounding box](-.39,-.36)rectangle(7.05,.86);
\draw(0,.63)--(7.00,.63);
\node[left,inner sep=2pt]at(0,.63){$D$};
\draw(.22,.57)--(.34,.69);
\draw(0,0)--(2.89,0);
\draw(3.90,0)--(7.00,0);
\node[left,inner sep=2pt]at(0,0){$a$};
\foreach \x in {.52,4.20}{
 \node[draw,fill=white,minimum width=.44cm,minimum height=.26cm,
 inner sep=1pt]at(\x,0){$|+\rangle$};
}
\foreach \x/\q in {1.55/{Q},5.23/{Q}}{
 \draw(\x,0)--(\x,.48);
 \fill(\x,0)circle(1.5pt);
 \node[draw,fill=white,minimum width=.55cm,minimum height=.30cm,
 inner sep=1pt]at(\x,.63){$\q$};
}
\foreach \x/\q in {2.58/{q_j},6.35/{q_{j+1}}}{
 \node[draw,fill=white,minimum width=.44cm,minimum height=.26cm,
 inner sep=1pt]at(\x,0){$M_X$};
 \draw[-{Latex[length=2pt,width=2pt]}](\x,-.17)--(\x,-.23);
 \node[below,inner sep=1pt]at(\x,-.23){$\q$};
}
\draw[->](3.04,0)--(3.69,0);
\node[above,inner sep=2pt]at(3.37,0){reset};
\end{tikzpicture}
\]
The ideal reset discards the readout's previous state and correlations while leaving the reduced state of $D$ unchanged. This repreparation is a preparation location and may be faulty. The reset does not remove errors already transferred to the code qubits. We finish each check-measurement circuit before beginning another that shares code qubits with it. Circuits on disjoint supports may run in parallel. The matter, gauge and completion qubits are retained between gauged rounds.

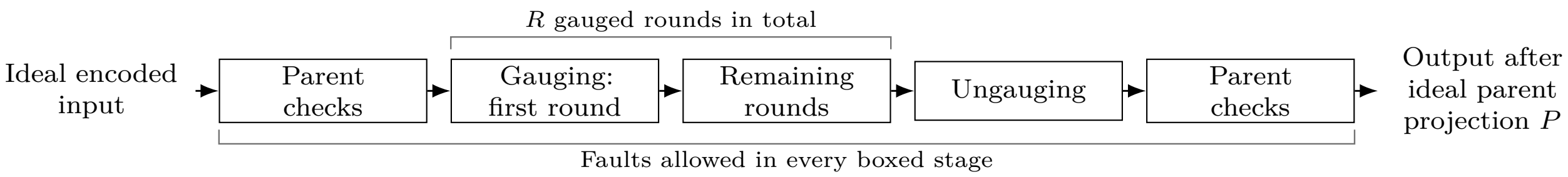
\begin{figure*}[tbp]
\centering
\begin{tikzpicture}[x=1cm,y=1cm,font=\footnotesize,>=Latex,
 stage/.style={draw=black,line width=.4pt,text width=2.05cm,
   minimum height=.62cm,align=center,inner sep=2pt},
 boundary/.style={text width=2.05cm,align=center,inner sep=2pt}]
\path[use as bounding box](0,-.62)rectangle(16.90,1.24);
\node[boundary](input)at(1.10,.33){Ideal encoded\\input};
\node[stage](pre)at(3.55,.33){Parent\\checks};
\node[stage](gauge)at(6.00,.33){Gauging:\\first round};
\node[stage](repeat)at(8.45,.33){Remaining\\rounds};
\node[stage](ungauge)at(10.90,.33){Ungauging};
\node[stage](post)at(13.35,.33){Parent\\checks};
\node[boundary](output)at(15.80,.33){Output after\\ideal parent\\projection $P$};
\foreach \a/\b in {input/pre,pre/gauge,gauge/repeat,
 repeat/ungauge,ungauge/post,post/output}
 {\draw[->,line width=.55pt](\a)--(\b);}
\draw[black!55,line width=.45pt](4.90,.77)--(4.90,.90)--(9.55,.90)--(9.55,.77);
\node[above,inner sep=1pt,font=\scriptsize]at(7.225,.92){$R$ gauged rounds in total};
\draw[black!55,line width=.45pt](2.45,-.08)--(2.45,-.23)--(14.45,-.23)--(14.45,-.08);
\node[below,inner sep=2pt,font=\scriptsize]at(8.45,-.23){Faults allowed in every boxed stage};
\end{tikzpicture}
\caption{Complete surgery circuit and its noise assumptions. Every boxed stage may be faulty, including parent-check extraction before gauging and after ungauging. Only the encoded input and final parent-code projection $P$ are ideal. The projection selects outcome $+1$ for every parent stabilizer. Gauging is the first of $R$ gauged rounds.}
\label{fig:v387-fault-model}
\end{figure*}

At a faulty native gate $G$, an arbitrary joint Kraus operator $E_\alpha$ may act on all its participants. For a three-participant gate, schematically,
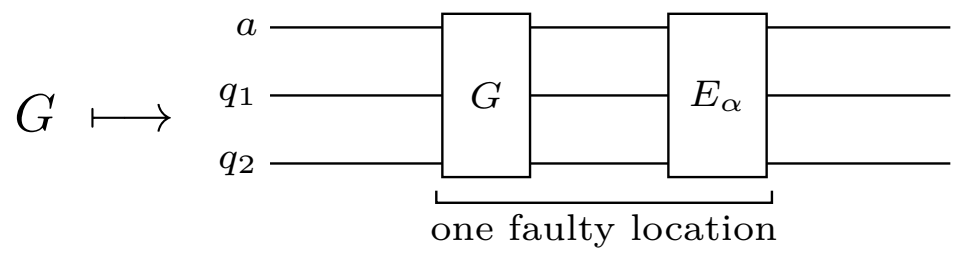
\[
G\ \longmapsto\
\begin{tikzpicture}[x=1cm,y=1cm,line width=.45pt,font=\scriptsize,
 baseline={([yshift=-.5ex]current bounding box.center)}]
\path[use as bounding box](-.42,-.50)rectangle(4.57,1.08);
\foreach \y/\name in {.90/a,.45/q_1,0/q_2}{
 \draw(0,\y)--(4.50,\y);
 \node[left,inner sep=2pt]at(0,\y){$\name$};
}
\node[draw,fill=white,minimum width=.58cm,minimum height=1.08cm,
 inner sep=1pt]at(1.43,.45){$G$};
\node[draw,fill=white,minimum width=.65cm,minimum height=1.08cm,
 inner sep=1pt]at(2.96,.45){$E_\alpha$};
\draw(1.10,-.18)--(1.10,-.26)--(3.32,-.26)--(3.32,-.18);
\node[below,inner sep=2pt]at(2.21,-.26){one faulty location};
\end{tikzpicture}
\]
This counts as one faulty location, even when $E_\alpha$ acts jointly on every participant. Preparations, readouts, basis changes and idles may also fail on matter, gauge, completion or measurement qubits. A measurement branch is the conditional circuit operation for one specified record of measurement outcomes. With noise, fixing the unobserved Kraus labels as well selects a linear branch operator. At a faulty location, the Kraus label is the index $\alpha$ in $E_\alpha$. Its expansion in participant Paulis is a proof device, without a Pauli-noise assumption. Classical processing is reliable. Figure~\ref{fig:v387-fault-model} summarizes the noisy surgery interval and its two ideal outer boundaries.

It is instructive to compare our setting with Ref.~\cite{Chang2026Magic}, which proves a fault-distance bound for raw gauged measurements while accounting for nondeterministic outcomes and both surgery interfaces. Its elementary faults are Pauli errors on matter blocks or single gauge qubits before measurement rounds, and errors in the recorded outcomes. Here, completion makes the measured checks commute on the full Hilbert space. This removes mutual measurement disturbance after an error and gives the bounded-propagation argument below. We also specify the extraction circuits and count faults at individual gates, including joint errors on the participants of a faulty entangling gate. Combining this propagation bound with the two interfaces gives a circuit fault-distance guarantee, sharpened to exact distance for our $D_4$ schedule in Section~\ref{sec:prx-d4}.

\subsubsection{Propagation through repeated measurements}
We bound the support of the effective error produced by each circuit fault. A fault may first spread through the gates of its own check circuit. Later ideal measurements can enlarge its support only to a fixed neighborhood of that circuit, independently of how many rounds follow. The next subsection combines this bound with both surgery interfaces and the parent-code distance.

Equation~\eqref{eq:prx273-joint-instrument} makes each complete ideal round a joint projective measurement on any input. Its outcomes repeat in subsequent ideal rounds, even if an earlier fault has changed some check values.

To isolate a fault, fix its circuit's recorded readout outcomes and unobserved Kraus labels. Contracting the ancilla preparation and readout leaves an operator $E$ on the code qubits. Let $B$ contain every code qubit used by that faulty circuit, including any qubits reached by the fault within the circuit.

An ideal measurement of $Q_i$ contributes the projector $(I+q_iQ_i)/2$. To expand a product of these projectors, choose either $I$ or $q_iQ_i$ from each factor. Each term contains a check product $\mathcal W=\prod_{i\in\mathcal I}Q_i$, where $\mathcal I$ is the index set of the selected checks. Scalar coefficients, including the outcomes $q_i$, are kept separately.

For each chosen product, let $\mathcal W_B$ be its subproduct containing only factors whose supports touch the original $B$. These are the only check factors that can fail to commute with $E$. A retained factor may still commute with $E$. Support overlap alone does not imply a changed measurement outcome. In each expansion term, the identity $\mathcal W E=(\mathcal W E\mathcal W^\dagger)\mathcal W$ moves the fault past that product. Every omitted factor commutes with $E$ by disjoint support and with all retained factors by global commutation. It therefore cancels from the conjugation, giving
\begin{equation}
 \mathcal{W}E\mathcal{W}^{\dagger}
 =\mathcal{W}_B E\mathcal{W}_B^{\dagger}.
 \label{eq:main-fault-dressing}
\end{equation}
All check measurements remain in the circuit. The cancellation concerns only the effective error. Every factor in $\mathcal W_B$ is a check touching the original $B$. Define the fixed neighborhood
\begin{equation}
 N(B)=B\cup\!\bigcup_{i:\operatorname{supp}Q_i\cap B\ne\varnothing}
 \operatorname{supp}Q_i.
 \label{eq:prx273-fault-neighborhood}
\end{equation}
Equation~\eqref{eq:main-fault-dressing} then gives the support bound
\[
\begin{aligned}
\operatorname{supp}(\mathcal W E\mathcal W^\dagger)
 &=\operatorname{supp}(\mathcal W_B E\mathcal W_B^\dagger)\\
 &\subseteq B\cup\operatorname{supp}(\mathcal W_B)
 \subseteq N(B).
\end{aligned}
\]
Only checks directly touching the original $B$ enter this neighborhood. A check that misses $B$ cancels from Eq.~\eqref{eq:main-fault-dressing} even if it overlaps $N(B)$. The neighborhood therefore does not grow from round to round. If $|B|\le w_0$, each check has weight at most $w$, and each code qubit belongs to at most $q_{\mathrm{inc}}$ checks, then $|N(B)|\le w_0[1+q_{\mathrm{inc}}(w-1)]$. Several faults contribute the union of their neighborhoods. This argument allows coherent and nonunitary faulty operators. Appendix~\ref{sec:locality-sm} also gives a direct proof by assigning each ideal projector to the latest fault whose original support it meets.

The observed record has a corresponding locality property. If two consecutive measurements of the same check are ideal, their outcomes agree unless an intervening faulty measurement circuit touches that check or another fault acts on its support. Every change between consecutive outcomes must therefore occur near a circuit fault. When check circuits have bounded support and round depth, and each qubit belongs to a bounded number of checks, one fault can account for only a bounded number of such changes. A fault may leave a check outcome at $-1$ for many rounds, but it cannot keep generating new changes during later ideal measurements. Appendix~\ref{sec:locality-sm} proves this statement.

\subsubsection{A circuit fault-distance theorem}\label{sec:prx-general-distance}
We now combine the propagation bound with gauging and ungauging in Protocol~\ref{alg:commuting-gauging}. Let $P$ project onto the parent qubit Pauli code of distance $d$. The measured symmetry $\mathcal U=\prod_vU_v$ preserves this code. Its unitary factors act on disjoint blocks of at most $b$ matter qubits and satisfy $U_v^2=I$. Each measured check and its measurement circuit must use a bounded number of qubits, and each code qubit must belong to a bounded number of checks. These bounds include all completed, central and pinning checks. Parent-check measurements, gauge readout and the physical corrections after ungauging may all be faulty. Only the encoded input and final parent-code projection are ideal, as in Ref.~\cite{WilliamsonYoder2026}.

Let $K_{F,m,\alpha}$ be the resulting circuit operator for a set $F$ of faulty native locations, recorded outcomes $m$ and unobserved Kraus labels $\alpha$, including the labels from discarding auxiliary qubits. Write $s(m)=\prod_v a_v$ for the logical sign reported from the initial Gauss outcomes $a_v$. Correct logical measurement requires
\begin{equation}
 PK_{F,m,\alpha}P\in\mathbb{C}\,\Pi_{s(m)}P,
 \label{eq:main-instrument-distance}
\end{equation}
where $\Pi_s=(I+s\mathcal{U})/2$. Thus, after the final projection, the circuit acts as the intended logical projection up to a scalar. This scalar may depend on $F,m,\alpha$, but not on the encoded state. The operation cannot otherwise change the state within the reported eigenspace. The circuit fault distance $d_{\mathrm{surg}}$ is the fewest faults for which a record passes the detection tests but an associated circuit operator violates Eq.~\eqref{eq:main-instrument-distance}.

For the proof, expand the propagated errors in Paulis. Let $\kappa$ bound, in each term, the number of affected matter blocks plus the number of flipped gauge qubits per native fault. This includes faults in the parent-check intervals and at both interfaces. Let $A=\{v:U_v\ne I\}$ denote the active vertices. For an edge cut $\delta y$, the graph parameter $h_A$ bounds the smaller number of active vertices on its two sides by $|\delta y|/h_A$. Let $\lambda$ be the fewest gauge flips that pass every measured flux check but cannot be written as $\delta y$ for any vertex assignment $y$. The construction must satisfy the gauging identity in Appendix~\ref{sec:v278-general-distance-parameters}, which also gives the precise graph definitions.

\begin{theorem}[Circuit fault distance of logical measurement]
\label{thm:end-distance}
Perform $R$ rounds of check measurements on the gauged code between the parent-check intervals, with ideal encoded input and an ideal final projection onto the parent code. Require outcome $+1$ for every parent check and every gauged check except the Gauss operators. Require the same full vector $a$ of Gauss outcomes in all $R$ rounds, and a final gauge record satisfying $z=\delta x$ on every edge. For $\gamma=b\kappa\max\{1,h_A^{-1}\}$,
\begin{equation}
 d_{\mathrm{surg}}\ge
 \min\!\left\{R,\left\lceil d/\gamma\right\rceil,
 \left\lceil\lambda/\kappa\right\rceil\right\}.
 \label{eq:graph-surgery-distance}
\end{equation}
In particular, $R=\Theta(d)$ gives $d_{\mathrm{surg}}=\Omega(d)$ when $b,\kappa=O(1)$, $h_A=\Omega(1)$ and $\lambda=\Omega(d)$.
\end{theorem}

For a fault count $f$ below the stated bound, $f<R$ guarantees a fault-free gauged round, although its input may still contain errors. The proof in Appendix~\ref{sec:v278-general-distance-parameters} converts the faults before and after that round into parent-code errors acting on fewer than $d$ qubits. The parent code detects these errors, giving Eq.~\eqref{eq:main-instrument-distance}. For our square-lattice circuit, tracking affected rows and columns gives the sharper result in Section~\ref{sec:prx-d4}, even though the graph expansion decreases with lattice size. Appendices~\ref{sec:sm-zn-measurement} and~\ref{sec:sm-zn-distance} give the $\mathbb{Z}_N$ logical-measurement protocol and its circuit-distance bound.

\subsection{Detecting faults with completion qubits}\label{sec:prx-dual-role}
Commuting checks prevent errors from spreading indefinitely through later ideal rounds. However, during a single check measurement, an error on the readout ancilla can still spread to several data qubits. This correlated error is known as a hook error. The same completion qubit that makes the checks commute can also serve as a partial flag. We show how two added CNOTs make its pin outcome detect these faults under the input and single-fault assumptions stated below.

Consider a Pauli central check $B=\prod_{j=1}^{w}Z_j$. Write $P_{s_B}(B)=(I+s_B B)/2$ for $s_B=\pm1$, so its pin is $T=P_-(B)+P_+(B)X_t$. Other Pauli labels are reduced to this form by single-qubit basis changes. Measure $B$ and $T$ jointly using one readout ancilla for each check. Prepare $a_B$ in $|1\rangle$ and $a_T$ in $|+\rangle$, apply $\operatorname{CNOT}(j\to a_B)$ in order $j=1,\ldots,w$ to accumulate the checked qubits' $Z$ parity in $a_B$, apply $\operatorname{Toffoli}(a_B,a_T\to t)$, and read $a_B$ in $Z$ and $a_T$ in $X$. For the $Z$-readout bit $q_B\in\{0,1\}$, report $s_B=2q_B-1$. Let $x$ be the $X$-readout outcome of $a_T$. The circuit implements $P_{s_B}(B)P_x(T)$, including when $s_B=-1$ and the pin is the identity.

Insert $\operatorname{CNOT}(t\to a_B)$ immediately before and after the CNOTs from the checked qubits to $a_B$. The two added gates commute with the CNOTs from the checked qubits to $a_B$ and cancel ideally, so the instrument is unchanged on every sector. If a $Z_{a_B}$ fault occurs during parity accumulation, it propagates $Z$ errors to the data qubits coupled to $a_B$ after the fault, and the second added CNOT adds $Z_t$. In the positive-$B$ sector, the pin measures $X_t$ and detects that change. When $B=-1$, the pin is the identity and supplies no such test.

\begin{lemma}[Completion-qubit hook flag]
\label{lem:prx273-completion-flag}
Suppose the joint gadget input satisfies $B=X_t=+1$. Allow one arbitrary native fault at a gate, preparation, readout, basis change or idle. Conditional on reported outcomes $B=T=+1$, every Kraus operator restricted to this input lies in the span of residual operators supported on at most one data qubit and the completion qubit.
\end{lemma}
\begin{proof}
Expand the fault in participant Paulis. After $\operatorname{CNOT}(k\to a_B)$, write it as $P_kX_{a_B}^{u}Z_{a_B}^{v}$, with $P_k$ on the data participant. Propagation through the remaining CNOTs from the checked qubits to $a_B$ and the second added CNOT gives
\begin{equation}
 P_k\left(\prod_{j>k}Z_j\right)^v
 X_{a_B}^{u}Z_{a_B}^{v}Z_t^v.
 \label{eq:prx273-flag-suffix}
\end{equation}
Before the final Toffoli, the ideal auxiliary state is $|1\rangle_{a_B}|+\rangle_{a_T}|+\rangle_t$. A nonzero $u$ changes the central readout. With $u=0$, a nonzero $v$ changes the pin readout. For the reported outcomes $B=T=+1$, only $P_k$ remains on the data. A fault at the first added CNOT can propagate only the full product $B$ to data, which fixes the assumed input. After the second added CNOT there are no further data couplings. Data idles and basis changes involve one data qubit, and auxiliary preparation and readout faults give the same cases. Linearity proves the claim for arbitrary local Kraus operators.
\end{proof}

The same qubit therefore cancels a check commutator and supplies a flag through its pin measurement, including when the fault acts on that completion qubit. For the bilayer toric circuit below, the full fault analysis includes repeated faults and intermediate negative-flux states and proves the same exact distance with or without the two added CNOTs. Appendix~\ref{sec:sm-prx-semidirect} gives the flag circuits and charge conditions for the dihedral reflection checks.

\section{\texorpdfstring{$D_4$}{D4} surgery: circuits, distance and recovery}\label{sec:prx-protocols}
\label{sec:sm-prx-protocols}
Having established the completion construction and its fault model, we now turn to our main example. We study a logical measurement obtained by gauging two toric codes. We specify its local measurement circuits, show how the measurement prepares a logical Toffoli state, and establish its circuit fault distance.

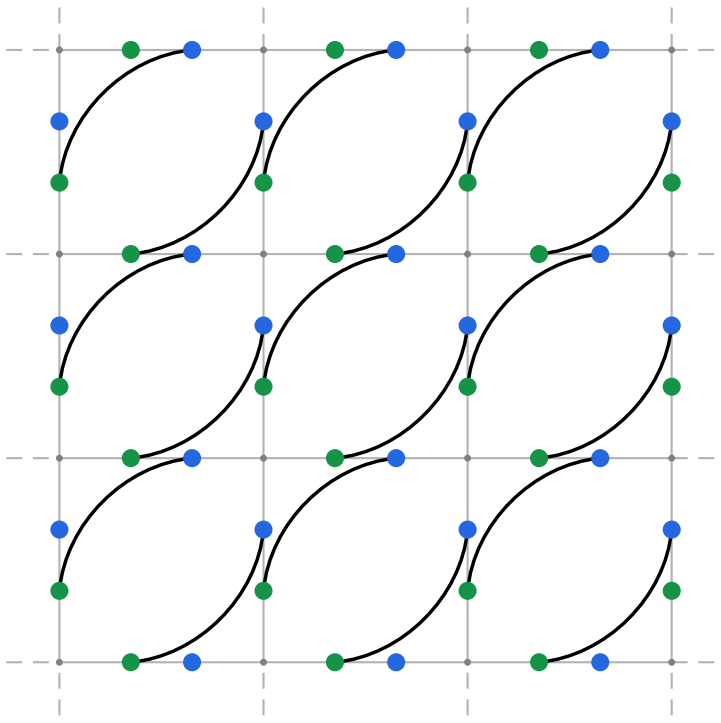
\begin{figure}[tbp]
\centering
\begin{tikzpicture}[x=1.35cm,y=1.35cm]
\path[use as bounding box](-.30,-.30)rectangle(3.30,3.30);
\draw[step=1,black!30,line width=.4pt](0,0)grid(3,3);
\foreach \k in {0,1,2,3}{
 \draw[black!30,densely dashed,line width=.4pt]
  (-.26,\k)--(0,\k) (3,\k)--(3.26,\k)
  (\k,-.26)--(\k,0) (\k,3)--(\k,3.26);
}
\foreach \i in {0,1,2}{\foreach \j in {0,1,2}{
 \draw[black,line width=.65pt]
  ({\i+.35},\j)..controls({\i+.68},{\j+.03})
  and({\i+.97},{\j+.32})..({\i+1},{\j+.65});
 \draw[black,line width=.65pt]
  (\i,{\j+.35})..controls({\i+.03},{\j+.68})
  and({\i+.32},{\j+.97})..({\i+.65},{\j+1});
}}
\foreach \i in {0,1,2}{\foreach \j in {0,1,2,3}{
 \fill[dFourGreen]({\i+.35},\j)circle(1.7pt);
 \fill[dFourBlue]({\i+.65},\j)circle(1.7pt);
 \fill[dFourGreen](\j,{\i+.35})circle(1.7pt);
 \fill[dFourBlue](\j,{\i+.65})circle(1.7pt);
}}
\foreach \i in {0,1,2,3}{\foreach \j in {0,1,2,3}{
 \fill[black!50](\i,\j)circle(.65pt);
}}
\end{tikzpicture}
\caption{Square-lattice layout of the two parent toric codes. Green and blue dots represent their physical qubits, with one of each color on every edge. Black curves are the disjoint CZ gates in Eq.~\eqref{eq:main-d4-physical}. The dashed lines indicate continuation of this local patch across the periodic lattice.}
\label{fig:d4-parent-symmetry}
\end{figure}

The two parent codes occupy an $L\times L$ periodic square lattice, with one green and one blue physical qubit on every edge (Fig.~\ref{fig:d4-parent-symmetry}). These colors label the two parent layers. Gauging their joint symmetry produces a single quantum-double code $D(D_4)$ of the order-eight dihedral group $D_4$. Ungauging returns to the two parent codes. The construction adds one red gauge qubit per edge and three completion qubits per face (Fig.~\ref{fig:d4-completed-checks}).

The surgery follows Protocol~\ref{alg:commuting-gauging}. We initialize the red gauge qubits in $|0\rangle$ and the completion qubits in $|+\rangle$. We then measure the completed checks for $R$ rounds, read the red qubits in $Z$ and apply the correction determined by that record to return to the parent codes. Parent-check rounds precede gauging and follow ungauging.

On the torus, each parent code encodes two logical qubits. We label them $1,3$ in the green layer and $2,4$ in the blue layer. The full surgery measures \mbox{$\overline{\mathcal{U}}=\overline{\operatorname{CZ}}_{12} \overline{\operatorname{CZ}}_{34}$}, where bars denote gates on the encoded qubits. Its recorded outcome $(-1)^{\eta}$, with $\eta\in\{0,1\}$, selects the logical projector
\begin{equation}
 \Pi_\eta=\frac{I+(-1)^{\eta}\overline{\mathcal{U}}}{2}.
 \label{eq:prx273-resource-projector}
\end{equation}
On the encoded stabilizer input $|+,+,1,+\rangle$, this projection prepares the Toffoli state in Eq.~\eqref{eq:main-d4-toffoli}, up to the known Pauli correction $\overline{X}_4^{\eta}$. The non-Clifford physical operations enter through the readout-controlled operations used to measure the completed checks.

For the native circuits and gate order specified in Section~\ref*{sec:prx-d4}, on an even-$L$ torus the fault distance is exactly $d_{\mathrm{surg}}=\min\{L,R\}$ under arbitrary local faults at native gates, preparations, readouts and idles. Theorem~\ref{thm:prx273-d4-surgery} specifies the complete schedule and detection tests, with ideal encoded input and final parent-code projection. At $R=L$, the circuit uses $18L^2$ allocated qubits for $3L$ extraction rounds. Table~\ref{tab:prx277-d4-resources} gives the costs.

Section~\ref{sec:prx-d4} constructs and analyzes this surgery. Section~\ref{sec:prx-d4-recovery} explains how retained completion qubits support recovery. Section~\ref{sec:prx278-t-source} uses a three-boundary square-lattice patch to prepare a logical $T$ state.

\subsection{A logical Toffoli source from two toric codes}\label{sec:prx-d4}
We first identify the physical symmetry and its action on the four encoded qubits. We then give measurement circuits for the completed checks and prove the circuit fault distance of the surgery. Take even $L\ge4$ for the periodic measurement schedule, with the qubit placement in Fig.~\ref{fig:d4-completed-checks}.

Let $e_x(i,j)$ and $e_y(i,j)$ be the edges extending right and up from vertex $(i,j)$, with coordinates modulo $L$. A face is indexed by its southwest vertex. Here $r_e$, $g_e$ and $b_e$ denote the red gauge, green matter and blue matter qubits on edge $e$, respectively. The physical symmetry consists of disjoint CZ pairs,
\begin{equation}
 \begin{aligned}
 \mathcal{U}=\prod_{i,j}&\operatorname{CZ}(g_{e_x(i,j)},b_{e_y(i+1,j)})\\
 &{}\times\operatorname{CZ}(g_{e_y(i,j)},b_{e_x(i,j+1)}).
 \end{aligned}
 \label{eq:main-d4-physical}
\end{equation}
Figure~\ref{fig:d4-parent-symmetry} shows the two parent codes and the CZ pairings in Eq.~\eqref{eq:main-d4-physical}. Each color has four-qubit $X$ checks at vertices and four-qubit $Z$ checks around faces. The same pairing pattern repeats over the periodic $L\times L$ lattice.

For each parent color $c=g,b$, $\alpha_c,\beta_c\in\{0,1\}$ label the encoded basis through the horizontal and vertical $Z$-loop eigenvalues $(-1)^{\alpha_c}$ and $(-1)^{\beta_c}$, respectively. Order the logical qubits as $(1,2,3,4)=(\alpha_g,\beta_b,\beta_g,\alpha_b)$. On the parent code, the physical symmetry in Eq.~\eqref{eq:main-d4-physical} acts on the four logical qubits as
\begin{equation}
 \overline{\mathcal{U}}=\overline{\operatorname{CZ}}_{12}\overline{\operatorname{CZ}}_{34}.
 \label{eq:main-d4-logical}
\end{equation}
\noindent The surgery measures the product in Eq.~\eqref{eq:main-d4-logical} as one observable, with the single outcome $(-1)^{\eta}$ in Eq.~\eqref{eq:prx273-resource-projector}. The torus protocol of Ref.~\cite{Davydova2026} measures this same logical observable. Building on that measurement, we give an explicit Toffoli-state preparation below and a logical Toffoli gate protocol in Appendix~\ref{sec:prx271-consumption}. To prepare the resource, initialize the encoded input $|+,+,1,+\rangle$. Because logical qubit 3 is fixed to $|1\rangle$, the restriction of $\overline{\mathcal U}$ to this input subspace is $\overline{\operatorname{CZ}}_{12}\overline{Z}_4$. Its eigenvalue on $|x,y,1,z\rangle$ is $(-1)^{xy+z}$, so the outcome $(-1)^{\eta}$ selects $z=\eta+xy\bmod2$. In the ideal circuit, each of the two outcomes $\eta=0,1$ occurs with probability $1/2$. For the recorded value $\eta$, the normalized state of logical qubits $1,2,4$ is
\begin{equation}
 \overline{X}_4^{\eta}|\mathrm{Tof}\rangle_{124},\qquad
 |\mathrm{Tof}\rangle_{124}=\frac12\sum_{x,y=0}^{1}|x,y,xy\rangle_{124}.
 \label{eq:main-d4-toffoli}
\end{equation}
Here $\eta$ is the outcome bit of the joint measurement of $\overline{\mathcal U}$. The operator $\overline{X}_4^{\eta}$ is a Pauli correction on blue logical qubit 4, not a measurement. For $\eta=0$, no correction is needed. For $\eta=1$, the logical bit flip changes the target value from $xy+1$ to $xy$ modulo two. Applying this correction or tracking it in the Pauli frame gives the same Toffoli resource for both logical outcomes. A Hadamard on target qubit 4 converts it to the CCZ magic state $\overline{\operatorname{CCZ}}|+,+,+\rangle$. With ideal encoded input and final parent-code projection, Theorem~\ref{thm:prx273-d4-surgery} ensures that fewer than $\min\{L,R\}$ arbitrary local native faults cannot give an incorrect resource while passing the stated error-detection tests.

The state in Eq.~\eqref{eq:main-d4-toffoli} is a resource for applying a logical Toffoli gate to arbitrary data. Its consumption circuit in Appendix~\ref{sec:prx271-consumption} uses logical CNOTs, Pauli measurements and outcome-dependent Clifford corrections.

\begin{figure*}[tp]
\centering
\begin{minipage}{.95\textwidth}\centering
\begin{minipage}[t]{.326\linewidth}\centering
\textbf{(a)}\; Qubit layout\par\smallskip
\scalebox{0.70}{\begin{tikzpicture}[x=2.5cm,y=2.5cm,font=\scriptsize]
\path[use as bounding box] (-.26,-.24) rectangle (1.26,1.30);
\fill[black!1] (0,0) rectangle (1,1);
\draw[black!55,line width=.45pt] (0,0) rectangle (1,1);
\fill[dFourRed] (0.280,0.000) circle[radius=2.35pt];
\fill[dFourRed] (0.280,1.000) circle[radius=2.35pt];
\fill[dFourRed] (0.000,0.280) circle[radius=2.35pt];
\fill[dFourRed] (1.000,0.280) circle[radius=2.35pt];
\node[diamond,aspect=1.05,draw=dFourRed,fill=white,line width=.9pt,minimum size=6.0mm,inner sep=.15pt,font=\normalsize,text=dFourRed!82!black] at (0.500,0.730) {$t_r$};
\fill[dFourGreen] (0.500,0.000) circle[radius=2.35pt];
\fill[dFourGreen] (0.500,1.000) circle[radius=2.35pt];
\fill[dFourGreen] (0.000,0.500) circle[radius=2.35pt];
\fill[dFourGreen] (1.000,0.500) circle[radius=2.35pt];
\node[diamond,aspect=1.05,draw=dFourGreen,fill=white,line width=.9pt,minimum size=6.0mm,inner sep=.15pt,font=\normalsize,text=dFourGreen!82!black] at (0.280,0.300) {$t_g$};
\fill[dFourBlue] (0.720,0.000) circle[radius=2.35pt];
\fill[dFourBlue] (0.720,1.000) circle[radius=2.35pt];
\fill[dFourBlue] (0.000,0.720) circle[radius=2.35pt];
\fill[dFourBlue] (1.000,0.720) circle[radius=2.35pt];
\node[diamond,aspect=1.05,draw=dFourBlue,fill=white,line width=.9pt,minimum size=6.0mm,inner sep=.15pt,font=\normalsize,text=dFourBlue!82!black] at (0.720,0.300) {$t_b$};
\foreach \x/\y in {0/0,1/0,0/1,1/1}{\fill[black!50] (\x,\y) circle[radius=.65pt];}
\node[anchor=south,inner sep=1.5pt,text=dFourRed!82!black,font=\normalsize] at (.28,1.04) {$r$};
\node[anchor=south,inner sep=1.5pt,text=dFourGreen!82!black,font=\normalsize] at (.5,1.04) {$g$};
\node[anchor=south,inner sep=1.5pt,text=dFourBlue!82!black,font=\normalsize] at (.72,1.04) {$b$};
\end{tikzpicture}}\par\smallskip
{\footnotesize Edge dots: gauge $r$, parent $g,b$.\par
Face diamonds: completion qubits.}
\end{minipage}
\hfill
\begin{minipage}[t]{.326\linewidth}\centering
\textbf{(b)}\; Flux check $B_{c,f}$\par\smallskip
\scalebox{0.70}{\begin{tikzpicture}[x=2.5cm,y=2.5cm,font=\small]
\path[use as bounding box] (-.26,-.24) rectangle (1.26,1.30);
\draw[black!48,line width=.38pt] (0,0) rectangle (1,1);
\draw[dFourGreen,line width=2pt] (0,0) rectangle (1,1);
\foreach \x/\y in {.5/0,.5/1,0/.5,1/.5}{\node[fill=white,inner sep=1.3pt,text=dFourGreen!82!black,font=\large] at (\x,\y) {$Z$};}
\node[font=\normalsize,text=black!80] at (.5,.5) {$f$};
\end{tikzpicture}}\par\smallskip
{\footnotesize $B_{c,f}=\displaystyle\prod_{e\in\partial f}Z_{c,e}$}\par
{\footnotesize $c=r,g,b$. One color shown.}
\end{minipage}
\hfill
\begin{minipage}[t]{.326\linewidth}\centering
\textbf{(c)}\; Pinning check $T_{c,f}$\par\smallskip
\scalebox{0.70}{\begin{tikzpicture}[x=2.5cm,y=2.5cm,font=\small,>=Latex]
\path[use as bounding box] (-.26,-.24) rectangle (1.26,1.30);
\fill[dFourGreen!5] (0,0) rectangle (1,1);
\draw[black!48,line width=.38pt] (0,0) rectangle (1,1);
\foreach \x/\y in {.5/0,.5/1,0/.5,1/.5}{
\draw[->,dFourGreen!85!black,line width=.7pt,shorten >=3.9mm] (\x,\y)--(.28,.30);
\fill[dFourGreen] (\x,\y) circle[radius=2.35pt];}
\node[diamond,aspect=1.05,draw=dFourGreen,fill=white,line width=.9pt,minimum size=6.0mm,inner sep=.15pt,font=\normalsize,text=dFourGreen!82!black] at (.28,.30) {$X$};

\end{tikzpicture}}\par\smallskip
{\footnotesize $T_{c,f}=X_{t_{c,f}}\displaystyle\prod_{e\in\partial f}\mathrm{CNOT}_{c,e\to t_{c,f}}$}\par
\makebox[\linewidth][c]{\footnotesize Arrows denote CNOTs, directed to the target.}
\end{minipage}
\par\medskip
\begin{minipage}[t]{.326\linewidth}\centering
\textbf{(d)}\; $\widehat{A}_{r,v}$\par\smallskip
\scalebox{0.70}{\begin{tikzpicture}[x=1.75cm,y=1.75cm,font=\small,>=Latex]
\path[use as bounding box] (-1.07,-1.075) rectangle (1.07,1.075);
\fill[dFourGreen!5] (0.000,0.000) rectangle (1.000,1.000);
\draw[step=1,black!48,line width=.38pt] (-1,-1) grid (1,1);
\draw[dFourRed,line width=2.1pt] (-1,0)--(1,0) (0,-1)--(0,1);
\draw[dFourGreen!85!black,densely dashed,line width=.65pt] (0.280,0.300)--(0.500,0.000);
\draw[dFourGreen!85!black,densely dashed,line width=.65pt] (0.280,0.300)--(0.500,1.000);
\draw[dFourGreen!85!black,densely dashed,line width=.65pt] (0.280,0.300)--(0.000,0.500);
\draw[dFourGreen!85!black,densely dashed,line width=.65pt] (0.280,0.300)--(1.000,0.500);
\draw[black!82,line width=.8pt] (0.500,0.000) .. controls (0.700,0.000) and (0.890,0.130) .. (0.925,0.300) .. controls (0.950,0.420) and (0.840,0.580) .. (1.000,0.720);
\draw[black!82,line width=.8pt] (0.000,0.500) .. controls (0.055,0.760) and (0.380,0.830) .. (0.720,1.000);
\fill[dFourGreen] (0.500,0.000) circle[radius=2.35pt];
\fill[dFourGreen] (0.500,1.000) circle[radius=2.35pt];
\fill[dFourGreen] (0.000,0.500) circle[radius=2.35pt];
\fill[dFourGreen] (1.000,0.500) circle[radius=2.35pt];
\fill[dFourGreen] (0.500,0.000) circle[radius=2.35pt];
\fill[dFourBlue] (1.000,0.720) circle[radius=2.35pt];
\fill[dFourGreen] (0.000,0.500) circle[radius=2.35pt];
\fill[dFourBlue] (0.720,1.000) circle[radius=2.35pt];
\node[fill=white,inner sep=.65pt,text=dFourRed!82!black,font=\large] at (0.280,0.000) {$X$};
\node[fill=white,inner sep=.65pt,text=dFourRed!82!black,font=\large] at (-0.720,0.000) {$X$};
\node[fill=white,inner sep=.65pt,text=dFourRed!82!black,font=\large] at (0.000,0.280) {$X$};
\node[fill=white,inner sep=.65pt,text=dFourRed!82!black,font=\large] at (0.000,-0.720) {$X$};
\node[diamond,aspect=1.05,draw=dFourGreen,fill=white,line width=.9pt,minimum size=6.0mm,inner sep=.15pt,font=\normalsize,text=dFourGreen!82!black] at (0.280,0.300) {$t_g$};
\node[diamond,aspect=1.05,draw=dFourBlue,fill=white,line width=.9pt,minimum size=6.0mm,inner sep=.15pt,font=\normalsize,text=dFourBlue!82!black] at (0.720,0.300) {$X$};
\fill[black!65] (0,0) circle[radius=.9pt];\node[anchor=north west,inner sep=1.2pt,font=\normalsize,text=black!85] at (.06,-.05) {$v$};
\end{tikzpicture}}\par\smallskip
{\footnotesize $\widehat{A}_{r,v}=A_{r,v}X_{t_{b,v}}V(B_{g,v},t_{g,v})$}
\end{minipage}
\hfill
\begin{minipage}[t]{.326\linewidth}\centering
\textbf{(e)}\; $\widehat{A}_{g,v}$\par\smallskip
\scalebox{0.70}{\begin{tikzpicture}[x=1.75cm,y=1.75cm,font=\small,>=Latex]
\path[use as bounding box] (-1.07,-1.075) rectangle (1.07,1.075);
\fill[dFourBlue!5] (0.000,0.000) rectangle (1.000,1.000);
\draw[step=1,black!48,line width=.38pt] (-1,-1) grid (1,1);
\draw[dFourGreen,line width=2.1pt] (-1,0)--(1,0) (0,-1)--(0,1);
\draw[dFourBlue!85!black,densely dashed,line width=.65pt] (0.720,0.300)--(0.720,0.000);
\draw[dFourBlue!85!black,densely dashed,line width=.65pt] (0.720,0.300)--(0.720,1.000);
\draw[dFourBlue!85!black,densely dashed,line width=.65pt] (0.720,0.300)--(0.000,0.720);
\draw[dFourBlue!85!black,densely dashed,line width=.65pt] (0.720,0.300)--(1.000,0.720);
\draw[black!82,line width=.8pt] (-0.720,0.000) .. controls (-0.790,0.430) and (-0.430,0.790) .. (0.000,0.720);
\draw[black!82,line width=.8pt] (0.000,-0.720) .. controls (0.430,-0.790) and (0.790,-0.430) .. (0.720,0.000);
\fill[dFourBlue] (0.720,0.000) circle[radius=2.35pt];
\fill[dFourBlue] (0.720,1.000) circle[radius=2.35pt];
\fill[dFourBlue] (0.000,0.720) circle[radius=2.35pt];
\fill[dFourBlue] (1.000,0.720) circle[radius=2.35pt];
\fill[dFourRed] (-0.720,0.000) circle[radius=2.35pt];
\fill[dFourBlue] (0.000,0.720) circle[radius=2.35pt];
\fill[dFourRed] (0.000,-0.720) circle[radius=2.35pt];
\fill[dFourBlue] (0.720,0.000) circle[radius=2.35pt];
\node[fill=white,inner sep=.65pt,text=dFourGreen!82!black,font=\large] at (0.500,0.000) {$X$};
\node[fill=white,inner sep=.65pt,text=dFourGreen!82!black,font=\large] at (-0.500,0.000) {$X$};
\node[fill=white,inner sep=.65pt,text=dFourGreen!82!black,font=\large] at (0.000,0.500) {$X$};
\node[fill=white,inner sep=.65pt,text=dFourGreen!82!black,font=\large] at (0.000,-0.500) {$X$};
\node[diamond,aspect=1.05,draw=dFourBlue,fill=white,line width=.9pt,minimum size=6.0mm,inner sep=.15pt,font=\normalsize,text=dFourBlue!82!black] at (0.720,0.300) {$t_b$};
\node[diamond,aspect=1.05,draw=dFourRed,fill=white,line width=.9pt,minimum size=6.0mm,inner sep=.15pt,font=\normalsize,text=dFourRed!82!black] at (-0.500,-0.270) {$X$};
\fill[black!65] (0,0) circle[radius=.9pt];\node[anchor=north west,inner sep=1.2pt,font=\normalsize,text=black!85] at (.06,-.05) {$v$};
\end{tikzpicture}}\par\smallskip
{\footnotesize $\widehat{A}_{g,v}=A_{g,v}X_{t_{r,v-\Delta}}V(B_{b,v},t_{b,v})$}
\end{minipage}
\hfill
\begin{minipage}[t]{.326\linewidth}\centering
\textbf{(f)}\; $\widehat{A}_{b,v}$\par\smallskip
\scalebox{0.70}{\begin{tikzpicture}[x=1.75cm,y=1.75cm,font=\small,>=Latex]
\path[use as bounding box] (-1.07,-1.075) rectangle (1.07,1.075);
\fill[dFourRed!5] (-1.000,-1.000) rectangle (0.000,0.000);
\draw[step=1,black!48,line width=.38pt] (-1,-1) grid (1,1);
\draw[dFourBlue,line width=2.1pt] (-1,0)--(1,0) (0,-1)--(0,1);
\draw[dFourRed!85!black,densely dashed,line width=.65pt] (-0.500,-0.270)--(-0.720,-1.000);
\draw[dFourRed!85!black,densely dashed,line width=.65pt] (-0.500,-0.270)--(-0.720,0.000);
\draw[dFourRed!85!black,densely dashed,line width=.65pt] (-0.500,-0.270)--(-1.000,-0.720);
\draw[dFourRed!85!black,densely dashed,line width=.65pt] (-0.500,-0.270)--(0.000,-0.720);
\draw[black!82,line width=.8pt] (-0.720,-1.000) .. controls (-0.120,-0.980) and (-0.030,-0.760) .. (0.000,-0.500);
\draw[black!82,line width=.8pt] (-1.000,-0.720) .. controls (-0.980,-0.120) and (-0.760,-0.030) .. (-0.500,0.000);
\fill[dFourRed] (-0.720,-1.000) circle[radius=2.35pt];
\fill[dFourRed] (-0.720,0.000) circle[radius=2.35pt];
\fill[dFourRed] (-1.000,-0.720) circle[radius=2.35pt];
\fill[dFourRed] (0.000,-0.720) circle[radius=2.35pt];
\fill[dFourRed] (-0.720,-1.000) circle[radius=2.35pt];
\fill[dFourGreen] (0.000,-0.500) circle[radius=2.35pt];
\fill[dFourRed] (-1.000,-0.720) circle[radius=2.35pt];
\fill[dFourGreen] (-0.500,0.000) circle[radius=2.35pt];
\node[fill=white,inner sep=.65pt,text=dFourBlue!82!black,font=\large] at (0.720,0.000) {$X$};
\node[fill=white,inner sep=.65pt,text=dFourBlue!82!black,font=\large] at (-0.280,0.000) {$X$};
\node[fill=white,inner sep=.65pt,text=dFourBlue!82!black,font=\large] at (0.000,0.720) {$X$};
\node[fill=white,inner sep=.65pt,text=dFourBlue!82!black,font=\large] at (0.000,-0.280) {$X$};
\node[diamond,aspect=1.05,draw=dFourRed,fill=white,line width=.9pt,minimum size=6.0mm,inner sep=.15pt,font=\normalsize,text=dFourRed!82!black] at (-0.500,-0.270) {$t_r$};
\node[diamond,aspect=1.05,draw=dFourGreen,fill=white,line width=.9pt,minimum size=6.0mm,inner sep=.15pt,font=\normalsize,text=dFourGreen!82!black] at (-0.720,-0.700) {$X$};
\fill[black!65] (0,0) circle[radius=.9pt];\node[anchor=north west,inner sep=1.2pt,font=\normalsize,text=black!85] at (.06,-.05) {$v$};
\end{tikzpicture}}\par\smallskip
{\footnotesize $\widehat{A}_{b,v}=A_{b,v}X_{t_{g,v-\Delta}}V(B_{r,v-\Delta},t_{r,v-\Delta})$}
\end{minipage}
\end{minipage}\par
\caption{Commuting checks of the $D(D_4)$ quantum double on the square lattice. (a) Each edge carries a red gauge qubit and green and blue matter qubits. Each face contains three completion qubits and is indexed by its southwest vertex. (b),(c) The flux $B_{c,f}$ controls the pinning check $T_{c,f}=T(B_{c,f},t_{c,f})$ on face $f$. (d)--(f) Raw stars $A_{c,v}$ and their completions $\widehat{A}_{c,v}$ for $c=r,g,b$, with $\Delta=(1,1)$. Colored edges carry $X$, solid curves denote the CZ factors in the raw stars, and dashed lines denote the CZs in $V(B,t)$. The $X$ diamond denotes the Pauli-$X$ operator on a completion qubit.}
\label{fig:d4-completed-checks}
\end{figure*}
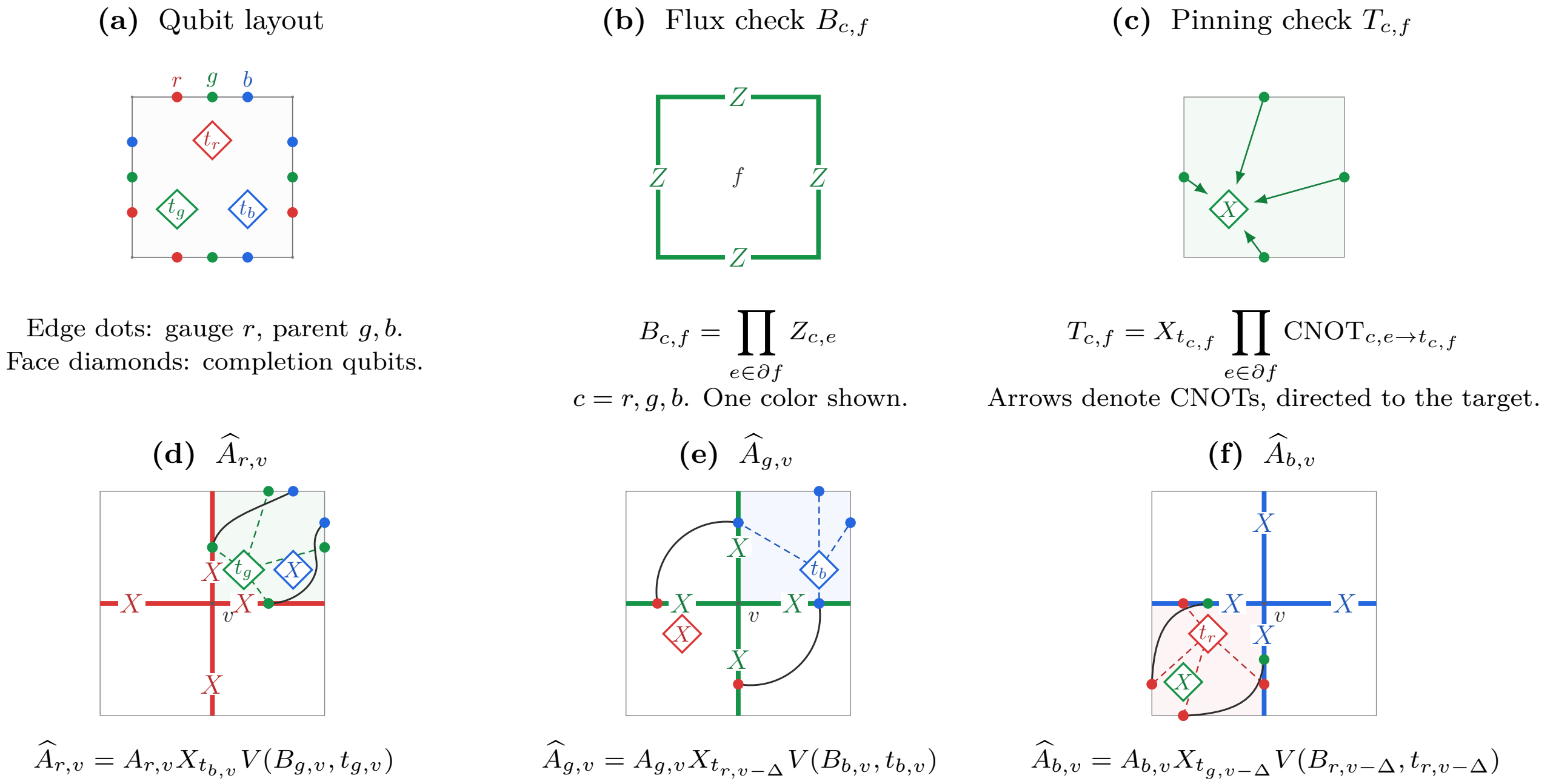
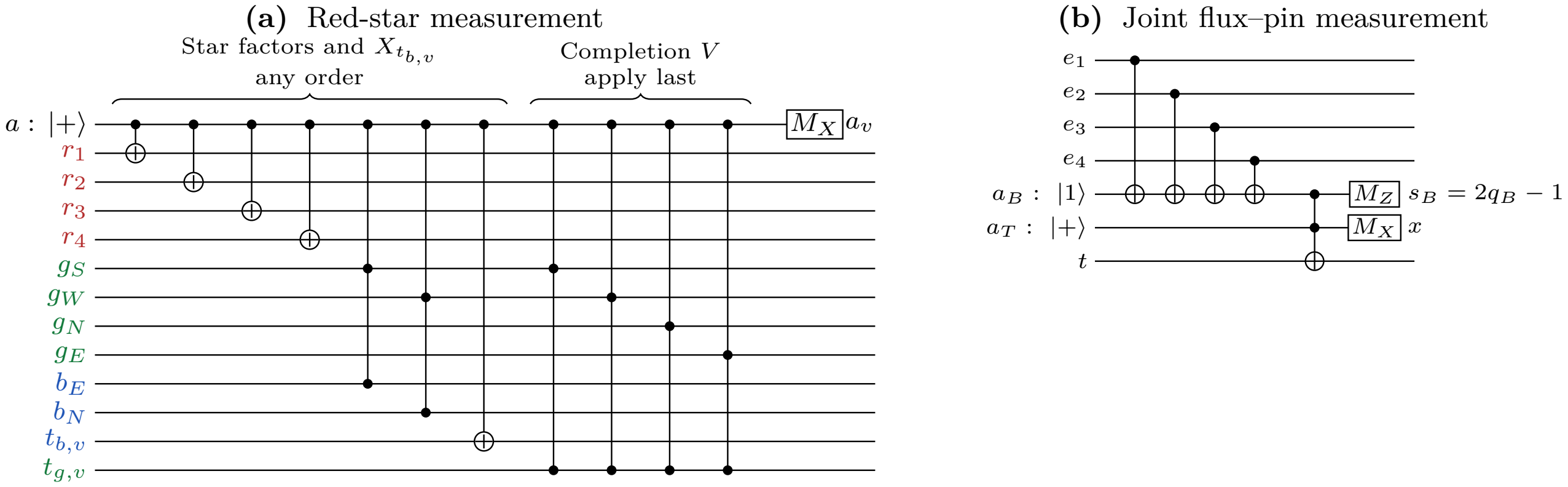
\begin{figure*}[tp]
\centering
\definecolor{dFourRed}{HTML}{DB3636}
\definecolor{dFourGreen}{HTML}{159447}
\definecolor{dFourBlue}{HTML}{2467DE}
\begin{minipage}[t]{.62\textwidth}\centering
\textbf{(a)}\; Red-star measurement\par\smallskip
\begin{tikzpicture}[x=1.25cm,y=1cm,line width=.45pt,font=\footnotesize,
  starcontrol/.style={circle,fill=black,inner sep=0pt,minimum size=3.0pt},
  startarget/.style={circle,draw,fill=white,inner sep=0pt,minimum size=5.8pt}]
\path[use as bounding box](-1.15,-3.92) rectangle (6.85,.83);
\draw (0,0)--(6.40,0);
\node[left,inner sep=2pt] at (0,0) {$a:\,|+\rangle$};
\foreach \y/\lab/\col in {
 -.31/{r_1}/dFourRed,-.62/{r_2}/dFourRed,
 -.93/{r_3}/dFourRed,-1.24/{r_4}/dFourRed,
 -1.55/{g_S}/dFourGreen,-1.86/{g_W}/dFourGreen,
 -2.17/{g_N}/dFourGreen,-2.48/{g_E}/dFourGreen,
 -2.79/{b_E}/dFourBlue,-3.10/{b_N}/dFourBlue,
 -3.41/{t_{b,v}}/dFourBlue,-3.72/{t_{g,v}}/dFourGreen}{
 \draw (0,\y)--(6.72,\y);
 \node[left,inner sep=2pt,text=\col!82!black] at (0,\y) {$\lab$};
}
\foreach \x/\y in {.35/-.31,.85/-.62,1.35/-.93,1.85/-1.24,3.35/-3.41}{
 \draw (\x,0)--(\x,\y);
 \node[starcontrol] at (\x,0) {};
 \node[startarget] at (\x,\y) {};
 \draw (\x-.075,\y)--(\x+.075,\y);
 \draw (\x,\y-.075)--(\x,\y+.075);
}
\foreach \x/\yg/\yb in {2.35/-1.55/-2.79,2.85/-1.86/-3.10}{
 \draw (\x,0)--(\x,\yb);
 \foreach \y in {0,\yg,\yb}{\node[starcontrol] at (\x,\y) {};}
}
\foreach \x/\yg in {3.95/-1.55,4.45/-1.86,4.95/-2.17,5.45/-2.48}{
 \draw (\x,0)--(\x,-3.72);
 \foreach \y in {0,\yg,-3.72}{\node[starcontrol] at (\x,\y) {};}
}
\node[draw,fill=white,minimum width=.40cm,minimum height=.27cm,
 inner sep=1pt] at (6.20,0) {$M_X$};
\node[right,inner sep=2pt] at (6.40,0) {$a_v$};
\draw[decorate,decoration={brace,amplitude=3pt}]
 (.15,.22)--(3.55,.22);
\node[anchor=south,align=center,font=\scriptsize,inner sep=1pt]
 at (1.85,.34) {Star factors and $X_{t_{b,v}}$\\any order};
\draw[decorate,decoration={brace,amplitude=3pt}]
 (3.75,.22)--(5.65,.22);
\node[anchor=south,align=center,font=\scriptsize,inner sep=1pt]
 at (4.70,.34) {Completion $V$\\apply last};
\end{tikzpicture}\end{minipage}\hfill
\begin{minipage}[t]{.36\textwidth}\centering
\textbf{(b)}\; Joint flux--pin measurement\par\smallskip
\begin{tikzpicture}[x=0.43cm,y=0.36cm,line width=0.45pt,font=\scriptsize]
\foreach \y/\name in {6/1,5/2,4/3,3/4}{
 \draw (0,\y)--(8.0,\y);
 \node[left,inner sep=2pt] at (0,\y) {$e_{\name}$};
}
\draw (0,2)--(7.05,2); \node[left,inner sep=2pt] at (0,2) {$a_B:\ |1\rangle$};
\draw (0,1)--(7.05,1); \node[left,inner sep=2pt] at (0,1) {$a_T:\ |+\rangle$};
\draw (0,0)--(8.0,0); \node[left,inner sep=2pt] at (0,0) {$t$};
\foreach \x/\y in {1/6,2/5,3/4,4/3}{
 \draw (\x,\y)--(\x,2); \fill (\x,\y) circle(1.5pt);
 \draw[fill=white] (\x,2) circle(2.8pt);\draw (\x,1.75)--(\x,2.25);
 \draw (\x-.22,2)--(\x+.22,2);
}
\draw (5.5,0)--(5.5,2);
\fill (5.5,2) circle(1.5pt);\fill (5.5,1) circle(1.5pt);
\draw[fill=white] (5.5,0) circle(2.8pt);\draw (5.5,-0.25)--(5.5,0.25);
\draw (5.28,0)--(5.72,0);
\node[draw,fill=white,minimum width=0.44cm,minimum height=0.26cm,inner sep=1pt] at (7,2) {$M_Z$};
\node[draw,fill=white,minimum width=0.44cm,minimum height=0.26cm,inner sep=1pt] at (7,1) {$M_X$};
\node[right,inner sep=2pt] at (7.65,2) {$s_B=2q_B-1$};
\node[right,inner sep=2pt] at (7.65,1) {$x$};
\end{tikzpicture}
\end{minipage}
\caption{Measurement circuits for completed checks. (a) Measurement of $\widehat A_{r,v}$ in Fig.~\ref{fig:d4-completed-checks}(d). The qubits $r_1,\ldots,r_4$ are the red edges incident to $v$. The labels $S,W,N,E$ on green and blue qubits denote the south, west, north and east sides, respectively, of the face whose southwest vertex is $v$. Three connected filled dots denote a CCZ. The first seven gates may be ordered arbitrarily, followed by the four controlled-$V$ CCZs. The green and blue circuits use the factors and completion placements in Eq.~\eqref{eq:prx273-d4-completed} and Fig.~\ref{fig:d4-completed-checks}(e),(f). (b) Local flux--pin measurement used on each face and color during a gauged round. It jointly measures $B=\prod_{j=1}^4Z_{e_j}$ and $T=P_-(B)+P_+(B)X_t$. Four CNOTs, each controlled by one checked edge qubit and targeting $a_B$, encode the eigenvalue of the flux check $B$ in $a_B$. Its $Z$-readout bit $q_B$ gives $s_B=2q_B-1\in\{\pm1\}$. The Toffoli enables an $X_t$ measurement when $s_B=+1$. When $s_B=-1$, the pin is identity and the $X$ readout of $a_T$ returns $x=+1$. Together the readouts implement $P_{s_B}(B)P_x(T)$.}
\label{fig:smd4-joint-circuit}
\end{figure*}
\subsubsection{Completed checks and their native circuits}\label{sec:prx-d4-circuits}
To implement this logical measurement, we use the raw checks from the gauging construction in Section~\ref{sec:prx-raw-gauging} and complete them as in Section~\ref{sec:prx-central-completion}. Appendix~\ref{sec:smd4-bilayer-gauging} derives their lattice expressions from the physical symmetry in Eq.~\eqref{eq:main-d4-physical}. The red stars are Gauss operators and the green and blue stars are modified parent checks. For a configuration $|z\rangle$, let $z_{c,e}\in\{0,1\}$ be the bit of the color-$c$ qubit on edge $e$, so $Z_{c,e}|z\rangle=(-1)^{z_{c,e}}|z\rangle$. Write $X^{s_{c,v}}$ for $X$ on the four color-$c$ edges incident to $v$, and $D_q|z\rangle=(-1)^{q(z)}|z\rangle$. The functions $q_{c,v}(z)$ below are the phase exponents in these diagonal signs. They are quadratic polynomials in the bits $z_{c,e}$, with arithmetic modulo two. The raw operators are
\begin{equation}
 A_{c,v}=X^{s_{c,v}}D_{q_{c,v}},\qquad
 B_{c,f}=\prod_{e\in\partial f}Z_{c,e},
 \label{eq:prx273-d4-raw}
\end{equation}
Each product $z_{c,e}z_{c',e'}$ in $q_{c,v}$ contributes a minus sign only when both bits are $1$, giving a CZ gate between the two indicated qubits. The corresponding quadratic polynomials are
\begin{align}
 q_{r,(i,j)}={}&z_{g,e_x(i,j)}z_{b,e_y(i+1,j)}\nonumber\\
 &{}+z_{g,e_y(i,j)}z_{b,e_x(i,j+1)},\nonumber\\
 q_{g,(i,j)}={}&z_{r,e_x(i-1,j)}z_{b,e_y(i,j)}\nonumber\\
 &{}+z_{r,e_y(i,j-1)}z_{b,e_x(i,j)},\nonumber\\
 q_{b,(i,j)}={}&z_{r,e_x(i-1,j-1)}z_{g,e_y(i,j-1)}\nonumber\\
 &{}+z_{r,e_y(i-1,j-1)}z_{g,e_x(i-1,j)}.
 \label{eq:prx273-d4-quadratics}
\end{align}
The only conflicts are the red--green pair at $v$, with label $B_{b,v}$, the green--blue pair at $v$, with label $B_{r,v-\Delta}$, and the blue star at $v+\Delta$ with the red star at $v$, with label $B_{g,v}$. Here, $\Delta=(1,1)$. Equation~\eqref{eq:main-completion} gives
\begin{align}
 \widehat{A}_{r,v}&=A_{r,v}X_{t_{b,v}}V(B_{g,v},t_{g,v}),\nonumber\\
 \widehat{A}_{g,v}&=A_{g,v}X_{t_{r,v-\Delta}}V(B_{b,v},t_{b,v}),\nonumber\\
 \widehat{A}_{b,v}&=A_{b,v}X_{t_{g,v-\Delta}}
 V(B_{r,v-\Delta},t_{r,v-\Delta}),
 \label{eq:prx273-d4-completed}
\end{align}
with $V(B_{c,f},t)=\prod_{e\in\partial f}\operatorname{CZ}(c_e,t)$. Measure these stars, all fluxes $B_{c,f}$ and all pins $T_{c,f}=P_-(B_{c,f})+P_+(B_{c,f})X_{t_{c,f}}$. The identity $X_tV=BVX_t$ cancels each conflict, while $B=T=+1$ fixes the completion register to its product state.

The common $+1$ eigenspace of these checks is the $D(D_4)$ code on the torus. Appendix~\ref{sec:smd4-ground-space} verifies its dimension $22$. To see why ungauging works globally on the torus, note that initializing the red gauge qubits in $|0\rangle$ fixes both red winding eigenvalues to $+1$. Here, the red winding operators are products of red $Z$ operators around horizontal and vertical noncontractible cycles. Ideal check measurements preserve these values for either logical outcome $\eta$. Together with positive red face fluxes, they ensure that the ideal final gauge record has the form $z=\delta x$. Fixing a reference vertex then determines the vertex bits $x$ and the ungauging correction $U(x)^\dagger$ in Protocol~\ref{alg:commuting-gauging}. Appendix~\ref{sec:smd4-bilayer-gauging} gives the corresponding boundary identities.

We now specify the physical check-measurement circuits and their gate order for the distance-preserving surgery of Theorem~\ref{thm:prx273-d4-surgery} below. The fault analysis below follows errors through these circuits and both the gauging and ungauging interfaces.

To measure a completed star $\widehat A_{c,v}$ in Fig.~\ref{fig:d4-completed-checks}(d)--(f), prepare a readout ancilla $a$ in $|+\rangle$ and use it to control each factor of the star. Figure~\ref{fig:smd4-joint-circuit}(a) shows the red-star circuit. First apply four CNOTs from $a$ to the four color-$c$ edge qubits incident to $v$, two CCZs on $a$ and the endpoint pairs of the solid CZ curves, and one CNOT from $a$ to the completion qubit marked $X$. These seven gates may be applied in any order.

Next implement the factor $V$ using four CCZs, each acting on $a$ and the two endpoints of one dashed CZ link. Finally, measure $a$ in $X$ to obtain the star outcome. The controlled-$V$ gates must follow these seven gates for the fault-distance proof in Appendix~\ref{sec:smd4-native-extraction}. Each star measurement therefore uses five CNOTs and six CCZs.

The remaining measurements in each gauged round are the face fluxes and their completion-qubit pins. For each face and color, the joint flux--pin circuit in Fig.~\ref{fig:smd4-joint-circuit}(b) uses one readout per check. Prepare $a_B$ in $|1\rangle$ and $a_T$ in $|+\rangle$. Apply four CNOTs, each controlled by one of the face's edge qubits and targeting $a_B$. They change the ancilla bit from $1$ to $1\oplus z_1\oplus z_2\oplus z_3\oplus z_4$, where $z_j\in\{0,1\}$ is the computational-basis bit of edge qubit $j$. Apply $\operatorname{Toffoli}(a_B,a_T\to t)$, then read $a_B$ in $Z$ and $a_T$ in $X$. The readout bit $q_B$ determines the flux sign $s_B=2q_B-1$, while $x=\pm1$ is the second readout. The two readout ancillas are reset after the measurement. The four edge qubits and the completion qubit $t$ remain part of the code. For an input $|\psi\rangle$ on these five qubits, $K_{s_B,x}|\psi\rangle$ is their unnormalized state conditioned on the two readouts, with
\begin{align}
 K_{s_B,x}&=P_{s_B}(B)\biggl[\frac{1+s_B}{2}\frac{I+xX_t}{2}
 +\frac{1-s_B}{2}\delta_{x,+1}I\biggr]\nonumber\\
 &=P_{s_B}(B)P_x(T).
 \label{eq:prx273-d4-joint-instrument}
\end{align}
Equation~\eqref{eq:prx273-d4-joint-instrument} has two cases, determined by the measured flux. If $B=+1$, the pinning check is $T=X_t$. The four CNOTs leave $a_B$ in $|1\rangle$, which enables the CNOT from $a_T$ to $t$. Reading $a_T$ in the $X$ basis then measures $X_t$. If $B=-1$, the pinning check is $T=I$. Now $a_B$ is in $|0\rangle$, the Toffoli does nothing, and $a_T$ remains in $|+\rangle$, giving $x=+1$ without measuring $X_t$. The same fixed circuit therefore measures the prescribed pinning check for either flux outcome.

A gauged round consists of the star measurements and these flux--pin measurements on all faces. With ideal operations, gauging, repeated rounds and corrected ungauging together implement the logical projector in Eq.~\eqref{eq:prx273-resource-projector}, which prepares the encoded Toffoli resource from the chosen input. Figure~\ref{fig:smd4-joint-circuit}(b) specifies one local part of this procedure. Its physical Toffoli acts on $a_B,a_T,t$. The next proof tracks errors through these local measurement circuits to establish the fault distance of the full surgery.

Schedule six star stages, one per color and vertex-checkerboard parity, and two face-checkerboard stages for joint flux--pin measurements. Serialize overlapping measurement circuits and parallelize disjoint ones. All have bounded support and use CNOT, CCZ and Toffoli gates, of hierarchy level at most three. Ungauging CZs are native two-qubit locations, and parent checks use four-CNOT extraction. Every gate, preparation, readout and idle obeys the local Kraus fault model of Section~\ref{sec:prx-ft}.

\subsubsection{Why the circuit retains the full distance}
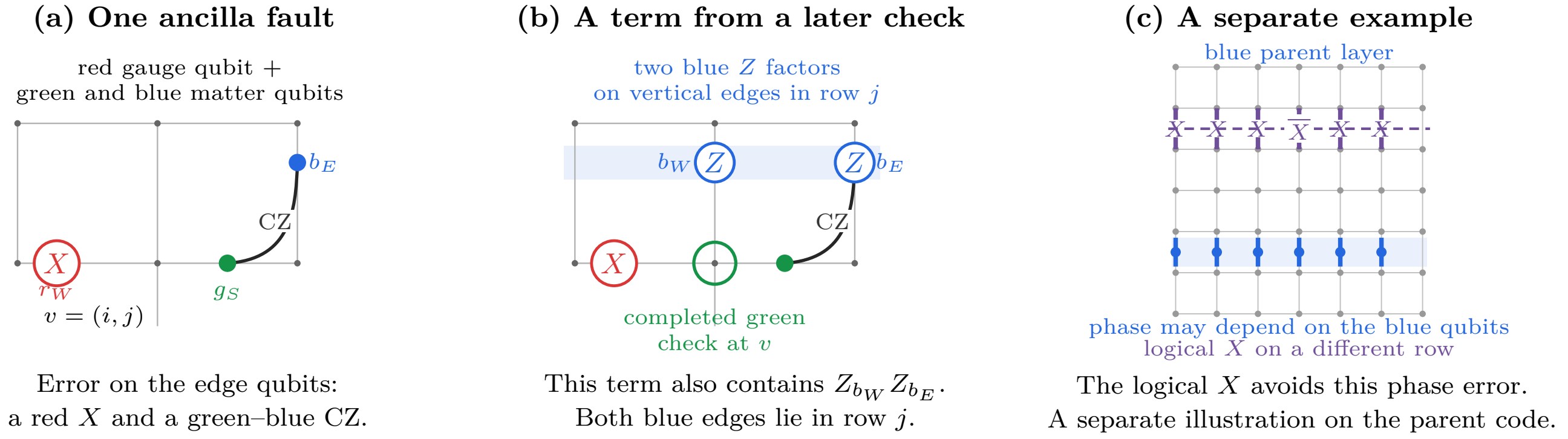
\begin{figure*}[tbp]
\centering
\begingroup
\definecolor{dFaultRed}{HTML}{DB3636}
\definecolor{dFaultGreen}{HTML}{159447}
\definecolor{dFaultBlue}{HTML}{2467DE}
\definecolor{dFaultLogical}{HTML}{6F4F99}
\begin{minipage}[t]{.32\textwidth}\centering
\textbf{(a) One ancilla fault}\par\smallskip
\begin{tikzpicture}[x=1.53cm,y=1.53cm,font=\scriptsize,text=black]
  \path[use as bounding box] (-1.20,-.70) rectangle (1.60,1.55);
  \draw[black!30,line width=.45pt]
   (-1,0)--(1,0) (-1,1)--(1,1)
   (-1,0)--(-1,1) (0,-.45)--(0,1) (1,0)--(1,1);
  \foreach \x/\y in {-1/0,0/0,1/0,-1/1,0/1,1/1}
   \fill[black!60] (\x,\y) circle (1pt);
  \draw[black!85,line width=1.0pt]
    (.5,0) .. controls (.94,.03) and (.99,.35) .. (1,.72);
  \node[fill=white,text=black!80,inner sep=1pt] at (.84,.30) {CZ};
  \node[circle,draw=dFaultRed,fill=white,inner sep=1.3pt,
    line width=.9pt,text=dFaultRed,font=\small] at (-.72,0) {$X$};
  \fill[dFaultGreen] (.5,0) circle (2.7pt);
  \fill[dFaultBlue] (1,.72) circle (2.7pt);
  \node[text=dFaultRed,below=3pt] at (-.72,-.02) {$r_W$};
  \node[text=dFaultGreen,below=3pt] at (.5,-.02) {$g_S$};
  \node[text=dFaultBlue,anchor=west,inner sep=2pt] at (1.03,.72) {$b_E$};
  \node[anchor=east,inner sep=2pt] at (-.04,-.38) {$v=(i,j)$};
  \node[align=center] at (.16,1.30) {red gauge qubit +\\green and blue matter qubits};
\end{tikzpicture}\par
{\footnotesize Error on the edge qubits:\\a red $X$ and a green--blue CZ.}
\end{minipage}\hfill
\begin{minipage}[t]{.32\textwidth}\centering
\textbf{(b) A term from a later check}\par\smallskip
\begin{tikzpicture}[x=1.53cm,y=1.53cm,font=\scriptsize,text=black]
  \path[use as bounding box] (-1.20,-.70) rectangle (1.60,1.55);
  \fill[dFaultBlue!10] (-1.08,.60) rectangle (1.18,.84);
  \draw[black!30,line width=.45pt]
   (-1,0)--(1,0) (-1,1)--(1,1)
   (-1,0)--(-1,1) (0,-.45)--(0,1) (1,0)--(1,1);
  \foreach \x/\y in {-1/0,0/0,1/0,-1/1,0/1,1/1}
   \fill[black!60] (\x,\y) circle (1pt);
  \draw[black!85,line width=1.0pt]
    (.5,0) .. controls (.94,.03) and (.99,.35) .. (1,.72);
  \node[fill=white,text=black!80,inner sep=1pt] at (.84,.30) {CZ};
  \node[circle,draw=dFaultRed,fill=white,inner sep=1.3pt,
    line width=.9pt,text=dFaultRed,font=\small] at (-.72,0) {$X$};
  \fill[dFaultGreen] (.5,0) circle (2.7pt);
  \foreach \x in {0,1}
    \node[circle,draw=dFaultBlue,fill=white,inner sep=1.0pt,
      line width=.9pt,text=dFaultBlue,font=\small] at (\x,.72) {$Z$};
  \node[text=dFaultBlue,anchor=east,inner sep=2pt] at (-.10,.72) {$b_W$};
  \node[text=dFaultBlue,anchor=west,inner sep=2pt] at (1.10,.72) {$b_E$};
  \draw[dFaultGreen,line width=1.0pt] (0,0) circle (.15);
  \node[text=dFaultGreen,align=center,anchor=north] at (0,-.25)
    {completed green\\check at $v$};
  \node[text=dFaultBlue,align=center] at (.16,1.30)
    {two blue $Z$ factors\\on vertical edges in row $j$};
\end{tikzpicture}\par
{\footnotesize This term also contains $Z_{b_W}Z_{b_E}$.\\Both blue edges lie in row $j$.}
\end{minipage}\hfill
\begin{minipage}[t]{.32\textwidth}\centering
\textbf{(c) A separate example}\par\smallskip
\begin{tikzpicture}[x=.45cm,y=.45cm,font=\scriptsize,text=black]
  \path[use as bounding box] (-.65,-1.2) rectangle (6.70,6.5);
  \fill[dFaultBlue!9] (-.10,1.15) rectangle (6.1,1.85);
  \draw[step=1,black!25,line width=.35pt] (0,0) grid (6,6);
  \foreach \x in {0,...,6} {
    \foreach \y in {0,...,6} {\fill[black!40] (\x,\y) circle (1pt);}
  }
  \foreach \x in {0,...,5} {
    \draw[dFaultBlue,line width=1.4pt] (\x,1.15)--(\x,1.85);
    \fill[dFaultBlue] (\x,1.5) circle (1.6pt);
  }
  \foreach \x in {0,...,5} {
    \draw[dFaultLogical,line width=1.4pt] (\x,4)--(\x,5);
    \node[fill=white,inner sep=.25pt,text=dFaultLogical] at (\x,4.5) {$X$};
  }
  \draw[dFaultLogical,densely dashed,line width=.8pt] (-.18,4.5)--(6.18,4.5);
  \node[fill=white,inner sep=1pt,text=dFaultLogical] at (3,4.5) {$\overline X$};
  \node[text=dFaultBlue,align=center] at (3,6.33) {blue parent layer};
  \node[text=dFaultBlue,align=center] at (3,-.35) {phase may depend on the blue qubits};
  \node[text=dFaultLogical,align=center] at (3,-.88) {logical $X$ on a different row};
\end{tikzpicture}\par
{\footnotesize The logical $X$ avoids this phase error.\\A separate illustration on the parent code.}
\end{minipage}
\endgroup
\caption{Fault propagation and the parent-code argument. (a) Apply $\operatorname{CNOT}(a\to r_W)$ and $\operatorname{CCZ}(a,g_S,b_E)$ first in an allowed red-star order. An $X_a$ fault immediately afterward gives the ideal star measurement followed by $F=X_{r_W}\operatorname{CZ}_{g_S,b_E}$, for either readout. (b) The expansion of a later green-check measurement includes the conjugated error $\widehat A_{g,v}F\widehat A_{g,v}=FZ_{b_W}Z_{b_E}$. The original blue phase involves $b_E$. The extra factors involve $b_W$ and $b_E$, both on vertical edges in row $j$, so no new blue phase row is reached. (c) A separate illustration of an error term reduced to the parent code for a measurement record that passes the detection tests in Theorem~\ref{thm:prx273-d4-surgery}. Blue marks indicate qubits on which a diagonal phase may depend, including correlated factors such as CZ. A logical $X$ on another row avoids that phase. Opposite sides are identified. The proof chooses such representatives in both directions and on both parent layers. This panel is not the output of (a,b).}
\label{fig:v435-square-ft}
\end{figure*}

To prove Theorem~\ref{thm:prx273-d4-surgery}, we track how faults spread through the measurement circuits. Figure~\ref{fig:v435-square-ft}(a) shows why bit flips alone are not enough: one ancilla $X$ fault produces $X_{r_W}\operatorname{CZ}_{g_S,b_E}$. Later ideal measurements can introduce further phase factors in the error expansion. In panel (b), the added blue $Z$ factors involve more qubits but remain in the same lattice row.

For vertex coordinates $(i,j)$, a column fixes $i$ and a row fixes $j$, for all three qubit colors. We will show that $f$ faulty locations affect at most $f$ coordinates in each of the row and column sets defined below. Thus $f<L$ leaves logical representatives that avoid the relevant bit-flip or phase error. On a parent toric code, logical $Z$ follows a winding lattice cycle, while its conjugate logical $X$ acts on edges crossed by a dual-lattice loop intersecting that cycle once. Choosing representatives away from an error is the idea behind the cleaning lemma~\cite{BravyiTerhal2009Cleaning}. Panel (c) illustrates this parent-code step separately from (a,b).

When $f<R$, at least one gauged round contains no new faults, though earlier errors may remain. For a measurement record satisfying the tests in Theorem~\ref{thm:prx273-d4-surgery}, the gauging and ungauging identities reduce the errors on its two sides to operators on the parent codes. Flux tests and winding cycles avoiding the flips make the surviving bit flips products of parent stars. After removing them inside the code projection, logical $X$ representatives avoiding the remaining phase exclude any logical relative phase. The projected error is therefore scalar, leaving the intended logical measurement. We now define the row and column sets used in this argument.

For each term $E=X^xD_q$ in the error expansion, $x$ specifies the bit flips and $q$ specifies the phase: $X^x|z\rangle=|z+x\rangle$ and $D_q|z\rangle=(-1)^{q(z)}|z\rangle$, with addition modulo two. For example, a CZ on qubits $u,v$ contributes $z_uz_v$ to $q$.

For color $c$, let $V_c$ list the columns containing flipped vertical edges and $H_c$ the rows containing flipped horizontal edges. For the phase already present, let $Q_c^{\mathrm h}$ list the columns of horizontal edge variables appearing in $q$, and $Q_c^{\mathrm v}$ the rows of its vertical edge variables.

A bit flip can also generate a phase on a neighboring qubit through the CZ factors in a check. We therefore include those neighbors in the phase sets. With colors ordered $r<g<b$, define
\begin{align}
 \mathcal{B}_c^{\mathrm h}(E)&=Q_c^{\mathrm h}\cup\!\bigcup_{d<c}V_d
                       \cup\!\bigcup_{d>c}(V_d-1),\nonumber\\
 \mathcal{B}_c^{\mathrm v}(E)&=Q_c^{\mathrm v}\cup\!\bigcup_{d<c}H_d
                       \cup\!\bigcup_{d>c}(H_d-1).
 \label{eq:prx273-d4-cut-sets}
\end{align}
Here $J-1=\{j-1\bmod L:j\in J\}$ shifts the listed coordinates by one lattice spacing, as required by the CZ pairs in the checks. Thus $\mathcal B_c^{\mathrm h}$ and $\mathcal B_c^{\mathrm v}$ include both the phase already present and the phase that the bit flips can generate. We bound four sets for each color: $V_c,H_c,\mathcal B_c^{\mathrm h},\mathcal B_c^{\mathrm v}$. The relevant rows and columns may differ between colors, edge orientations and terms in the error expansion.

Two facts give the bound. First, one fault, propagated through the remaining gates of its check measurement, contributes at most one row or column to each set, up to factors generated by
\begin{equation}
 \mathsf{S}_0=\langle B_{c,f},X_{t_{c,f}}\rangle.
 \label{eq:prx273-d4-trivial-subgroup}
\end{equation}
These factors act as $+1$ on the initialized state, including its gauge and completion qubits, and on the ideal joint-check projector with positive flux and pin outcomes. The proof removes them at those boundaries; it does not assume that an intermediate faulty state has positive flux.

For a star measurement, an ancilla $X$ fault spreads to the remaining star edges and creates CZ errors between other colors. The vertical star edges share one column, and the horizontal edges share one row. The CZ pairs align the additional phase contributions with the rows and columns already included in Eq.~\eqref{eq:prx273-d4-cut-sets}. Faults in the final controlled-$V$ gates can instead use a neighboring coordinate. At that point, all CNOTs to star edges and CCZs from the raw star have finished, so the fault cannot also spread through those earlier gates and add a second coordinate to the same set. This is why their order matters.

A $Z$ fault on the flux readout ancilla $a_B$ spreads to the face edges whose CNOTs have not yet occurred. Their horizontal edges share one column and their vertical edges one row. Toffoli faults act only on the readout and completion qubits. Faults involving completion qubits obey the same row/column bound. Lemma~\ref{lem:smd4-native} checks each type of gate.

Second, later ideal measurements do not enlarge these sets. Expanding their projectors brings in conjugations by the checks. For a raw star $A=A_{c,v}=X^sD_{q_{c,v}}$, with $s=s_{c,v}$,
\begin{equation}
 AEA=X^xD_{q(z+s)+q_{c,v}(z)+q_{c,v}(z+x)}.
 \label{eq:prx273-d4-cut-closure}
\end{equation}
The bit flips remain $x$. Replacing $z$ by $z+s$ introduces no new variables into $q$, and the two quadratic star phases add only the CZ neighbors already included in Eq.~\eqref{eq:prx273-d4-cut-sets}. Completed checks have the same effect on these sets up to factors in $\mathsf S_0$. For a product of errors, each set is contained in the union of the corresponding sets of the factors. Thus every term produced by $f$ faults satisfies
\begin{equation}
 |V_c|,|H_c|,|\mathcal{B}_c^{\mathrm h}|,|\mathcal{B}_c^{\mathrm v}|\le f,
 \label{eq:prx273-d4-fault-cuts}
\end{equation}
independently of the number of ideal rounds. When $f<L$, each set leaves at least one row or column available for the corresponding logical operator.

\begin{theorem}[Distance-preserving bilayer surgery]
\label{thm:prx273-d4-surgery}
For even $L\ge4$, use the circuits and gate order above, $L$ parent rounds before gauging, $R$ gauged rounds, and $L$ parent rounds after ungauging. Assume ideal encoded input and ideal final parent projection. Require positive parent, flux, pin, green-star and blue-star outcomes, a repeated red-star vector $a$, and an exact final red-link record. Under arbitrary local native faults,
\begin{equation}
 d_{\mathrm{surg}}=\min\{L,R\}.
 \label{eq:main-d4-distance}
\end{equation}
The same statement holds after adding the two CNOTs of Lemma~\ref{lem:prx273-completion-flag} around each flux extraction.
\end{theorem}
\begin{proof}
For $f<\min(L,R)$, at least one gauged round contains no faulty locations. Earlier errors may still be present. This round applies the ideal joint-check projector. As in the general proof, expand the errors before and after this projector into terms satisfying Eq.~\eqref{eq:prx273-d4-fault-cuts}.

At input, the flux projection keeps only flip patterns with even parity around every face. Since $f<L$, each color also has a horizontal and a vertical winding cycle containing no flipped edges. Their flip parities are therefore zero. Together these conditions imply $x_{c,(uv)}=\phi_c(u)+\phi_c(v)\bmod2$ for bits $\phi_c$ assigned to vertices. In other words, the flips are produced by stars at the vertices with $\phi_c=1$. On every row outside $H_c$ and every column outside $V_c$, $\phi_c$ is constant. These lines intersect, so one can choose this constant to be zero on all of them.

The gauging identity now expresses the error as an operator on the parent codes. The red flips become a product of local symmetry operators. Its CZ phases involve only the rows and columns already included in Eq.~\eqref{eq:prx273-d4-cut-sets}. At output, the gauge-readout test requires the reported red-link bits to have the form $\widehat z_{uv}=\widehat y(u)+\widehat y(v)\bmod2$. The ungauging identity then forces the outgoing red flip pattern to have the same vertex form. Applying the recorded CZ correction can add only the neighbor phases included in that equation. The final parent projection imposes the face-parity condition on the green and blue flips. Their available winding cycles again imply that these patterns are products of parent stars. Lemma~\ref{lem:smd4-transfer} gives both identities explicitly.

On either parent code, the remaining bit flips are a product of its star stabilizers and can be removed inside the code projection. The remaining error is diagonal. For each of the four parent logical $X$ operators, choose a representative on a row or column absent from the corresponding phase set. The original phase and the phases introduced by gauging and ungauging all avoid these edges, by the bounds above. Hence each chosen logical $X$ commutes with the diagonal error by disjoint support. The projected error also commutes with every logical $Z$ because it is diagonal. It therefore acts as a scalar on the parent code. This step compares equivalent operators in the proof; no additional star operation is performed.

Thus the errors on both sides of the fault-free round reduce to scalars, leaving the intended logical projector. This proves Eq.~\eqref{eq:main-instrument-distance} for each term. Summing the terms covers arbitrary coherent local faults.

The two optional CNOTs preserve this argument. A fault on the first can add only a full flux operator $B\in\mathsf S_0$ on the edge qubits; the second has no later edge interaction in that measurement. Faults in the original flux circuit reach the same edge qubits, and the added gates contribute only further readout/completion errors. Equation~\eqref{eq:prx273-d4-fault-cuts} therefore still holds, including for multiple faults and intermediate negative-flux states.

Finally, $L$ faults can place $Z$ on a length-$L$ noncontractible loop of matter edges after their last gates. The resulting logical $Z$ passes all tests and acts nontrivially within the measured logical eigenspace. Alternatively, $R$ faults can reverse the same red-star readout in every gauged round. The reported vector still repeats, but its product gives the wrong logical sign. These examples give the matching upper bounds for both circuits.
\end{proof}

Theorem~\ref{thm:smd4-memory} also gives $d(P')=d_{\mathrm{mem}}=L$ for the full 22-dimensional code, where $P'$ is its code projector and $d_{\mathrm{mem}}$ is the memory circuit fault distance defined there. The surgery uses joint-check eigenspaces with both red winding eigenvalues $+1$ and outcome-dependent completed red-star signs. For records passing its detection tests, its fault-distance guarantee protects the reported logical sign and the state within the corresponding logical eigenspace.

\subsubsection{Resource cost and completion-qubit flags}
The distance result fixes the required number of gauged rounds, $R\ge L$, to attain distance $L$. We now count the qubits, gates and extraction rounds used by that circuit. The $3L^2$ stars use one readout each, and the $3L^2$ joint flux--pin measurement circuits use two. These readouts reset each round, while completion qubits are kept throughout the gauged interval. This gives $9L^2$ allocated readout qubits, in addition to the $9L^2$ matter, gauge and completion qubits.

Both versions in Table~\ref{tab:prx277-d4-resources} use completion qubits. The original circuit is shown in Fig.~\ref{fig:smd4-joint-circuit}(b). The second version adds $\operatorname{CNOT}(t\to a_B)$ immediately before and after the four CNOTs that record the flux. These two gates let the existing completion qubit and pin readout detect certain errors spreading from $a_B$. They add $6L^2$ CNOTs per gauged round and no qubits or readouts. The prescribed initialization gives $B=+1$ and $|+\rangle_t$, hence $B=X_t=+1$. Ideal check measurements preserve these eigenvalues. Lemma~\ref{lem:prx273-completion-flag} proves the single-fault flag guarantee from this initialized subspace. The full distance proof also covers earlier faults, repeated faults and intermediate negative-flux states. Both versions have the same exact distance $\min\{L,R\}$.

\begin{table}[!ht]
\caption{Resources for the bilayer Toffoli source on an even-$L$ torus. All entries are in units of $L^2$; gates are counted per gauged round. Both versions use completion qubits and the same completed checks. The second adds two flag CNOTs per joint flux--pin measurement circuit. Theorem~\ref{thm:prx273-d4-surgery} gives the same exact distance for both.}
\label{tab:prx277-d4-resources}
\centering
\begin{tabular}{@{}lrr@{}}
\toprule
Resource & \shortstack{Original\\circuit} & \shortstack{With flag\\CNOTs}\\
\midrule
Matter qubits & 4 & 4\\
Gauge qubits & 2 & 2\\
Completion qubits & 3 & 3\\
Readout qubits & 9 & 9\\
Total allocated qubits & 18 & 18\\
\midrule
CNOTs per gauged round & 27 & 33\\
CCZs per gauged round & 18 & 18\\
Toffolis per gauged round & 3 & 3\\
\bottomrule
\end{tabular}
\end{table}

Each gauged round uses $9L^2$ native $X$- or $Z$-basis preparations and $9L^2$ readouts. Counts exclude decomposed basis changes and idles. Reserving all qubits for $2L+R$ extraction rounds gives $18L^2(2L+R)$ qubit-rounds, or $54L^3$ at distance $L$ with $R=L$. The counts also exclude encoded-input preparation and resource consumption.

\subsection{Recovery using the completion qubits}
\label{sec:prx-d4-recovery}
Completion qubits make the check measurements compatible and retain quantum information for coherent recovery before ungauging. We show explicitly how they repair a green-$X$ error at the input to gauging, with all subsequent operations ideal. The repair restores the intended logical measurement on the original input while preserving superpositions within each outcome.

\subsubsection{A detected error at the gauging interface}
Let $|\psi\rangle$ be an arbitrary parent-code state, and let $J$ initialize the red gauge qubits in $|0\rangle$ and the completion qubits in $|+\rangle$. Suppose $D=X_{g,e}$ occurs after the last parent round and before gauged extraction. This is the temporal boundary where the measured checks switch from the parent code to the gauged code. The two faces next to $e$ have negative green flux. Their pins are therefore identity, leaving the completion qubits $t_{g,v}$ and $t_{g,w}$ free to store quantum information. Here $v=v_e$ labels the symmetry block containing $e$, and $b(e)$ is the blue qubit paired with $g_e$ by a CZ in Eq.~\eqref{eq:main-d4-physical}.

To see what is stored, write $U_v$ for the two CZ factors indexed by vertex $v$ in Eq.~\eqref{eq:main-d4-physical}. Let $V$ be the vertex set, $y\in\{0,1\}^V$, $U(y)=\prod_fU_f^{y_f}$ and $a^y=\prod_fa_f^{y_f}$, where $a_f=\pm1$ are the completed red-star outcomes. Their joint projector is
\begin{equation}
 G_a=\prod_{f\in V}\frac{I+a_f\widehat A_{r,f}}2.
 \label{eq:prx412-recovery-gauss}
\end{equation}
The green and blue completed stars initially remain positive. Expanding the red projectors gives
\begin{equation}
 \begin{split}
 G_aDJ|\psi\rangle={}&2^{-|V|}\sum_y a^y
 U(y)D|\psi\rangle|\delta y\rangle_r\,|\tau_y\rangle_t,\\
 |\tau_y\rangle_t={}&|y_v\rangle_{X,t_{g,v}}|y_w\rangle_{X,t_{g,w}},
 \end{split}
 \label{eq:prx412-recovery-state}
\end{equation}
where $(\delta y)_{fh}=y_f\oplus y_h$ is the red gauge bit on the link between neighboring vertices $f,h$. We omit the other completion qubits, which remain in $|+\rangle$, and use $|0\rangle_X=|+\rangle$, $|1\rangle_X=|-\rangle$. The variables $y_f$ label terms in a coherent sum, not recorded measurement outcomes. The completion qubits retain these variables without measuring them.

The error and the local symmetry satisfy $U(y)D=D Z_{b(e)}^{y_v}U(y)$. Removing $D$ alone would therefore leave a blue-$Z$ error controlled by $y_v$. Instead, apply the local Clifford repair
\begin{equation}
 R_e=X_{g,e}\operatorname{CNOT}_{b(e)\to t_{g,v}}.
 \label{eq:prx412-recovery-repair}
\end{equation}
The CNOT acts first. Since its target is $|y_v\rangle_X$, it applies $Z_{b(e)}^{y_v}$ to its blue control and cancels the conditional error without reading $y_v$. The following $X_{g,e}$ removes the green error. Thus, the completion qubit supplies the coherent control needed to recover the matter state.

Now measure both active completion qubits in $Z$, with signs $q_v,q_w$, and reset them to $|+\rangle$. Denote this measurement-and-reset operator by $M_q$. With $P_{\rm par}$ the parent-code projector, the exact recovery identity is
\begin{equation}
 \begin{gathered}
 M_qR_eG_aDJP_{\rm par}=\frac12G_{a'}JP_{\rm par},\\
 a'_f=\begin{cases}a_fq_f,&f\in\{v,w\},\\a_f,&f\notin\{v,w\}.
 \end{cases}
 \end{gathered}
 \label{eq:prx412-recovery-identity}
\end{equation}
The $Z$ readout does not reveal the stored $X$-basis bit $y_f$. Its phase $q_f^{y_f}$ is absorbed into the updated Gauss sign $a'_f$. The right-hand side is ideal gauging of the original encoded input. Ordinary ungauging therefore implements the logical projector in Eq.~\eqref{eq:prx273-resource-projector}, with $(-1)^\eta=\prod_fa'_f$. The negative-flux branches are recovered without rejection. Alternatively, after reset, the next ideal completed-check round measures $a'$ directly, so the logical answer need not rely on the completion-readout bits.

The stored information is quantum. Tracing out the active completion qubits in Eq.~\eqref{eq:prx412-recovery-state} removes coherence between terms with different stored $y_v,y_w$. Keeping them until the repair avoids this loss. Appendix~\ref{sec:sm297-coherent-recovery} extends the repair to every green-$X$ boundary pattern of weight below $L/2$, assuming all subsequent operations are ideal.

\subsection{A \texorpdfstring{$D_4$}{D4} route to a logical \texorpdfstring{$T$}{T} state}
\label{sec:prx278-t-source}

\begin{figure*}[t]
\centering
\begin{minipage}[t]{.39\textwidth}
\centering\textbf{(a) Three boundary segments}\par\smallskip
\begin{tikzpicture}[x=.82cm,y=.82cm,font=\small,>=Latex,
 completion/.style={diamond,aspect=1.05,fill=white,line width=.55pt,
 minimum size=2.3mm,inner sep=0pt}]
\path[use as bounding box](-1.10,-1.33)rectangle(5.35,5.10);
\draw[step=1,black!35,line width=.4pt](0,0)grid(4,4);
\foreach \i in {0,1,2,3}{
 \foreach \j in {0,1,2,3,4}{
  \fill[dFourRed]({\i+.28},\j)circle(1.15pt);
  \fill[dFourGreen]({\i+.50},\j)circle(1.15pt);
  \fill[dFourBlue]({\i+.72},\j)circle(1.15pt);
  \fill[dFourRed](\j,{\i+.28})circle(1.15pt);
  \fill[dFourGreen](\j,{\i+.50})circle(1.15pt);
  \fill[dFourBlue](\j,{\i+.72})circle(1.15pt);
 }
}
\draw[dFourRed,line width=1.8pt](-.17,0)--(-.17,4);
\draw[dFourBlue,line width=1.8pt](0,4.17)--(4,4.17);
\draw[violet!85!black,line width=1.8pt](0,-.17)--(4.17,-.17)--(4.17,4);
\foreach \i in {1,2,3}{
 \foreach \j in {1,2,3}{
  \node[completion,draw=dFourBlue]at({\i+.72},{\j+.30}){};
 }
 \node[completion,draw=dFourBlue]at({\i+.72},4.30){};
}
\foreach \j in {1,2,3}{
 \node[completion,draw=dFourRed]at(-.50,{\j-.27}){};
 \foreach \i in {1,2,3}{
  \node[completion,draw=dFourRed]at({\i-.50},{\j-.27}){};
 }
}
\foreach \i in {1,2,3}{\foreach \j in {0,1,2}{
 \node[completion,draw=dFourGreen]at({\i+.28},{\j+.30}){};
}}
\foreach \i in {1,2,3}{
 \node[completion,draw=dFourGreen]at({\i+.28},-.70){};
}
\foreach \j in {0,1,2}{
 \node[completion,draw=dFourGreen]at(4.28,{\j+.30}){};
}
\foreach \x/\y in {0/0,0/4,4/4}{
 \fill[black](\x,\y)circle(2.3pt);
}
\node[text=dFourRed,rotate=90]at(-.84,2){$\mathcal L_r:\ Z_r$};
\node[text=dFourBlue]at(2,4.86){$\mathcal L_b:\ Z_b$};
\node[text=violet!85!black]at(2,-1.12){$\mathcal L_{rb}:\ Z_rZ_b,\ Z_g$};
\node[text=violet!85!black,rotate=90]at(4.86,2){same $\mathcal L_{rb}$};
\end{tikzpicture}\par
{\footnotesize Dots: edge qubits. Diamonds: completion qubits.\par
\textcolor{dFourBlue}{$\alpha_v$},\quad \textcolor{dFourRed}{$\beta_v$},\quad \textcolor{dFourGreen}{$\gamma_v$} (faces),\quad \textcolor{dFourGreen}{$u_e$} (boundary).}
\end{minipage}\hfill
\begin{minipage}[t]{.28\textwidth}
\centering\textbf{(b) Logical gate}\par\smallskip
\begin{tikzpicture}[x=1.15cm,y=1.15cm,font=\small,>=Latex]
\path[use as bounding box](-.28,-.70)rectangle(2.75,3.27);
\draw[black!50](0,0)rectangle(2,2);
\foreach \x/\y/\n/\pos in {0/0/0/below left,2/0/1/below right,0/2/2/above left,2/2/3/above right}{
 \fill[black!65](\x,\y)circle(1pt);
 \node[\pos]at(\x,\y){$\n$};
}
\draw[line width=.85pt](.55,0)..controls(1.30,.10)and(1.90,.70)..(2,1.45);
\draw[line width=.85pt](0,.55)..controls(.10,1.30)and(.70,1.90)..(1.45,2);
\fill[dFourRed](.55,0)circle(2pt);
\fill[dFourBlue](2,1.45)circle(2pt);
\fill[dFourRed](0,.55)circle(2pt);
\fill[dFourBlue](1.45,2)circle(2pt);
\node[fill=white,inner sep=1pt]at(1.60,.52){$\mathrm{CS}$};
\node[fill=white,inner sep=1pt]at(.52,1.48){$\mathrm{CS}^{\dagger}$};
\node[align=center]at(1,2.85){$V:\ \mathrm{CNOT}(r\!\to\!g)$\\[-1pt]
 $\phantom{V:\ }\mathrm{CNOT}(b\!\to\!g)$};
\node[align=center]at(1,-.60){$T_r^{\dagger}$ on $\mathcal L_{rb}$};
\end{tikzpicture}\par
{\footnotesize $U=WV$ acts as $\overline{T}^{\dagger}$.}
\end{minipage}\hfill
\begin{minipage}[t]{.29\textwidth}
\centering\textbf{(c) State preparation}\par\smallskip
\begin{tikzpicture}[font=\small,>=Latex,node distance=.83cm,
 box/.style={draw=black!55,rounded corners=2pt,fill=black!2,
 text width=3.65cm,align=center,minimum height=.85cm,inner sep=4pt}]
\node[box](parent){Folded surface code\\$|+\rangle_L$};
\node[box,below=of parent](gauged){Completed $D(D_4)$ code\\apply $U^{\dagger}$};
\node[box,below=of gauged](output){Folded surface code\\$|T\rangle_L$};
\draw[->,line width=.8pt](parent)--node[right,font=\scriptsize]{gauge}(gauged);
\draw[->,line width=.8pt](gauged)--node[right,font=\scriptsize]{ungauge}(output);
\end{tikzpicture}\par\smallskip
{\footnotesize Correct the measured branch at each interface.}
\end{minipage}
\caption{A logical $T$ source on the square lattice. (a) Boundary conditions of Ref.~\cite{KobayashiZhuHsin2026}. The boundary labels specify $+1$ checks on each edge $e$ of the indicated segment: $Z_{r,e}$ on the left, $Z_{b,e}$ on top, and both $Z_{r,e}Z_{b,e}$ and $Z_{g,e}$ on the bottom and right. Here $Z_{c,e}$ acts on the color-$c$ qubit of that edge, and the labels suppress $e$. These local boundary checks are imposed together with the face fluxes $B_{c,f}=\prod_{e\in\partial f}Z_{c,e}$ for all three colors on every face. Diamonds show all completion qubits for $L=4$, using the color positions of Fig.~\ref{fig:d4-completed-checks}(a) and the boundary families in Appendix~\ref{sec:v386-boundary-catalogue}. The lower-right bend does not change the boundary type. The star generators are specified in Table~\ref{tab:v385-boundaries}. The completed boundary generators and pinning checks are drawn in Fig.~\ref{fig:v386-boundary-stars}. (b) The two controlled-phase factors of $W$ on a face with vertices $0,1,2,3$, the $T^\dagger$ gate on each red qubit along $\mathcal L_{rb}$, and the edgewise CNOTs of $V$. (c) The completed version of the code-switching protocol uses the inverse gate to prepare the standard $T$ state. Appendix~\ref{sec:v385-three-boundary} gives the boundary completion and the ideal switching maps.}
\label{fig:v385-square-t}
\end{figure*}
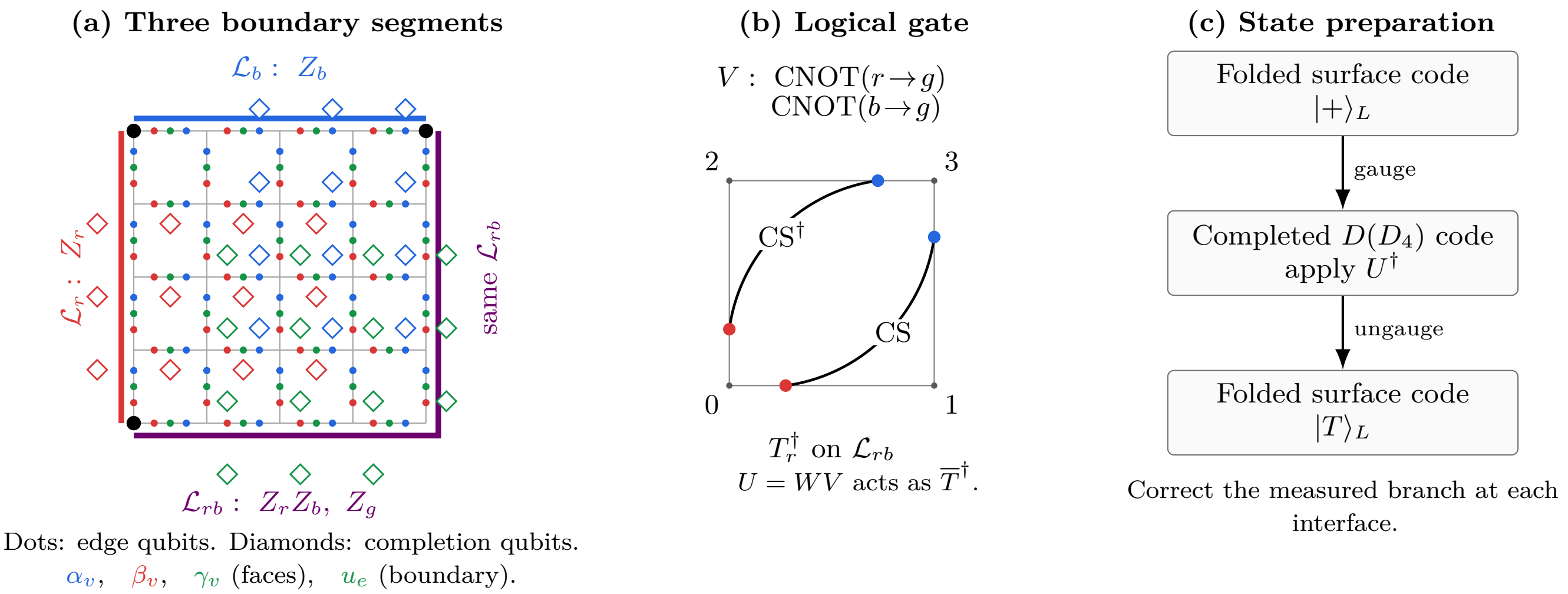

We prepare a logical $T$ state by changing from a one-qubit surface-code parent to a completed $D(D_4)$ patch, applying a local circuit there, and returning to the parent code. We use the three-boundary construction of Ref.~\cite{KobayashiZhuHsin2026}. The left side has boundary type $\mathcal L_r$, the top has $\mathcal L_b$, and the bottom and right together form $\mathcal L_{rb}$. Thus, ``triangular'' refers to the three boundary segments and their junctions; the underlying lattice remains square. Figure~\ref{fig:v385-square-t} shows the boundaries and the preparation sequence.

The boundary stars must preserve the Pauli boundary checks. Each color star flips its own color on the incident edges, while its CZ factors commute with every $Z$ check. Thus $Z_r$ on $\mathcal L_r$ excludes a red star, and $Z_b$ on $\mathcal L_b$ excludes a blue star. On each edge of $\mathcal L_{rb}$, we impose $Z_g$ and $Z_rZ_b$ separately. The first forbids a green flip. The second allows red and blue to flip together, but forbids either alone. We therefore retain $M_v=A_{r,v}A_{b,v}$ there, giving the generators in Table~\ref{tab:v385-boundaries}. At a junction impose both adjoining conditions. This leaves $A_g$ at the upper-left corner and no star at the lower-left and upper-right corners. The lower-right bend retains $A_rA_b$ because the boundary type does not change there.

To obtain the complete raw operator, take the indicated bulk star or same-vertex product, retain its $X$ factors on existing edges, and omit a CZ factor only when one of its edge qubits is absent. All face fluxes are retained. The factors $A_{r,v}$ and $A_{b,v}$ commute at the same vertex on the full Hilbert space, so the retained raw generators are Hermitian Clifford involutions. Figure~\ref{fig:v386-boundary-stars} shows representatives of every boundary and corner star type on one square-lattice layout, together with the central checks and completion pins. Appendix~\ref{sec:v386-boundary-catalogue} gives the full family and endpoint exceptions.

\begin{table}[tb]
\caption{Boundary checks for the square-lattice $T$ source. Stars are truncated to existing edges. The bulk retains all three colors, and every face retains its three flux checks.}
\label{tab:v385-boundaries}
\centering\renewcommand{\arraystretch}{1.15}
\begin{tabular}{@{}lll@{}}
\toprule
Segment & Pauli boundary checks & Retained stars\\
\midrule
$\mathcal L_r$ & $Z_r$ & $A_g,\ A_b$\\
$\mathcal L_b$ & $Z_b$ & $A_r,\ A_g$\\
$\mathcal L_{rb}$ & $Z_rZ_b,\ Z_g$ & $A_rA_b$\\
\bottomrule
\end{tabular}
\end{table}

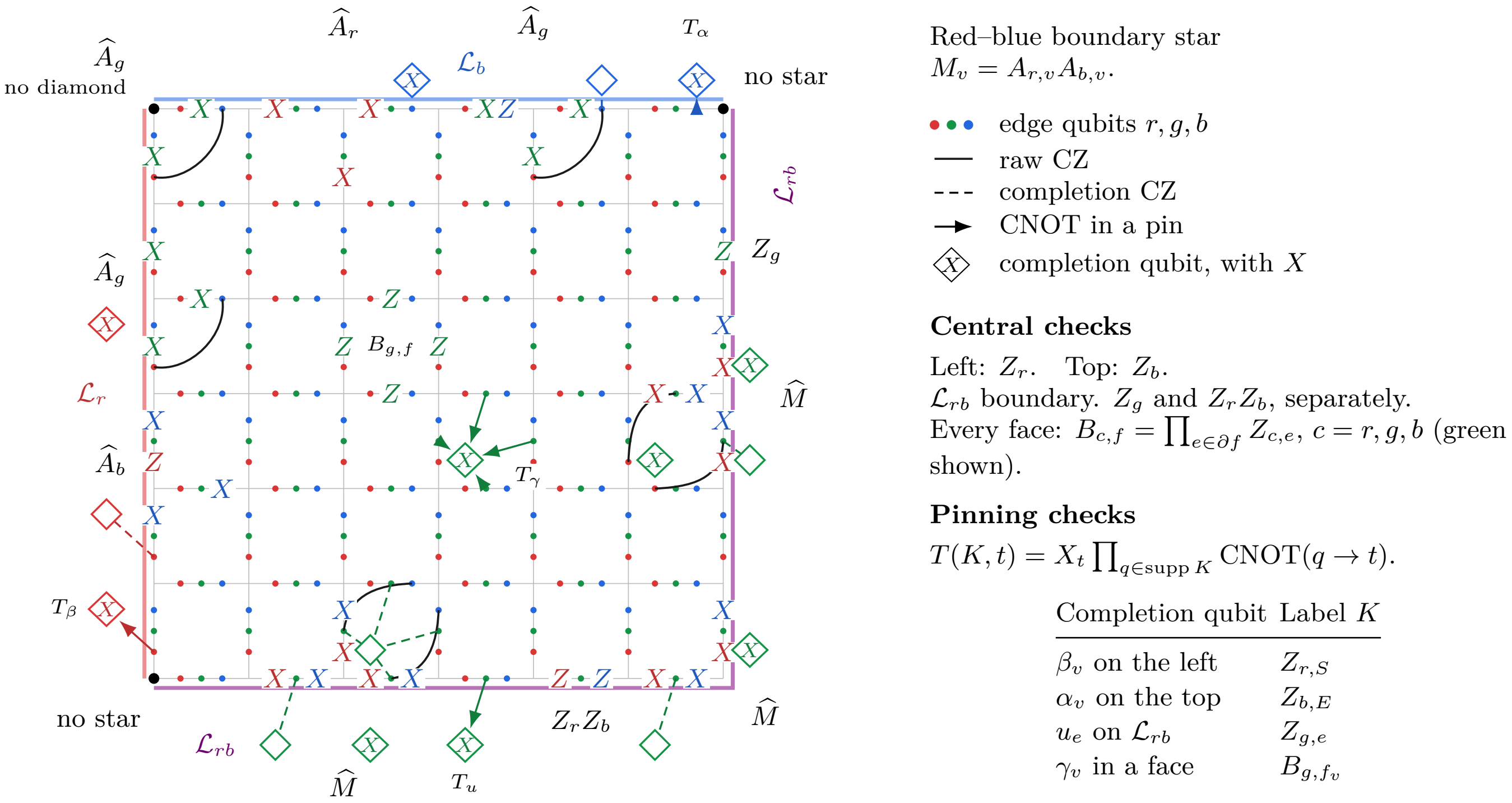
\begin{figure*}[t]
\centering
\begingroup
\newcommand{\bndX}[3]{\node[fill=white,inner sep=.3pt,font=\small,text=#1!85!black] at (#2,#3) {$X$};}
\newcommand{\bndZ}[3]{\node[fill=white,inner sep=.3pt,font=\small,text=#1!85!black] at (#2,#3) {$Z$};}
\newcommand{\bndT}[4]{\node[diamond,aspect=1.05,draw=#1,fill=white,line width=.65pt,minimum size=3.6mm,inner sep=0pt,font=\scriptsize,text=#1!85!black] at (#2,#3) {#4};}
\begin{minipage}[c]{.59\textwidth}\centering
\begin{tikzpicture}[x=1.12cm,y=1.12cm,font=\small,>=Latex,
 raw/.style={black!90,line width=.7pt},
 comp/.style={densely dashed,line width=.65pt},
 pin/.style={->,line width=.65pt,shorten >=2.0mm},
 lab/.style={fill=white,inner sep=1pt,font=\small}]
\path[use as bounding box](-1.15,-1.35)rectangle(7.1,7.1);
\draw[step=1,black!25,line width=.32pt](0,0)grid(6,6);
\draw[dFourRed!55,line width=1.3pt](-.10,0)--(-.10,6);
\draw[dFourBlue!55,line width=1.3pt](0,6.10)--(6,6.10);
\draw[violet!55,line width=1.3pt](0,-.10)--(6.10,-.10)--(6.10,6);
\foreach \i in {0,...,5}{\foreach \j in {0,...,6}{
 \fill[dFourRed]({\i+.28},\j)circle(1.05pt);
 \fill[dFourGreen]({\i+.50},\j)circle(1.05pt);
 \fill[dFourBlue]({\i+.72},\j)circle(1.05pt);
 \fill[dFourRed](\j,{\i+.28})circle(1.05pt);
 \fill[dFourGreen](\j,{\i+.50})circle(1.05pt);
 \fill[dFourBlue](\j,{\i+.72})circle(1.05pt);
}}
\draw[raw](0,3.28)..controls(.38,3.22)and(.78,3.62)..(.72,4);
\foreach \x/\y in {0/4.5,.5/4,0/3.5}{\bndX{dFourGreen}{\x}{\y}}
\bndT{dFourRed}{-.50}{3.73}{$X$}
\node[lab,anchor=east]at(-.25,4.30){$\widehat A_g$};
\foreach \x/\y in {0/2.72,.72/2,0/1.72}{\bndX{dFourBlue}{\x}{\y}}
\draw[comp,dFourRed!85!black](0,1.28)--(-.50,1.73);
\bndT{dFourRed}{-.50}{1.73}{}
\node[lab,anchor=east]at(-.25,2.30){$\widehat A_b$};
\foreach \x/\y in {1.28/6,2.28/6,2/5.28}{\bndX{dFourRed}{\x}{\y}}
\bndT{dFourBlue}{2.72}{6.30}{$X$}
\node[lab]at(2,6.90){$\widehat A_r$};
\draw[raw](4,5.28)..controls(4.38,5.22)and(4.78,5.62)..(4.72,6);
\foreach \x/\y in {3.5/6,4.5/6,4/5.5}{\bndX{dFourGreen}{\x}{\y}}
\draw[comp,dFourBlue!85!black](4.72,6)--(4.72,6.30);
\bndT{dFourBlue}{4.72}{6.30}{}
\node[lab]at(4,6.90){$\widehat A_g$};
\draw[raw](2.5,0)..controls(2.76,.04)and(2.98,.18)..(3,.72);
\draw[raw](2,.5)..controls(2.04,.76)and(2.18,.98)..(2.72,1);
\foreach \x/\y in {1.28/0,2.28/0,2/.28}{\bndX{dFourRed}{\x}{\y}}
\foreach \x/\y in {1.72/0,2.72/0,2/.72}{\bndX{dFourBlue}{\x}{\y}}
\draw[comp,dFourGreen!85!black](1.5,0)--(1.28,-.70);
\bndT{dFourGreen}{1.28}{-.70}{}
\bndT{dFourGreen}{2.28}{-.70}{$X$}
\foreach \x/\y in {2.5/0,3/.5,2.5/1,2/.5}{
 \draw[comp,dFourGreen!85!black](\x,\y)--(2.28,.30);
}
\bndT{dFourGreen}{2.28}{.30}{}
\node[lab]at(2,-1.10){$\widehat M$};
\draw[raw](5.28,2)..controls(5.78,2.02)and(5.98,2.16)..(6,2.5);
\draw[raw](5,2.28)..controls(5.02,2.78)and(5.16,2.98)..(5.5,3);
\foreach \x/\y in {5.28/3,6/3.28,6/2.28}{\bndX{dFourRed}{\x}{\y}}
\foreach \x/\y in {5.72/3,6/3.72,6/2.72}{\bndX{dFourBlue}{\x}{\y}}
\bndT{dFourGreen}{5.28}{2.30}{$X$}
\draw[comp,dFourGreen!85!black](6,2.5)--(6.28,2.30);
\bndT{dFourGreen}{6.28}{2.30}{}
\bndT{dFourGreen}{6.28}{3.30}{$X$}
\node[lab,anchor=west]at(6.55,3){$\widehat M$};
\draw[raw](0,5.28)..controls(.38,5.22)and(.78,5.62)..(.72,6);
\bndX{dFourGreen}{.5}{6}\bndX{dFourGreen}{0}{5.5}
\node[lab,anchor=east,align=right]at(-.25,6.45){$\widehat A_g$\\[-1pt]\scriptsize no diamond};
\foreach \x/\y in {5.28/0,6/.28}{\bndX{dFourRed}{\x}{\y}}
\foreach \x/\y in {5.72/0,6/.72}{\bndX{dFourBlue}{\x}{\y}}
\draw[comp,dFourGreen!85!black](5.5,0)--(5.28,-.70);
\bndT{dFourGreen}{5.28}{-.70}{}
\bndT{dFourGreen}{6.28}{.30}{$X$}
\node[lab,anchor=west]at(6.25,-.35){$\widehat M$};
\foreach \x/\y in {0/0,0/6,6/6}{\fill[black](\x,\y)circle(1.8pt);}
\node[lab,anchor=north east]at(-.10,-.30){no star};
\node[lab,anchor=south west]at(6.18,6.22){no star};
\bndZ{dFourRed}{0}{2.28}
\bndZ{dFourBlue}{3.72}{6}
\bndZ{dFourGreen}{6}{4.5}
\bndZ{dFourRed}{4.28}{0}\bndZ{dFourBlue}{4.72}{0}
\node[lab,anchor=west]at(6.25,4.5){$Z_g$};
\node[lab]at(4.5,-.45){$Z_rZ_b$};
\draw[pin,dFourBlue!85!black](5.72,6)--(5.72,6.30);
\bndT{dFourBlue}{5.72}{6.30}{$X$}
\node[lab,font=\scriptsize]at(5.72,6.85){$T_\alpha$};
\draw[pin,dFourRed!85!black](0,.28)--(-.50,.73);
\bndT{dFourRed}{-.50}{.73}{$X$}
\node[lab,anchor=east,font=\scriptsize]at(-.75,.73){$T_\beta$};
\draw[pin,dFourGreen!85!black](3.5,0)--(3.28,-.70);
\bndT{dFourGreen}{3.28}{-.70}{$X$}
\node[lab,font=\scriptsize]at(3.28,-1.10){$T_u$};
\foreach \x/\y in {2.5/3,3/3.5,2.5/4,2/3.5}{\bndZ{dFourGreen}{\x}{\y}}
\node[lab,font=\scriptsize]at(2.5,3.5){$B_{g,f}$};
\foreach \x/\y in {3.5/2,4/2.5,3.5/3,3/2.5}{
 \draw[pin,dFourGreen!85!black](\x,\y)--(3.28,2.30);
}
\bndT{dFourGreen}{3.28}{2.30}{$X$}
\node[lab,font=\scriptsize]at(3.95,2.12){$T_\gamma$};
\node[text=dFourRed!85!black,lab]at(-.65,3.0){$\mathcal L_r$};
\node[text=dFourBlue!85!black,lab]at(3.35,6.48){$\mathcal L_b$};
\node[text=violet!85!black,lab,rotate=90]at(6.65,5.2){$\mathcal L_{rb}$};
\node[text=violet!85!black,lab]at(.65,-.70){$\mathcal L_{rb}$};
\end{tikzpicture}
\end{minipage}\hfill
\begin{minipage}[c]{.38\textwidth}
\small\raggedright
Red--blue boundary star\par
$M_v=A_{r,v}A_{b,v}$.
\par\medskip
\begin{tikzpicture}[x=1cm,y=1cm,font=\small,>=Latex]
\fill[dFourRed](0,1.45)circle(1.6pt);
\fill[dFourGreen](.2,1.45)circle(1.6pt);
\fill[dFourBlue](.4,1.45)circle(1.6pt);
\node[anchor=west]at(.65,1.45){edge qubits $r,g,b$};
\draw[black!90,line width=.7pt](0,1.05)--(.45,1.05);
\node[anchor=west]at(.65,1.05){raw CZ};
\draw[densely dashed,line width=.65pt](0,.65)--(.45,.65);
\node[anchor=west]at(.65,.65){completion CZ};
\draw[->,line width=.65pt](0,.25)--(.45,.25);
\node[anchor=west]at(.65,.25){CNOT in a pin};
\node[diamond,aspect=1.05,draw,minimum size=3.6mm,inner sep=0pt,font=\scriptsize]at(.2,-.2){$X$};
\node[anchor=west]at(.65,-.2){completion qubit, with $X$};
\end{tikzpicture}
\par\medskip
\textbf{Central checks}\par\smallskip
Left: $Z_r$.\quad Top: $Z_b$.\par
$\mathcal L_{rb}$ boundary. $Z_g$ and $Z_rZ_b$, separately.\par
Every face: $B_{c,f}=\prod_{e\in\partial f}Z_{c,e}$,
$c=r,g,b$ (green shown).
\par\medskip
\textbf{Pinning checks}\par\smallskip
$T(K,t)=X_t\prod_{q\in\operatorname{supp}K}
 \operatorname{CNOT}(q\to t).$
\par\medskip
\centering
\renewcommand{\arraystretch}{1.1}
\begin{tabular}{@{}ll@{}}
Completion qubit & Label $K$\\
\midrule
$\beta_v$ on the left & $Z_{r,S}$\\
$\alpha_v$ on the top & $Z_{b,E}$\\
$u_e$ on $\mathcal L_{rb}$ & $Z_{g,e}$\\
$\gamma_v$ in a face & $B_{g,f_v}$
\end{tabular}

\end{minipage}
\endgroup
\caption{Completed boundary checks on one square-lattice layout. The $\widehat M$ are completed red--blue boundary stars on $\mathcal L_{rb}$, the bottom and right sides. Representative translates of every boundary and corner star type are drawn with all their factors, together with boundary constraints, a face flux and pin examples. Diamonds are completion qubits. Blank diamonds participate through the incident dashed CZs. The pin arrows belong to separate checks from the displayed stars. The empty junctions retain their edge and face constraints. Near a junction, omit a completion factor when its conflicting partner is absent, as specified in Appendix~\ref{sec:v386-boundary-catalogue}. That appendix gives the full translated family and all index ranges.}
\label{fig:v386-boundary-stars}
\label{fig:v386-boundary-central}
\end{figure*}

Completion is then derived from the raw commutators. For example, at a left-side vertex the raw pair is
\begin{align}
 A_{g,v}&=X_{g,N}X_{g,E}X_{g,S}
          \operatorname{CZ}(r_S,b_E),\nonumber\\
 A_{b,v}&=X_{b,N}X_{b,E}X_{b,S},\nonumber\\
 A_{g,v}A_{b,v}&=Z_{r,S}A_{b,v}A_{g,v}.
 \label{eq:v386-left-boundary-conflict}
\end{align}
Here $N,E,S$ are the existing edges incident to $v$. Add one completion qubit $\beta_v$. The green star receives $X_{\beta_v}$, the blue star receives $\operatorname{CZ}(r_S,\beta_v)$, and the pin is $X_{\beta_v}\operatorname{CNOT}(r_S\!\to\!\beta_v)$. The added factors contribute the same commutator $Z_{r,S}$, canceling the raw sign for either eigenvalue of the boundary check. The other conflicts are treated in the same way. Their labels are products of retained fluxes and Pauli boundary checks and commute with every retained star. Appendix~\ref{sec:v385-three-boundary} gives the complete conflict list and derives the factors in the figures. If $P_{\triangle}$ denotes the raw projector, including these boundary constraints, its completed projector is
\begin{equation}
 \widehat P_{\triangle}=P_{\triangle}\otimes
 (|+\rangle\langle+|)^{\otimes c_L}.
 \label{eq:v385-triangle-projector-main}
\end{equation}
Here $c_L=3L^2-2L-1$ for an $L\times L$ patch. Appendix~\ref{sec:v385-three-boundary} proves the boundary algebra and the one-qubit dimension. Completion preserves this logical qubit, the code distance and its gapped boundary conditions. The $D_4$ surface-code patch with one boundary type, used to display individual completed edge and corner checks, is given in Appendix~\ref{sec:sm279-d4-boundary}.

For a face $f=(0123)$, $c_{ab}$ denotes the color-$c$ qubit on the edge joining vertices $a$ and $b$. The gate $\operatorname{CS}$ is controlled $S=\operatorname{diag}(1,i)$. In the orientation of Figure~\ref{fig:v385-square-t}(b), the boundary-preserving gate of Ref.~\cite{KobayashiZhuHsin2026} is
\begin{align}
 V&=\prod_e\operatorname{CNOT}(r_e\!\to\!g_e)
                   \operatorname{CNOT}(b_e\!\to\!g_e),
 \label{eq:v385-square-t-V}\\
 W&=\left[\prod_{f=(0123)}
       \operatorname{CS}(r_{01},b_{13})
       \operatorname{CS}^{\dagger}(r_{02},b_{23})\right]\nonumber\\
 &\quad\times\prod_{e\in\mathcal L_{rb}}T_{r_e}^{\dagger},
 \label{eq:v385-square-t-W}\\
 U&=WV,\qquad
 U|m\rangle_{\triangle}=e^{-i\pi m/4}|m\rangle_{\triangle}.
 \label{eq:v385-square-t-action}
\end{align}
Here $|m\rangle_{\triangle}$, $m\in\{0,1\}$, is the encoded basis of the three-boundary patch, with eigenvalue $(-1)^m$ under $\overline{Z}=\prod_{e\in\mathcal L_{rb}}Z_{r,e}$. All displayed gates have bounded support and admit a constant-depth schedule. By Eq.~\eqref{eq:v385-triangle-projector-main}, $U\otimes I_t$ preserves the completed code and has the same logical action. No gate on a completion qubit is needed for this ideal logical operation.

For state preparation, use the red and blue surface-code layers joined along $\mathcal L_{rb}$, with the parent checks specified in Appendix~\ref{sec:v385-square-switch}. Prepare their encoded qubit in $|+\rangle_L$, add green qubits in $|0\rangle$, and initialize the completion qubits in $|+\rangle$. Measure the completed checks and apply green $Z$ corrections to set the green-star outcomes to $+1$. This changes the code without measuring the encoded qubit. Apply $U^{\dagger}\otimes I_t$, then measure the green qubits in $Z$. Their outcomes specify a product of red--blue CZ gates to undo. Appendix~\ref{sec:v385-square-switch} proves the preparation map
\begin{equation}
 |+\rangle_L\longmapsto
 |T\rangle_L=\frac{|0\rangle_L+e^{i\pi/4}|1\rangle_L}{\sqrt2}
 \label{eq:v385-square-t-output}
\end{equation}
for every corrected ideal record. Green qubits are the gauge register here. The preceding torus protocol gauges red. The non-Clifford operations enter through both check extraction and the explicit gate in Eqs.~\eqref{eq:v385-square-t-V}--\eqref{eq:v385-square-t-W}.

For the noisy $T$-preparation circuit, Appendix~\ref{sec:v445-three-boundary-distance} proves $d_{\rm circ}\ge \min\{R_{\rm pre},R_{\rm post},\lceil L/\kappa\rceil\}$. Here $R_{\rm pre}$ and $R_{\rm post}$ count repeated check rounds before and after the inserted gate, and $\kappa$ is a constant fixed by the local circuits. The bound includes faults in both code switches, the gate and the physical corrections, with ideal encoded input, final parent projection and the stated tests on the measurement record. Choosing both repeated intervals proportional to $L$ gives a linear lower bound. Completion separately preserves the raw patch's static code distance. Appendix~\ref{sec:v278-t-details} gives an alternative color-code protocol with a linear circuit-distance bound.

\section{Completion beyond central commutators}
\label{sec:prx-semidirect}
The construction in Section~\ref{sec:prx-central} requires the commutator of each conflicting pair to commute with every raw check. This condition lets us cancel each conflict with an added qubit. More general code symmetries can change the commutators themselves. We now extend completion to such cases. For example, a dihedral reflection exchanges a rotation check $R$ with its inverse $R^{-1}$. We add a qubit and replace $R$ by a joint check that acts as $R$ on the qubit's $|0\rangle$ component and $R^{-1}$ on its $|1\rangle$ component. The completed reflection check also flips this qubit. The two exchanges cancel, leaving the joint check unchanged. These two joint checks therefore commute.

Theorem~\ref{thm:prx-semidirect} applies whenever the raw checks represent $G=A\rtimes_\phi H$ with finite abelian $A$ and $H$. This is a split extension, with short exact sequence $1\to A\to G\to H\to1$ and action $\phi$ of $H$ on $A$. The splitting identifies $H$ with a subgroup of $G$. This includes the non-dihedral example $A_4\simeq(\mathbb Z_2\times\mathbb Z_2)\rtimes\mathbb Z_3$. The construction applies independently of geometric locality. For a qLDPC implementation, check weight and qudit dimension must be bounded, and each qudit must participate in a bounded number of checks. The two-gross-code example in Appendix~\ref{sec:sm-gross} already has central Pauli commutators and uses Section~\ref{sec:prx-central}.

The completed code is the original code tensored with a fixed product state, so its code distance is unchanged. The dihedral circuits below also have a circuit-distance lower bound. Appendix~\ref{sec:prx278-d8} gives the square-lattice $D_8$ checks, qubit circuits and gate counts.

\subsection{Completion with a qudit}
\label{sec:prx273-frame}
Let $G=A\rtimes_\phi H$, where $A$ and $H$ are finite abelian groups, and suppose the raw operators form a unitary representation
\begin{equation}
 \begin{gathered}
 R_aR_b=R_{a+b},\qquad G_hG_k=G_{hk},\\
 G_hR_aG_h^\dagger=R_{\phi_h(a)}.
 \end{gathered}
 \label{eq:prx-semidirect-input}
\end{equation}
We write $A$ additively. Add one $|H|$-dimensional completion qudit with basis $|x\rangle$, $x\in H$. Its value controls which raw check $R_{\phi_x(a)}$ is measured. Accompany each symmetry operation $G_h$ by the qudit shift $L_h|x\rangle=|hx\rangle$, so the check label changes together with the data. Let $D$ be the subgroup of $A$ generated by all differences $\phi_h(a)-a$ with $a\in A$ and $h\in H$. The projector
\begin{equation}
 E_D=\frac{1}{|D|}\sum_{d\in D}R_d
 \label{eq:prx273-frame-enable}
\end{equation}
selects the states on which every $R_d$, $d\in D$, has eigenvalue one. On these states, $R_{\phi_h(a)}$ and $R_a$ agree. In this subspace, the pinning checks reduce to shifts of the completion qudit. Their common $+1$ eigenstate is the uniform superposition $|+_H\rangle$ defined below. The outcome $+1$ is a check eigenvalue, not the basis label $1$. For a completion qubit it selects $|+\rangle$, not $|1\rangle$. The original quantum-double code is recovered on the original qudits when the vertex and flatness checks also have outcome $+1$.

\begin{theorem}[Completion for semidirect-product checks]
\label{thm:prx-semidirect}
For finite abelian $A,H$, the following unitary checks commute:
\begin{align}
 \widehat{R}_a&=\sum_{x\in H}R_{\phi_x(a)}\otimes|x\rangle\langle x|,
 \label{eq:prx-frame-r}\\
 \widehat{G}_h&=G_h\otimes L_h,\nonumber\\
 T_h&=(I-E_D)\otimes I+E_D\otimes L_h.
 \label{eq:prx-frame-g}
\end{align}
Their joint eigenvalue-one space is the original $G$-invariant space tensored with $|+_H\rangle=|H|^{-1/2}\sum_x|x\rangle$. More generally, impose $\widehat{R}_a=T_h=1$ and $\widehat{G}_h=\chi(h)$ for a character $\chi$ of $H$. The resulting projector is $P_{A,\chi}\otimes|+_H\rangle\langle+_H|$, where $P_{A,\chi}$ imposes $R_a=1$ and $G_h=\chi(h)$ on the original system. For quantum-double vertex actions and gauge-invariant flatness checks, one completion qudit per vertex gives
\begin{equation}
 P_{\mathrm{comp}}=P_{D(A\rtimes_\phi H)}
 \otimes|+_H\rangle\langle+_H|^{\otimes |V|},
 \label{eq:prx-semidirect-projector}
\end{equation}
and preserves static code distance in the same physical weight convention.
\end{theorem}

\begin{proof}
The subgroup $D$ is invariant under $H$, so $E_D$ commutes with all $R_a,G_h$. Since $\phi_x(a)-a\in D$,
\begin{equation}
 R_{\phi_x(a)}E_D=R_aE_D.
 \label{eq:prx273-frame-neutral}
\end{equation}
The $\widehat{R}_a$ commute for every qudit value $x$, and the $\widehat{G}_h$ commute because $H$ is abelian. Conjugating a completed check by a completed symmetry gives
\begin{align}
 \widehat{G}_h\widehat{R}_a\widehat{G}_h^\dagger
 &=\sum_x R_{\phi_{hx}(a)}\otimes|hx\rangle\langle hx|\nonumber\\
 &=\widehat{R}_a.
 \label{eq:prx273-frame-commutation}
\end{align}
The qudit shift therefore compensates for the symmetry's change of check label. On the range of $E_D$, the check is independent of the qudit value by Eq.~\eqref{eq:prx273-frame-neutral}, so the pinning shifts cannot disturb it. The other commutations follow from $[E_D,G_h]=[L_h,L_k]=0$.

Let $P_A=|A|^{-1}\sum_aR_a$. For every qudit value $x$, imposing all $\widehat{R}_a=1$ gives the same projector $P_A$, because $\phi_x$ permutes $A$. On $P_A\otimes I$, the pins reduce to $L_h$. Their common eigenvalue-one space is spanned by $|+_H\rangle$. The remaining conditions reduce to $G_h=\chi(h)$ on the original system. This also proves the identity when the prescribed sector is absent, in which case both projectors vanish. Generators of $A$ and $H$ suffice.

At distinct vertices, left and right group actions on a shared edge commute. Gauge-invariant flatness checks commute with all vertex actions, so the local construction proves Eq.~\eqref{eq:prx-semidirect-projector}. Taking the matrix element of an error in the fixed completion state leaves a data operator supported only where the error already acted. Conversely, every original logical operator remains a logical operator after adjoining the completion qudits. These two directions establish equality of static distances.
\end{proof}

Section~\ref{sec:sm-semidirect-without-completion} compares this construction with commuting checks on the original qudits.

\subsection{Completed dihedral checks}
\label{sec:prx273-dihedral-checks}
For $D_m=\mathbb{Z}_m\rtimes\mathbb{Z}_2$, of order $2m$, reflection acts by inversion. Write an edge group element as $(a,b)$, with $a\in\mathbb{Z}_m$ and $b\in\mathbb{Z}_2$, and use the group law
\begin{equation}
 (a,b)(c,d)=(a+(-1)^b c,b+d).
 \label{eq:prx273-dihedral-law}
\end{equation}
The rotation and reflection generators are $r=(1,0)$ and $s=(0,1)$. We write $g_e=(a_e,b_e)$ for the computational-basis label on edge $e$, and $Z_{b_e}$ acts on the gauge qubit storing $b_e$. The action $A_v(g)$ multiplies outgoing edges by $g$ on the left and incoming edges by $g^{-1}$ on the right. Set $R_v=A_v(r)$ and $S_v=A_v(s)$. Here $A=\mathbb Z_m$ and $H=\mathbb Z_2$, so the completion qudit is a qubit. The completed reflection flips this qubit, while the completed rotation uses its value to select $R_v$ or $R_v^{-1}$. The completed checks are
\begin{align}
 \widehat{R}_v&=R_v\otimes|0\rangle\langle0|_{t_v}
              +R_v^{-1}\otimes|1\rangle\langle1|_{t_v},\nonumber\\
 \widehat{S}_v&=S_vX_{t_v},\qquad E_v=\Pi(R_v^2=1),\nonumber\\
 T_v&=I-E_v+E_vX_{t_v}.
 \label{eq:prx-dihedral-completion}
\end{align}
The sum defining $\widehat R_v$ specifies a controlled unitary. Equivalently, $\widehat R_v=(R_v\otimes I)\operatorname{ctrl}_{t_v}(R_v^{-2})$ is a product of gates. The projector $E_v$ specifies a condition and is not an additional unitary check. To apply $T_v$, coherently compute whether $R_v^2=1$ into a clean work qubit, use it to control $X_{t_v}$, and reverse the computation. This resets the temporary qubits and qudits without measuring the condition separately.

Here, $D=2\mathbb{Z}_m$ and $E_v$ averages $R_v^{2j}$ over this subgroup. This covers every $D_{4N}$ and also $S_3=D_3$. For $m=3$ or $m>4$, the commutator $S_vR_vS_v^{-1}R_v^{-1}=R_v^{-2}$ is noncentral, so these cases extend beyond Equation~\eqref{eq:main-central}.

Retain the plaquette checks
\begin{equation}
 B_p=\prod_{e\in\partial p}Z_{b_e},\qquad
 F_p=2\Pi(h_p=1)-I,
 \label{eq:prx273-dihedral-flux}
\end{equation}
Here $h_p$ is the ordered product of edge group labels around $p$. The check $B_p$ is a product of Pauli $Z$ gates. To apply $F_p$, coherently compute whether $h_p\ne1$ into a clean work qubit, apply $Z$ to it, and reverse the computation. This gives phase $+1$ when $h_p=1$ and $-1$ otherwise. The surgery protocol also measures the $\mathbb{Z}_2$ gauge-flux check $B_p$ separately, although full flatness $F_p=1$ implies $B_p=1$. The family supports higher-level logical operations~\cite{WarmanSchaferNameki2026}. We now give the completed-check extraction circuits and prove the distance bound for charge-conjugation surgery.

\subsection{Measurement circuits}
\label{sec:prx273-dihedral-instruments}
The circuits below measure each check without resolving additional observables. Circuits sharing matter, gauge or completion qudits run sequentially. Any temporary work registers are used by one circuit at a time and reset before reuse. Appendix~\ref{sec:sm-prx-dihedral-circuits} gives the ordered edge operations.

For rotation, let $\omega_m=e^{2\pi i/m}$ and $\Pi_k=\Pi(\widehat{R}_v=\omega_m^k)$. Prepare an $m$-level readout qudit $y$ in its uniform state and apply
\begin{equation}
 U_{\mathrm{phase}}=\sum_{y=0}^{m-1}|y\rangle\langle y|
                       \otimes\widehat{R}_v^y.
 \label{eq:prx273-phase-query}
\end{equation}
This controlled operation consists of four signed edge additions in the order $N,E,S,W$. An inverse Fourier transform records the charge $k$. A separate pin readout $u$, prepared in $|+\rangle$, controls $X_t$ only when $2k=0\pmod m$. Raw reflection sends a rotation charge $k$ to $-k$, so this condition selects precisely the charges unchanged by reflection. Reading $y$ in its computational basis and $u$ in $X$ gives
\begin{equation}
 K_{k,x}=\Pi_k\frac{I+xT_v}{2},\qquad x=\pm1.
 \label{eq:prx273-rotation-pin-instrument}
\end{equation}
The pin measures $X_t$ when $2k=0$ and otherwise returns $x=+1$, without an additional measurement of the condition. For $m=2^q$, charge and pin readout use $q+1$ qubits. Reflection uses one readout $a$ controlling the four $S_v$ edge factors, followed by $\operatorname{CNOT}(a\to t_v)$ and an $X$ measurement. The outgoing factors are $K_mX_{b_e}$ and the incoming factors $X_{b_e}$, where $K_m|j\rangle=|-j\rangle$.

The flatness circuit computes $h_p=g_Sg_Eg_N^{-1}g_W^{-1}$ in temporary qudits, records only whether $h_p=1$, and reverses the computation to reset those qudits. Its two branch operators are
\begin{equation}
 K_{p,+}=\Pi(h_p=1),\qquad K_{p,-}=I-\Pi(h_p=1).
 \label{eq:prx273-flatness-instrument}
\end{equation}
Reading the full group element $h_p$ would measure additional information and change the postmeasurement state. To extract $B_p$, prepare $a_B$ in $|0\rangle$, apply four CNOTs from the edge gauge qubits $b_e$ to $a_B$, and measure $a_B$ in $Z$. The extra flag gates in Appendix~\ref{sec:sm-prx-dual-role} preserve these measurements and detect certain faults that spread to several edges, under the stated charge and pin conditions.

\begin{figure}[tbp]
\centering
\begin{tikzpicture}[x=.92cm,y=.92cm,font=\footnotesize,
 >={Latex[length=4pt,width=3pt]},line width=.65pt]
\path[use as bounding box](-1.70,-1.45)rectangle(6.05,1.68);
\draw[->,black!35](-1.45,0)--(-.08,0);
\draw[->,black!35](0,-1.35)--(0,-.08);
\draw[->,line width=.95pt](.08,0)--(1.45,0);
\draw[->,line width=.95pt](0,.08)--(0,1.35);
\foreach \x/\y in {-1.45/0,1.45/0,0/-1.35,0/1.35}
 {\draw[fill=white](\x,\y)circle(1.5pt);}
\draw[fill=white,line width=.8pt](0,0)circle(2.4pt);
\node[above left,inner sep=3pt]at(0,0){$v$};
\fill[black!35](-.72,0)circle(2.6pt);
\fill[black!35](0,-.68)circle(2.6pt);
\fill(.72,0)circle(2.6pt);
\fill(0,.68)circle(2.6pt);
\node[below,inner sep=4pt,text=black!50]at(-.72,0){$W(v)$};
\node[left,inner sep=4pt,text=black!50]at(0,-.68){$S(v)$};
\node[below,inner sep=4pt]at(.72,0){$E(v)$};
\node[left,inner sep=4pt]at(0,.68){$N(v)$};
\node[align=center,text width=3.75cm]at(3.80,.37)
 {The north and east matter qudits\\form the block at $v$:\\[3pt]
 $U_v=K_{m,N(v)}K_{m,E(v)}$};
\end{tikzpicture}
\caption{Meaning of the outgoing matter qudits. Arrows orient the lattice edges east and north. The two highlighted edge qudits belong to the symmetry block at $v$. Each matter qudit belongs to exactly one such block, so the factors $U_v$ act on disjoint qudits. Only matter qudits are shown.}
\label{fig:v466-oriented-edges}
\end{figure}
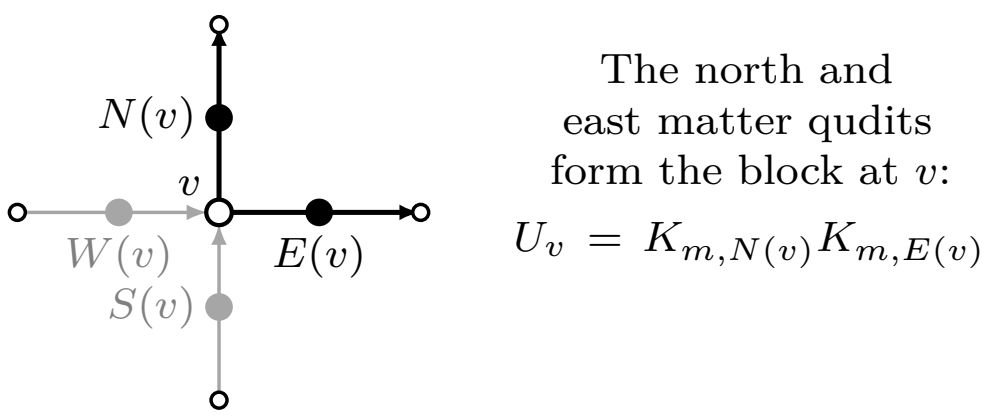

\subsection{A circuit-distance lower bound}
\label{sec:prx273-dihedral-distance}
For every $D_{4N}$, use an $L\times L$ square torus with $L\ge4$ and the native circuits of Section~\ref{sec:prx273-dihedral-instruments} and Appendix~\ref{sec:sm-prx-dihedral-circuits}. We use the schedule and local Kraus fault model of Section~\ref{sec:prx-ft} for qudits, including faults on completion and temporary qudits and in resets. For fixed $m$, each check circuit has depth and temporary workspace bounded independently of $L$.

For surgery, follow the gauging and ungauging steps of Protocol~\ref{alg:commuting-gauging} with the finite-order check readouts and additional work registers specified above. Start with the $\mathbb{Z}_m$ toric code and gauge charge conjugation. Orient horizontal edges east and vertical edges north, as in Fig.~\ref{fig:v466-oriented-edges}. The matter qudits on the north and east edges starting at $v$ form one symmetry block, with
\begin{equation}
 U_v=K_{m,N(v)}K_{m,E(v)},\qquad \mathcal{U}=\prod_vU_v.
 \label{eq:prx273-dihedral-symmetry}
\end{equation}
Write $U(x)=\prod_vU_v^{x_v}$ for a binary vertex assignment $x$. Initialize the gauge qubits in $|0\rangle$ and the completion qubits in $|+\rangle$. Extract the completed checks for $R$ rounds, measure the gauge qubits in $Z$, and apply $U(x)^\dagger$ when the final record is $z=\delta x$. Parent-check circuits stay within one star or face, and correction gates within individual matter edges. Both parent-check intervals and the physical corrections may be faulty. The input is encoded and the final parent-code projection $P$ is ideal. Require the recorded parent, flux and pin outcomes to be positive, the recorded rotation charge to be neutral, the reflection vector $\boldsymbol{s}=(s_v)$ to agree across all gauged rounds, and the final link record to have the form $z=\delta x$. The reported symmetry outcome is $s=\prod_vs_v$.

\begin{theorem}[Dihedral circuit distance]
\label{thm:prx273-dihedral-distance}
For every $m=4N$, the specified circuits have
\begin{equation}
 d_{\mathrm{surg}}\geq\min\{R,\lceil L/4\rceil\}.
 \label{eq:prx-dihedral-distance}
\end{equation}
For fewer faults than this bound, every measurement record $\mathsf{m}$ satisfying the conditions just stated and every associated noise branch $\mu$ obey
\begin{equation}
 P K_{F,\mathsf{m},\mu}P
 =c_{F,\mathsf{m},\mu}\frac{I+s\mathcal{U}}{2}P.
 \label{eq:prx273-dihedral-branch}
\end{equation}
With ideal initial and final code boundaries, the same extraction circuits have memory fault distance at least $\lceil L/4\rceil$. The geometric constant four is independent of $N$, although native gate counts and round duration depend on $N$.
\end{theorem}

\begin{proof}
The commuting-check reduction confines each fault to its original star or face and the checks meeting that support. Repeated ideal rounds do not enlarge it further. Index each edge by its starting vertex. Each resulting neighborhood occupies at most four rows and four columns of starting vertices, including the optional completion-qubit flag. Thus $4f<L$ leaves a full row and column outside the support of the effective errors. On the parent toric code, these support a complete set of logical $X$ and $Z$ operators.

For surgery, $f<R$ also leaves a fault-free gauged round. By Theorem~\ref{thm:prx-semidirect}, the projector for its recorded outcomes is the raw gauging projector with those outcomes, tensored with the completion plus state. The gauge-flux tests $B_p=+1$ retain only incoming gauge-bit flip patterns with even parity around every face. Since the unused row and column contain no flipped edges, these patterns also have zero parity around both noncontractible cycles. Each pattern therefore has the form $\eta_{uv}=y_u+y_v\pmod2$ for bits $y_v$ assigned to vertices, which is the exactness condition. The final link test similarly selects outgoing patterns of this form. The vertex bits can be chosen zero on the unused row and column. The resulting matter errors commute with all logical Pauli operators and hence are scalar after parent-code projection. Applying this argument on both sides of the fault-free round gives Eq.~\eqref{eq:prx273-dihedral-branch}.

For memory, the unused row and column retain both winding cycles. Appendix~\ref{sec:sm-prx-dihedral-distance} proves that every operator on the complementary region is scalar after code projection. It also gives the surgery contractions and extends the argument by linearity to arbitrary local Kraus operators.
\end{proof}

For fixed $N$, taking $R=\Theta(L)$ gives linear distance in $O(L)$ rounds. The bound allows faults to spread across several edges and does not use the optional flag gates. The bilayer $D_4$ circuit has the sharper exact-distance proof.

\section{Compatible measurements without completion}
\label{sec:sm-without-completion}
The preceding sections make measurements compatible by adding completion qubits or qudits. Here we construct compatible checks using only the existing matter and gauge qudits. Section~\ref{sec:sm-without-completion-construction} gives the construction. Section~\ref{sec:prx277-local-comparison} explains what changes when completion qubits are omitted, both in the circuits and in the measurements after errors. Section~\ref{sec:sm-semidirect-without-completion} then gives explicit lattice checks and circuits for a family of quantum-double codes.

On encoded inputs with the prescribed gauge initialization, these constructions implement the same ideal gauging map as completion, with the fixed completion-qudit product state omitted (Appendix~\ref{sec:sm-explicit-gauged-projector}).

\subsection{Check construction}
\label{sec:sm-without-completion-construction}
\paragraph{Central commutators.}
We call the modified observables below \emph{controlled checks}. The central-check eigenvalues determine whether each raw check acts. With the central relations~\eqref{eq:main-central}, assign each conflict $e=(i,j)$ to one endpoint. If $\mathcal{A}_i$ is the set assigned to $i$, use
\begin{equation}
 E_i=\prod_{e\in\mathcal{A}_i}\frac{I+\mathsf{K}_e}{2},\qquad
 C_i^{\mathrm{ctrl}}=(I-E_i)+E_iC_i,
 \label{eq:conditional-central-completion}
\end{equation}
and retain every central check. The superscript $\mathrm{ctrl}$ denotes this coherent control by $E_i$. The resulting unitary acts as $C_i$ on $\operatorname{im}E_i$ and as $I$ on its complementary eigenspaces. Both subspaces remain in the physical Hilbert space. The circuit measures the modified check without separately measuring $E_i$ or postselecting its value. For a conflicting pair,
\[
 [C_i^{\mathrm{ctrl}},C_j^{\mathrm{ctrl}}]
 =E_iE_j(\mathsf{K}_{ij}-I)C_jC_i=0,
\]
because the assigned endpoint contributes $(I+\mathsf{K}_{ij})/2$. The checks are Hermitian unitary operators. In the common $+1$ sector of the central checks they equal the raw checks, so they preserve $P_g$, its distance, and its compatible measurement branches. Bounded raw-check weight and incidence give bounded weight and incidence for the controlled checks. In gauging, assign each Gauss--parent conflict to the parent check, leaving the Gauss checks unchanged.

The controlled checks obtained from Clifford raw checks need not be Clifford. Their hierarchy levels can depend on which check receives each conflict, as the examples in Appendix~\ref{sec:v460-comparison-details} show.

\paragraph{Local symmetric constraints.}
The following construction applies without requiring central commutators. It uses the $\mathbb{Z}_N$ gauging notation and projector identity of Appendix~\ref{sec:sm-zn-proof}.

\begin{proposition}[Gauging local symmetric checks]
\label{prop:zn-patch-checks}
Let the $U_v$ act on disjoint blocks, with identity factors on blocks where the symmetry does not act. Suppose the parent code is the common $+1$ space of commuting Hermitian unitary operators $J_i$ on patches $D_i$ of whole blocks. Each $J_i$ commutes with $\prod_{v\in A_i}U_v$, where $A_i$ lists its active blocks. Then, the gauged parent constraints can be chosen as globally commuting Hermitian unitaries compatible with $A_v,B_f$ and satisfying Eq.~\eqref{eq:sm-zn-projector}. Bounded patch size, tree size and patch overlap bound their weight. Bounded vertex degree and face size bound the Gauss and flux weights.
\end{proposition}
\begin{proof}
Choose a tree connecting all vertices in $A_i$, possibly through other vertices, and a reference vertex $r_i$ on that tree. Set $y_i(r_i)=0$ and determine each $y_i(v)$ along the tree path from $r_i$ to $v$, using $y_i(s(e))-y_i(t(e))=z_e$ modulo $N$. Put $W_i=\prod_{v\in A_i}U_v^{y_i(v)}$, with $W_i=I$ for inactive patches. For each overlapping pair $(i,j)$, test whether $y_i-y_j$ is constant on $A_i\cap A_j$. Assign this test to one of the two checks. Enable a check when every test assigned to it passes, including when no test is assigned. Empty intersections pass. Use
\[
 C_i(z)=
 \begin{cases}
 W_i(z)J_iW_i(z)^\dagger,&i\text{ enabled},\\
 I,&i\text{ disabled}.
 \end{cases}
\]
The check is diagonal in the link basis and commutes with fluxes. Use the gauge qudits to control the compatibility tests without measuring their outcomes separately. Record only the eigenvalue of $C_i$. Under $z\mapsto z+\delta u$, $y_i(v)$ shifts by $u_v-u_{r_i}$. The test is unchanged and patch invariance removes the shift at the reference block, so $C_i(z+\delta u)=U(u)C_i(z)U(u)^\dagger$ and $C_i$ commutes with the Gauss's law operators. If two neighboring checks are enabled, their pair test must have passed because its assigned check is enabled. We can therefore shift $y_j$ to match $y_i$ on their common active blocks. A single unitary then conjugates both commuting checks. A disabled check acts as identity.

At $z=\delta y$ all checks are enabled and select $U(y)\operatorname{im}P$. The Gauss outcomes away from the reference vertex fix the phases in Eq.~\eqref{eq:sm-zn-gamma}, and the product of all Gauss outcomes fixes $\mathcal{U}=\omega^\alpha$. The checks preserve the exact sector, giving Eq.~\eqref{eq:sm-zn-projector}.
\end{proof}

\paragraph{Pauli stabilizer parents.}
Enlarge each stabilizer support to a patch $D_i$ of whole symmetry blocks. Let $P_i$ project onto the common $+1$ space of all parent stabilizers supported in $D_i$, and set $J_i=2P_i-I$. The $J_i$ commute and together test every parent generator. The stabilizer group average gives $\operatorname{Tr}_{D_i^c}P\propto P_i$, since every Pauli operator not proportional to identity is traceless, including in composite dimension. Factoring $\mathcal{U}$ across $D_i$ and its complement in $\mathcal{U}P\mathcal{U}^\dagger=P$ then proves patch invariance.

The circuit for $C_i$ first applies $W_i^\dagger$ and diagonalizes the parent stabilizers on $D_i$. It applies a phase $-1$ when the check is enabled and at least one stabilizer has eigenvalue different from $+1$, then reverses these basis changes. Bounded-depth extraction requires bounded incidence and routing congestion. The hierarchy cost depends on the controlled symmetry powers and patch reflections.

For $\mathbb{Z}_2$ logical measurement, use Protocol~\ref{alg:commuting-gauging}, including ungauging, with completion initialization omitted and $D_v=I$. For $\mathbb{Z}_N$, use the procedure in Appendix~\ref{sec:sm-zn-measurement} with the full commuting family and omit completion initialization. The construction also applies to twisted quantum doubles obtained by gauging local symmetric constraints, including the double-semion model~\cite{Tantivasadakarn2024LongRange}.

\subsection{Bilayer surgery with and without completion}
\label{sec:prx277-local-comparison}
For bilayer surgery on the even-$L$ torus, completion keeps the star checks Clifford and supplies coherent controls for boundary-error recovery.

\paragraph{Repairing a detected boundary error.}
Consider green-$X$ errors at the input to gauging, with all later operations ideal. For one erroneous edge, $D=X_{g,e}$, let $b(e)$ be its blue edge partner under the symmetry and let $y_v$ be the bit labeling a term in the coherent gauging sum. The relation $U(y)D=D Z_{b(e)}^{y_v}U(y)$ shows why removing $D$ alone leaves a conditional blue-$Z$ error. At a negative green flux, the pin releases its completion qubit, which stores $|y_v\rangle_X$. The CNOT from $b(e)$ to that qubit cancels $Z_{b(e)}^{y_v}$ coherently. Removing $D$, measuring and resetting the active completion qubits, and updating the Gauss record then restore ideal gauging before ungauging (Section~\ref{sec:prx-d4-recovery}).

For every green-$X$ boundary pattern of weight below $L/2$, this repair implements the full ideal measurement when all later operations are ideal. The retained completion qubits supply the coherent controls and avoid the extra matter measurements used in the controlled-check recovery of Appendix~\ref{sec:sm297-no-completion-recovery}.

\paragraph{Clifford checks and explicit circuits.}
With $3L^2$ persistent completion qubits and $3L^2$ dedicated pin readouts, every completed star is Clifford. For the controlled assignment in Eq.~\eqref{eq:sm297-conditional-assignment}, the blue and green stars have hierarchy levels three and four. The corresponding readout-controlled unitaries have hierarchy level three for the completed stars, compared with levels four and five for these controlled stars (Appendix~\ref{sec:v460-comparison-d4}). The completed extraction uses CNOT, CCZ and Toffoli gates, with its resources given in Table~\ref{tab:prx277-d4-resources}. For the specified gate order and detection protocol, Theorem~\ref{thm:prx273-d4-surgery} establishes exact fault distance $\min\{L,R\}$.

\subsection{Semidirect-product quantum doubles}

We now give explicit lattice checks and measurement circuits without completion for a family that includes noncentral commutators. This provides a concrete application of Section~\ref{sec:sm-without-completion-construction}, together with a circuit-distance lower bound.

\label{sec:sm-semidirect-without-completion}
On a $\mathbb{Z}_m$ toric code ($m,N\ge2$), gauge $K_q|a\rangle=|qa\rangle$ for $q$ invertible modulo $m$ and $q^N=1\bmod m$. The edge labels $(a,b)\in\mathbb{Z}_m\times\mathbb{Z}_N$ form the group
\begin{equation}
 (a,b)(c,d)=(a+q^bc,b+d),\qquad
 G=\mathbb{Z}_m\rtimes_q\mathbb{Z}_N.
 \label{eq:em-semidirect}
\end{equation}
Orient edges north and east. Assign the two outgoing matter qudits at each vertex $v$ to $U_v=K_{q,\mathrm{N}(v)}K_{q,\mathrm{E}(v)}$. Subscripts $d\in\{\mathrm{N},\mathrm{E},\mathrm{S},\mathrm{W}\}$ denote edges of a star or face carrying matter and gauge qudits with group labels $g_d=(a_d,b_d)$. Use $X_m,Z_m$ as for the gauge qudits, the Fourier transform $\mathsf{F}_mX_m\mathsf{F}_m^\dagger=Z_m$, and $R_m=\operatorname{diag}(1,-1,\ldots,-1)$. The gate $\mathrm{C}_b(K_q^{-1})$ sends $|b,a\rangle$ to $|b,q^{-b}a\rangle$, and $\mathrm{CX}_m$ adds its control modulo $m$. On pairs of matter and gauge qudits $CL|g,h\rangle=|g,gh\rangle$ and $\mathrm{Inv}|g\rangle=|g^{-1}\rangle$. The coordinate changes are
\begin{equation}
 \begin{aligned}
 Q_v&=\mathrm{C}_{b_{\mathrm{S}}}(K_{q,\mathrm{S}}^{-1})\mathrm{C}_{b_{\mathrm{W}}}(K_{q,\mathrm{W}}^{-1}),\\
 {L_v}&=\mathrm{CX}_m(\mathrm{N}\to\mathrm{W})\mathrm{CX}_m(\mathrm{N}\to\mathrm{S})
       \mathrm{CX}_m(\mathrm{N}\to\mathrm{E})^{-1},\\
 V_v&=\mathsf{F}_{m,\mathrm{N}}{L_v}Q_v,\\
 W_p&=CL_{\mathrm{S}\to\mathrm{W}}CL_{\mathrm{E}\to\mathrm{W}}CL_{\mathrm{N}\to\mathrm{W}}^{-1}\mathrm{Inv}_{\mathrm{W}}.
 \end{aligned}
 \label{eq:em-zn-words}
\end{equation}
Here, $W_p$ writes the ordered group product around the face, $h_p=g_{\mathrm{S}}g_{\mathrm{E}}g_{\mathrm{N}}^{-1}g_{\mathrm{W}}^{-1}$, into the west-edge matter and gauge qudits alone. Vertex checks test $\mathbb{Z}_N$ Gauss's law and rotation neutrality. The $\mathbb{Z}_N$ gauge-flux check $H_p$ has eigenvalue one exactly when the oriented sum $b_{\mathrm{S}}+b_{\mathrm{E}}-b_{\mathrm{N}}-b_{\mathrm{W}}$ vanishes modulo $N$. The check $F_p$ tests rotation flux in this sector and acts as the identity otherwise.
\begin{equation}
 \begin{aligned}
 G_v&=X_{N,b_{\mathrm{N}}}X_{N,b_{\mathrm{E}}}X_{N,b_{\mathrm{S}}}^{-1}X_{N,b_{\mathrm{W}}}^{-1}
       K_{q,\mathrm{N}}K_{q,\mathrm{E}},\\
 E_v&=V_v^\dagger R_{m,\mathrm{N}}V_v,\\
 H_p&=Z_{N,b_{\mathrm{S}}}Z_{N,b_{\mathrm{E}}}Z_{N,b_{\mathrm{N}}}^{-1}Z_{N,b_{\mathrm{W}}}^{-1},\\
 F_p&=W_p^\dagger\mathrm{C}_{b_{\mathrm{W}}=0}(R_{m,\mathrm{W}})W_p.
 \end{aligned}
 \label{eq:em-zn-generators}
\end{equation}
The last factor applies $R_m$ only when $H_p=1$.

\begin{proposition}[Commuting checks for the semidirect product]
The checks in Eq.~\eqref{eq:em-zn-generators} commute globally and have as their common $+1$ eigenspace the untwisted $D(G)$ code space without completion qudits.
\end{proposition}
\begin{proof}
For vertex action $A_v(h)$, multiply outgoing edges on the left by $h$ and incoming edges on the right by $h^{-1}$. With $r=(1,0)$ and $s=(0,1)$, $V_v$ sends $A_v(r)$ to $Z_{m,\mathrm{N}}$, so $E_v$ tests rotation neutrality. Since $srs^{-1}=r^q$, $G_v=A_v(s)$ preserves neutrality. Distinct vertex actions commute, and both flux checks are conjugacy-invariant functions of holonomy. The joint $+1$ eigenspace imposes vertex invariance and flatness, giving the quantum-double code space.
\end{proof}
At $z=\delta x$, conjugation by $U(x)^\dagger$ restores the parent star and plaquette constraints, and $\mathbb{Z}_N$ Gauss's law operators fix the phases relative to the reference vertex. This proves Eq.~\eqref{eq:sm-zn-projector}, since every check preserves the exact sector. Examples $(m,N,q)=(3,2,2)$ and $(7,3,2)$ give $S_3$ and the non-Abelian group of order $21$, respectively.

\paragraph{Fault tolerance.}
For fixed $(m,N,q)$ on an $L\times L$ square lattice on the torus with $L\ge4$, measure each check within one star or plaquette, serializing overlaps. Under the detection tests and assumptions of ideal encoded input and ideal final parent-code projection of Theorem~\ref{thm:zn-accepted-distance},
\begin{equation}
 d_{\mathrm{surg}}\ge\min\{R,\lceil L/4\rceil\}.
 \label{eq:em-zn-distance}
\end{equation}
Index each edge by its source vertex $(i,j)$. Here $\mathcal{F}$ denotes the full check family in Eq.~\eqref{eq:em-zn-generators}. The residual support of a fault in a circuit on $S$ lies in
\begin{equation}
 \mathcal{N}(S)=S\cup\!\bigcup_{C\in\mathcal{F}:\,\operatorname{supp}C\cap S\ne\varnothing}
 \operatorname{supp}C.
 \label{eq:sm-s3-ft-neighborhood}
\end{equation}
For a star at $(i,j)$ this neighborhood occupies columns $i-2,\ldots,i+1$ and rows $j-2,\ldots,j+1$. A face with southwest corner $(i,j)$ occupies $i-1,\ldots,i+2$ and $j-1,\ldots,j+2$. Moving a residual Pauli operator through the parent-check projectors changes only their outcome labels, without enlarging its support, so each fault costs at most four rows and four columns. For $f=|F|<\min\{R,\lceil L/4\rceil\}$ native faults, choose a fault-free gauged round. On either side, the union of residual supports occupies at most $4f<L$ rows and columns, leaving a common row and column untouched. An incoming shift that passes the flux checks is flat and has zero winding around both cycles, so it is exact, also for composite $N$. The exact final link record also forces each surviving outgoing shift to be exact. Choose each potential to vanish on the unused row and column. Since $U_v$ acts only on outgoing matter qudits, the byproduct and direct matter error avoid the supports of the two logical $X$ operators and two logical $Z$ operators on that row and column. Their products span all operators on the encoded states, including for composite $m$, so multiplying the effective parent operators in Eqs.~\eqref{eq:sm-zn-input} and~\eqref{eq:sm-zn-output} by $P$ on both sides gives scalar multiples of $P$. For a record $\mathsf{m}$ satisfying these conditions with reported label $\alpha(\mathsf{m})$ and noise branch $\mu$ this gives
\begin{equation}
 P K_{F,\mathsf{m},\mu}P
 \in\mathbb{C}\Pi_{\alpha(\mathsf{m})}(\mathcal{U})P,
 \label{eq:sm-zn-instrument-distance}
\end{equation}
including for inputs entangled with a reference.

Theorem~\ref{thm:s3-completion-nogo} identifies an encoding-dependent obstruction to finite-hierarchy completion. For the $D(S_3)$ code in its fixed qubit--qutrit encoding, no product-state completion preserving the same code space can give globally commuting checks at finite levels of the hierarchy defined by that qubit--qutrit Pauli structure.

\section{Discussion}
\label{sec:prx-discussion}
We construct compatible local checks for gauged non-Pauli codes and give simple, explicit measurement circuits. The fault analysis treats individual circuit locations and both code switches. For the specified bilayer $D_4$ surgery, it establishes exact circuit distance. Completion qubits also retain quantum information that supports recovery.

The explicit surgery circuits and completed measurement records provide a concrete setting for decoder development and numerical studies of logical-error rates. Existing work provides decoding methods and performance estimates for non-Abelian codes~\cite{MagdalenaD4Decoding2026,Jing2026Intrinsic,Sajith2026Fusion,Jing2026Integer}, and threshold results for specified non-Abelian protocols~\cite{Davydova2026,Ren2026CircuitThresholds}. In a separate work, we will address how our construction enables efficient decoding of non-Abelian topological codes for quantum memory.

Further directions include simultaneous logical measurements and gauging higher-form symmetries~\cite{Williamson2026HigherForm,Cowtan2026Parallel}, as well as the additional logical operations supported by dihedral codes~\cite{WarmanSchaferNameki2026}. These settings offer applications for compatible checks and for fault analysis that follows errors through shared circuits and code-switching interfaces.

\begin{acknowledgments}
Z.W. thanks Yu-An Chen, Margarita Davydova, Arpit Dua, and Nathanan Tantivasadakarn for many inspiring discussions. Z.W. is supported by the National Natural Science Foundation of China (Grant No.~12474491) and the Central University Fundamental Research Funds (Peking University).
\end{acknowledgments}

\clearpage

\appendix
\makeatletter
\@addtoreset{theorem}{section}
\makeatother
\renewcommand{\thetheorem}{\thesection\arabic{theorem}}
\renewcommand{\theHsection}{appendix.\Alph{section}}
\renewcommand{\theHsubsection}{appendix.\Alph{section}.\arabic{subsection}}
\renewcommand{\theHsubsubsection}{appendix.\Alph{section}.\arabic{subsection}.\arabic{subsubsection}}
\renewcommand{\theHequation}{appendix.\Alph{section}.\arabic{equation}}
\renewcommand{\theHtheorem}{appendix.\Alph{section}.\arabic{theorem}}
\setcounter{secnumdepth}{2}
\section*{Appendices}

The appendices give the detailed proofs, gate sequences and additional examples supporting the main text.

\pdfbookmark[1]{Appendix contents}{appendix-contents}
\subsection*{Appendix contents}
\begingroup
\makeatletter
\def\AppendixReferencesEntry{\numberline {}References}
\def\AppendixContentsLine#1#2#3#4{\def\AppendixEntryTitle{#2}\ifx\AppendixEntryTitle\AppendixReferencesEntry\else
    \savedcontentsline{#1}{#2}{#3}{#4}\fi}
\def\StartAppendixContents{\let\contentsline\AppendixContentsLine}
\addtocontents{toc}{\string\StartAppendixContents}
\let\savedcontentsline\contentsline
\let\contentsline\@gobblefour
\hypersetup{linktoc=all,linkcolor=black}
\def\@pnumwidth{2.5em}
\def\@tocrmarg{3.5em}
\def\l@section#1#2{\addvspace{0.4\baselineskip}{\bfseries\@dottedtocline{1}{0em}{2.5em}{#1}{#2}}}
\def\l@subsection#1#2{\@dottedtocline{2}{1.3em}{2.0em}{#1}{#2}}
\def\l@subsubsection{\@dottedtocline{3}{2.6em}{3.8em}}
\setcounter{tocdepth}{2}
\let\appendix\appendix@toc
\@starttoc{toc}
\makeatother
\endgroup

\section{Construction and resources for \texorpdfstring{$\mathbb{Z}_2$}{Z2} Clifford gauging on qubits}
\label{sec:general-sm}

This appendix supplies the graph conventions, extraction circuits and phase factors for a qubit Pauli parent with disjoint bounded-support Clifford symmetry factors $U_v^2=I$. The completion theorem has broader scope. It applies to any nonzero raw code space with the check relations in Equation~\eqref{eq:main-central} of the main text. The raw construction here uses the standard symmetry-enrichment and gauging procedure~\cite{Verresen2021Enrichment,Tantivasadakarn2024LongRange}.
\subsection{Choosing the gauging graph and its flux checks}
\label{sec:sm-choose-graph}
The graph conditions below ensure a collective symmetry measurement and bounded raw-check weight and incidence for an LDPC parent.

We call a matter-qubit block with $U_v\ne I$ an active block and assign it a vertex of $G$. For each parent stabilizer, paths connect the active blocks it meets and specify the gauge qubits in its modified check. Vertices with $U_v=I$ may be added for routing and connectedness.

The chosen flux faces, edges and vertices span the chain spaces $C_2,C_1,C_0$ over $\mathbb{Z}_2$, respectively, with boundary maps $\partial_2,\partial_1$ and vertex coboundary $\delta$ given by
\begin{equation}
 C_2\xrightarrow{\partial_2}C_1\xrightarrow{\partial_1}C_0,
 \qquad \partial_1\partial_2=0,\quad\delta=\partial_1^T.
 \label{eq:sm-chain}
\end{equation}
The maps take an edge to its endpoints and a face to its boundary, modulo two. Closed face boundaries give $\partial_1\partial_2=0$. Only $C_2,C_1,C_0$ are needed, irrespective of spatial dimension.

For $x\in C_0$, $(\delta x)_{uv}=x_u+x_v\bmod2$. An edge vector $c\in C_1$ selects the edges with $c_e=1$. Write $X_g(c)=\prod_eX_{g_e}^{c_e}$ and $Z_g(c)=\prod_eZ_{g_e}^{c_e}$. For a vertex basis vector $v$, $\delta v$ selects its incident edges, so a vertex with $U_v=I$ has Gauss operator $X_g(\delta v)$.

\emph{Connectedness.} We choose a connected graph so that the protocol measures only the collective symmetry $\mathcal{U}$. On a disconnected graph, products of Gauss outcomes separately determine the symmetry eigenvalue of each connected component. This can destroy coherence within a fixed eigenspace of $\mathcal{U}$. For example, two isolated vertices for $\mathcal{U}=Z_1Z_2$ measure $Z_1$ and $Z_2$ separately, destroying coherence between $|00\rangle$ and $|11\rangle$ even though both have $\mathcal{U}=+1$. Join them by one link initialized in $|0\rangle_g$. For vertex outcomes $a_1,a_2$ and final link bit $z$,
\[
 \langle z|\frac{I+a_1 Z_1X_g}{2}\,
              \frac{I+a_2 Z_2X_g}{2}|0\rangle
  =\frac{a_2^z}{2}Z_2^z\,\Pi_{a_1a_2}(\mathcal{U}).
\]
Correcting by $Z_2^z$ measures only the joint parity. Equation~\eqref{eq:em-zn-instrument} gives the corresponding identity on any connected graph.

\emph{Edges and faces.} Require bounded vertex degree, path length and paths per edge, together with bounded face boundary length and faces per edge. Face boundaries must span the commutator cycles in Equation~\eqref{eq:sm-commutator-cycle}, making every raw commutator a product of retained flux checks. Other cycles may remain outside this span. Their parity is fixed by gauge initialization and tested by $z=\delta x$ at ungauging. Modified parent and invariant checks use only their original blocks and paths (Appendix~\ref{sec:sm-clifford-derivation}). For an LDPC parent with bounded blocks, these conditions bound all raw weights and incidences. Corollary~\ref{cor:parsimonious-completion} supplies such a graph.

\emph{Ungauging.} The ungauging procedure is the same as in Refs.~\cite{WilliamsonYoder2026,Chang2026Magic}. Choose a reference vertex $v_0$ and an arbitrary edge-path from $v_0$ to each vertex $v$. Set $x_{v_0}=0$ and obtain $x_v$ by summing the measured bits along that path modulo two. Test $z=\delta x$ on every edge, which also checks parity around winding cycles (Appendix~\ref{sec:graph-distance}). On an ideal measurement branch, changing the reference vertex changes the corrected state only by a known overall sign.

\subsection{Completion-qubit placement and extraction circuits}
\label{sec:sm-place-ancillas}

Use the central relations in Equation~\eqref{eq:main-central} of the main text and a nonzero projector $P_g$ onto the common $+1$ eigenspace of the raw and central checks. Original qudits may have any finite dimension. Assign one completion qubit to each oriented conflict edge, including edges with repeated labels, and measure $\mathcal{F}=\{\widehat{C}_i,\mathsf{K}_a,T_e\}$ as defined in Equations~\eqref{eq:prx273-completion-factor}--\eqref{eq:main-pinning} of the main text. Theorem~\ref{thm:prx273-central-completion} of the main text proves commutation, preservation of compatible measurement outcomes and their conditional states, and unchanged code distance, including completion-qubit errors. Write $P_\pm(\mathsf{K})=(I\pm\mathsf{K})/2$.

\begin{center}
\small
\begin{tabularx}{\columnwidth}{@{}p{.31\columnwidth}X@{}}
\toprule
Qubit role & Preparation, use and release \\
\midrule
Gauge qubit $g_e$ & Prepare $|0\rangle$ on edge $e$. Keep the qubit through all syndrome extraction rounds, then measure it in $Z$ for ungauging. \\[3pt]
Completion qubit $t_e$ & Prepare $|+\rangle$ for edge $e$ of $\mathcal{H}$. Couple it to both endpoint checks, measure $T_e$ in each round, and discard the qubit at ungauging. \\[3pt]
Measurement ancilla & Assign one to each check. Record its measurement outcome and reset it for the next round. \\
\bottomrule
\end{tabularx}
\end{center}

\paragraph{Circuit for gauging and for one syndrome extraction round.}
For $C_{i(v)}=A_v$ and $V_\ell=\operatorname{ctrl}_{t_\ell}(\mathsf{K}_\ell)$ the completion factor is
\begin{equation}
 D_v=\prod_{\ell:i(v)\to j}X_{t_\ell}
           \prod_{\ell:j\to i(v)} V_\ell,\qquad
 \widehat{A}_v=A_vD_v .
 \label{eq:sm-entry-dressing}
\end{equation}
On outgoing edges the vertex ancilla controls $X_{t_\ell}$ by a CNOT. On incoming edges it controls $V_\ell$. Centrality gives $[D_v,A_u]=0$ for every $u,v$. No edge joins two Gauss operators, so the $D_v$ commute and the controlled operators in Equation~\eqref{eq:main-circuit-gauging} of the main text decompose into these factors. For the $D(D_4)$ gate order see Appendix~\ref{sec:smd4-native-extraction}. Define the initialization map $J$ for the gauge and completion qubits by
\begin{equation}
 J|\psi\rangle_m=|0\cdots0\rangle_g\otimes|\psi\rangle_m
                    \otimes|+\cdots+\rangle_t,
 \label{eq:sm-initialization-map}
\end{equation}
with $|E(G)|$ gauge qubits and $|E(\mathcal{H})|$ completion qubits. With $a^x=\prod_va_v^{x_v}$, vertex measurement gives
\begin{equation}
 {}_\mu\!\langle a^X|\mathcal{G}_a|+\rangle_\mu J=2^{-|V|}\!\sum_{x\in\mathbb{F}_2^V}a^x
           \left(\prod_v\widehat{A}_v^{\,x_v}\right)J .
 \label{eq:sm-entry-measure}
\end{equation}
On $JP$ the gauge configuration is zero and the invariant checks reduce to parent stabilizers. The Gauss operators preserve all central outcomes $+1$, where $V_\ell=I$ and $X_{t_\ell}$ fixes $|+\rangle_{t_\ell}$. Thus, the encoded branch reduces to Equation~\eqref{eq:prx273-gauging-sum} of the main text. Pairing $x$ with $x+\mathbf{1}$ gives the isometry and record probability in Equations~\eqref{eq:main-flat-certificate} and~\eqref{eq:prx273-record-probability} of the main text. For a full round let $r_Q$ be the measurement ancilla for $Q\in\mathcal{F}$, $b_Q$ its outcome and $\langle b_X|_r$ the corresponding product bra. The round has Kraus operator
\begin{equation}
 \begin{aligned}
 M_b&={}_r\!\langle b_X|
       \left[\prod_{Q\in\mathcal{F}}\operatorname{ctrl}_{r_Q}(Q)\right]
           |+\rangle_r\\
    &=\prod_{Q\in\mathcal{F}}\frac{I+b_Q Q}{2}.
 \end{aligned}
 \label{eq:sm-entry-full-sweep}
\end{equation}
Global commutation makes $M_b$ a joint projector. On encoded input $M_bJP={}_\mu\!\langle a^X|\mathcal{G}_a|+\rangle_\mu JP$ for a compatible record and vanishes otherwise. Compatibility requires Gauss outcomes $a$ and all other outcomes $+1$. Reset the measurement ancillas between rounds, while retaining the matter, gauge and completion qubits. Appendix~\ref{sec:smd4-native-extraction} gives the joint $D(D_4)$ flux and pinning circuit.

\emph{Resources and scheduling.} The weight and incidence bounds follow from Equation~\eqref{eq:prx273-completed-weights} of the main text and the derivation following it. Each conflict adds one completion qubit and one measurement ancilla. For qubit Clifford checks, use the controlled-gate decomposition and native gates specified in Section~\ref{sec:prx-spatial-costs}. This gives bounded depth on the check support and its assigned readout ancilla. Form the overlap graph of these circuits, joining circuits that share a qubit. Bounded check weight and incidence give this graph bounded degree. A constant coloring then schedules disjoint circuits in parallel and overlapping circuits serially, giving constant-depth rounds.

\subsection{Relation to the physical gauging map}
\label{sec:sm269-physical-gauging}
To implement the operator correspondences in Eq.~\eqref{eq:prx336-kw-map}, we couple each gauge qubit to the two auxiliary vertex qubits at the endpoints of its edge. The gauge qubits start with $Z_{g_e}=+1$. For each edge $e=(uv)$, apply two CNOT gates with controls $\mu_u,\mu_v$ and common target $g_e$. These gates extend each auxiliary-qubit operator $X_{\mu_v}$ onto the incident gauge qubits, producing the factors needed for the Gauss check. Their product over the graph is
\begin{equation}
 \bm{E}_{\mathrm{link}}=\prod_{e=(uv)}
 \operatorname{CNOT}_{\mu_u\to g_e}
 \operatorname{CNOT}_{\mu_v\to g_e}.
 \label{eq:prx273-link-circuit}
\end{equation}
Conjugating the enriched vertex constraint and the initial edge polarization gives
\begin{align}
 \bm{E}_{\mathrm{link}}(X_{\mu_v}U_v)\bm{E}_{\mathrm{link}}^\dagger
 &=X_{\mu_v}A_v,\label{eq:prx335-gauss-dressing}\\
 \bm{E}_{\mathrm{link}}Z_{g_e}\bm{E}_{\mathrm{link}}^\dagger
 &=Z_{\mu_u}Z_{g_e}Z_{\mu_v}.
 \label{eq:prx335-link-constraint}
\end{align}
The transformed vertex constraint is therefore $X_{\mu_v}A_v=+1$. The second equation transforms the initial edge polarization into the constraint $Z_{\mu_u}Z_{g_e}Z_{\mu_v}=+1$. Thus, $Z_{g_e}=Z_{\mu_u}Z_{\mu_v}$ on the coupled subspace, so each gauge link represents the relative $Z$ parity of the auxiliary vertex qubits at its endpoints. These dressed edge constraints commute with the enlarged Gauss operators $X_{\mu_v}A_v$, as in the usual lattice-gauging construction~\cite{Shirley2019Gauging}. Multiplying them around a closed path cancels every vertex factor,
\begin{equation}
 \prod_{e=(uv)\in\partial p}\!\bigl(Z_{\mu_u}Z_{g_e}Z_{\mu_v}\bigr)
 =\prod_{e\in\partial p}Z_{g_e}=B_p.
 \label{eq:prx335-flux-cancellation}
\end{equation}
Hence, the initial edge constraints ensure $B_p=+1$.

The auxiliary vertex qubits are temporary readout qubits in this implementation. Measuring their commuting operators $X_{\mu_v}$ with outcomes $a_v=\pm1$ turns the constraints $X_{\mu_v}A_v=+1$ into the Gauss-check conditions,
\begin{equation}
 X_{\mu_v}A_v=+1,\quad X_{\mu_v}=a_v
 \quad\Longrightarrow\quad A_v=a_v.
 \label{eq:prx335-gauss-reduction}
\end{equation}
In each outcome branch, the auxiliary qubits are fixed in product $X$ eigenstates and can be reset or reused. The remaining state is carried by the matter and gauge qubits. Measuring the auxiliary qubits is the nonunitary step of this implementation. The individual dressed edge constraints are not retained, but their closed-loop products $B_p$ commute with every $X_{\mu_v}$ and retain eigenvalue $+1$.

The factorization in Equation~\eqref{eq:prx273-physical-factorization} of the main text identifies the temporary vertex ancillas with the Gauss readouts above. It specifies the ideal instrument. Fault-distance bounds additionally use the stated gate order and schedule.

With the two added CNOTs, the completion qubits also supply a hook-fault flag under the input condition $B=X_t=+1$ of Lemma~\ref{lem:prx273-completion-flag} of the main text. Appendix~\ref{sec:sm269-completion-flag} derives this propagation through the central-check extraction circuit. The algebraic completion and encoded instrument apply independently of the spatial layout.

\subsection{Derivation from a blockwise Clifford operator}
\label{sec:sm-clifford-derivation}
Use the chain complex of Eq.~\eqref{eq:sm-chain} and the Clifford gauging construction of Ref.~\cite{Chang2026Magic}. The disjoint local operators $U_v$, with $U_v^2=I$, generate a symmetry $\mathcal{U}$ preserving $\mathcal{S}=\langle S_i\rangle$. Qubits outside the active blocks retain their original check factors. Write $S_{iv}$ for the local Pauli factor of $S_i$ on block $v$ and $A_i=\{v:U_v\ne I,\ S_{iv}\ne I\}$ for the active blocks it meets. Denote its modified check by $\widetilde{S}_i$. If $A_i$ is empty, set $\widetilde{S}_i=S_i$ and $\mathcal{N}_i=I$, with no paths. Otherwise, choose $r_i\in A_i$ and paths $p_i(v)$ for $v\in A_i$, with $\partial_1p_i(v)=v+r_i$ and $p_i(r_i)=0$. The nontrivial factors in $W_i(z)$ below come from $v\in A_i\setminus\{r_i\}$. For sums over all vertices, set $p_i(v)=0$ when $v\notin A_i$, where $0$ denotes the zero binary edge vector. For an edge configuration $z\in\mathbb{F}_2^E$, let $p_i(v)\cdot z$ be its parity along this path. The matter operation $W_i(z)$ controlled by the gauge configuration $z$, its action on $S_i$, the Gauss's law operator $A_v$ and the flux check $B_p$ of face $p$ are
\begin{align}
 a_{iv}(z)&=p_i(v)\cdot z,\qquad
 W_i(z)=\prod_{v\ne r_i}U_v^{a_{iv}(z)},\nonumber\\
 \widetilde{S}_i&=\sum_z|z\rangle\langle z|\otimes W_i(z)S_iW_i(z)^\dagger,
 \nonumber\\
 A_v&=U_vX_g(\delta v),\qquad
 B_p=Z_g(\partial_2p).
 \label{eq:general-raw-gauging}
\end{align}
Conditional conjugation preserves Hermiticity and unitarity.

The symmetry-invariant parent stabilizer is $N_i=S_i\mathcal{U}S_i\mathcal{U}\in\mathcal{S}$. Its invariant check $\mathcal{N}_i$ is obtained by replacing $S_i$ with $N_i$ in Eq.~\eqref{eq:general-raw-gauging}, using the same paths. Decompose $N_i$ into Hermitian local Pauli factors as $N_i=\bigotimes_vN_{iv}$. Define $\mu_{iv}\in\mathbb{Z}_4$ by $S_{iv}U_vS_{iv}U_v=i^{\mu_{iv}}N_{iv}$, choosing factor signs so that $\sum_v\mu_{iv}=0\pmod4$. The factors $i^{\mu_{iv}}$ arise from multiplying operators within block $v$. They determine the controlled phase gates in the modified check. With $\mathsf{S}(B)=P_+(B)+iP_-(B)$, Eq.~\eqref{eq:general-raw-gauging} becomes
\begin{align}
 \widetilde{S}_i={}&S_i\prod_{v\ne r_i}\mathsf{S}(Z_g(p_i(v)))^{\mu_{iv}}
 \nonumber\\[-2pt]
 &\times\prod_{v\ne r_i}\prod_{e\in p_i(v)}
                    \operatorname{ctrl}_{g_e}(N_{iv}),\label{eq:general-clifford-form}\\
 \mathcal{N}_i={}&N_i\prod_{v\ne r_i}Z_g(p_i(v))^{\mu_{iv}}.
 \label{eq:general-invariant-checks}
\end{align}
All displayed factors of $\widetilde{S}_i$ are Clifford. The invariant checks $\mathcal{N}_i$ are Paulis. The stated operator order in Eq.~\eqref{eq:general-clifford-form} retains the phase factors.

The commutation relations below are those of Corollary~II.11 of Ref.~\cite{Chang2026Magic}, expressed in our chain-complex notation.

\begin{lemma}[Central algebra of order-two gauging]
The $B_p$ and $\mathcal{N}_i$ commute with all raw checks. The only possible nonidentity group commutators, with $[D,E]_{\rm grp}=DED^\dagger E^\dagger$, are
\begin{equation}
 [\widetilde{S}_i,A_{r_i}]_{\rm grp}=\mathcal{N}_i,
 \qquad [\widetilde{S}_i,\widetilde{S}_j]_{\rm grp}=Z_g(\gamma_{ij}),
 \label{eq:general-raw-commutators}
\end{equation}
where $\partial_1\gamma_{ij}=0$.
\end{lemma}
\begin{proof}
Applying the raw Gauss's law operator at an active block other than the reference block changes only the corresponding path parity and is compensated by $U_v$. At the reference block it flips every path parity, replacing $S_i$ by $\mathcal{U}S_i\mathcal{U}=S_iN_i$. Since $U_vN_{iv}U_v=(-1)^{\mu_{iv}}N_{iv}$, the resulting invariant check is $\mathcal{N}_i$.

For Paulis $D,E$ on block $v$, let $\langle D,E\rangle$ be their binary symplectic pairing (one for anticommutation) and $J_{\rm sp}$ its matrix. Let $F_v$ be the binary symplectic action of $U_v$. Then, $F_v^2=I$ gives $(I+F_v)^TJ_{\rm sp}(I+F_v)=0$ and $(I+F_v)^TJ_{\rm sp}=J_{\rm sp}(I+F_v)$. Consequently, the $N_{iv}$ commute locally, and $\eta_{ij,v}=\langle S_{iv},N_{jv}\rangle =\langle N_{iv},S_{jv}\rangle$ has even global sum. The conditional Pauli commutator therefore has phase $(-1)^{\sum_v\eta_{ij,v}[a_{iv}(z)+a_{jv}(z)]}$, or
\begin{equation}
 \gamma_{ij}=\sum_v\eta_{ij,v}[p_i(v)+p_j(v)].
 \label{eq:sm-commutator-cycle}
\end{equation}
Its boundary vanishes by the even sum of $\eta_{ij,v}$. Each $\mathcal{N}_i$ commutes with every $\widetilde{S}_j$ because $N_i$ is a parent stabilizer and the local $N$ factors commute. In its commutator with $A_v$, the matter sign $(-1)^{\mu_{iv}}$ cancels the path-$Z$ sign. The same holds at the reference block because $\sum_v\mu_{iv}$ is even.
\end{proof}

Expressing every $\gamma_{ij}$ as a sum of face boundaries and retaining all $\mathcal{N}_i$ gives Equation~\eqref{eq:main-central} of the main text. Edge assignments with even parity around each chosen face are called flat. Those of the form $\delta x$ are called exact. Flat assignments modulo exact ones form $H^1=\ker\partial_2^T/\operatorname{im}\delta$. Initialization selects exact assignments. Local flux checks alone do not test parity around winding cycles that are not sums of the chosen face boundaries.

\paragraph{Paramagnet and two-copy examples.}
\label{ex:paramagnet}
For the paramagnet $S_v=X_v$, $\mathcal{U}=\prod_vX_v$, the conflict graph has no edges. The checks are $X_v$, $A_v=X_v\prod_{e\ni v}X_{g_e}$ and plaquette $Z$ flux checks. Fixing and discarding the matter $X_v=+1$ qubits leaves the toric code on a toric cellulation. For two identical Calderbank--Shor--Steane (CSS) blocks $a,b$, take transversal CNOT $\mathcal{U}=\prod_v\operatorname{CNOT}(a_v\to b_v)$. Let $x,z$ be supports of rows of the $X$- and $Z$-check matrices, respectively. Write $X_a(x)=\prod_{v\in x}X_{a_v}$, and use the same convention for $b$ and for $Z$. For the chosen reference blocks $r_x,r_z$ and paths $p_x(v),p_z(v)$ one may use
\begin{align}
 C_x^X&=X_a(x)\prod_{v\in x\setminus\{r_x\}}
                \prod_{e\in p_x(v)}\operatorname{CNOT}(g_e\to b_v),\nonumber\\
 C_z^Z&=Z_b(z)\prod_{v\in z\setminus\{r_z\}}
                \prod_{e\in p_z(v)}\operatorname{CZ}(g_e,a_v),
 \label{eq:general-css-checks}
\end{align}
with invariant checks $X_b(x),Z_a(z)$. The commutator of $C_x^X$ and $C_z^Z$ is the product of gauge-qubit $Z$ operators on $\sum_{v\in x\cap z}[p_x(v)+p_z(v)]$, a cycle by CSS orthogonality. Their commutators with the Gauss's law operators at the respective reference blocks give the two invariant checks. This applies to any CSS family with a suitable gauging graph and measures the collective transversal logical CNOT eigenvalue. Two copies of any Pauli code similarly admit blockwise SWAP, with invariant $N_i=S_i^aS_i^b$. The controlled gates in Eq.~\eqref{eq:general-clifford-form} then apply the paired local Pauli product $S_{iv}^aS_{iv}^b$, controlled by each gauge qubit along the chosen path to block $v$.

\subsection{Implementation algorithm}
\label{sec:sm-implementation}
This algorithm constructs and implements Protocol~\ref{alg:commuting-gauging} of the main text for order-two Clifford symmetries.

\emph{Input.} Parent stabilizer generators $S_i$, an encoded state $\rho=P\rho P$, and disjoint bounded-support Clifford factors $U_v$ satisfying $U_v^2=I$, whose product $\mathcal{U}$ preserves $P$.

\emph{Output.} A symmetry outcome $s=\pm1$ and the corresponding postmeasurement state on the parent qubits.

\begin{protocolsteps}
\item \emph{Construct the raw checks.} Choose a connected gauging graph $G$, a reference vertex $v_0$, and an arbitrary edge-path from $v_0$ to each vertex. For each $S_i$, choose a reference block and paths as in Eq.~\eqref{eq:general-raw-gauging}. A generator disjoint from the active vertices is unchanged. Construct the modified generators $\widetilde{S}_i$ and invariant checks $\mathcal{N}_i$ using Eqs.~\eqref{eq:general-clifford-form} and~\eqref{eq:general-invariant-checks}, retaining their phase factors. Compute the cycles in Eq.~\eqref{eq:sm-commutator-cycle} and choose flux faces whose boundaries span them. Impose the locality conditions of Appendix~\ref{sec:sm-choose-graph}.

\item \emph{Complete the checks.} Enumerate the $A_v$ and $\widetilde{S}_i$ as $C_i$, and include every flux check $B_p$ and invariant check $\mathcal{N}_i$ among the central checks $\mathsf{K}_a$. For each $\mathsf{K}_{ij}=C_iC_jC_iC_j\ne I$, orient one edge of $\mathcal{H}$ and assign a separate completion qubit $t_e$. Form the completed checks and pinning checks using Equations~\eqref{eq:prx273-completion-factor}, \eqref{eq:main-completion} and~\eqref{eq:main-pinning} of the main text. The commuting-check family is $\mathcal{F}=\{\widehat{C}_i,\mathsf{K}_a,T_e\}$.

\item \emph{Fix the extraction schedule.} Assign one measurement ancilla $r_Q$ to each $Q\in\mathcal{F}$ and compile its controlled-$Q$ circuit. Preserve the operator order in Eq.~\eqref{eq:general-clifford-form}. Serialize extraction circuits that overlap. Disjoint circuits may run in parallel. Fix the schedule before collecting measurement outcomes.

\item \emph{Initialize and measure.} Prepare every gauge qubit in $|0\rangle$ and every completion qubit in $|+\rangle$. For each check $Q$ in the chosen schedule, prepare $r_Q$ in $|+\rangle$, apply $\operatorname{ctrl}_{r_Q}(Q)$, and measure $r_Q$ in $X$, recording $b_Q=\pm1$. Denote the completed Gauss outcomes by $a_v$ and set $s=\prod_va_v$.

\item \emph{Ungauge and correct.} Measure every gauge qubit in $Z$, obtaining bits $z_e$. Set $x_{v_0}=0$ and sum $z_e$ modulo two along the chosen path from $v_0$ to each vertex $v$ to obtain $x_v$. Check $z_e=x_u+x_v\bmod2$ on every edge $e=uv$. Fault-free outcomes satisfy this condition. Apply $U_v^\dagger$ at each vertex with $x_v=1$, discard the gauge and completion qubits, and retain $s$ and the parent qubits.
\end{protocolsteps}

In the absence of faults, summing the other measurement records at fixed $s$ gives the unnormalized output $\Pi_s(\mathcal{U})\rho\Pi_s(\mathcal{U})$, where $\Pi_s(\mathcal{U})=(I+s\mathcal{U})/2$.

For the circuit-distance analysis, insert parent-check intervals before initialization and after correction, and repeat the check measurements in step~4 for $R$ rounds. Reset the measurement ancillas while retaining the gauge and completion qubits. Use the assumptions and detection tests of Theorem~\ref{thm:end-distance} of the main text, proved in Appendix~\ref{sec:graph-distance}. Appendix~\ref{sec:sm297-boundary-recovery} gives quantum recovery operations for detected boundary errors. Proposition~\ref{prop:zn-patch-checks} supplies alternative commuting checks for non-Clifford symmetries, and Appendix~\ref{sec:sm-zn-proof} gives the general $\mathbb{Z}_N$ protocol.

\section{\texorpdfstring{Qudit completion and $\mathbb{Z}_N$ gauging}{Qudit completion and ZN gauging}}
\label{sec:sm-zn-proof}

This appendix extends completion to finite-order central commutators, bounds the Clifford-hierarchy levels of the completed checks and their extraction gates, and gives the $\mathbb{Z}_N$ logical-measurement protocol. Appendix~\ref{sec:sm-zn-distance} proves its circuit-distance bound. The protocol and distance bound also apply to the constructions without completion in Section~\ref{sec:sm-without-completion}.

\subsection{Completion qudits for finite-order checks}
\label{sec:sm-qudit-completion}
The completion construction extends to finite-order checks whose commutators are central. Let $C_i$ be unitaries with $C_i^{m_i}=I$ and let $\mathsf{K}_a$ be mutually commuting finite-order unitaries that commute with every $C_i$. Assume
\begin{equation}
 C_iC_j=\mathsf{K}_{ij}C_jC_i,\qquad
 \mathsf{K}_{ij}\in\langle\mathsf{K}_a\rangle,
 \label{eq:qudit-central-input}
\end{equation}
and that the common $+1$ space of all $C_i$ and $\mathsf{K}_a$ has nonzero projector $P_g$.

Orient each nontrivial commutator as $e:i\to j$ and put $\mathsf{K}_e=\mathsf{K}_{ij}$. Let $n_e=\operatorname{ord}(\mathsf{K}_e)$ be its operator order, the smallest positive integer for which $\mathsf{K}_e^{n_e}=I$. Add one $n_e$-level completion qudit $t_e$, initialized in $|+_{n_e}\rangle=n_e^{-1/2}\sum_{r=0}^{n_e-1}|r\rangle$. Write $X_{n_e,t_e}|r\rangle=|r+1\bmod n_e\rangle$. Let $E_e=n_e^{-1}\sum_{r=0}^{n_e-1}\mathsf{K}_e^r$ be the projector onto the eigenvalue-one sector of $\mathsf{K}_e$. The pinning check is
\begin{align}
 V_e&=\sum_{r=0}^{n_e-1}\mathsf{K}_e^r\otimes|r\rangle\langle r|_{t_e},\nonumber\\
 T_e&=E_eX_{n_e,t_e}+(I-E_e),\nonumber\\
 \widehat{C}_i&=C_i\prod_{e:i\to j}X_{n_e,t_e}
                    \prod_{e:j\to i}V_e.
 \label{eq:qudit-completion}
\end{align}
Both $V_e$ and $T_e$ are unitary. The first applies powers of $\mathsf{K}_e$ controlled by $t_e$. The pinning check $T_e$ shifts $t_e$ only in the eigenvalue-one sector of $\mathsf{K}_e$. To write it as a gate product, choose a unitary $D_e$ on the support of $\mathsf{K}_e$ with $D_e\mathsf{K}_eD_e^\dagger=\sum_x\kappa_x|x\rangle\langle x|$. Then
\begin{equation}
 T_e=D_e^\dagger
 \left[\prod_{x:\kappa_x=1}
 \operatorname{ctrl}_x(X_{n_e,t_e})\right]D_e,
 \label{eq:qudit-pinning-gate-product}
\end{equation}
Here $\operatorname{ctrl}_x$ applies the shift only when the control qudits have computational-basis label $x$. This product uses no additional qudits. For the prime-dimensional Pauli labels considered below, $D_e$ can be Clifford, sending $\mathsf{K}_e$ to $Z$ on one qudit. The product then reduces to a single shift controlled on that qudit being in $|0\rangle$. The measurement checks are $\widehat{C}_i$, $T_e$, and $\mathsf{K}_a$.

\begin{theorem}[Completion with qudits]
\label{thm:qudit-completion}
The measured checks are globally commuting unitaries, with $\widehat{C}_i^{m_i}=I$ and $T_e^{n_e}=I$. Their common $+1$ projector obeys
\begin{equation}
 \begin{aligned}
 \widehat{P}_g&=P_g\otimes\bigotimes_e|+_{n_e}\rangle\langle+_{n_e}|,\\
 d(\widehat{P}_g)&=d(P_g).
 \end{aligned}
 \label{eq:qudit-completion-projector}
\end{equation}
In the common $+1$ sector of the central checks with completion qudits in this product state, measurements preserve every compatible joint raw-check outcome and its conditional state.
\end{theorem}
\begin{proof}
Centrality implies $\mathsf{K}_e^{m_i}=\mathsf{K}_e^{m_j}=I$, so $n_e$ divides both endpoint orders. The identity $X_{n_e,t_e}V_e=\mathsf{K}_e^{-1}V_eX_{n_e,t_e}$ cancels the raw endpoint commutator. The remaining commutation and order statements follow as in the proof of Theorem~\ref{thm:prx273-central-completion} of the main text. In the common $+1$ sector of the central checks, $V_e=I$ and $T_e=X_{n_e,t_e}$, so pinning fixes the completion register to $|\Omega\rangle=\bigotimes_e|+_{n_e}\rangle$. The projector, compatible-outcome and distance conclusions then follow from the same restriction and error-compression arguments.
\end{proof}

\emph{Application to $\mathbb{Z}_N$ gauging.} If all raw checks have order dividing $N$, then $n_e\mid N$. An $N$-level completion qudit can also be used on every conflict by replacing $n_e$ with $N$ throughout Eq.~\eqref{eq:qudit-completion} and preparing $|+_N\rangle$. Since $\mathsf{K}_e^N=I$, the averaged operator $E_e$ is unchanged, and the completed checks have orders dividing $N$. The construction applies when the raw family obeys Eq.~\eqref{eq:qudit-central-input} and its joint projectors in the common $+1$ sector of the central checks realize the intended gauged parent code, including every allowed Gauss outcome. Completion then preserves these projectors with a fixed product state appended. The ideal measurement and circuit fault-distance analysis below apply to this family. The required projector identity for each Gauss outcome is given in Eq.~\eqref{eq:sm-zn-projector}, and circuit fault distance still requires the stated bound on fault propagation.

\emph{Resources.} Let $w,q$ bound raw check weight and incidence, including central checks. Each check has at most $\Delta=w(q-1)$ conflicts. Completed checks have weight at most $w+(2w+1)\Delta$ and pinning checks at most $2w+1$. Incidence remains bounded by the same counting as in Appendix~\ref{sec:sm-place-ancillas}. Each conflict adds one completion qudit and one measurement ancilla for its pinning check. An order-$m$ check is measured using one $m$-level measurement ancilla, controlled powers and Fourier readout. Bounded orders of the raw and central checks keep the added dimensions bounded when each completion qudit has dimension $n_e$. The uniform $N$-level choice requires bounded $N$. For prime-dimensional qudits with Pauli central checks, we bound the Clifford-hierarchy cost in the next subsection.

\subsection{Hierarchy costs for prime-dimensional completion}
\label{sec:sm-qudit-hierarchy}
Completion can trade additional qudits for a hierarchy bound that does not grow with the number of conflicts at a check.

Let every qudit have the same prime dimension $p$, write $\omega=e^{2\pi i/p}$, and let $\mathcal{C}_k$ denote hierarchy level $k$, with $\mathcal{C}_1$ the Pauli operators up to phase.

\begin{theorem}[Completion with bounded hierarchy level]
\label{thm:qudit-hierarchy-control}
Assume the central algebra of Eq.~\eqref{eq:qudit-central-input}, with Pauli central checks and a nonzero projector $P_g$ onto the common $+1$ eigenspace of all raw and central checks. Suppose that $C_i^p=I$ and $C_i\in\mathcal{C}_h$ for every raw check, where $h\ge2$. Add one $p$-level completion qudit per nontrivial commutator and use Eq.~\eqref{eq:qudit-completion}. Then the measured family is globally commuting and preserves the projector, distance and compatible measurement branches of Theorem~\ref{thm:qudit-completion}. Moreover,
\begin{equation}
 \widehat{C}_i\in\mathcal{C}_h,\qquad
 T_e\in\mathcal{C}_p\setminus\mathcal{C}_{p-1}.
 \label{eq:sm-qudit-hierarchy-cap}
\end{equation}
Thus, every measured check has hierarchy level at most $\max\{h,p\}$, independently of the number of conflicts at a raw check.

If the raw controlled-power operators admit exact circuits whose gates have hierarchy level at most $h_{\rm raw}$, the completed extraction circuits use gates at hierarchy level at most $\max\{h_{\rm raw},p+1\}$.
\end{theorem}
\begin{proof}
The algebraic conclusions follow from Theorem~\ref{thm:qudit-completion}. A nontrivial commutator $K=\mathsf{K}_e$ is a nonscalar Pauli of order $p$. A nonidentity scalar label would contradict the nonzero common $+1$ eigenspace. Each added shift $X_t$ is Pauli, and $V_e=\sum_r K^r\otimes|r\rangle\langle r|_t$ is Clifford. Hence, the product of completion factors multiplying each $C_i$ is Clifford. Every level $\mathcal{C}_h$ with $h\ge2$ is invariant under left and right multiplication by Clifford operators, proving the first inclusion without using closure of higher hierarchy levels under general multiplication.

A Clifford change of basis sends $K$ to $Z$ on one qudit. Fourier-transforming the completion qudit sends $X_t$ to $Z_t$. In these bases, with $k,u,m\in\mathbb{F}_p$, let $\omega^k$ be the eigenvalue of $K$, and let $u$ and $m$ be the basis labels of the transformed completion qudit and control ancilla, respectively. The pinning check and its controlled powers have phases
\begin{equation}
 \begin{aligned}
 T:\quad&\omega^{u(1-k^{p-1})},\\
 \sum_{m=0}^{p-1}|m\rangle\langle m|\otimes T^m:\quad
 &\omega^{mu(1-k^{p-1})}.
 \end{aligned}
 \label{eq:sm-qudit-pin-hierarchy}
\end{equation}
The factor $1-k^{p-1}$ is one exactly when $K$ has eigenvalue one. These reduced polynomials have degrees $p$ and $p+1$. The diagonal-hierarchy classification~\cite{Cui2017Diagonal} therefore puts the pinning check exactly in $\mathcal{C}_p\setminus\mathcal{C}_{p-1}$ and its controlled powers exactly in $\mathcal{C}_{p+1}\setminus\mathcal{C}_p$.

Measure each check with a $p$-level measurement ancilla, controlled powers and Fourier readout. The factors within a completed check $\widehat{C}_i$ commute, so its controlled powers factor into those of $C_i$ and its completion factors. Controlled powers of $X_t$ and $V_e$ have levels at most two and three, respectively. Central-check extraction is Clifford. Combining these factors with the assumed raw circuits and the pinning-check extraction gives the stated gate bound.
\end{proof}

\paragraph{Clifford raw checks.}
When every raw check $C_i$ is Clifford, every corresponding completed check $\widehat{C}_i$ is Clifford. The pinning checks have levels two for qubits and three for qutrits, with direct controlled-power extraction at levels three and four, respectively. For odd $p$ and raw checks on at most $w$ qudits, a sufficient bound is
\begin{equation}
 h_{\rm raw}\le p(2w+2)+1.
 \label{eq:sm-qudit-raw-hierarchy-bound}
\end{equation}
To prove this, let $\mathcal{P}_w$ be the Pauli group with $p$th-root phases. The group $H=\langle\mathcal{P}_w,C_i\rangle$ has order at most $p^{2w+2}$ because $C_i$ normalizes $\mathcal{P}_w$ and $C_i^p=I$. The controlled-power operator $\sum_m|m\rangle\langle m|\otimes C_i^m$, all target Paulis, and both measurement-qudit Paulis lie in $H^p\rtimes\mathbb{Z}_p$. The base group acts independently in the $p$ computational sectors of the measurement qudit, and $\mathbb{Z}_p$ cyclically permutes those sectors. This is a finite $p$-group of order at most $p^{p(2w+2)+1}$. Its lower central series terminates within $p(2w+2)+1$ steps. The group contains every physical Pauli on these qudits, and an operator commuting with all of them is scalar. Successive Pauli commutators therefore give Eq.~\eqref{eq:sm-qudit-raw-hierarchy-bound} by the recursive hierarchy definition.

For $p=2$, the sharper bound $h_{\rm raw}\le3$ follows directly from the Clifford property and $C_i^2=I$. Conjugation by controlled-$C_i$ sends control $X$ to $X\otimes C_i$ and each target Pauli to a target Pauli times a controlled Pauli. These images are Clifford. Thus, fixed $p$ and bounded raw-check support give an extraction bound independent of code size.

\paragraph{Dependence on the conflict assignment.}
The following local family illustrates the conflict-independent bound of Theorem~\ref{thm:qudit-hierarchy-control} and the hierarchy growth for one particular construction without completion. Take $C_0=X_a$ and $C_j=\operatorname{CZ}_{a,b_j}$ for $1\le j\le m$, where $\operatorname{CZ}|x,y\rangle=\omega^{xy}|x,y\rangle$. Include the independent central checks $K_j=Z_{b_j}^{-1}$. The only nontrivial raw commutators are $[C_0,C_j]_{\rm grp}=K_j$.

One way to make this family commute is to replace $C_0$ by an operator that acts as $C_0$ only when all $K_j$ have eigenvalue one and acts as identity otherwise. After Fourier transformation on $a$, let $u\in\mathbb{F}_p$ label its transformed basis and let $\omega^{k_j}$ be the eigenvalue of $K_j$, with $k_j\in\mathbb{F}_p$. This replacement has phase
\begin{equation}
 \omega^{u\prod_{j=1}^{m}(1-k_j^{p-1})}.
 \label{eq:sm-qudit-disabling-cost}
\end{equation}
Its hierarchy level is exactly $1+m(p-1)$, and its direct controlled-power gate has level $2+m(p-1)$. Completion adds $m$ completion qudits and gives Clifford checks $\widehat{C}_i$ together with one level-$p$ pinning check per conflict. All measured checks then have level at most $\max\{2,p\}$, and their extraction uses gates at level at most $p+1$. This local comparison uses the one-dimensional common $+1$ space spanned by $|+\rangle_a|0\rangle_{b_1}\cdots|0\rangle_{b_m}$. For this assignment of all controls to $C_0$, completion keeps the measured-check hierarchy bounded as the number of conflicts grows. These are the hierarchy levels of the full check and its direct controlled-power operator.

\subsection{Gauging with the completed checks}
\label{sec:sm-zn-measurement}

Let $\mathcal{U}=\prod_{v\in V}U_v$ preserve the parent projector $P$, with commuting unitary factors satisfying $U_v^N=I$ on a finite connected graph $G=(V,E)$. We gauge the resulting $\mathbb{Z}_N$ action, $N\ge2$. A faithful action has symmetry order $N$. The ideal measurement permits overlapping factors~\cite{WilliamsonYoder2026,Chang2026Magic}. Matter qudits can have arbitrary finite dimensions $d_{\rm m}$. Each edge carries an $N$-level gauge qudit.

Choose a reference vertex $v_0$ and orient each edge from $s(e)$ to $t(e)$. For vertex labels $y\in\mathbb{Z}_N^V$, put $U(y)=\prod_vU_v^{y_v}$ and $(\delta y)_e=y_{s(e)}-y_{t(e)}$. Write $\omega=e^{2\pi i/N}$, $X_N|b\rangle=|b+1\rangle$, $Z_N|b\rangle=\omega^b|b\rangle$, and $X_g(k)=\prod_eX_{N,e}^{k_e}$, likewise $Z_g(k)$. Let $\partial_2$ map each chosen oriented face to its signed edge boundary, with coefficients modulo $N$. For the vertex basis vector $v$ and oriented face boundary $\partial_2f$, the Gauss's law operators and flux checks obey
\begin{equation}
 A_v=U_vX_g(\delta v),\qquad B_f=Z_g(\partial_2f),
 \qquad\prod_v A_v=\mathcal{U} .
 \label{eq:em-zn-gauss}
\end{equation}
Commutation follows from $[U_v,U_w]=0$ and $\partial_2^T\delta=0$. Opposite edge shifts cancel in the product over vertices.

For an encoded input $\rho=P\rho P$, prepare each gauge qudit in $|0\rangle$ and each completion qudit $t_e$ in $|+_{n_e}\rangle$. Measure the full completed family from Theorem~\ref{thm:qudit-completion}, using one measurement ancilla per check. The completed Gauss outcomes $\omega^{a_v}$ give the reported eigenvalue $\omega^\alpha$, with $\alpha=\sum_v a_v\bmod N$. In the absence of faults, all other checks give $+1$. To ungauge, measure the gauge qudits in $Z_N$ with outcomes $\omega^{z_e}$. Set $y_{v_0}=0$ and determine each $y_v$ along an arbitrary edge-path from $v_0$ to $v$, using $y_{s(e)}-y_{t(e)}=z_e$ modulo $N$. Then test this equation on every edge. Apply $U(y)^\dagger$ and discard the gauge and completion qudits. If measurements are repeated, retain the gauge and completion qudits through all rounds and reset the measurement ancillas.

Let $P_{\rm ex}$ project onto link strings $z\in\operatorname{im}\delta$. The gauging isometry is
\begin{equation}
 \Gamma_a|\psi\rangle=N^{-(|V|-1)/2}
 \sum_{x:x_{v_0}=0}\omega^{-a\cdot x}
 |\delta x\rangle_g U(x)|\psi\rangle.
 \label{eq:sm-zn-gamma}
\end{equation}
Distinct $x$ with the reference value fixed have distinct gauge strings. Include in $\Gamma_a$ the fixed completion product state $|\mathrm{aux}\rangle$. Let $\Pi_\alpha(\mathcal{U})$ project onto the $\omega^\alpha$ eigenspace of $\mathcal{U}$. For a globally commuting family $\mathcal{F}$, let $\widetilde{Q}_a$ be the joint projector for Gauss's law outcomes $a$ and outcome $+1$ on every other check. We require
\begin{equation}
 \widetilde{Q}_a P_{\rm ex}=Q_a
     =\Gamma_a\Pi_\alpha(\mathcal{U})P\Gamma_a^\dagger.
 \label{eq:sm-zn-projector}
\end{equation}
Taking adjoints gives $P_{\rm ex}\widetilde{Q}_a=Q_a$. This identity is a condition on the raw gauging construction. Theorem~\ref{thm:qudit-completion} preserves it when the raw checks realize it in the common $+1$ sector of the central checks.

To obtain the ideal action, let $J$ append the initialized gauge and completion qudits and set $c_V=N^{-(|V|-1)/2}$. Equation~\eqref{eq:sm-zn-gamma} gives $\Gamma_a^\dagger JP=c_VP$ and ${}_g\!\langle\delta y|\Gamma_a=c_V\omega^{-a\cdot y}U(y)\otimes|\mathrm{aux}\rangle$. Substituting these contractions into Eq.~\eqref{eq:sm-zn-projector} and applying $U(y)^\dagger$ gives the corrected matter branch for $z=\delta y$,
\begin{equation}
 K_{a,z}P=N^{1-|V|}\omega^{-a\cdot y}\Pi_\alpha(\mathcal{U})P.
 \label{eq:em-zn-instrument}
\end{equation}
Nonexact $z$ has zero amplitude. The completion qudits remain in $|\mathrm{aux}\rangle$, so discarding them does not change this branch. At fixed $\alpha$ there are $N^{|V|-1}$ vectors $a$ and the same number of exact records $z$. Summing their branches therefore gives the unnormalized output $\Pi_\alpha(\mathcal{U})\rho\Pi_\alpha(\mathcal{U})$. Summing over $\alpha$ exhausts probability one, so all other checks have outcome $+1$ on ideal encoded input.

The same argument applies to the commuting families without completion in Section~\ref{sec:sm-without-completion}, with the completion state omitted. The circuit-distance analysis in Appendix~\ref{sec:sm-zn-distance} uses the same projector condition.

\subsection{Examples and scope of the measurement protocol}
On the Steane code $\mathcal{U}=S^{\otimes7}$, $S=\operatorname{diag}(1,i)$, has order four and acts as $\overline{S}^\dagger$~\cite{Gottesman1997Steane}. Basis strings have weights $0,4$ in $|0_L\rangle$ and $3,7$ in $|1_L\rangle$. The gauge links have four levels. On encoded inputs the logical measurement returns $1$ or $-i$. On three suitably stacked 3D surface codes, a product of CCZs on disjoint vertex triples implements logical CCZ~\cite{VasmerBrowne2019CCZ}. Gauging measures its sign with qubit links, while direct control of each CCZ gives a fourth-level unitary. Thus, symmetry order sets the link dimension used here, while the local symmetry factors set the hierarchy level of their direct controls.

\paragraph{A spatial constraint on gauging.}
The qubit toric code admits a depth-three Clifford circuit of order four that exchanges $e$ and $m$. Its square fixes the anyon labels but acts nontrivially on microscopic operators. This symmetry cannot be made onsite even after adding onsite ancillas~\cite{Shirley2025Onsite}.

\begin{proposition}[Onsite symmetry after adding ancillas]
\label{prop:stable-onsite}
An action $\mathcal{U}=\prod_v U_v$ with mutually commuting factors of bounded spatial range and bounded overlap, each satisfying $U_v^N=I$, becomes onsite after adding onsite regular representations and conjugating by a finite-depth local circuit.
\end{proposition}
\begin{proof}
Append one $N$-level ancilla $a_v$ per factor and let $V=\prod_vX_{a_v}$. The circuit $W=\prod_v(\sum_k|k\rangle\langle k|_{a_v}\otimes U_v^k)$ has bounded depth by coloring the overlap graph. Incrementing $k$ gives $WVW^\dagger=V\mathcal{U}$, including at wraparound since $U_v^N=I$.
\end{proof}
The toric-code symmetry therefore lacks the spatially local commuting factorization used in Appendix~\ref{sec:sm-zn-measurement}.

\Needspace{5\baselineskip}
\section{Fault propagation under commuting checks}
\label{sec:locality-sm}
Let $S_i$ be local, mutually commuting finite-order unitary checks with $S_iP=P$. Here $P$ denotes the code projector protected by this fixed check family. Their code qudits exclude measurement ancillas and, during gauging, include matter, gauge and completion qudits. We bound fault propagation for this fixed family, then apply the result to surgery in Appendix~\ref{sec:graph-distance}.

Fix a measurement record $m$, the noise Kraus labels and any unobserved ancillary discard labels, with branch operator $K_m$. Each check circuit starts with fresh or reset ancillary qudits. Serialize circuits sharing code qudits. Preparation and reset faults may alter the prepared state, and readout faults may alter the quantum-to-classical map. No auxiliary quantum output is carried into the next circuit. After contracting all ancillary qudits, represent each faulty circuit, including its preparation, reset and readout, by $F_a$ on its full code-qudit support $R_a$. Include standalone preparation and idle faults among the $F_a$, in time order. Let $f$ count these fault operators $F_a$. It does not exceed the number of faulty native locations.

For a set $R$ of qudits of the code define
\begin{equation}
 N(R)=R\cup\!\bigcup_{i:\,\operatorname{supp}S_i\cap R\ne\varnothing}
                         \operatorname{supp}S_i,
 \qquad C=\max_a|N(R_a)|.
\end{equation}
Here, $C$ is uniform over all allowed faulty-operation supports.

\begin{theorem}[Support of propagated faults]
On every nonzero branch there exist propagated fault operators $O_a$, one for each $F_a$, such that
\begin{equation}
 K_mP=O_f\cdots O_1P,\qquad
 \operatorname{supp}O_a\subseteq N(R_a).
 \label{eq:local-dressing}
\end{equation}
The $O_a$ may be nonunitary. Their support is independent of the number of intervening fault-free rounds.
\end{theorem}
\begin{proof}
Let $\Pi_i(s)$ be the spectral projector of $S_i$ for the recorded eigenvalue $s$. For Hermitian checks, $\Pi_i(s)=(I+sS_i)/2$ with $s=\pm1$. Assign each ideal projector $\Pi_i(s)$ to the latest preceding $F_a$ whose original support meets $S_i$, or to the initial code if there is none. Move it backwards to that operator. It commutes with all intervening ideal projectors and with every intervening $F_b$ by disjointness. The projectors assigned to $F_a$ form a commuting product $Q_a$. Identical projectors are idempotent, while distinct outcomes of a check give zero. Initial projectors act as identity or zero on $P$. Taking $O_a=Q_aF_a$ proves the claim. Assignment uses the original $R_a$, never a recursively enlarged support, because all projector factors commute.
\end{proof}

Thus, errors from at most $t$ faulty operations lie in the span of operators on at most $Ct$ qudits of the code. If $P$ detects every operator of support below $d$, then $PE^\dagger FP$ is a scalar multiple of $P$ for any two such errors $E,F$ whenever $2Ct<d$, proving the Knill--Laflamme conditions\cite{KnillLaflamme1997}. These conditions guarantee the existence of ideal recovery for the fixed code. Appendix~\ref{sec:graph-distance} extends the analysis through gauging and ungauging to protect the logical measurement.

\begin{proposition}[Locality of temporal detection events]
For a fixed repeated schedule with bounded check-circuit support, check incidence and round depth in native circuit time steps, let $d_{i,r}$ be one if the measurement outcomes of $S_i$ in rounds $r-1$ and $r$ differ, and zero otherwise. Start the comparison in the second round. Write $m$ for the measurement record and $F_0$ for the set of faulty native locations. Each location $a$ has a bounded spacetime neighborhood of detectors $N_{\rm det}(a)$ such that every nonzero branch satisfies
\begin{equation}
 \{(i,r):d_{i,r}=1\}\subseteq\bigcup_{a\in F_0}N_{\rm det}(a).
 \label{eq:local-detectors}
\end{equation}
If each neighborhood contains at most $C_{\rm det}$ detectors and each detector has at most $D_{\rm det}$ candidate fault locations, consider local stochastic faults of parameter $p$, so $\Pr[T\subseteq F_0]\le p^{|T|}$ for every set $T$ of locations. For any detector set $Q$, the set of detection events $D(m)=\{(i,r):d_{i,r}=1\}$ obeys
\begin{equation}
 \Pr[Q\subseteq D(m)]
 \le\min\{1,(D_{\rm det} p^{1/C_{\rm det}})^{|Q|}\}.
 \label{eq:detector-stochastic-bound}
\end{equation}
\end{proposition}
\begin{proof}
If two consecutive measurements of $S_i$ are fault-free and neither an intervening faulty check circuit nor a standalone faulty operation meets its support, the first projector commutes with every intervening ideal check projector and fault operator $F_a$. Different measurement outcomes then have zero joint probability. A detection event thus requires a fault in either endpoint measurement circuit, an intervening faulty check circuit whose support meets $S_i$, or an intervening standalone fault on its support. Bounded circuit support, check incidence and round depth bound both the number of detectors affected by one fault and the number of candidate faults for each detector. A flipped measurement outcome may affect both adjacent detectors. For a set $Q$, assign each detection event one of its at most $D_{\rm det}$ candidate faults. There are at most $D_{\rm det}^{|Q|}$ assignments, each requiring at least $|Q|/C_{\rm det}$ distinct faults. The local-stochastic assumption and a union bound prove Eq.~\eqref{eq:detector-stochastic-bound}.
\end{proof}

\section{Fault tolerance of logical measurement}
\label{sec:graph-distance}

We prove Theorem~\ref{thm:end-distance} of the main text, including both surgery interfaces. Appendix~\ref{sec:sm-zn-distance} extends the bound to $\mathbb{Z}_N$. Let $P$ be the distance-$d$ qubit Pauli parent projector and $\mathcal{U}=\prod_{v\in A}U_v$ a code-preserving symmetry with disjoint blocks of at most $b$ qubits and $U_v^2=I$. The connected graph $G=(V,E)$ may include vertices $v\in V\setminus A$ with $U_v=I$. Count each spectator matter qubit as a one-qubit block, without requiring a graph vertex. For $Y\subseteq V$, let $\delta Y$ be its edge cut. Define the relative expansion $h_A$ and the minimum weight $\lambda$ of a flat gauge pattern that is not exact by

\begin{align}
 h_A&=\min_{0<|Y\cap A|<|A|}
 \frac{|\delta Y|}{\min\{|Y\cap A|,|A\setminus Y|\}},
 \label{eq:graph-expansion}\\
 \lambda&=\min\{|\eta|:\partial_2^T\eta=0,\
                  \eta\notin\operatorname{im}\delta\}.
 \label{eq:gauge-holonomy-distance}
\end{align}
An empty minimum is $\infty$. The complete logical measurement uses the detection tests, ideal encoded input and ideal final parent-code projection of Theorem~\ref{thm:end-distance} of the main text. All prescribed fluxes are measured. The measured family must satisfy Equation~\eqref{eq:em-gauged-projector} and have bounded support and incidence, including the measurement circuits. For the patch construction this requires bounded patches, tree sizes, overlap and routing congestion. For Clifford completion the prescribed faces must also span all raw commutator cycles.

\subsection{Graph parameters and proof of the circuit-distance theorem}
\label{sec:v278-general-distance-parameters}
Let $P_{\mathrm{ex}}$ project onto gauge strings $z=\delta x$, where $z_{uv}=x_u+x_v\bmod2$. Write $U(x)=\prod_vU_v^{x_v}$ and let $\widetilde{Q}_a$ be the joint projector for Gauss outcomes $a=(a_v)$ and positive outcomes of every other gauged check. With $s=\prod_va_v$, the required gauging identity is
\begin{equation}
 Q_a\coloneqq\widetilde{Q}_aP_{\mathrm{ex}}
 =\Gamma_a\Pi_sP\Gamma_a^{\dagger}.
 \label{eq:em-gauged-projector}
\end{equation}
The isometry $\Gamma_a$ is given in Equation~\eqref{eq:prx273-rooted-isometry} of the main text. The identity says that on an exact gauge string the matter state belongs to $U(x)\operatorname{im}P$. It holds for the completed gauging construction and the alternatives in Section~\ref{sec:sm-without-completion}.

The parameters in Eqs.~\eqref{eq:graph-expansion} and~\eqref{eq:gauge-holonomy-distance} control two different errors. The inequality
\begin{equation}
 |\delta Y|\ge h_A\min\{|Y\cap A|,|A\setminus Y|\}
 \label{eq:em-relative-expansion}
\end{equation}
bounds the smaller matter region enclosed by a gauge error. Fewer than $\lambda$ gauge flips that pass the flux checks must have the form $\delta y$.

Let $\kappa$ bound, per native fault, the sum of affected matter blocks and flipped gauge qubits after the propagation reduction above. Enlarge it to include faults in both parent intervals, initialization, link readout and the physical correction. The bounded circuit supports and the blockwise form of the correction make $\kappa$ independent of the number of repeated rounds. For a gadget touching at most $w_0$ code qubits, check weight $w$ and incidence $q_{\mathrm{inc}}$ bound the propagated support by $w_0[1+q_{\mathrm{inc}}(w-1)]$ (Appendix~\ref{sec:locality-sm}). Set $\gamma=b\kappa\max\{1,h_A^{-1}\}$.

\begin{proof}[Proof of Theorem~\ref{thm:end-distance}]
Fix a measurement record satisfying the detection tests of Theorem~\ref{thm:end-distance} of the main text and its noise Kraus labels, with $f$ faults below the right-hand side of Equation~\eqref{eq:graph-surgery-distance} of the main text. Since $f<R$, some gauged round is fault-free and applies exactly $\widetilde{Q}_a$. Expand the other ideal projectors and move their check factors to this round, where they act by the recorded eigenvalues. Equation~\eqref{eq:main-fault-dressing} of the main text controls the remaining conjugated faults. Expanding these local operators in Paulis gives a sum of terms
\begin{equation}
 P V_{z,\xi}E_+\widetilde{Q}_aE_-JP.
 \label{eq:prx273-two-interface}
\end{equation}
Here, $J$ appends the initialized gauge and completion qubits. The map $V_{z,\xi}$ contracts the gauge qubits against their measurement record and the completion qubits against a discard bra $\langle\xi|$, then applies $U(x)^{\dagger}$. The discard label $\xi$ is unobserved, and $x$ is the vertex assignment satisfying $z=\delta x$ and $x_{v_0}=0$. Each residual Pauli has the form $E=M X_g(\eta)Z_g(\zeta)E_t$. Here $M$ and $E_t$ act on the matter and completion qubits, respectively, and $\eta,\zeta\in\mathbb{F}_2^E$ specify the gauge-$X$ and gauge-$Z$ factors. Let $f_-$ and $f_+$ count faults before and after the chosen round, respectively. Writing $v(M)$ for the number of matter blocks touched by $M$, including spectators, gives
\begin{equation}
 v(M_\pm)+|\eta_\pm|\le\kappa f_\pm,
 \qquad f_-+f_+\le f.
 \label{eq:em-fault-support}
\end{equation}
Moving a residual Pauli past a parent-check projector changes only the projector's outcome label and does not enlarge the Pauli's support. Gauge-readout faults are included in $\eta_+$, and conjugating a correction fault by $U(x)$ preserves its bounded block support.

The incoming pattern passes the flux checks and has weight below $\lambda$, so $\eta_-=\delta y_-$. The selected round therefore lies in the exact-string sector of Eq.~\eqref{eq:em-gauged-projector}. For an exact residual shift $\eta=\delta y$, contracting the gauging isometry gives
\begin{align}
 Q_aEJP&\propto\Gamma_a\Pi_sP U(y)MP,
 \label{eq:prx273-main-input-transfer}\\
 PV_{z,\xi}E\Gamma_a\Pi_sP
 &\propto P U(x)^{\dagger}M U(x)U(y)\Pi_sP.
 \label{eq:prx273-main-output-transfer}
\end{align}
Gauge phases and completion contractions contribute scalars. In the output equation the final exact record forces the outgoing shift to be exact as well. These identities show how gauge flips become a product of symmetry factors on a matter region.

The vertex assignments $y$ and $y+\mathbf{1}$ give matter operations differing by the measured global symmetry, which is scalar on the selected branch. Choose the representative with fewer active vertices. Equation~\eqref{eq:em-relative-expansion} bounds this number by $|\eta|/h_A$. Each effective parent error in the two transfer equations consequently occupies at most
\begin{equation}
 b\bigl[v(M)+|\eta|/h_A\bigr]\le\gamma f<d
 \label{eq:em-parent-error-bound}
\end{equation}
parent qubits. The parent error-detection identity $PEP\in\mathbb{C}P$ makes the operators on both sides scalar~\cite{KnillLaflamme1997}. Every term in Eq.~\eqref{eq:prx273-two-interface}, and therefore their coherent sum, obeys Equation~\eqref{eq:main-instrument-distance} of the main text.
\end{proof}

Relative expansion bounds the support of the symmetry correction induced by gauge errors, and $\kappa$ bounds fault spread within the measurement circuits. For a symmetry on $W$ bounded blocks of a Pauli LDPC parent, the cellulated graph of Ref.~\cite{Yuan2026Parsimonious} has $h_A=\Omega(1)$ and $\lambda=\infty$, with $O(W\log(2+W))$ auxiliary qubits. Completion adds $O(W)$ qubits and pin readouts, preserving this peak spatial bound and giving distance $\Omega(d)$ in $O(d)$ rounds with unrestricted bounded-degree connectivity. Readouts are reused between rounds. Corollary~\ref{cor:parsimonious-completion} gives the graph construction. The square torus has decreasing expansion, and its exact distance uses the directional circuit argument in Section~\ref{sec:prx-d4} of the main text.

\subsection{An explicit gauged projector}\label{sec:sm-explicit-gauged-projector}
For Gauss's law outcomes $a_v=\pm1$, let $a=(a_v)$, $s=\prod_v a_v$, $\Pi_s=(I+s\mathcal{U})/2$, $a^x=\prod_v a_v^{x_v}$ and $U(x)=\prod_vU_v^{x_v}$. With reference vertex $v_0$, the binary isometry of Eq.~\eqref{eq:sm-zn-gamma} is
\begin{equation}
 \Gamma_a|\psi\rangle=2^{-(|V|-1)/2}\sum_{x:x_{v_0}=0}a^x|\delta x\rangle_g\,
 U(x)|\psi\rangle_m\otimes|+\rangle_t^{\otimes c}.
 \label{eq:explicit-gamma}
\end{equation}
Here, $c$ counts completion qubits and is zero when the checks use only the matter and gauge qubits.

\begin{lemma}[Projector for exact gauge configurations]
The completion construction of Appendix~\ref{sec:general-sm} and the constructions without completion qudits in Section~\ref{sec:sm-without-completion} preserve the subspace spanned by gauge configurations $z=\delta x$, with projector $P_{\rm ex}$. The signed joint projector $\widetilde{Q}_a$ for outcomes $a$ and all other checks positive satisfies
\begin{equation}
 Q_a=\widetilde{Q}_aP_{\rm ex}=\Gamma_a\Pi_sP\Gamma_a^\dagger.
 \label{eq:explicit-gauged-projector}
\end{equation}
If $H^1=0$, $Q_a=\widetilde{Q}_a$.
\end{lemma}
\begin{proof}
Proposition~\ref{prop:zn-patch-checks} proves the patch case. In the Clifford completion the Gauss's law operators shift links by cuts and the remaining checks preserve each gauge configuration. On $z=\delta x$, the path from $v$ to its reference block $r_i$ has parity $x_v+x_{r_i}$. For a parent generator $S_i$ meeting active blocks,
\[
 \widetilde{S}_i(z)=U(x)\mathcal{U}^{x_{r_i}}S_i\mathcal{U}^{x_{r_i}}U(x)^\dagger .
\]
The invariant check is $\mathcal{N}_i(z)=U(x)N_iU(x)^\dagger$. Since $\mathcal{U}S_i\mathcal{U}=S_iN_i$ and $N_i$ is a parent stabilizer, the common $+1$ eigenspace of the modified parent stabilizers and invariant checks is $U(x)\operatorname{im}P$ even when different generators have different reference blocks. A generator disjoint from active blocks is unchanged. Central outcomes $+1$ set $V=I$ and the pinning checks fix every completion qubit in $|+\rangle$. The Gauss outcomes at vertices other than the reference vertex fix the phase factor $a^x$ in Eq.~\eqref{eq:explicit-gamma}, and the remaining vertex imposes $\mathcal{U}=s$. This gives Eq.~\eqref{eq:explicit-gauged-projector}. For the construction in Eq.~\eqref{eq:conditional-central-completion}, all central outcomes are likewise $+1$, so every modified check equals its raw counterpart. The same argument applies with $c=0$. Both constructions inherit this identity whenever the family of raw checks has the required signed projector in the common $+1$ sector of the central checks, because both preserve its measurement branches there.
\end{proof}

On an even $L\times L$ square torus with all vertices active, half the columns give $|\delta Y|=2L$ and $|Y|=L^2/2$, so $h_A\le4/L$. A seam of $L$ flipped horizontal links passes all local flux checks but has odd parity around each horizontal winding loop. Conversely, any flat gauge pattern with odd parity around a horizontal winding loop meets every horizontal row. The same argument applies in the vertical direction. Thus, $\lambda=L$. The final check that $z=\delta x$ on every edge detects the seam (Appendix~\ref{sec:sm-choose-graph}). The directional proof for $D(D_4)$ gives $\min(L,R)$ despite the small $h_A$, including faults in the parent rounds (Appendix~\ref{sec:smd4}).

\subsection{\texorpdfstring{Logical measurement of $\mathbb{Z}_N$ symmetries}{Logical measurement of ZN symmetries}}
\label{sec:sm-zn-distance}

Use a fixed commuting family of finite-order checks satisfying Eq.~\eqref{eq:sm-zn-projector} and containing the prescribed fluxes $B_f$. Serialize overlapping extraction circuits. The detection tests require that every parent-check outcome is $+1$, the record repeats $a$ for all $R$ gauged rounds, gives $+1$ on the other gauged checks, and ends with an exact link record. Assume ideal encoded input and final parent projection, and reliable classical processing. For a shift mask $\eta$, write $|\eta|$ for its support size. Define
\begin{align}
 \lambda_N&=\min\{|\eta|:\partial_2^T\eta=0,
                 \ \eta\notin\operatorname{im}\delta\},\nonumber\\
 \beta_N&=\max_{\delta y\ne0}
 \frac{\min_{c\in\mathbb{Z}_N}|\operatorname{supp}(y-c\mathbf{1})|}
      {|\operatorname{supp}\delta y|}.
 \label{eq:sm-zn-graph-parameters}
\end{align}
Set $\lambda_N=\infty$ if every flat mask is exact. For edge expansion $h$ (minimum cut size divided by its smaller vertex side), $\beta_2=1/h$ and $\beta_N\le2/h$. Omit a most frequent label of $y$ and apply expansion to each other label class. Each contains at most $|V|/2$ vertices, and every cut edge is counted at most twice. For a single vertex put $\beta_N=0$, $h=\infty$.

\begin{theorem}[Circuit fault distance of $\mathbb{Z}_N$ logical measurement]
\label{thm:zn-accepted-distance}
Suppose the disjoint matter blocks contain at most $b$ matter qudits each, and $P$ detects every operator on fewer than $d$ matter qudits. For any fault-free gauged round, let $f_-$ and $f_+$ count faults before and after it. Assume every residual product term obeys $v(M_\pm)+|\eta_\pm|\le\kappa_N f_\pm$, where $v(M_\pm)$ counts blocks touched by the matter part, including spectator qudits counted as one-qudit blocks, and $\eta_\pm$ is the gauge shift. Include all native locations, from preparation through both parent intervals, link measurements, physical correction and idles. Then,
\begin{equation}
 d_{\rm surg}\ge
 \min\!\left\{R,
 \left\lceil\frac{d}{b\kappa_N\max\{1,\beta_N\}}\right\rceil,
 \left\lceil\frac{\lambda_N}{\kappa_N}\right\rceil\right\}.
 \label{eq:sm-zn-graph-distance}
\end{equation}
This distance is the least fault count required for a branch $K$ whose measurement outcomes satisfy all detection conditions to violate $PKP\in\mathbb{C}\,\Pi_\alpha(\mathcal{U})P$. Here, $K$ fixes the noise Kraus labels and the discard label $\xi$.
\end{theorem}
\begin{proof}
For $f$ below the bound, $f<R$ ensures a fault-free round projecting onto $\widetilde{Q}_a$. Expand intervening projectors in powers of their checks and move these products to that round. Disjoint factors cancel in conjugation by commutation. Only checks meeting the original faulty support can enlarge it, bounding spread in the gauged interval for bounded support and incidence.

Write each tensor-product residual term as $D=M X_g(\eta)Z_g(\zeta)E_t$. Here $E_t$ acts on the completion qudits. Let $J$ append the initialized gauge and completion qudits. A surviving incoming shift is flat by the flux checks and exact since $|\eta|<\lambda_N$. Thus, $\widetilde{Q}_a$ acts as $Q_a$. For $\eta=\delta y$, taking gauge matrix elements in Eq.~\eqref{eq:sm-zn-gamma} gives
\begin{equation}
 Q_aDJP\ \propto\
 \Gamma_a\Pi_\alpha(\mathcal{U})P U(y)^\dagger M P.
 \label{eq:sm-zn-input}
\end{equation}
An exact final record $\widehat{z}=\delta\widehat{x}$ requires an exact outgoing shift. Contracting the gauge and completion qudits gives
\begin{equation}
 \begin{aligned}
 &P U(\widehat{x})^\dagger\,{}_{g,t}\!\langle\widehat{z},\xi|
 D\Gamma_a\Pi_\alpha(\mathcal{U})P\\
 &\quad\propto P U(\widehat{x})^\dagger M U(\widehat{x})
 U(y)^\dagger\Pi_\alpha(\mathcal{U})P.
 \end{aligned}
 \label{eq:sm-zn-output}
\end{equation}
The completion factors $\langle\mathrm{aux}|E_t|\mathrm{aux}\rangle$ and $\langle\xi|E_t|\mathrm{aux}\rangle$, at input and output respectively, are scalars, as are gauge phases. Gauge readout errors enter the outgoing shift mask. Replacing $y$ by $y-c\mathbf{1}$ contributes only the selected eigenvalue of $\mathcal{U}^c$. Since conjugation by $U(\widehat{x})$ preserves block support, each effective parent operator occupies at most $b[v(M)+\beta_N|\eta|]<d$ matter qudits. Multiplying either effective parent operator by $P$ on both sides gives a scalar multiple of $P$. Summing the residual terms therefore gives the claimed logical-measurement condition.
\end{proof}

\subsection{Compatibility with low-overhead graph constructions}
\begin{lemma}[Routing and relative expansion on active vertices]
\label{lem:relative-routing}
Let a graph $G_0$ on $A$ have edge expansion $h_0$. If its edges are routed between their endpoints in $G$, with at most $c_{\rm route}$ routed paths using any edge of $G$, then $h_A(G)\ge h_0/c_{\rm route}$. In particular, $h_A(G)\ge h_0$ for $G_0\subseteq G$.
\end{lemma}
\begin{proof}
Each base edge crossing $(Y\cap A,A\setminus Y)$ has a route crossing $\delta_GY$. Consequently,
\[
 h_0\min\{|Y\cap A|,|A\setminus Y|\}
 \le |\delta_{G_0}(Y\cap A)|\le c_{\rm route}|\delta_G Y|.
\]
\end{proof}

\begin{corollary}[Low-overhead commuting-check logical measurement]
\label{cor:parsimonious-completion}
Let a distance-$d$ qubit Pauli stabilizer family have bounded generator weight and incidence. Suppose a symmetry preserving the stabilizer group is a product of Hermitian Clifford operators on $W\ge1$ disjoint active blocks of bounded size. On an unrestricted bounded-degree interaction graph, logical measurement by commuting-check gauging and ungauging uses $O(W\log(2+W))$ additional qubits including measurement ancillas, with extraction gates of level at most three. With bounded-size measurement circuits and the boundary conditions and detection tests of Theorem~\ref{thm:end-distance} of the main text, $R=\Theta(d)$ gives $d_{\rm surg}=\Omega(d)$.
\end{corollary}
\begin{proof}
For $W\ge2$, augment a constant-degree expander on $A$ with a tree joining the active blocks of each parent generator. Bounded generator weight and incidence bound the vertex degree, the lengths of paths to each reference vertex, and the number of such paths using any edge. For $W=1$ a single vertex suffices.

The cellulated cone of Ref.~\cite{Yuan2026Parsimonious}, Theorem~IV.23, retains this graph as a subgraph, has $O(W\log(2+W))$ cells and trivial first homology, and bounds face length, degree and edge--face incidence. The original paths remain available, all commutator cycles are filled, and Lemma~\ref{lem:relative-routing} gives $h_A=\Omega(1)$. Over $\mathbb{F}_2$ the vanishing first homology gives $\lambda=\infty$.

The gauge qubits and new Gauss and flux measurement ancillas cost $O(W\log(2+W))$ qubits. Each conflict involves a modified parent check meeting an active block. There are $O(W)$ such checks, each of bounded degree in the conflict graph. Completion and pinning therefore cost $O(W)$ qubits. All resulting checks have bounded weight and incidence and admit bounded-size Clifford-check measurement circuits. Thus, $\kappa=O(1)$ and Equation~\eqref{eq:graph-surgery-distance} of the main text applies. The $O(W\log(2+W))$ bound counts the peak number of additional qubits for one logical measurement. Measurement ancillas are reused across rounds. Total space including an $n$-qubit parent memory and its measurement ancillas is $O(n+W\log(2+W))$.
\end{proof}

\section{Square-lattice \texorpdfstring{$D(D_4)$}{D(D4)}: checks, explicit circuits, and distance}
\label{sec:smd4}

We use $D_4$ for the order-eight dihedral group and the square-lattice raw checks of Ref.~\cite{KobayashiZhuHsin2026} on the $L\times L$ torus $\Lambda$, with $L\ge3$. The proofs below use the specified native schedule for memory on all ground sectors and for bilayer surgery.

\subsection{Lattice convention and completed checks}
For a vertex $v=(i,j)$, let $e_x(i,j)$ and $e_y(i,j)$ be its outgoing horizontal and vertical edges. Identify each face $f$ with its lower-left vertex. Coordinates are modulo $L$. Write $c_e$ for the qubit of color $c\in\{r,g,b\}$ on edge $e$, with colors ordered red, green, blue. The red qubit is the gauge qubit, and the green and blue qubits belong to the parent toric codes. The computational-basis bit of $c_e$ is $z_{c,e}$. All vertex and edge assignments take values in $\mathbb{Z}_2$. For a mask $x$ and a polynomial $q$, write $X^x|z\rangle=|z+x\rangle$ and $D_q|z\rangle=(-1)^{q(z)}|z\rangle$. Let $s_{c,v}$ mark the four color-$c$ edges incident to $v$. A raw star and a flux check are
\begin{equation}
 A_{c,v}=X^{s_{c,v}}D_{q_{c,v}},\qquad
 B_{c,f}=\prod_{e\in\partial f}Z_{c,e},
 \label{eq:smd4-raw}
\end{equation}
where $\partial f$ is the face boundary. The red star is the Gauss's law operator and the green and blue stars are the modified checks of the two parent toric codes. The quadratic phase $q_{c,v}$ specifies two CZ gates per star, one for each monomial in
\begin{align}
 q_{r,(i,j)}&=z_{g,e_x(i,j)}z_{b,e_y(i+1,j)}\nonumber\\
 &\quad+z_{g,e_y(i,j)}z_{b,e_x(i,j+1)},\nonumber\\
 q_{g,(i,j)}&=z_{r,e_x(i-1,j)}z_{b,e_y(i,j)}\nonumber\\
 &\quad+z_{r,e_y(i,j-1)}z_{b,e_x(i,j)},\nonumber\\
 q_{b,(i,j)}&=z_{r,e_x(i-1,j-1)}z_{g,e_y(i,j-1)}\nonumber\\
 &\quad+z_{r,e_y(i-1,j-1)}z_{g,e_x(i-1,j)}.
 \label{eq:smd4-Q}
\end{align}
The $X$ factors and CZ gates act on different colors, so every $A_{c,v}$ is a Hermitian Clifford operator. The flux checks commute with all stars. With the diagonal displacement $\Delta=(1,1)$, the only nontrivial star commutators are
\begin{align}
 A_{r,v}A_{g,v}&=B_{b,v}A_{g,v}A_{r,v},\nonumber\\
 A_{g,v}A_{b,v}&=B_{r,v-\Delta}A_{b,v}A_{g,v},\nonumber\\
 A_{b,v+\Delta}A_{r,v}&=B_{g,v}A_{r,v}A_{b,v+\Delta}.
 \label{eq:smd4-conflicts}
\end{align}
These identities follow directly by translating the quadratic polynomials in Eq.~\eqref{eq:smd4-Q} by the corresponding star masks.

Orient the first conflict in Equation~\eqref{eq:smd4-conflicts} from red to green at $v$. Its label $B_{b,v}$ places $t_{b,v}$ on face $v$. The red star gains $X_{t_{b,v}}$ and the green star gains four CZs to the blue face edges. The product of these four CZs is $Z_{t_{b,v}}$ when $B_{b,v}=-1$. Its anticommutation with the added $X_{t_{b,v}}$ cancels the raw sign. When $B_{b,v}=+1$, the four CZs multiply to identity and the positive pin constraint fixes $X_{t_{b,v}}=+1$. The other conflicts give the placements in Figure~\ref{fig:d4-completed-checks}(a) of the main text. For $B=B_{c,f}$ and $t=t_{c,f}$, write $P_\pm(B)=(I\pm B)/2$. Then

\begin{align}
 V(B,t)&=P_+(B)+P_-(B)Z_t
       =\prod_{e\in\partial f}\mathrm{CZ}(c_e,t),\nonumber\\
 T(B,t)&=P_-(B)+P_+(B)X_t
       =X_t\prod_{e\in\partial f}\mathrm{CNOT}(c_e\to t).
 \label{eq:smd4-VT}
\end{align}
The flux and pinning checks are shown in Figure~\ref{fig:d4-completed-checks}(b),(c) of the main text. Measure the flux checks $B_{c,f}$, the pinning checks $T_{c,f}=T(B_{c,f},t_{c,f})$, and the completed stars shown in Figure~\ref{fig:d4-completed-checks}(d)--(f) of the main text:
\begin{align}
 \widehat{A}_{r,v}&=A_{r,v}X_{t_{b,v}}V(B_{g,v},t_{g,v}),\nonumber\\
 \widehat{A}_{g,v}&=A_{g,v}X_{t_{r,v-\Delta}}V(B_{b,v},t_{b,v}),\nonumber\\
 \widehat{A}_{b,v}&=A_{b,v}X_{t_{g,v-\Delta}}
                         V(B_{r,v-\Delta},t_{r,v-\Delta}).
 \label{eq:smd4-completed}
\end{align}
The identity $X_tV(B,t)=BV(B,t)X_t$ cancels each commutator in Eq.~\eqref{eq:smd4-conflicts}. The pinning checks commute with $X_t$ and $V(B,t)$. In the negative-flux sector $T=I$, whereas in the positive sector $V=I$. Hence, the entire measured family commutes on the full enlarged Hilbert space.

Let $P_0$ project onto the common $+1$ space of the raw stars and flux checks and let $P'$ project onto the common $+1$ space of the completed family. There $B=T=+1$ imposes $X_t=+1$ on each completion qubit, so
\begin{equation}
 P'=P_0\otimes|+\rangle\langle+|^{\otimes3L^2}.
 \label{eq:smd4-productcode}
\end{equation}

\subsection{Ground-space degeneracy}
\label{sec:smd4-ground-space}
In a flat configuration the parity of $z_{c,e}$ around each face is zero for every color. Define $\alpha_c$ and $\beta_c$ as its parities around a horizontal and a vertical winding cycle. For each color, the $L^2$ flux checks have $L^2-1$ independent constraints, leaving $2^{3L^2+3}$ configurations across the three colors. The star-flip masks have total binary rank $3(L^2-1)$, so these configurations split into $64$ orbits, each of size $2^{3L^2-3}$. The vectors of winding parities $\alpha,\beta\in\mathbb{F}_2^3$ label the star-flip orbits. On flat configurations all stars commute. The product of all color-$c$ stars acts by the sign $(-1)^{\chi_c}$, where
\begin{equation}
 \chi_c=\alpha_d\beta_e+\beta_d\alpha_e,
 \qquad \{c,d,e\}=\{r,g,b\}.
 \label{eq:smd4-chi}
\end{equation}
For example, $\sum_v q_{r,v}$ equals the cubical cup product of the green and blue edge cochains evaluated on the torus, giving $\chi_r$. For each color the only nonempty product with zero star mask is the product over all vertices. Thus, an orbit supports a common $+1$ eigenstate exactly when all three $\chi_c$ vanish. In that case, the state is unique and has uniform computational probabilities. Equivalently, $\alpha,\beta$ are linearly dependent, so
\begin{equation}
 \dim P_0=8+7\times2=22.
\end{equation}
This count holds for both odd and even $L\ge3$. Every length-$L$ product of single-color Pauli $Z$ operators around a horizontal or vertical cycle is nonscalar on the code space, because both zero and single-bit winding parities occur among these sectors.

\subsection{Error detection}
\label{sec:smd4-error-detection}
For a monomial error $E=X^xD_q$, let $\operatorname{var}(q)$ denote the variables on which its reduced Boolean polynomial depends. For an index set $J$, write $J-1=\{j-1:j\in J\}$ modulo $L$. The sets $V_c,H_c$ track columns and rows of the $X$ mask. The auxiliary sets $Q_c^{\mathrm h},Q_c^{\mathrm v}$ track the phase variables on horizontal and vertical edges. Including the shifts from the CZ factors gives the phase-obstruction sets $\mathcal{B}_c^{\mathrm h},\mathcal{B}_c^{\mathrm v}$. Explicitly,
\begin{align}
 V_c(x)&=\{i:\exists j, x_{c,e_y(i,j)}=1\},\nonumber\\
 H_c(x)&=\{j:\exists i, x_{c,e_x(i,j)}=1\},\nonumber\\
 Q_c^{\mathrm h}(q)&=\{i:\exists j, z_{c,e_x(i,j)}\in\operatorname{var}(q)\},\nonumber\\
 Q_c^{\mathrm v}(q)&=\{j:\exists i, z_{c,e_y(i,j)}\in\operatorname{var}(q)\},\nonumber\\
 \mathcal{B}_c^{\mathrm h}(E)&=Q_c^{\mathrm h}(q)\cup\bigcup_{d<c}V_d(x)
                             \cup\bigcup_{d>c}(V_d(x)-1),\nonumber\\
 \mathcal{B}_c^{\mathrm v}(E)&=Q_c^{\mathrm v}(q)\cup\bigcup_{d<c}H_d(x)
                             \cup\bigcup_{d>c}(H_d(x)-1).
 \label{eq:smd4-cuts}
\end{align}
For each of the three colors, the four sets $V_c,H_c,\mathcal{B}_c^{\mathrm h},\mathcal{B}_c^{\mathrm v}$ record the affected rows and columns. An omitted coordinate of $H_c$ or $V_c$ gives a winding cycle that does not meet the $X$ support. If $k\notin\mathcal{B}_c^{\mathrm h}$, the horizontal edges $\{e_x(k,j):j\in\mathbb{Z}_L\}$ contain no phase variables, and the relevant columns contain no bit flips that could generate phase on these edges. These are the edges crossed by a vertical dual-lattice loop wrapping around the torus. Similarly, a row omitted from $\mathcal{B}_c^{\mathrm v}$ gives vertical edges crossed by a horizontal dual loop.

\begin{lemma}[Closure of obstruction sets]
\label{lem:smd4-closure}
Multiplication gives union bounds for all sets in Eq.~\eqref{eq:smd4-cuts}. Conjugation by a raw check preserves the $X$ sets and does not enlarge the phase-obstruction sets.
\end{lemma}
\begin{proof}
The product rule is
\begin{equation}
 (X^{x_1}D_{q_1})(X^{x_2}D_{q_2})
 =X^{x_1+x_2}D_{q_1(z+x_2)+q_2(z)}.
\end{equation}
Translation introduces no new polynomial variables. For $A=A_{c,v}=X^sD_{q_{c,v}}$, with $s=s_{c,v}$, the $X^s$ and $D_{q_{c,v}}$ factors act on disjoint colors, giving
\begin{equation}
 AEA=X^xD_{q(z+s)+q_{c,v}(z)+q_{c,v}(z+x)}.
 \label{eq:smd4-conj}
\end{equation}
For qubits $k,\ell$, a quadratic phase term $z_kz_\ell$ adds only $x_kz_\ell+x_\ell z_k+x_kx_\ell$. Its new variables are CZ neighbors of $X$ support. Write $\mathbf{1}_{d>c}$ for the indicator that color $d$ follows $c$. In Eq.~\eqref{eq:smd4-Q}, a horizontal color-$c$ edge at column $k$ has its color-$d$ vertical neighbors at column $k+\mathbf{1}_{d>c}$. The row relation is identical. These are exactly the shifts in Eq.~\eqref{eq:smd4-cuts}. Conjugation by a flux check changes only a scalar sign.
\end{proof}

\begin{lemma}[Error detection on the full code]
\label{lem:smd4-clean}
If every set $V_c,H_c,\mathcal{B}_c^{\mathrm h},\mathcal{B}_c^{\mathrm v}$ has size less than $L$, then $P_0EP_0\in\mathbb{C}P_0$.
\end{lemma}
\begin{proof}
If $x$ is nonflat, $X^x$ anticommutes with a flux check, so $P_0EP_0=0$. Otherwise, the empty cycles force both winding parities to vanish for each color. Hence, $x_c=\delta\tau_c$ for a binary vertex potential $\tau_c$, where $(\delta\tau_c)_{vw}=\tau_c(v)+\tau_c(w)$ on edge $vw$. On each column outside $V_c$ and row outside $H_c$, the potential $\tau_c$ is constant. Their intersections force these constants to agree. Subtract this one constant so that $\tau_c$ vanishes on every such row and column simultaneously.

Let $R_\tau$ be the ordered product of stars with $\tau_c(v)=1$ and $q_\tau$ its phase polynomial. Then, $R_\tau=X^xD_{q_\tau}$ and $R_\tau P_0=P_0$. The remaining phase polynomial $\rho$ satisfies
\begin{equation}
 EP_0=D_\rho P_0,\qquad \rho(z)=q(z+x)+q_\tau(z).
\end{equation}
Fix $k\notin\mathcal{B}_c^{\mathrm h}(E)$. The original $q$ contains no variables on the horizontal edges $e_x(k,*)$ crossed by the vertical dual loop. Every color-$d$ star that could add phase on these edges lies on column $k+\mathbf{1}_{d>c}$, where $\tau_d=0$. Thus, $q_\tau$ also contains no variables on these edges. The same argument with rows gives a horizontal dual loop whose crossed vertical edges contain no remaining phase variables. The same fixed potential and the same $\rho$ work for the two loops of each of the three colors.

Choose a flat representative $z_\eta$ for each pair of winding parities $\eta=(\alpha,\beta)$, including pairs without a ground state. Let $\mathcal{V}_0=C^0(\Lambda,\mathbb{F}_2)^3$ be the space of three-color vertex potentials. Define the orbit average
\begin{equation}
 \Phi_\rho(\eta)=\frac{1}{|\mathcal{V}_0|}
 \sum_{u\in\mathcal{V}_0}(-1)^{\rho(z_\eta+\delta u)}.
\end{equation}
Flipping the edge bits crossed by any one of these six loops adds a flat cochain and toggles one winding parity. Because $\rho$ contains none of these edge variables, the flip leaves the average unchanged. Hence, $\Phi_\rho$ is constant over all 64 pairs of winding parities, and in particular over the 22 allowed sectors. The diagonal matrix elements of $D_\rho$ are those averages, while exact $x$ cannot connect different winding sectors. Thus, $P_0EP_0$ is a scalar multiple of $P_0$ on the full code.
\end{proof}

\subsection{Check measurement circuits and obstruction bounds}
\label{sec:smd4-native-extraction}

For the circuit analysis below, assume even $L\ge4$ so that the periodic checkerboard schedule is well defined. The factors within each check in Figure~\ref{fig:d4-completed-checks} of the main text commute, but constituent gates belonging to different checks need not. Reordering extraction gates that commute in the absence of faults can also change fault propagation. We therefore use the order below for the circuit fault-distance proof.

To measure a completed star, prepare its measurement ancilla in $|+\rangle$. For the red star at $v$, this ancilla controls $U_v$, an $X$ on each of the four incident gauge qubits, and $D_v=X_{t_{b,v}}\prod_{e\in\partial f_v}\mathrm{CZ}(g_e,t_{g,v})$, where $f_v$ has southwest vertex $v$ and $g_e$ is the green qubit on edge $e$. Here $U_v=D_{q_{r,v}}$ is the pair of CZ gates in the raw red star. The completion adds one CNOT to the blue completion qubit and four CCZs involving the green completion qubit. Apply the four CNOTs to star edges, the two CCZs from the raw star, and the outgoing completion CNOT first, in any order. Apply the four CCZs implementing controlled $V$ last, then measure the ancilla in $X$. Measure each flux check $B$ jointly with its pinning check $T$ using their assigned measurement ancillas $a_B$ and $a_T$. Prepare $a_B$ in $|1\rangle$ and $a_T$ in $|+\rangle$, apply one CNOT from each of the four checked edge qubits to $a_B$, encoding the eigenvalue of the flux check $B$ in $a_B$, apply $\mathrm{Toffoli}(a_B,a_T\to t)$, then measure $a_B$ in $Z$ and $a_T$ in $X$. Let $q_B\in\{0,1\}$ be the measurement outcome of $a_B$ and $x\in\{+1,-1\}$ that of $a_T$. The flux check outcome is $s_B=2q_B-1$. The joint Kraus operator for outcomes $s_B,x$ is
\begin{equation}
 \begin{aligned}
 K_{s_B,x}&=P_{s_B}(B)\left[\frac{1+s_B}{2}\frac{I+xX_t}{2}
 +\frac{1-s_B}{2}\delta_{x,+1}I\right]\\
 &=P_{s_B}(B)P_x(T).
 \end{aligned}
 \label{eq:smd4-jointBT}
\end{equation}
This joint measurement uses a fixed circuit. Complete each check measurement circuit before beginning another on overlapping qubits. Circuits with disjoint supports may run in parallel.

A native fault is an arbitrary error on the participants of one CNOT, CCZ or Toffoli, or an error at a preparation, measurement, basis change or idle location. Each two-qubit CZ applied during ungauging is also one native fault location. A fault may act jointly on both data qubits. We expand arbitrary local errors in participant Paulis. To compare errors on the edge qubits, we work modulo the subgroup
\begin{equation}
 \mathsf{S}_0=\langle B_{c,f},X_{t_{c,f}}\rangle
\end{equation}
which fixes $P'$ and is normalized by every completed check. In particular, conjugating an incoming completion $X_t$ by its completed star multiplies it by the flux check $B$ in $V(B,t)$. Modulo $\mathsf{S}_0$, conjugation by $T$ preserves the obstruction sets of the edge qubits, and conjugation by $\widehat{A}$ acts on those sets like conjugation by the raw $A$. Completion $Z$ factors may remain but do not contribute to the obstruction sets on the edge qubits. The subgroup also fixes the initialized bilayer code space and the ideal joint-check projector with positive flux and pin outcomes. Its factors are removed at these ideal boundaries. An intermediate faulty state need not have positive flux.

\begin{lemma}[Obstruction bound for one circuit fault]
\label{lem:smd4-native}
Propagate one native fault through the remaining fault-free gates of its check measurement circuit, then expand the operator on the measurement ancillas and qubits of the code in Paulis. Modulo $\mathsf{S}_0$, for each of the twelve coordinate sets in Eq.~\eqref{eq:smd4-cuts}, at most one coordinate suffices for all resulting terms on the edge qubits. The same coordinate remains sufficient after conjugation by any completed check. Products of $f$ faulty-location contributions occupy at most $f$ coordinates in each set.
\end{lemma}
\begin{proof}
For a raw star of color $a$ at $(i,j)$, let $S_{\rm rem}$ be the set of remaining star targets and $R_{\rm rem}$ the set of remaining raw CZ pairs. Write $X_{S_{\rm rem}}$ for the product of $X$ on these targets. Let $\epsilon\in\{0,1\}$ indicate whether the ancilla Pauli has an $X$ component. Its response is a direct participant Pauli times
\begin{equation}
 X_{S_{\rm rem}}^\epsilon
 \prod_{p\in R_{\rm rem}}\mathrm{CZ}_p^\epsilon.
\end{equation}
All star subsets have vertical edges at column $i$ and horizontal edges at row $j$. For target color $c\ne a$, every CZ phase from the remaining raw gates and the shifted star-$X$ contribution align at
\begin{equation}
 (i-\mathbf{1}_{a>c},\ j-\mathbf{1}_{a>c}).
\end{equation}
At a CNOT fault the direct data Pauli is on one color-$a$ edge. Its phase adds at most one coordinate to each phase-obstruction set for color $a$. At a CCZ fault the two data participants have different complementary colors and opposite orientations. For each target color their direct phases, transverse $X$ contributions and CZ phases from the remaining gates align at the displayed coordinate. Correlated participant errors therefore contribute at most one coordinate to each obstruction set.

The outgoing completion CNOT adds no support on the edge qubits. The final four CCZs implementing controlled $V(B,t)$ introduce phase variables only on the plaquette of $B$. Its color and lower-left vertex are $(g,(i,j))$, $(b,(i,j))$, or $(r,(i-1,j-1))$ for red, green, or blue stars, respectively. These coincide with the raw phase coordinates. An $X$ on an edge qubit of a raw CZ factor propagates through one of these CCZs by adding only a CZ between the measurement ancilla and the completion qubit. A fault at one of these four CCZs directly affects at most one edge qubit and creates phase variables only on the same plaquette. The direct edge bit flip can occupy a neighboring coordinate in another color's phase-obstruction set. For example, an $X$ on the right green vertical edge of a red-star controlled-$V$ gate contributes $i+1$ to $\mathcal B_b^{\mathrm h}$, rather than $i$. All CNOTs to star edges and CCZs from the raw star have already finished, so the fault cannot also spread through those earlier gates and add a second coordinate to the same set. Thus, neither case creates a second coordinate in any family.

In the joint measurement of $B,T$, a $Z$ on the measurement ancilla $a_B$ propagates to the remaining plaquette edges through the CNOTs from the checked qubits to $a_B$ that follow it. A fault at one of these CNOTs directly affects at most one edge qubit. Horizontal perimeter edges share one anchor column and vertical perimeter edges one anchor row. The final Toffoli acts on $a_B,a_T$ and the completion qubit $t$. An $X$ fault on $a_B$ propagates through the Toffoli by a CNOT from $a_T$ to $t$, whose Pauli expansion adds only $I$ and $X_t$ factors. Thus, the Toffoli does not enlarge the propagated error on the edge qubits. A fault at the Toffoli has support only on the measurement ancillas and the completion qubit, even when the flux parity stored in $a_B$ is wrong.

The same coordinate sets contain every term in the coherent Pauli expansion. A CZ expansion uses only endpoint $Z$ factors. Completion errors propagate as described above, and Lemma~\ref{lem:smd4-closure} proves closure on the edge qubits. Idle faults between check measurement circuits obey the one-qubit bound. An idle fault inside a circuit can be absorbed into its next gate involving that qubit, or treated as a terminal error if no such gate remains. Preparation and measurement faults give the same local Pauli expansions. For several faults in one check measurement circuit, multiply their propagated operators in circuit order before taking matrix elements on the measurement ancillas. Taking the union of these coordinate sets counts each original faulty location once.
\end{proof}

For a memory history, let $F_0$ be its faulty locations, $m$ its measurement record and $\ell$ its noise branch. Expand its Kraus operator $K_{F_0,m,\ell}$ by moving products of ideal checks to an ideal memory boundary. The resulting monomial terms $E_\nu$, including their completion Paulis, have complex coefficients $\lambda_\nu$ and obey
\begin{equation}
 \begin{gathered}
 K_{F_0,m,\ell}P'=\sum_\nu\lambda_\nu E_\nu P',\\
 |V_c|,|H_c|,|\mathcal{B}_c^{\mathrm h}|,|\mathcal{B}_c^{\mathrm v}|\le|F_0|.
 \end{gathered}
 \label{eq:smd4-history}
\end{equation}
In each term remove $\mathsf{S}_0$ factors with their scalar Pauli commutation phases. This is justified by normalization of $\mathsf{S}_0$ and its action on the boundary state. A final $Z$ on a completion qubit gives zero when multiplied by $P'$ on both sides. Otherwise, Lemma~\ref{lem:smd4-clean} applies.

\begin{theorem}[Full-code memory distance]
\label{thm:smd4-memory}
Define the circuit fault distance of the memory, $d_{\rm mem}$, as the fewest native faults producing a nonscalar memory action with all check outcomes positive, ideal input in $P'$ and ideal final projection onto $P'$. For even $L\ge4$ and the specified schedule, the circuit fault distance equals the code distance,
\begin{equation}
 d(P')=d_{\rm mem}=L.
\end{equation}
The memory error span for histories of at most $t$ faults and arbitrary records admits ideal recovery whenever $2t<L$.
\end{theorem}
\begin{proof}
For fewer than $L$ faults, Eq.~\eqref{eq:smd4-history} and Lemma~\ref{lem:smd4-clean} prove the circuit lower bound, including errors that act nonunitarily on the encoded state. Every weight-$w$ Pauli obeys the same obstruction bound with $f=w$, so the code distance is at least $L$. These length-$L$ logical operators give both upper bounds. For the memory circuit insert one $Z$ per loop edge after its last participating gate or at a terminal idle. These are distinct fault locations and their product commutes with the final projection. Equation~\eqref{eq:smd4-productcode} shows that adding the completion qubits preserves this code distance. Finally, two memory histories with at most $t$ faults have pairwise products costing at most $2t$ in every family. Multiplying these products by $P'$ on both sides gives scalar multiples of $P'$, which are the Knill--Laflamme conditions on the full 22-dimensional code.
\end{proof}

Fix six star stages using the three colors and checkerboard parity. Joint measurements of $B,T$ for all colors fit into two plaquette-checkerboard stages, with overlapping gates serialized. Figure~\ref{fig:smd4-joint-circuit}(b) of the main text gives the joint circuit.

\begin{lemma}[Completion flags preserve the bilayer obstruction bound]
\label{lem:smd4-wrapper-cuts}
Insert $\operatorname{CNOT}(t\to a_B)$ immediately before and after the four CNOTs from the plaquette edges to $a_B$ in each joint flux--pin circuit of Appendix~\ref{sec:smd4-native-extraction}. Preserve the remaining gate order and serialize overlapping gadgets as before. The modified circuit obeys the same one-coordinate bound in Lemma~\ref{lem:smd4-native}. Consequently, Theorems~\ref{thm:smd4-memory} and~\ref{thm:smd4-surgery} retain their stated distances under the same boundary and fault assumptions.
\end{lemma}
\begin{proof}
The two added CNOTs commute with the intervening CNOTs from the plaquette edges to $a_B$ and cancel ideally, so the joint measurement instrument is unchanged in every flux sector. Expand a fault at the first added CNOT in participant Paulis on $t,a_B$. Its $Z_{a_B}$ component propagates through all four CNOTs from the plaquette edges to $a_B$ to the full plaquette $B$. This factor belongs to $\mathsf{S}_0=\langle B_{c,f},X_{t_{c,f}}\rangle$. All its other propagated factors act on auxiliary qubits. Thus, a fault at the first added CNOT has no edge-qubit cost modulo $\mathsf{S}_0$. A fault at the second added CNOT has no subsequent edge-qubit interaction within this gadget. Faults at the original CNOTs from the plaquette edges to $a_B$ produce the same propagated errors on the edge qubits as before, and the second added CNOT adds only auxiliary factors. An idle error on $t$ also creates no additional edge support. The final Toffoli acts only on auxiliaries. These observations cover all new fault locations and preserve the original bound at the old ones. Subsequent ideal completed checks normalize $\mathsf{S}_0$, and conjugation of a completion Pauli adds only flux factors or a scalar sign. The original closure under ideal checks and the input and output transfer identities therefore apply without change. The lower bounds follow from the preceding distance proofs. The length-$L$ terminal matter loop and the $R$ repeated red-star readout flips remain available, proving the same upper bounds.
\end{proof}
In the positive flux sector with the completion qubit initially in $|+\rangle$, these two CNOTs also give the local hook-flag guarantee of Lemma~\ref{lem:prx273-completion-flag} of the main text. The exact bilayer distance above follows from the obstruction bound, including multiple faults and all intermediate flux sectors.

\subsection{Bilayer gauging and parent boundary identities}
\label{sec:smd4-bilayer-gauging}
Gauging two toric codes gives the raw $D(D_4)$ checks~\cite{Tantivasadakarn2023Shortest,Tantivasadakarn2023Hierarchy,Christos2026NonAbelian}. Place a green and a blue data qubit on each edge of a square lattice. The parent checks, with $c=g,b$, are
\begin{equation}
 S_{c,v}=\prod_{e\ni v}X_{c,e},\qquad
 B_{c,f}=\prod_{e\in\partial f}Z_{c,e}.
 \label{eq:prx332-parent-tc}
\end{equation}
Write $e_x(v)$ and $e_y(v)$ for the edges extending right and up from vertex $v$, and label each face by its lower-left vertex. Here $\hat x=(1,0)$ and $\hat y=(0,1)$ are the lattice unit displacements. Choose $\mathcal U=\prod_v U_v$ with
\begin{equation}
 U_v=\operatorname{CZ}_{g_{e_x(v)},b_{e_y(v+\hat x)}}
     \operatorname{CZ}_{g_{e_y(v)},b_{e_x(v+\hat y)}}.
 \label{eq:smd4-U}
\end{equation}
Each block contains the four data qubits in these two disjoint CZ pairs. The symmetry leaves the parent flux checks unchanged and obeys
\begin{align}
 \mathcal U S_{g,v}\mathcal U^\dagger&=S_{g,v}B_{b,v},\nonumber\\
 \mathcal U S_{b,v}\mathcal U^\dagger&=S_{b,v}B_{g,v-\Delta},
 \qquad \Delta=\hat x+\hat y.
 \label{eq:prx332-parent-symmetry}
\end{align}
Thus, it preserves the parent code space, where both flux families have eigenvalue $+1$.

For this symmetry, enrichment uses two CCZ operators per block. With an auxiliary vertex qubit polarized by $X_{\mu_v}=+1$, the local enrichment operator is
\begin{equation}
 \begin{aligned}
 \operatorname{ctrl}_{\mu_v}(U_v)
 & =\operatorname{CCZ}_{\mu_v,g_{e_x(v)},b_{e_y(v+\hat x)}}\\[-2pt]
 &\quad\times\operatorname{CCZ}_{\mu_v,g_{e_y(v)},b_{e_x(v+\hat y)}}.
 \end{aligned}
 \label{eq:prx334-d4-enrichment}
\end{equation}
Its action follows from the conjugation identity
\begin{equation}
 \operatorname{CCZ}_{\mu,a,b}X_\mu\operatorname{CCZ}_{\mu,a,b}
 =X_\mu\operatorname{CZ}_{a,b}.
 \label{eq:prx335-ccz-dressing}
\end{equation}
Thus, the two CCZ operators dress $X_{\mu_v}$ with the two green--blue CZ factors in $U_v$, turning the auxiliary constraint into $X_{\mu_v}U_v=+1$.

Take the square lattice as the gauging graph and add a red gauge qubit $r_e$ with $Z_{r_e}=+1$ to each edge. Here, $r_e$ denotes the gauge qubit called $g_e$ in the general construction. The link coupling attaches $X$ operators on the four incident red qubits, transforming the vertex constraint into $X_{\mu_v}A_{r,v}=+1$, where $A_{r,v}=U_v\prod_{e\ni v}X_{r_e}$. Measuring $X_{\mu_v}$ on the auxiliary vertex qubits with outcomes $a_v$ gives the red Gauss laws $A_{r,v}=a_v$. The red flux checks satisfy $B_{r,f}=+1$ by the same closed-loop argument.

We now derive the two modified parent stars from Eq.~\eqref{eq:prx273-rooted-check}. For the green star $S_{g,v}$, the active blocks are $v$, $v-\hat x$ and $v-\hat y$. Choose $v$ as the reference block. The paths to the other two blocks are the single edges $e_x(v-\hat x)$ and $e_y(v-\hat y)$, respectively. Thus,
\begin{align}
 \bm W_{g,v}
 &=\operatorname{ctrl}_{r_{e_x(v-\hat x)}}(U_{v-\hat x})\nonumber\\
 &\quad\times\operatorname{ctrl}_{r_{e_y(v-\hat y)}}(U_{v-\hat y}),\nonumber\\
 A_{g,v}&=\bm W_{g,v}S_{g,v}\bm W_{g,v}^\dagger\nonumber\\
 &=S_{g,v}\operatorname{CZ}_{r_{e_x(v-\hat x)},b_{e_y(v)}}\nonumber\\
 &\quad\times\operatorname{CZ}_{r_{e_y(v-\hat y)},b_{e_x(v)}}.
 \label{eq:smd4-green-path-dressing}
\end{align}
Each controlled $U$ contains two CCZ gates. Only one meets the green star. Conjugating its $X$ produces the displayed red--blue CZ by Eq.~\eqref{eq:prx335-ccz-dressing}, with the qubit labels permuted.

For the blue star $S_{b,v}$, the active blocks are $v-\Delta$, $v-\hat x$ and $v-\hat y$. Choose $v-\Delta$ as the reference. The paths to $v-\hat y$ and $v-\hat x$ are the single edges $e_x(v-\Delta)$ and $e_y(v-\Delta)$, respectively. The same conjugation gives
\begin{align}
 \bm W_{b,v}
 &=\operatorname{ctrl}_{r_{e_x(v-\Delta)}}(U_{v-\hat y})\nonumber\\
 &\quad\times\operatorname{ctrl}_{r_{e_y(v-\Delta)}}(U_{v-\hat x}),\nonumber\\
 A_{b,v}&=\bm W_{b,v}S_{b,v}\bm W_{b,v}^\dagger\nonumber\\
 &=S_{b,v}\operatorname{CZ}_{r_{e_x(v-\Delta)},g_{e_y(v-\hat y)}}\nonumber\\
 &\quad\times\operatorname{CZ}_{r_{e_y(v-\Delta)},g_{e_x(v-\hat x)}}.
 \label{eq:smd4-blue-path-dressing}
\end{align}
These are exact operator identities on the full Hilbert space. They do not require a flux constraint. Together with $A_{r,v}=U_v\prod_{e\ni v}X_{r_e}$, they give precisely the quadratic phases in Eq.~\eqref{eq:smd4-Q} and the three star diagrams in Eqs.~\eqref{eq:prx332-raw-r}--\eqref{eq:prx332-raw-b} of the main text. The parent green and blue flux checks remain unchanged because the path dressing is diagonal. With the red flux checks already derived above, this gives the full raw check family in Eqs.~\eqref{eq:prx332-raw-r}--\eqref{eq:prx332-raw-flux}.

Let $P_{\rm par}$ project onto two ordinary toric codes on green and blue edges. Initialize the gauge qubits in $|0\rangle$ and completion qubits in $|+\rangle$. Denote the embedding by $J$. Use the parent Clifford symmetry in Eq.~\eqref{eq:smd4-U}. Its disjoint four-qubit supports partition the parent matter qubits. On the parent's winding parities,
\begin{equation}
 \mathcal{U}=(-1)^{\alpha_g\beta_b+\beta_g\alpha_b}.
\end{equation}
Thus, its two eigenspaces have dimensions ten and six. Gauging fixes both red winding parities to zero. The surgery uses joint eigenspaces with outcome-dependent completed red-star signs. On a $|+\rangle_L^{\otimes4}$ input, its two measurement branches have logical computational-basis supports of sizes ten and six and are therefore nonstabilizer states.

The red winding operators have eigenvalues $(-1)^{\alpha_r}$ and $(-1)^{\beta_r}$:
\begin{equation*}
 W_{r,\mu}=\prod_{e\in C_\mu}Z_{r,e},\qquad \mu=x,y,
\end{equation*}
where $C_x$ and $C_y$ wrap once around the torus horizontally and vertically. Each eigenvalue is the product of the red $Z$ outcomes along that loop. These operators commute with every completed check. Gauge initialization fixes both values to $+1$, so a flat ideal gauge record is exact, $z=\delta x$, for either logical outcome $\eta$.

The raw red star is $X^{s_{r,v}}U_v$. When all gauge bits are zero the other stars reduce to the parent stars. Let $V$ be the torus vertex set and fix a reference vertex $v_0\in V$. For the vector $a=(a_v)_{v\in V}$ of completed red-star measurement outcomes, write $s=\prod_v a_v$ and $\Pi_s=(I+s\mathcal{U})/2$. For a binary vertex potential $y$, set $U(y)=\prod_v U_v^{y(v)}$ and $a^y=\prod_v a_v^{y(v)}$. Write $|+\rangle_t$ for the product state of all completion qubits. The normalized gauging isometry $\Gamma_a$ is
\begin{equation}
 \begin{gathered}
 \Gamma_a|\psi\rangle=c_V\sum_{y:y(v_0)=0}
 a^yU(y)|\psi\rangle|\delta y\rangle_r|+\rangle_t,\\
 c_V=2^{-(|V|-1)/2},
 \end{gathered}
 \label{eq:smd4-gamma}
\end{equation}
Here, $|\delta y\rangle_r$ is the gauge-qubit configuration. Let $\widetilde{Q}_a$ project onto $B=T=+1$, positive completed green and blue stars, and completed red-star outcomes $a$. Its initialized-boundary identity is
\begin{equation}
 \widetilde{Q}_aJ=c_V\Gamma_a\Pi_sP_{\rm par}.
 \label{eq:smd4-idealboundary}
\end{equation}
It holds on arbitrary parent matter states. Central flux projection acts first, green/blue stars then impose the parent electric constraints, and the remaining red-star product gives the sum with the reference value fixed. Its restriction to zero red winding parities is $Q_a=\Gamma_a\Pi_sP_{\rm par}\Gamma_a^\dagger$.

For the parent version of Lemma~\ref{lem:smd4-clean}, write $z_m=(z_g,z_b)$ for the green/blue matter configuration and $x_m=(x_g,x_b)$ for its flip mask. If a parent monomial $M=X^{x_m}D_p$ has flat $x_g,x_b$ with an empty cycle in each direction, and, for each of the four parent logical $X$ operators, there is a representative whose support is disjoint from that of $D_p$, then
\begin{equation}
 P_{\rm par}MP_{\rm par}\in\mathbb{C}P_{\rm par}.
 \label{eq:smd4-parentclean}
\end{equation}
Indeed, $x_c=\delta\tau_c$, so its $X$ part is a parent stabilizer and $P_{\rm par}MP_{\rm par}=P_{\rm par}D_pP_{\rm par}$. Choose representatives of the four parent logical $X$ operators whose supports avoid the variables in $p$. Each commutes with $D_p$ by disjoint support. The projected diagonal operator also commutes with every logical $Z$. It thus commutes with the full logical operator algebra of the two parent toric codes and must be scalar.

\begin{lemma}[Input and output transfer]
\label{lem:smd4-transfer}
Let $E$ be a monomial on the edge qubits, possibly tensored with a completion Pauli. If all its obstruction sets have size below $L$, then for some scalar $\lambda_E$,
\begin{equation}
 \widetilde{Q}_aEJP_{\rm par}=\lambda_E \widetilde{Q}_aJP_{\rm par}.
 \label{eq:smd4-input}
\end{equation}
For an exact string of gauge-qubit measurement outcomes $\widehat{z}=\delta\widehat{y}$, absorb any gauge-qubit measurement fault into $E$. For each completion basis bra $\langle\xi|$ resolving the discard, there is a scalar $\mu_E$, possibly depending on the record and bra, such that
\begin{equation}
 \begin{split}
 P_{\rm par}U(\widehat{y})^\dagger\,{}_r\!\langle\widehat{z}|
 \,{}_t\!\langle\xi|E\Gamma_a\Pi_sP_{\rm par}
 =\mu_E\Pi_sP_{\rm par}.
 \end{split}
 \label{eq:smd4-output}
\end{equation}
\end{lemma}
\begin{proof}
Completion Pauli contraction is scalar, possibly zero, so it suffices to consider the edge qubits. For the input, flux projection forces all $x_c$ flat. The empty cycles make them exact. Write $x_r=\delta y$ and normalize $y$ to vanish on every row/column outside $H_r,V_r$, as in Lemma~\ref{lem:smd4-clean}. This also forces both red winding parities to vanish. Setting the initial gauge bits to zero in the phase gives $q_0(z_m)=q(0,z_m)$. The gauging sum reduces the input to
\begin{equation}
 \Gamma_a\Pi_sP_{\rm par}U(y)X^{x_m}D_{q_0}P_{\rm par},
\end{equation}
up to a scalar. Fixing the reference value only multiplies by global $\mathcal{U}$, scalar on the selected branch. If $U(y)=D_{\Phi_y}$, the reduced phase is $\Phi_y(z_m+x_m)+q_0(z_m)$. A selected red vertex at $(i,j)$ contributes horizontal phase variables only at column $i$ and vertical variables only at row $j$. These coordinates lie in $V_r,H_r$, already included in the green/blue phase obstructions. Translation introduces no further variables. The four chosen parent logical-$X$ representatives therefore still avoid the phase error, and Eq.~\eqref{eq:smd4-parentclean} proves Eq.~\eqref{eq:smd4-input}.

At output let $w$ denote the vertex potential in the selected term of the gauging sum. It obeys $\delta w=\widehat{z}+x_r$. A nonzero term requires $x_r=\delta y$, with $y=0$ on every row outside $H_r$ and column outside $V_r$, so $w=\widehat{y}+y$ up to a constant. After applying $U(\widehat{y})^\dagger$, the parent monomial phase is
\begin{align}
 p(z_m)&=q(\widehat{z}+x_r,z_m)+\Phi_y(z_m)\nonumber\\
 &\quad+\Phi_{\widehat{y}}(z_m+x_m)+\Phi_{\widehat{y}}(z_m).
\end{align}
Substituting the gauge bits introduces no new matter variables. The $\Phi_y$ term has the bounds just proved. The last two terms are precisely the linear CZ-neighbor variables induced by matter $X$. Green vertical $X$ at column $i$ adds blue horizontal phase at $i$, while blue vertical $X$ at $i$ adds green horizontal phase at $i-1$. The row statement is identical. These are the shifts in Eq.~\eqref{eq:smd4-cuts}. Final parent projection annihilates nonflat matter masks. For surviving terms the masks are exact and Eq.~\eqref{eq:smd4-parentclean} gives Eq.~\eqref{eq:smd4-output}.
\end{proof}

Each fault during preparation of the gauge or completion qubits, or during measurement of the gauge qubits, has a local Pauli expansion and contributes at most one coordinate to each of the twelve sets. The byproduct correction uses a subset of the disjoint CZ pairs in Eq.~\eqref{eq:smd4-U}. A correlated participant fault on such a pair has the same coordinate sets on the edge qubits as a red-star CCZ fault. Moving it through the byproduct correction adds only the already-counted neighbor phases. Parent stars use four CNOTs controlled by a $|+\rangle$ ancilla, followed by $X$ readout. Parent plaquettes use four CNOTs targeting a $|0\rangle$ ancilla, followed by $Z$ readout. Serialize overlapping circuits. A star $X$ hook stays within one parent star, a plaquette $Z$ hook within one parent plaquette, and no parent gate joins the two colors.

\begin{proposition}[Bilayer preparation of a logical Toffoli resource]
\label{prop:smd4-toffoli}
Order the parent logical coordinates as $(1,2,3,4)=(\alpha_g,\beta_b,\beta_g,\alpha_b)$. Apply the bilayer symmetry measurement to $|+,+,1,+\rangle$, and let $s=(-1)^{\eta}$ be its outcome. Each ideal outcome has probability $1/2$. A known logical $\overline{X}_4^{\eta}$ correction gives $|1\rangle_3|\mathrm{Tof}\rangle_{124}$, where
\begin{equation}
 |\mathrm{Tof}\rangle_{124}=\frac{1}{2}\sum_{x,y=0}^{1}|x,y,xy\rangle_{124}.
\end{equation}
With the circuit and boundary conditions of Theorem~\ref{thm:smd4-surgery}, fewer than $\min(L,R)$ native faults give a scalar multiple of this intended corrected state for every measurement record passing that theorem's detection tests and every associated noise branch.
\end{proposition}
\begin{proof}
The logical symmetry is $\overline{\operatorname{CZ}}_{12}\overline{\operatorname{CZ}}_{34}$. On the input, qubit 3 is fixed to $1$, so its eigenvalue on a basis vector $|x,y,1,z\rangle$ is $(-1)^{xy+z}$. The outcome $(-1)^{\eta}$ retains precisely the four terms with $z=\eta+xy$ modulo two. Their common initial amplitude is $1/\sqrt{8}$, giving probability $1/2$ and normalized output $\frac{1}{2}\sum_{x,y}|x,y,1,\eta+xy\rangle$. The correction $\overline{X}_4^{\eta}$ gives the claimed state. Its support on the three active qubits is not an affine binary subspace, so it is a nonstabilizer state. Equivalently, $\overline{H}_4|\mathrm{Tof}\rangle=\overline{\operatorname{CCZ}}_{124}|+,+,+\rangle$. The fault statement follows by applying the scalar branch identity in Theorem~\ref{thm:smd4-surgery} to this fixed encoded input and then updating the known logical Pauli frame. It introduces no additional physical gate location.
\end{proof}

\subsection{Fault tolerance of bilayer logical measurement}
Let $S$ count the syndrome extraction rounds in each parent interval and $R$ the rounds in the gauged code. Run $S$ parent rounds, prepare the gauge and completion qubits, and run $R$ rounds in the gauged code. Measure the gauge qubits in $Z$, apply the byproduct correction, and run $S$ parent rounds. We assume ideal encoded input and an ideal final projection onto the parent code. The detection tests require positive outcomes for all parent checks and for the gauged flux checks, pinning checks, and completed green and blue stars. The completed red-star outcomes form the same vector $a$ in all $R$ rounds, and the string of gauge-qubit measurement outcomes is exact. The logical measurement outcome is $s=\prod_v a_v$.

\begin{theorem}[Circuit fault distance of bilayer logical measurement]
\label{thm:smd4-surgery}
For even $L\ge4$, this circuit and the stated native fault model give
\begin{equation}
 d_{\rm surg}=\min\{L,R\}.
\end{equation}
For a set $F_0$ of native fault locations, each record $m$ satisfying these conditions with reported sign $s(m)$ and each noise branch $\ell$, there is a scalar $\lambda_{F_0,m,\ell}$ such that
\begin{equation}
 P_{\rm par}K_{F_0,m,\ell}P_{\rm par}
 =\lambda_{F_0,m,\ell}\Pi_{s(m)}P_{\rm par}
\end{equation}
whenever $|F_0|<\min(L,R)$.
\end{theorem}
\begin{proof}
Fewer than $R$ faults leave a fault-free round in the gauged code. Because the checks commute, that round applies the joint projector $\widetilde{Q}_a$. Commute the products of ideal checks on each side up to the projector. In the resulting operator product, the residual errors from earlier faults stand to its right and those from later faults to its left. The subgroup $\mathsf{S}_0$ fixes both $JP_{\rm par}$ and the branch selected by the fault-free round, so removing its factors contributes only scalar phases. Lemma~\ref{lem:smd4-native} bounds the number of occupied coordinates in each input and output coordinate set by $|F_0|$. The initial and final parent-code measurement intervals reduce to sums of $EP_{\rm par}$ and $P_{\rm par}E'$ by commuting their Pauli errors through parent syndrome projectors. Their native faults also affect at most one coordinate in each coordinate set. Equations~\eqref{eq:smd4-input} and \eqref{eq:smd4-output} now give the claimed scalar multiple of the intended projector for every coherent term.

A product of matter Pauli $Z$ operators around a length-$L$ noncontractible loop, inserted after its supporting qubits' last participating gates, gives a nontrivial surviving action while all detection tests pass. It is nonscalar on either parent symmetry branch. Alternatively, flipping the same red-star measurement outcome in each of the $R$ gauged rounds reverses the logical measurement outcome while passing the temporal and parent checks. These prove the matching upper bounds.
\end{proof}

At $S=R=L$ the operation has $3L$ syndrome extraction rounds and circuit fault distance $L$.

\subsection{\texorpdfstring{A $D_4$ surface-code patch}{A D4 surface-code patch}}
\label{sec:sm279-d4-boundary}

\label{sec:prx279-d4-boundary}
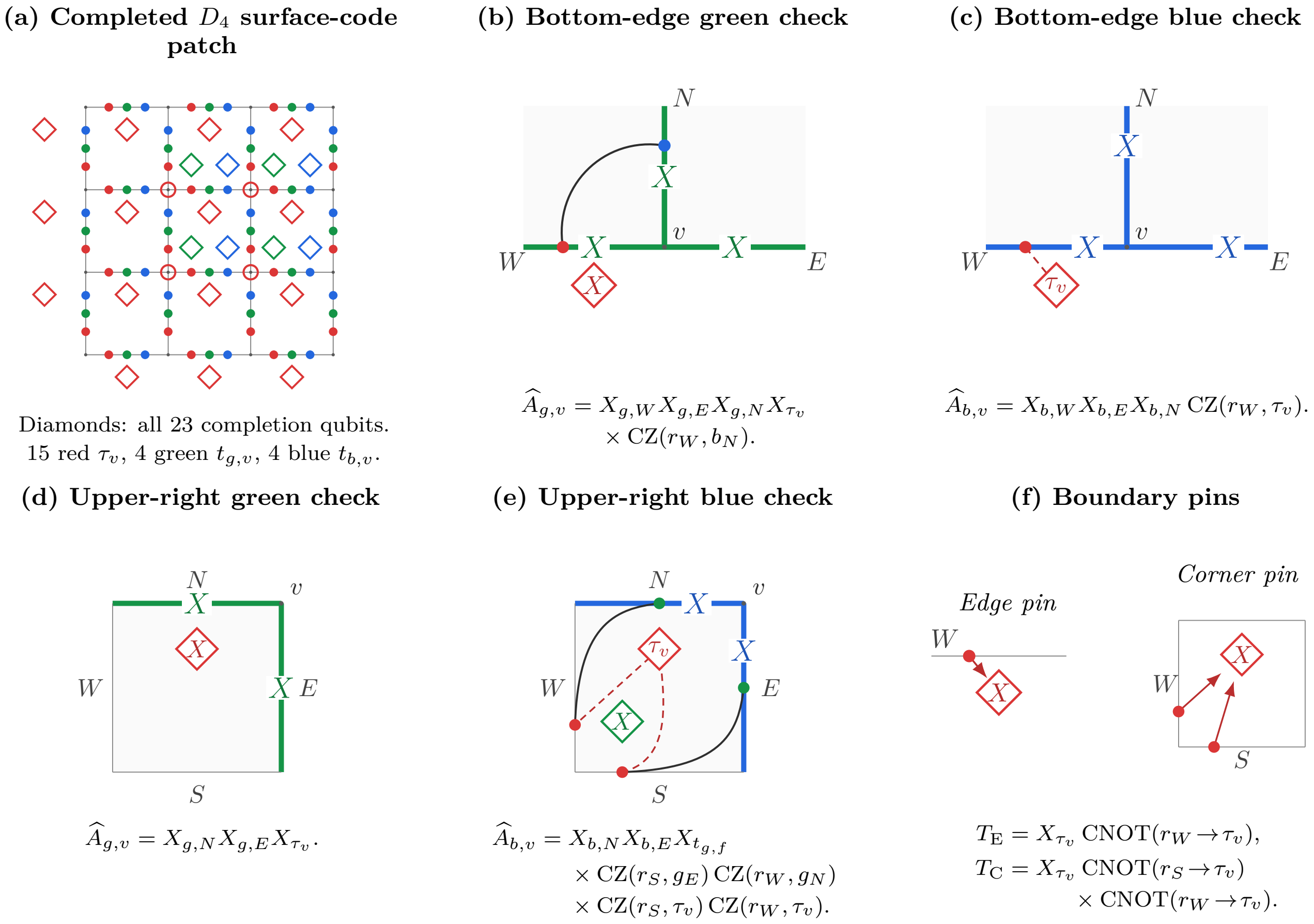
\begin{figure*}[tp]
\centering
\begin{minipage}{.98\textwidth}\centering
\begin{minipage}[t]{.326\linewidth}\centering\textbf{(a) Completed $D_4$ surface-code patch}\par\smallskip
\begin{minipage}[c][4.3cm][c]{\linewidth}\centering
\scalebox{0.96}{\begin{tikzpicture}[x=1.10cm,y=1.10cm,font=\small,>=Latex]
\path[use as bounding box] (-.68,-.55) rectangle (3.50,3.40);
\foreach \i in {0,1,2}{
  \foreach \j in {0,1,2,3}{
    \draw[black!48,line width=.38pt] (\i,\j)--({\i+1},\j);
    \fill[dFourRed] ({\i+.28},\j) circle[radius=1.65pt];
    \fill[dFourGreen] ({\i+.50},\j) circle[radius=1.65pt];
    \fill[dFourBlue] ({\i+.72},\j) circle[radius=1.65pt];
  }
}
\foreach \i in {0,1,2,3}{
  \foreach \j in {0,1,2}{
    \draw[black!48,line width=.38pt] (\i,\j)--(\i,{\j+1});
    \fill[dFourRed] (\i,{\j+.28}) circle[radius=1.65pt];
    \fill[dFourGreen] (\i,{\j+.50}) circle[radius=1.65pt];
    \fill[dFourBlue] (\i,{\j+.72}) circle[radius=1.65pt];
  }
}
\foreach \i in {1,2,3}{
  \foreach \j in {0,1,2,3}{
    \node[diamond,aspect=1.05,draw=dFourRed,fill=white,line width=.75pt,
          minimum size=3.1mm,inner sep=0pt] at ({\i-.50},{\j-.27}) {};
  }
}
\foreach \j in {1,2,3}{
  \node[diamond,aspect=1.05,draw=dFourRed,fill=white,line width=.75pt,
        minimum size=3.1mm,inner sep=0pt] at (-.50,{\j-.27}) {};
}
\foreach \i in {1,2}{
  \foreach \j in {1,2}{
    \node[diamond,aspect=1.05,draw=dFourGreen,fill=white,line width=.75pt,
          minimum size=3.1mm,inner sep=0pt] at ({\i+.28},{\j+.30}) {};
    \node[diamond,aspect=1.05,draw=dFourBlue,fill=white,line width=.75pt,
          minimum size=3.1mm,inner sep=0pt] at ({\i+.72},{\j+.30}) {};
  }
}
\foreach \i in {0,1,2,3}{
  \foreach \j in {0,1,2,3}{
    \fill[black!65] (\i,\j) circle[radius=.6pt];
  }
}
\foreach \i in {1,2}{
  \foreach \j in {1,2}{
    \draw[dFourRed,line width=.8pt] (\i,\j) circle[radius=2.7pt];
  }
}
\end{tikzpicture}}
\end{minipage}\par\smallskip
{\footnotesize Diamonds: all 23 completion qubits.\par
$15$ red $\tau_v$, $4$ green $t_{g,v}$, $4$ blue $t_{b,v}$.}
\end{minipage}
\hfill
\begin{minipage}[t]{.326\linewidth}\centering\textbf{(b) Bottom-edge green check}\par\smallskip
\begin{minipage}[c][4.3cm][c]{\linewidth}\centering
\scalebox{0.95}{\begin{tikzpicture}[x=1.9cm,y=1.9cm,font=\small,>=Latex]
\path[use as bounding box] (-1.16,-0.62) rectangle (1.16,1.15);
\fill[black!2] (-1,0) rectangle (1,1);
\draw[dFourGreen,line width=2.1pt] (-1,0)--(0,0);
\draw[dFourGreen,line width=2.1pt] (0,0)--(1,0);
\draw[dFourGreen,line width=2.1pt] (0,0)--(0,1);
\draw[black!82,line width=.8pt] (-.72,0) .. controls (-.79,.43) and (-.43,.79) .. (0,.72);
\fill[dFourRed] (-0.72,0) circle[radius=2.35pt];
\fill[dFourBlue] (0,0.72) circle[radius=2.35pt];
\node[fill=white,inner sep=.65pt,text=dFourGreen!82!black,font=\large] at (-0.5,0) {$X$};
\node[fill=white,inner sep=.65pt,text=dFourGreen!82!black,font=\large] at (0.5,0) {$X$};
\node[fill=white,inner sep=.65pt,text=dFourGreen!82!black,font=\large] at (0,0.5) {$X$};
\node[diamond,aspect=1.05,draw=dFourRed,fill=white,line width=.9pt,minimum size=6mm,inner sep=.15pt,font=\normalsize,text=dFourRed!82!black] at (-0.5,-0.27) {$X$};
\fill[black!65] (0,0) circle[radius=.9pt];
\node[anchor=south west,inner sep=1pt,font=\normalsize,text=black!85] at (0.035,0.035) {$v$};
\node[font=\normalsize,text=black!75,inner sep=1pt] at (-1.08,-0.1) {$W$};
\node[font=\normalsize,text=black!75,inner sep=1pt] at (1.08,-0.1) {$E$};
\node[font=\normalsize,text=black!75,inner sep=1pt] at (0.14,1.06) {$N$};
\end{tikzpicture}}
\end{minipage}\par\smallskip
{\footnotesize $\begin{aligned}\widehat{A}_{g,v}={}&X_{g,W}X_{g,E}X_{g,N}X_{\tau_v}\\[-1pt]&{}\times\operatorname{CZ}(r_W,b_N).\end{aligned}$}
\end{minipage}
\hfill
\begin{minipage}[t]{.326\linewidth}\centering\textbf{(c) Bottom-edge blue check}\par\smallskip
\begin{minipage}[c][4.3cm][c]{\linewidth}\centering
\scalebox{0.95}{\begin{tikzpicture}[x=1.9cm,y=1.9cm,font=\small,>=Latex]
\path[use as bounding box] (-1.16,-0.62) rectangle (1.16,1.15);
\fill[black!2] (-1,0) rectangle (1,1);
\draw[dFourBlue,line width=2.1pt] (-1,0)--(0,0);
\draw[dFourBlue,line width=2.1pt] (0,0)--(1,0);
\draw[dFourBlue,line width=2.1pt] (0,0)--(0,1);
\draw[dFourRed!85!black,densely dashed,line width=.7pt] (-.72,0)--(-.5,-.27);
\fill[dFourRed] (-0.72,0) circle[radius=2.35pt];
\node[fill=white,inner sep=.65pt,text=dFourBlue!82!black,font=\large] at (-0.28,0) {$X$};
\node[fill=white,inner sep=.65pt,text=dFourBlue!82!black,font=\large] at (0.72,0) {$X$};
\node[fill=white,inner sep=.65pt,text=dFourBlue!82!black,font=\large] at (0,0.72) {$X$};
\node[diamond,aspect=1.05,draw=dFourRed,fill=white,line width=.9pt,minimum size=6mm,inner sep=.15pt,font=\normalsize,text=dFourRed!82!black] at (-0.5,-0.27) {$\tau_v$};
\fill[black!65] (0,0) circle[radius=.9pt];
\node[anchor=south west,inner sep=1pt,font=\normalsize,text=black!85] at (0.035,0.035) {$v$};
\node[font=\normalsize,text=black!75,inner sep=1pt] at (-1.08,-0.1) {$W$};
\node[font=\normalsize,text=black!75,inner sep=1pt] at (1.08,-0.1) {$E$};
\node[font=\normalsize,text=black!75,inner sep=1pt] at (0.14,1.06) {$N$};
\end{tikzpicture}}
\end{minipage}\par\smallskip
{\footnotesize $\widehat{A}_{b,v}=X_{b,W}X_{b,E}X_{b,N}\operatorname{CZ}(r_W,\tau_v).$}
\end{minipage}
\par\medskip
\begin{minipage}[t]{.326\linewidth}\centering\textbf{(d) Upper-right green check}\par\smallskip
\begin{minipage}[c][3.7cm][c]{\linewidth}\centering
\scalebox{0.9}{\begin{tikzpicture}[x=2.4cm,y=2.4cm,font=\small,>=Latex]
\path[use as bounding box] (-1.26,-1.2) rectangle (0.32,0.45);
\fill[black!2] (-1,-1) rectangle (0,0);
\draw[dFourGreen,line width=2.1pt] (-1,0)--(0,0);
\draw[dFourGreen,line width=2.1pt] (0,-1)--(0,0);
\draw[black!48,line width=.38pt] (-1,-1)--(0,-1);
\draw[black!48,line width=.38pt] (-1,-1)--(-1,0);
\node[fill=white,inner sep=.65pt,text=dFourGreen!82!black,font=\large] at (-0.5,0) {$X$};
\node[fill=white,inner sep=.65pt,text=dFourGreen!82!black,font=\large] at (0,-0.5) {$X$};
\node[diamond,aspect=1.05,draw=dFourRed,fill=white,line width=.9pt,minimum size=6mm,inner sep=.15pt,font=\normalsize,text=dFourRed!82!black] at (-0.5,-0.27) {$X$};
\fill[black!65] (0,0) circle[radius=.9pt];
\node[anchor=south west,inner sep=1pt,font=\normalsize,text=black!85] at (0.035,0.035) {$v$};
\node[font=\normalsize,text=black!75,inner sep=1pt] at (-0.5,0.14) {$N$};
\node[font=\normalsize,text=black!75,inner sep=1pt] at (0.16,-0.5) {$E$};
\node[font=\normalsize,text=black!75,inner sep=1pt] at (-0.5,-1.13) {$S$};
\node[font=\normalsize,text=black!75,inner sep=1pt] at (-1.13,-0.5) {$W$};
\end{tikzpicture}}
\end{minipage}\par\smallskip
{\footnotesize $\widehat{A}_{g,v}=X_{g,N}X_{g,E}X_{\tau_v}.$}
\end{minipage}
\hfill
\begin{minipage}[t]{.326\linewidth}\centering\textbf{(e) Upper-right blue check}\par\smallskip
\begin{minipage}[c][3.7cm][c]{\linewidth}\centering
\scalebox{0.9}{\begin{tikzpicture}[x=2.4cm,y=2.4cm,font=\small,>=Latex]
\path[use as bounding box] (-1.26,-1.2) rectangle (0.32,0.45);
\fill[black!2] (-1,-1) rectangle (0,0);
\draw[dFourBlue,line width=2.1pt] (-1,0)--(0,0);
\draw[dFourBlue,line width=2.1pt] (0,-1)--(0,0);
\draw[black!48,line width=.38pt] (-1,-1)--(0,-1);
\draw[black!48,line width=.38pt] (-1,-1)--(-1,0);
\draw[black!82,line width=.8pt] (-.72,-1) .. controls (-.12,-.98) and (-.03,-.76) .. (0,-.5);
\draw[black!82,line width=.8pt] (-1,-.72) .. controls (-.98,-.12) and (-.76,-.03) .. (-.5,0);
\draw[dFourRed!85!black,densely dashed,line width=.7pt] (-.5,-.27) .. controls (-.48,-.52) and (-.4,-.92) .. (-.72,-1);
\draw[dFourRed!85!black,densely dashed,line width=.7pt] (-.5,-.27)--(-1,-.72);
\fill[dFourRed] (-0.72,-1) circle[radius=2.35pt];
\fill[dFourRed] (-1,-0.72) circle[radius=2.35pt];
\fill[dFourGreen] (0,-0.5) circle[radius=2.35pt];
\fill[dFourGreen] (-0.5,0) circle[radius=2.35pt];
\node[fill=white,inner sep=.65pt,text=dFourBlue!82!black,font=\large] at (-0.28,0) {$X$};
\node[fill=white,inner sep=.65pt,text=dFourBlue!82!black,font=\large] at (0,-0.28) {$X$};
\node[diamond,aspect=1.05,draw=dFourRed,fill=white,line width=.9pt,minimum size=6mm,inner sep=.15pt,font=\normalsize,text=dFourRed!82!black] at (-0.5,-0.27) {$\tau_v$};
\node[diamond,aspect=1.05,draw=dFourGreen,fill=white,line width=.9pt,minimum size=6mm,inner sep=.15pt,font=\normalsize,text=dFourGreen!82!black] at (-0.72,-0.7) {$X$};
\fill[black!65] (0,0) circle[radius=.9pt];
\node[anchor=south west,inner sep=1pt,font=\normalsize,text=black!85] at (0.035,0.035) {$v$};
\node[font=\normalsize,text=black!75,inner sep=1pt] at (-0.5,0.14) {$N$};
\node[font=\normalsize,text=black!75,inner sep=1pt] at (0.16,-0.5) {$E$};
\node[font=\normalsize,text=black!75,inner sep=1pt] at (-0.5,-1.13) {$S$};
\node[font=\normalsize,text=black!75,inner sep=1pt] at (-1.13,-0.5) {$W$};
\end{tikzpicture}}
\end{minipage}\par\smallskip
{\footnotesize $\begin{aligned}\widehat{A}_{b,v}={}&X_{b,N}X_{b,E}X_{t_{g,f}}\\[-1pt]&{}\times\operatorname{CZ}(r_S,g_E)\operatorname{CZ}(r_W,g_N)\\[-1pt]&{}\times\operatorname{CZ}(r_S,\tau_v)\operatorname{CZ}(r_W,\tau_v).\end{aligned}$}
\end{minipage}
\hfill
\begin{minipage}[t]{.326\linewidth}\centering\textbf{(f) Boundary pins}\par\smallskip
\begin{minipage}[c][3.7cm][c]{\linewidth}\centering
\begin{minipage}{.47\linewidth}\centering
\textit{Edge pin}\par\smallskip
\scalebox{0.96}{\begin{tikzpicture}[x=1.8cm,y=1.8cm,font=\small,>=Latex]
\path[use as bounding box] (-0.98,-0.62) rectangle (0.12,0.22);
\draw[black!48,line width=.38pt] (-1,0)--(0,0);
\node[diamond,aspect=1.05,draw=dFourRed,fill=white,line width=.9pt,minimum size=6mm,inner sep=.15pt,font=\normalsize,text=dFourRed!82!black] (pinEdgeTarget) at (-0.5,-0.27) {$X$};
\draw[->,dFourRed!85!black,line width=.7pt,shorten >=.5pt] (-.72,0)--(pinEdgeTarget);
\fill[dFourRed] (-0.72,0) circle[radius=2.35pt];
\node[font=\normalsize,text=black!75,inner sep=1pt] at (-0.9,0.13) {$W$};
\end{tikzpicture}}\end{minipage}\hfill
\begin{minipage}{.50\linewidth}\centering
\textit{Corner pin}\par\smallskip
\scalebox{0.9}{\begin{tikzpicture}[x=1.8cm,y=1.8cm,font=\small,>=Latex]
\path[use as bounding box] (-1.17,-1.17) rectangle (0.12,0.18);
\draw[black!48,line width=.38pt] (-1,0)--(0,0);
\draw[black!48,line width=.38pt] (0,-1)--(0,0);
\draw[black!48,line width=.38pt] (-1,-1)--(0,-1);
\draw[black!48,line width=.38pt] (-1,-1)--(-1,0);
\draw[->,dFourRed!85!black,line width=.7pt,shorten >=3.9mm] (-0.72,-1)--(-.5,-.27);
\fill[dFourRed] (-0.72,-1) circle[radius=2.35pt];
\draw[->,dFourRed!85!black,line width=.7pt,shorten >=3.9mm] (-1,-0.72)--(-.5,-.27);
\fill[dFourRed] (-1,-0.72) circle[radius=2.35pt];
\node[diamond,aspect=1.05,draw=dFourRed,fill=white,line width=.9pt,minimum size=6mm,inner sep=.15pt,font=\normalsize,text=dFourRed!82!black] at (-0.5,-0.27) {$X$};
\node[font=\normalsize,text=black!75,inner sep=1pt] at (-0.5,-1.1) {$S$};
\node[font=\normalsize,text=black!75,inner sep=1pt] at (-1.1,-0.5) {$W$};
\end{tikzpicture}}\end{minipage}
\end{minipage}\par\smallskip
{\footnotesize $\begin{aligned}T_{\mathrm{E}}&=X_{\tau_v}\operatorname{CNOT}(r_W\!\to\!\tau_v),\\T_{\mathrm{C}}&=X_{\tau_v}\operatorname{CNOT}(r_S\!\to\!\tau_v)\\[-1pt]&\phantom{=X_{\tau_v}}{}\times\operatorname{CNOT}(r_W\!\to\!\tau_v).\end{aligned}$}
\end{minipage}
\end{minipage}
\caption{Boundary checks of the completed $D_4$ surface-code patch, in the qubit convention of Figure~\ref{fig:d4-completed-checks}. Filled dots are edge qubits and colored diamonds are completion qubits. Thick colored edges and $X$ labels denote Pauli factors. Solid black curves are raw CZs, dashed curves are completion CZs, and arrows are CNOTs. (a) The full completed $L=3$ patch. The boundary condition is $Z_{r,e}=+1$ on red perimeter qubits. Red stars occur at circled interior vertices, green and blue stars at every vertex, and all face fluxes are retained. Red completion diamonds also lie outside the bottom and left boundaries. No completion qubit is needed at the lower-left corner, where the green and blue stars commute. (b),(c) The paired checks at a bottom-edge vertex. (d),(e) The paired checks at the upper-right corner. Here, $N,E,S,W$ label the inward face $f=v-\Delta$, and the green $X$ diamond is $X_{t_{g,f}}$. (f) The associated pins. The formulas list every factor in the displayed checks.}
\label{fig:prx279-d4-boundary}
\end{figure*}

Figure~\ref{fig:prx279-d4-boundary} shows the completed $D_4$ surface-code patch using the bulk qubit placement and gate notation. The perimeter $Z_{r,e}$ checks specify the boundary condition. Completion makes the measurements compatible also when these checks have outcome $-1$.

The central green--blue commutator $K_v$ is the red face flux in the bulk, a one-edge red-$Z$ product on the left and bottom sides, a three-edge product on the top and right, and a two-edge product at the upper-right corner. All are generated by retained flux and perimeter checks. The lower-left pair commutes. Assign $\tau_v$ to each nontrivial green--blue pair, and $t_{b,v},t_{g,v}$ only to interior red stars. Then
\begin{align}
 \widehat{A}_{r,v}&=A_{r,v}X_{t_{b,v}}V(B_{g,v},t_{g,v}),
 \label{eq:prx279-d4-boundary-red}\\
 \widehat{A}_{g,v}&=A_{g,v}X_{\tau_v}V(B_{b,v},t_{b,v}),
 \label{eq:prx279-d4-boundary-green}\\
 \widehat{A}_{b,v}&=A_{b,v}X_{t_{g,v-\Delta}}V(K_v,\tau_v).
 \label{eq:prx279-d4-boundary-blue}
\end{align}
Omit factors without an assigned qubit. In the bulk, $\tau_v=t_{r,v-\Delta}$. Include every pin $T(K,t)$. Centrality and $X_tV=KVX_t$ prove global commutation, including all boundary and corner checks. The derivation below proves the code-space identity. For example, let $W$ be the edge west of the lower-boundary vertex in Figure~\ref{fig:prx279-d4-boundary}. The raw $\operatorname{CZ}(r_W,b_N)$ gives commutator $Z_{r,W}$. The completion factors $X_{\tau_v}$ and $\operatorname{CZ}(r_W,\tau_v)$ contribute the same sign, giving
\begin{equation}
 \widehat{A}_{g,v}\widehat{A}_{b,v}
 =Z_{r,W}^{2}\widehat{A}_{b,v}\widehat{A}_{g,v}
 =\widehat{A}_{b,v}\widehat{A}_{g,v}.
 \label{eq:prx280-boundary-cancellation}
\end{equation}
This holds for either eigenvalue of $Z_{r,W}$, with pin $X_{\tau_v}\operatorname{CNOT}(r_W\to\tau_v)$, and extends to every central boundary conflict.

A gapped boundary to vacuum specifies which bulk anyons condense and how they fuse, described by a Lagrangian algebra~\cite{KitaevKong2012Boundaries,Kong2014Condensation}. The checks in Figure~\ref{fig:prx279-d4-boundary} implement the boundary that fixes red gauge variables and permits green and blue boundary gauge transformations. The derivation below identifies its Lagrangian algebra explicitly. In the color-variable description, the red electric charge and three bosonic flux types can be absorbed at the boundary. Equation~\eqref{eq:sm280-boundary-gap} below shows that completion preserves this boundary while making its measurements compatible.

This surface-code patch is a vacuum disk. The three-boundary logical patch in Section~\ref{sec:prx278-t-source} uses a different termination. The preceding exact-distance theorem concerns the torus circuit.

We prove the boundary commutation and code-space statements illustrated in Figure~\ref{fig:prx279-d4-boundary}. The patch has vertices $(i,j)$ for $0\le i,j\le L$, horizontal edges $e_x(i,j)$ for $i<L$, and vertical edges $e_y(i,j)$ for $j<L$. There are three qubits on each edge. Write $I_L=\{1,\ldots,L-1\}^2$ for the interior vertices. Keep the raw green and blue stars at every vertex, red stars at $I_L$, all elementary face fluxes, and $Z_{r,e}$ for every perimeter edge. Use the quadratic polynomials in Equation~\eqref{eq:prx273-d4-quadratics} of the main text, deleting a monomial exactly when one of its two edge qubits is absent.

The raw conflicts involving a red star are unchanged:
\begin{align}
 A_{r,v}A_{g,v}&=B_{b,v}A_{g,v}A_{r,v},
 \label{eq:sm279-boundary-rg}\\
 A_{b,v+\Delta}A_{r,v}&=B_{g,v}A_{r,v}A_{b,v+\Delta},
 \qquad v\in I_L.
 \label{eq:sm279-boundary-br}
\end{align}
The only green--blue conflicts occur at the same vertex, $A_{g,v}A_{b,v}=K_vA_{b,v}A_{g,v}$. For $v=(i,j)$ their labels are
\begin{align}
 K_{(0,0)}&=I,
 \label{eq:sm279-boundary-k00}\\
 K_{(i,0)}&=Z_{r,e_x(i-1,0)},\qquad i\ge1,
 \label{eq:sm279-boundary-kbottom}\\
 K_{(0,j)}&=Z_{r,e_y(0,j-1)},\qquad j\ge1,
 \label{eq:sm279-boundary-kleft}\\
 K_{(i,j)}&=B_{r,(i-1,j-1)}
       Z_{r,e_x(i-1,L)}^{[j=L]}
       Z_{r,e_y(L,j-1)}^{[i=L]},\qquad i,j\ge1.
 \label{eq:sm279-boundary-kpositive}
\end{align}
An exponent in square brackets is one when its condition holds and zero otherwise. Thus, the full four-edge flux loses its north edge at the top boundary, its east edge at the right boundary, and both at their common corner. These identities follow by taking the discrete difference of each quadratic phase under the other star's $X$ shift. There are no additional off-site conflicts.

Every face flux commutes with every retained star because an incident star meets a full face boundary in two edges. A red perimeter $Z$ commutes with the red stars because no retained red star touches that edge. It commutes with green and blue shifts by color. Therefore, every $K_v$ and every other commutator label is central. Blindly truncating all three colors of boundary stars would not have this property. For example, the lower-edge commutator $K_{(1,0)}=Z_{r,e_x(0,0)}$ anticommutes with the omitted red star at $(1,0)$.

Complete the three classes of conflicts with the orientations of Equations~\eqref{eq:prx279-d4-boundary-red}--\eqref{eq:prx279-d4-boundary-blue}. The identity $X_tV(K,t)=KV(K,t)X_t$ cancels each raw commutator. All other pairs remain commuting because their labels are central. The pin
\begin{equation}
 T(K,t)=\frac{I-K}{2}+\frac{I+K}{2}X_t
 \label{eq:sm279-boundary-pin}
\end{equation}
also commutes with all completed stars and central checks. This follows from $KV=VK$, $[K,A]=0$ and the same single-qubit identity. Pins on different completion qubits commute. The argument applies to bulk--bulk, bulk--boundary and boundary--boundary pairs on the full Hilbert space.

Let $P_{\rm disk}$ be the common positive raw projector, including the perimeter $Z$ checks. All conflict labels equal $+1$ on this space, so the pins fix the completion qubits to $|+\rangle$. Conversely, positive central and pin checks force this product state and reduce the completed stars to the raw stars. Hence,
\begin{equation}
 P_{\rm completed,disk}=P_{\rm disk}\otimes
 (|+\rangle\langle+|)^{\otimes(3L^2-2L+2)}.
 \label{eq:sm279-boundary-projector}
\end{equation}

This is a nonempty vacuum termination. A flat green or blue edge string is a vertex coboundary, with its potential defined modulo a constant. A flat red string satisfying the perimeter checks is a coboundary of a potential that vanishes on the boundary. The retained stars act transitively on these configurations. Their only shift dependencies are the products of all green stars and all blue stars. Both products have zero phase on the constrained flat configurations. Substitution of the vertex potentials into the summed quadratic phases cancels the bulk terms, and the remaining terms vanish with the red boundary potential. The orbit average of the zero configuration is therefore a nonzero common positive state, and transitivity makes it unique.

We can identify the boundary condensation directly in the type-III $G=\mathbb Z_2^r\times\mathbb Z_2^g\times\mathbb Z_2^b$ description of Appendix~\ref{sec:prx313-typeiii}. For group elements $a,b,c\in G$, subscripts $r,g,b$ denote their color components. The perimeter checks restrict boundary edge values to the subgroup $H_\partial$, and the green and blue stars generate its boundary gauge transformations:
\begin{equation}
 \begin{gathered}
 H_\partial=\{(0,g,b):g,b\in\mathbb Z_2\},\\
 \omega(a,b,c)=(-1)^{a_r b_g c_b},\qquad
 \omega|_{H_\partial}=1.
 \end{gathered}
 \label{eq:v378-boundary-subgroup}
\end{equation}
The subgroup alone does not specify the boundary phases. To determine them, restrict the sum in Eq.~\eqref{eq:prx313-typeiii-phase} to the existing faces and set $\phi_r=0$ on the boundary. For every retained star, the finite difference of this phase under its vertex flip is exactly $q_{c,v}(\delta\phi)$, with absent-edge monomials deleted. Thus, on these flat configurations, the ground-state amplitudes are proportional to $(-1)^{F(\phi)}$, with no additional boundary phase. This identifies the boundary with $(H_\partial,\beta_\partial=1)$, where $\beta_\partial$ is the two-cochain specifying the boundary phases in the standard subgroup construction~\cite{BullivantHuWan2017Boundaries}.

These data specify a Lagrangian algebra~\cite{DavydovSimmons2017Lagrangian}. Its decomposition into bulk anyons is
\begin{equation}
 \begin{gathered}
 \mathcal A_\partial=\mathbf1\oplus e_r\oplus m_g^+
       \oplus m_b^+\oplus m_{g+b}^+,\\
 \dim\mathcal A_\partial=1+1+2+2+2=8.
 \end{gathered}
 \label{eq:v378-boundary-condensate}
\end{equation}
Here, $e_r$ is the red electric charge, with character $(-1)^{x_r}$, and $m_h^+$ is the dimension-two flux-$h$ anyon for which the gauge transformation $h$ acts as $+I$. These labels refer to the type-III variables. To check the decomposition, write $h=(0,u,v)\ne0$ in $H_\partial$. The projective multiplication phase and a corresponding irreducible representation are
\begin{equation}
 \begin{aligned}
 \alpha_h(x,y)&=(-1)^{x_r(vy_g+uy_b)},\\
 \rho_h(x)&=Z^{v x_g+u x_b}X^{x_r}.
 \end{aligned}
 \label{eq:v378-boundary-flux-rep}
\end{equation}
The Pauli matrices here act on the two-dimensional representation space. They obey $\rho_h(x)\rho_h(y)=\alpha_h(x,y)\rho_h(x+y)$. Moreover, $\rho_h(h)=I$, so the anyon is bosonic, and its restriction to $H_\partial$ has a one-dimensional invariant subspace. At zero flux, only $\mathbf1$ and $e_r$ are trivial on $H_\partial$. Fluxes outside $H_\partial$ do not occur in the condensate. The subgroup construction also fixes the algebra multiplication. The Lagrangian algebra describes the boundary realized by the checks collectively, rather than an algebra of individual stabilizer operators.

For the boundary gap, let $H_0^{\partial}$ sum the penalties $(I-Q)/2$ for the raw stars, fluxes and perimeter checks, together with $(I-X_t)/2$ on each completion qubit. Let $H_1^{\partial}$ use the completed stars and conditional pins instead. Their common ground projector is given in Eq.~\eqref{eq:sm279-boundary-projector}. Any negative central-check sector of $H_0^{\partial}$ costs at least one; in the positive central sector, its penalties commute. The penalties of $H_1^{\partial}$ commute globally. Both endpoints therefore have gap at least one, and for $0\le u\le1$,
\begin{equation}
 (1-u)H_0^{\partial}+uH_1^{\partial}\ge
 I-P_{\rm disk}\otimes(|+\rangle\langle+|)^{\otimes(3L^2-2L+2)}.
 \label{eq:sm280-boundary-gap}
\end{equation}
All terms have bounded range, including at the corners, so the path preserves the boundary gap uniformly in $L$. Thus, completion preserves the boundary phase and its condensation algebra.

\subsection{A square lattice with three boundary types}
\label{sec:v385-three-boundaries}
\label{sec:v385-three-boundary}
The triangular boundary condition of Ref.~\cite{KobayashiZhuHsin2026} can be realized on the same square cellulation. Here ``triangular'' refers to three boundary types and three junctions. We give the raw generators and their completion on $\Lambda_L=\{0,\ldots,L\}^2$, with $L\ge2$. The left side is $\mathcal L_r$, the top is $\mathcal L_b$, and the bottom and right sides together form $\mathcal L_{rb}$. Thus the lower-right geometric corner lies within one boundary segment.

Retain all three face fluxes. On boundary edges retain
\begin{equation}
 \begin{array}{c|c}
 \mathcal L_r&Z_{r,e}\\
 \mathcal L_b&Z_{b,e}\\
 \mathcal L_{rb}&Z_{g,e},\ Z_{r,e}Z_{b,e}.
 \end{array}
 \label{eq:v385-three-boundary-pins}
\end{equation}
Define $A_{c,v}$ from the shifts and quadratic polynomials of Eqs.~\eqref{eq:prx273-d4-raw} and \eqref{eq:prx273-d4-quadratics}, retaining existing edges and deleting a quadratic monomial only if one of its edges is absent. The retained raw stars are
\begin{equation}
 \begin{array}{c|c}
 \text{interior}&A_{r,v},A_{g,v},A_{b,v}\\
 \mathcal L_r\text{ interior}&A_{g,v},A_{b,v}\\
 \mathcal L_b\text{ interior}&A_{r,v},A_{g,v}\\
 \mathcal L_{rb}\text{ interior}&A_{r,v}A_{b,v}.
 \end{array}
 \label{eq:v385-three-boundary-stars}
\end{equation}
Retain $A_{g,(0,L)}$ at the upper-left junction and no star at the lower-left or upper-right junction. At $(L,0)$ retain $A_{r,(L,0)}A_{b,(L,0)}$, as on the rest of $\mathcal L_{rb}$. The two factors of each red--blue boundary star commute on the full Hilbert space. Neither quadratic phase contains a bit shifted by the other factor at that vertex. Every retained star is therefore a Hermitian Clifford involution.

For completeness, the nontrivial commutators can be listed without an assumption on the flux sector. The interior red--green and green--blue pairs have labels $B_{b,v}$ and $B_{r,v-\Delta}$, respectively. At a left-side vertex $(0,j)$, the green--blue label is $Z_{r,e_y(0,j-1)}$. At a top-side vertex $(i,L)$, the red--green label is $Z_{b,e_x(i,L)}$. A red factor at $v$ and a blue factor at $v+\Delta$ have label $B_{g,v}$ whenever both corresponding retained stars exist. Finally, adjacent red--blue boundary stars on the bottom or right side have label $Z_{g,e}$ on their common boundary edge. All other retained pairs commute. These relations follow by taking the binary difference of each quadratic phase under the other star's shift. There are $3L^2-2L-1$ nontrivial pairs.

Each label is generated by the retained flux and boundary checks. Those checks commute with every retained star. Face fluxes meet a star shift on an even number of edges. Boundary shifts preserve the pinned color or color parity in Eq.~\eqref{eq:v385-three-boundary-pins}. Hence all commutators are central on the full Hilbert space, including at the three junctions. In particular, no projection onto the positive boundary sector is needed for completion.

Index the retained raw stars by $i$ and write $C_iC_j=K_{ij}C_jC_i$. Independently orient each nontrivial conflict $i\to j$ and assign it a qubit $t_{ij}$. Define
\begin{align}
 \widehat C_i&=C_i
 \prod_{i\to j}X_{t_{ij}}
 \prod_{j\to i}V(K_{ji},t_{ji}),
 \label{eq:v385-three-boundary-completion}\\
 T_{ij}&=\frac{I-K_{ij}}{2}
       +\frac{I+K_{ij}}{2}X_{t_{ij}}.
 \label{eq:v385-three-boundary-completion-pin}
\end{align}
Together with the fluxes and boundary checks, these are globally commuting Clifford involutions. The same local cancellation $X_tV(K,t)=KV(K,t)X_t$ applies at bulk sites, boundary sites and junctions. There are $c_L=3L^2-2L-1$ completion qubits, all within a bounded neighborhood of their conflicting stars.

Let $P_\triangle$ be the raw positive code projector. The completed projector is exactly
\begin{equation}
 \widehat P_\triangle=P_\triangle\otimes
 (|+\rangle\langle+|)^{\otimes c_L}.
 \label{eq:v385-three-boundary-projector}
\end{equation}
Indeed, positive central checks set every $K_{ij}=1$, and the pins then fix every completion qubit to $|+\rangle$. Conversely, on that product state the completed stars reduce to their raw counterparts. The completion theorem therefore preserves the static code distance and the boundary code space exactly.

We can count this code space directly. There are $6L(L+1)$ edge qubits. The $3L^2$ face-flux constraints and $6L$ boundary constraints are independent. One way to see the latter statement is to express each flat color field as the coboundary of a vertex potential. Fix the red and blue potentials at the upper-left vertex and the green potential at the lower-left vertex. The left and top constraints successively fix the red and blue potentials there. Along the connected bottom--right path, the green potential is constant and the red-plus-blue potential is constant. These are $6L$ independent conditions, leaving $3L^2$ binary degrees of freedom.

There are $s=3L^2-1$ retained star shifts, and they are independent. A vanishing sum of these shifts would give a constant potential in each color. The potential at a junction with no star must be zero, so all three constants vanish. In the positive central sector the stars commute and square to the identity. Their independent shifts therefore give a free action with $2^{3L^2-1}$ configurations per orbit. Each orbit supports exactly one common positive state. There are two orbits, so $\operatorname{rank}P_\triangle=2$. This is the one-qubit boundary code used in the logical $T$ construction. The same central-sector argument used for Eq.~\eqref{eq:sm280-boundary-gap} gives a uniformly gapped local interpolation for this boundary family. Appendix~\ref{sec:v445-three-boundary-distance} analyzes the circuit fault distance for this patch.

\subsubsection{Every boundary generator and its completion}
\label{sec:v386-boundary-catalogue}
For $L\ge2$, we give the local operators for every boundary position, including the three changes of boundary type. This also fixes one explicit orientation of every completion conflict. At a vertex $v=(i,j)$ write
\begin{align*}
 E&=e_x(i,j),&N&=e_y(i,j),\\
 W&=e_x(i-1,j),&S&=e_y(i,j-1).
\end{align*}
Only existing edges are used. The other two edges of the northeast face $f_v$ are $E^+=e_y(i+1,j)$ and $N^+=e_x(i,j+1)$. The other two edges of the southwest face $f_{v-\Delta}$ are $S^-=e_x(i-1,j-1)$ and $W^-=e_y(i-1,j-1)$. Thus $\partial f_v=\{E,N,E^+,N^+\}$ and $\partial f_{v-\Delta}=\{W,S,S^-,W^-\}$. Write $M_v=A_{r,v}A_{b,v}$ for a red--blue boundary star.

\emph{Raw operators.} On the open left side, $v=(0,j)$ with $1\le j<L$,
\begin{align}
 A_{g,v}&=X_{g,E}X_{g,N}X_{g,S}
             \operatorname{CZ}(r_S,b_E),\nonumber\\
 A_{b,v}&=X_{b,E}X_{b,N}X_{b,S}.
 \label{eq:v386-left-raw}
\end{align}
On the open top side, $v=(i,L)$ with $1\le i<L$,
\begin{align}
 A_{r,v}&=X_{r,W}X_{r,E}X_{r,S},\nonumber\\
 A_{g,v}&=X_{g,W}X_{g,E}X_{g,S}
             \operatorname{CZ}(r_S,b_E).
 \label{eq:v386-top-raw}
\end{align}
On the open bottom side, $v=(i,0)$ with $1\le i<L$,
\begin{equation}
 \begin{split}
 M_v={}&\left(\prod_{e\in\{W,E,N\}}X_{r,e}X_{b,e}\right)\\
 &{}\times\operatorname{CZ}(g_E,b_{E^+})
          \operatorname{CZ}(g_N,b_{N^+}).
 \end{split}
 \label{eq:v386-bottom-raw}
\end{equation}
On the open right side, $v=(L,j)$ with $1\le j<L$,
\begin{equation}
 \begin{split}
 M_v={}&\left(\prod_{e\in\{W,N,S\}}X_{r,e}X_{b,e}\right)\\
 &{}\times\operatorname{CZ}(r_{S^-},g_S)
          \operatorname{CZ}(r_{W^-},g_W).
 \end{split}
 \label{eq:v386-right-raw}
\end{equation}
At the upper-left junction $v=(0,L)$, retain only
\begin{equation}
 A_{g,v}=X_{g,E}X_{g,S}\operatorname{CZ}(r_S,b_E).
 \label{eq:v386-upper-left-raw}
\end{equation}
There is no star at $(0,0)$ or $(L,L)$. At the lower-right geometric corner $v=(L,0)$, which lies within $\mathcal L_{rb}$, retain
\begin{equation}
 M_v=X_{r,W}X_{r,N}X_{b,W}X_{b,N}.
 \label{eq:v386-lower-right-raw}
\end{equation}
All face fluxes and all Pauli boundary checks remain as specified above, including the edges incident to junctions.

These formulas follow directly from the bulk quadratic phases. For example, the left green star loses the monomial containing its absent west edge, leaving $z_{r,S}z_{b,E}$. Both monomials of the left blue star contain absent southwest edges, so that star has no CZ factor. At the bottom only the CZ factors from the red star survive, whereas at the right only those from the blue star survive. At the upper-left junction the remaining CZ acts on two existing boundary edges and must be retained, even though it is trivial in the positive boundary sector. At the lower-right corner neither star has CZ factors because their required northeast or southwest edges are absent.

\emph{Completion orientation and placement.} Use four kinds of completion qubits:
\begin{enumerate}[leftmargin=1.5em,itemsep=2pt]
\item $\alpha_v$ for the red--green pair at $1\le i<L$, $1\le j\le L$, directed from red to green.
\item $\beta_v$ for the green--blue pair at $0\le i<L$, $1\le j<L$, directed from green to blue.
\item $\gamma_v$ for the pair consisting of the blue factor at $v+\Delta$ and the red factor at $v$, where $1\le i<L$, $0\le j<L-1$, directed from blue to red. A red or blue factor here belongs to $M_v$ on $\mathcal L_{rb}$.
\item $u_e$ for neighboring red--blue boundary stars, directed along the bottom from left to right and then along the right from bottom to top. Such a qubit exists precisely on $e_x(i,0)$, $1\le i<L$, and $e_y(L,j)$, $0\le j<L-1$.
\end{enumerate}
Place $\gamma_v$ inside $f_v$ and $u_e$ beside its boundary edge. Place $\alpha_v$ near the red--green pair, on its northeast side, and $\beta_v$ near the green--blue pair, on its southwest side. The latter placements can lie just outside the top and left edges. Every assignment has bounded range, including at a junction. The four counts are $L(L-1)$, $L(L-1)$, $(L-1)^2$ and $2(L-1)$, whose sum is $3L^2-2L-1$.

On an arrow with central label $K$, attach $X_t$ to its source and $V(K,t)$ to its target. Independent arrow orientations are allowed. A common ordering of all generators is unnecessary. With the arrows above, the completed left-side operators are
\begin{align}
 \widehat A_{g,v}&=A_{g,v}X_{\beta_v},\nonumber\\
 \widehat A_{b,v}&=A_{b,v}\operatorname{CZ}(r_S,\beta_v),
 \qquad v=(0,j),\ 1\le j<L.
 \label{eq:v386-left-completed}
\end{align}
The completed top-side operators are
\begin{align}
 \widehat A_{r,v}&=A_{r,v}X_{\alpha_v},\nonumber\\
 \widehat A_{g,v}&=A_{g,v}\operatorname{CZ}(b_E,\alpha_v),
 \qquad v=(i,L),\ 1\le i<L.
 \label{eq:v386-top-completed}
\end{align}
For the bottom side,
\begin{equation}
 \begin{split}
 \widehat M_v={}&M_vX_{u_E}
   \operatorname{CZ}(g_W,u_W)^{[i>1]}\\
 &{}\times\operatorname{CZ}(g_E,\gamma_v)
          \operatorname{CZ}(g_N,\gamma_v)\\
 &{}\times\operatorname{CZ}(g_{E^+},\gamma_v)
          \operatorname{CZ}(g_{N^+},\gamma_v),\\[-2pt]
 &\hspace{2cm}v=(i,0),\quad1\le i<L.
 \end{split}
 \label{eq:v386-bottom-completed}
\end{equation}
Here $u_E$ and $u_W$ mean the completion qubits assigned to those edges. An exponent in square brackets is one when its condition holds and zero otherwise. A factor with exponent zero is omitted. In particular, there is no $u_W$ for the first bottom star. The four CZs involving $\gamma_v$ are precisely $V(B_{g,f_v},\gamma_v)$, with every face edge displayed. For the right side,
\begin{equation}
 \begin{split}
 \widehat M_v={}&M_vX_{\gamma_{v-\Delta}}
      X_{u_N}^{[j<L-1]}\operatorname{CZ}(g_S,u_S),\\[-2pt]
 &\hspace{1cm}v=(L,j),\quad1\le j<L.
 \end{split}
 \label{eq:v386-right-completed}
\end{equation}
The last right-side star has no $u_N$. At the lower-right corner,
\begin{equation}
 \widehat M_{(L,0)}=M_{(L,0)}X_{u_N}
                    \operatorname{CZ}(g_W,u_W).
 \label{eq:v386-lower-right-completed}
\end{equation}
The upper-left star in Eq.~\eqref{eq:v386-upper-left-raw} needs no completion factor. No star is introduced at either empty junction. Equations~\eqref{eq:v386-left-completed}--\eqref{eq:v386-lower-right-completed} therefore include every completion factor on every boundary star.

\emph{All pin operators.} The central label for each completion qubit is
\begin{equation}
 \begin{array}{c|c}
 t&K_t\\ \hline
 \alpha_{(i,j)},\ j<L&B_{b,(i,j)}\\
 \alpha_{(i,L)}&Z_{b,e_x(i,L)}\\
 \beta_{(i,j)},\ i>0&B_{r,(i-1,j-1)}\\
 \beta_{(0,j)}&Z_{r,e_y(0,j-1)}\\
 \gamma_{(i,j)}&B_{g,(i,j)}\\
 u_e&Z_{g,e}.
 \end{array}
 \label{eq:v386-completion-labels}
\end{equation}
Each row retains the index range specified above. Every label is a product of $Z$ operators of one color on either one edge or one full face. If $K_t=\prod_{e\in F_t}Z_{c,e}$, the complete pin is
\begin{equation}
 T_t=X_t\prod_{e\in F_t}\operatorname{CNOT}(c_e\to t)
 =\frac{I-K_t}{2}+\frac{I+K_t}{2}X_t.
 \label{eq:v386-all-pins}
\end{equation}
Thus the pins on the left, top and $\mathcal L_{rb}$ boundaries each have one CNOT, while face pins have four. The positive $Z_{r,e}Z_{b,e}$ boundary checks are retained directly. They need no extra completion qubit.

As an example of the cancellation, the raw left-side pair obeys $A_{g,v}A_{b,v}=Z_{r,S}A_{b,v}A_{g,v}$. Its two completion factors obey $X_{\beta_v}\operatorname{CZ}(r_S,\beta_v) =Z_{r,S}\operatorname{CZ}(r_S,\beta_v)X_{\beta_v}$, so the two signs cancel. Adjacent red--blue boundary stars use the same identity with $Z_{g,e}$, and the diagonal red--blue pair uses it with the green face flux. Since these labels commute with every retained star, the added factors create no new conflict. This proves compatibility on the full Hilbert space, including negative-flux and negative-boundary sectors.

\subsubsection{Ideal switching and logical action}
\label{sec:v385-three-arc-switch}
\label{sec:v385-square-switch}
The three-boundary patch also gives an explicit ideal switching map. Here the green qubits are the gauge qubits, while red and blue form the folded surface-code parent. Let $V_g$ be the vertices where the boundary prescription retains a green star, and put $M=|V_g|$. Extend a binary assignment $x\in\mathbb F_2^{V_g}$ by zero to the boundary $\mathcal L_{rb}$. Its coboundary $(\delta x)_{uv}=x_u+x_v$ is a green edge string. Since the patch is a disk and $\mathcal L_{rb}$ is connected, every flat green string that vanishes on that boundary has a unique such potential. In particular, $\delta x=0$ implies $x=0$.

Write the retained green stars as
\begin{equation}
 A_{g,v}=X_g(\delta v)Q_v,\qquad
 Q(x)=\prod_{v\in V_g}Q_v^{x_v},
 \label{eq:v385-green-factor}
\end{equation}
where $Q_v$ is the product of red--blue CZ factors obtained by retaining the existing faces in Eq.~\eqref{eq:smd4-Q}. These diagonal Clifford operators commute. Let $P_{\mathrm{fold}}$ be the common positive projector of the red and blue face fluxes, the red and blue boundary constraints in Eq.~\eqref{eq:v385-three-boundary-pins}, and the Pauli stars obtained by restricting every retained non-green star to $g=0$. These are the truncated $\prod_{e\ni v}X_{r,e}$ and $\prod_{e\ni v}X_{b,e}$ wherever the corresponding individual stars are retained, and $\prod_{e\ni v}X_{r,e}X_{b,e}$ at each site with a retained red--blue boundary star. Their CZ factors act as identity when all green bits vanish. The normalized switching map is
\begin{equation}
 \Gamma|\psi\rangle
 =2^{-M/2}\sum_{x\in\mathbb F_2^{V_g}}
 Q(x)|\psi\rangle\otimes|\delta x\rangle_g,
 \qquad |\psi\rangle\in\operatorname{im}P_{\mathrm{fold}}.
 \label{eq:v385-three-arc-isometry}
\end{equation}
Extend $\Gamma$ by zero outside the parent code. Distinct potentials give orthogonal green configurations, proving $\Gamma^\dagger\Gamma=P_{\mathrm{fold}}$. To identify its image, put $J_0|\psi\rangle=|\psi\rangle\otimes|0\rangle_g$. Then
\begin{equation}
 \Gamma=2^{M/2}\left[\prod_{v\in V_g}\frac{I+A_{g,v}}2\right]
 J_0P_{\mathrm{fold}}.
 \label{eq:v405-folded-switch-projectors}
\end{equation}
The initial image satisfies all central constraints and retained non-green stars. In this central sector the green stars commute with those checks, so $P_\triangle\Gamma=\Gamma$. Conversely, every state in $\operatorname{im}P_\triangle$ has green support only on strings $\delta x$, with unique $x$ because the boundary $\mathcal L_{rb}$ is connected. The green-star constraints fix its matter coefficients to $Q(x)|\psi_0\rangle$, and its green-zero coefficient satisfies the defining checks of $P_{\mathrm{fold}}$. Thus $\Gamma\Gamma^\dagger=P_\triangle$.

Starting with green qubits in $|0\rangle$, measurement outcomes $(-1)^{m_v}$ for the green stars insert $(-1)^{m\cdot x}$ in the sum. Each outcome has probability $2^{-M}$, independent of the logical state. Choose a green edge chain $\eta$ whose boundary agrees with $m$ at the vertices in $V_g$. Its other endpoints may lie on $\mathcal L_{rb}$. Then $\eta\cdot\delta x=m\cdot x$, so applying $Z_g(\eta)$ removes this sign. Thus every ideal outcome gives the same isometry after its known correction.

The parent has one logical qubit. To see this directly, write its flat red and blue fields as $a_r=\delta\phi_r$ and $a_b=\delta\phi_b$. Fix their additive constants so that $\phi_r=0$ on $\mathcal L_r$ and $\phi_b=0$ on $\mathcal L_b$. The boundary $\mathcal L_{rb}$ imposes $\phi_r+\phi_b=\ell$ there, with $\ell\in\mathbb F_2$ constant. The allowed star operations connect all configurations with the same $\ell$. The two remaining classes are distinguished by $\overline{Z}=\prod_{e\in\mathcal L_{rb}}Z_{r,e}$, with eigenvalue $(-1)^\ell$.

\paragraph{Square-lattice phase calculation.}
The logical action of the physical gate $U$ from Ref.~\cite{KobayashiZhuHsin2026} can be checked directly for every patch size in this convention. For a flat red--blue configuration, write $r_v=\phi_r(v)$, $b_v=\phi_b(v)$ and $s_v=r_v+b_v+\ell\pmod 2$, so $s=0$ on $\mathcal L_{rb}$. The CNOT part of $U$ changes the green potential in Eq.~\eqref{eq:v385-three-arc-isometry} from $x$ to $x+s$. Put $Q(s)=(-1)^{q(s)}$ on this configuration. For a face with vertices $(0,1,2,3)=(\mathrm{SW},\mathrm{SE},\mathrm{NW},\mathrm{NE})$, define $r_{uv}=r_u\mathbin\oplus r_v$ and $b_{uv}=b_u\mathbin\oplus b_v$, with $r_e=r_{uv}$ for $e=(u,v)$. The total phase exponent in units of $\pi/4$, including the change of amplitude in the sum defining $\Gamma$, is
\begin{align}
 E&=2\sum_f(r_{01}b_{13}-r_{02}b_{23})
   -\sum_{e\in\mathcal L_{rb}}r_e+4q(s),\nonumber\\
 q(s)&=\sum_f(s_1r_{01}b_{13}+s_2r_{02}b_{23})\pmod2.
 \label{eq:v385-square-T-phase}
\end{align}
Orient $\mathcal L_{rb}$ from the lower-left corner along the bottom and then upward along the right side. Define the integer edge expression $F_\ell(u,v)=2(-1)^{s_u}r_u(b_u\mathbin\oplus b_v)$ modulo eight. Expanding binary sums gives
\begin{equation}
 F_\ell(v,w)-F_\ell(u,w)+F_\ell(u,v)
 =2(-1)^{s_v}r_{uv}b_{vw}\pmod8.
 \label{eq:v385-square-T-coboundary}
\end{equation}
Summing this identity over the two oriented triangles in every square reduces the first and third terms of $E$ to the boundary sum of $F_\ell$. Its left and top contributions vanish. On $\mathcal L_{rb}$, $s_u=0$ and $b_{uv}=r_{uv}$, so the remaining summand, including the second term of $E$, is $2r_u(r_u\mathbin\oplus r_v)-(r_u\mathbin\oplus r_v)=r_u-r_v$. It telescopes from $r=0$ at the lower-left corner to $r=\ell$ at the upper-right corner. Hence $E=-\ell\pmod8$, and therefore

\begin{equation}
 U\Gamma=\Gamma\overline{T}^{\dagger},\qquad
 U^{\dagger}\Gamma=\Gamma\overline{T}.
 \label{eq:v385-three-arc-logical-map}
\end{equation}
After $U^{\dagger}$, green $Z$ readout gives some $\delta x$. The unnormalized matter branch is $2^{-M/2}Q(x)\overline{T}|\psi\rangle$. Applying the known Clifford $Q(x)^{\dagger}$ returns $\overline{T}|\psi\rangle$ to the folded parent. In particular, input $|+\rangle_L$ gives the standard $T$ state in every ideal branch.

Equation~\eqref{eq:v385-three-boundary-projector} gives $\widehat\Gamma=\Gamma\otimes|+\rangle^{\otimes c_L}$. Thus $U\otimes I_t$ has the same logical action on the completed code, and the completion qubits can be discarded after ungauging. We now specify the measurements and record tests that protect this protocol against circuit faults.

\subsubsection{Faults in the three-boundary $T$ protocol}
\label{sec:v445-three-boundary-distance}

The three-boundary protocol protects the logical gate by repeated verification before and after its application. The following single-fault example shows the role of these measurements. Let $v_*=(0,L)$ be the upper-left junction. If a single readout fault flips its reported green-star sign, the prescribed correction differs by $Z_g(\eta)$, where $\eta$ joins $v_*$ to $\mathcal L_{rb}$. The inserted gate gives
\begin{align}
 U^\dagger Z_g(\eta)U&=Z_g(\eta)Z_r(\eta)Z_b(\eta),\nonumber\\
 Z_r(\eta)Z_b(\eta)P_{\mathrm{fold}}
 &=\overline ZP_{\mathrm{fold}}.
 \label{eq:v445-corner-report-obstruction}
\end{align}
The second equality follows by summing the flat red and blue vertex potentials along $\eta$: their sum is zero at $v_*$ and is the logical bit $\ell$ on $\mathcal L_{rb}$. Green $Z$ readout removes the first factor as a branch scalar, leaving an undetected logical $Z$ on the output. Thus one gauging round followed by the gate and direct ungauging, without the post-gate verification below, has fault distance one, although its ideal map is correct.

For the verified protocol, write $P=P_{\mathrm{fold}}$ and $\widehat P=\widehat\Gamma\widehat\Gamma^\dagger$ for the positive completed patch projector.  Let $\widehat P_m$ have green-star signs $(-1)^{m_v}$ and positive signs for all other checks.  Choose a green edge chain $\eta_m$ whose boundary on $V_g$ is $m$, and set $C_m=Z_g(\eta_m)$.  Its other endpoints may lie on $\mathcal L_{rb}$. Let $J|\psi\rangle=|\psi\rangle|0\rangle_g |+\rangle_t^{\otimes c_L}$ denote preparation of the gauge and completion qubits.

Use the following detection protocol.  After initializing these qubits, measure the full completed family for $R_{\rm pre}$ rounds. Require positive flux, boundary, non-green-star and pin outcomes, and require a common green-star vector $m$ in every round.  Apply $C_m$ and then $U^\dagger$.  Measure the full completed family for $R_{\rm post}$ further rounds, requiring every outcome to be positive. Finally measure the green edges in $Z$ and require the reported string $z$ to be flat and zero on $\mathcal L_{rb}$.  Thus $z=\delta x$ with a unique $x$ vanishing there.  Apply $Q(x)^\dagger$ to the parent and perform the final parent checks.  Parent-check rounds before the switch and after the correction may also be included; their required outcomes are positive.  The encoded input and final parent projection are ideal. We use the local Kraus fault model of Section~\ref{sec:prx-ft}. All physical preparations, native gates, measurements, resets and idles between these outer boundaries may be faulty.  Classical processing of the reported record is ideal, as in the other circuit-distance statements.

The completed checks, boundary checks and pins are measured by exact bounded-size circuits on their supports and assigned readout qubits. A central check and its pin share a joint measurement circuit with two readout qubits. Readout qubits are prepared afresh or reset before reuse. The gauge and completion qubits persist throughout the switch. Overlapping gadgets are serialized in a fixed bounded-depth round schedule.  Implement $U^\dagger=VW^\dagger$ by the disjoint face $\operatorname{CS}^{\dagger}/\operatorname{CS}$ pairs, the boundary $T$ gates, all red-to-green CNOTs, and then all blue-to-green CNOTs. The known exit correction is a subset of the disjoint red--blue CZ pairs.  The Pauli correction $C_m$ uses single-qubit gates.

For an explicit native extraction, prepare a star readout $a$ in $|+\rangle$.  First apply every CNOT controlled by $a$ for the raw star's $X$ factors and its outgoing completion-$X$ factors.  Next apply $\operatorname{CCZ}(a,u,v)$ for each raw CZ factor on $u,v$. Finally apply the incoming controlled-$V$ factors: if $K=\prod_{e\in F}Z_e$, these are $\operatorname{CCZ}(a,e,t)$ for $e\in F$.  Read $a$ in $X$. This order also applies to red--blue boundary stars.  Measure each central $Z$ check $K$ and its pin $T=P_-(K)+P_+(K)X_t$ jointly by preparing $a_B$ in $|1\rangle$ and $a_T$ in $|+\rangle$, applying the CNOTs from the checked edge qubits to $a_B$, then $\operatorname{Toffoli}(a_B,a_T\to t)$, and reading $a_B$ in $Z$ and $a_T$ in $X$.  A $Z$-readout bit $q_B$ reports $K=2q_B-1$; the $X$ readout reports the pin sign.  Other retained $Z$ checks use ordinary CNOT parity extraction.  Parent CSS $Z$ checks use a $|0\rangle$ readout targeted by data-controlled CNOTs and a final $Z$ readout; parent $X$ checks use a $|+\rangle$ readout controlling CNOTs onto the data and a final $X$ readout.

\begin{lemma}[A common unused row and column]
\label{lem:v445-folded-coordinate-compression}
Let $S$ be a set of edge qubits of the three colors, and let $I_0(S)$ and $J_0(S)$ contain their first and second edge coordinates, respectively.  Put
\begin{equation}
 I(S)=I_0(S)\cup(I_0(S)-1),\qquad
 J(S)=J_0(S)\cup(J_0(S)-1).
 \label{eq:v445-coordinate-cover}
\end{equation}
Coordinates are ordinary integers; there is no periodic identification. Suppose $|I(S)|<L$ and $|J(S)|<L$. Any operator supported on $S$ and the completion qubits acts as a scalar after projection onto the completed code. The same conclusion holds after either code switch, with parent projection on both sides, provided the recorded outgoing green string is exact.  Extra parent operators supported in $S$ can be included in either interface.
\end{lemma}

\begin{proof}
We place logical $X$ and $Z$ operators away from the error, including the extra support created by the code switches. An error that survives projection then commutes with both logical operators, so it acts as a scalar on the encoded qubit.

First take the matrix element on the completion qubits in their prescribed product state, or in the chosen Kraus branch for their final discard. Expansion in a Pauli basis on the completion qubits leaves a sum of edge operators supported in $S$. It is enough to consider $E=X_m^aX_g^hD_q$, where $D_q$ is diagonal. Initially $D_q$ can be chosen as a Pauli $Z$ product. The code switches can add other diagonal phases, so we keep this more general notation.

Every nonzero projected term has a green shift $h$ that is flat and zero on $\mathcal L_{rb}$.  Hence $h=\delta\xi$, uniquely, with $\xi=0$ on that boundary.  At input, this follows by projecting $EJ$ onto the positive green constraints.  In the middle it follows because both code states have exact green strings.  At output it follows because the reported string and the green string in the incoming code state are both exact; faults in the green readout are included in $h$.

A column not containing a flipped vertical green edge has constant $\xi$ down to the bottom boundary, and therefore has $\xi=0$. Likewise a row not containing a flipped horizontal green edge has $\xi=0$, since it reaches the right boundary.  Thus a nonzero $\xi(i,j)$ requires $i\in I_0(S)$ and $j\in J_0(S)$.  Consequently $Q(\xi)$ has red-horizontal phase variables only in columns $I_0(S)-1$, blue-horizontal phase variables only in columns $I_0(S)$, red-vertical phase variables only in rows $J_0(S)-1$, and blue-vertical phase variables only in rows $J_0(S)$.

The contractions can be written explicitly.  Suppressing completion scalars and writing $J_0|\psi\rangle=|\psi\rangle|0\rangle_g$,
\begin{align}
 \Gamma^\dagger E J_0P
 &=2^{-M/2}P Q(\xi)^\dagger X_m^aD_{q(\cdot,0)}P,
 \label{eq:v445-entry-local}\\
 \Gamma^\dagger E\Gamma
 &=2^{-M}\sum_y
 P Q(y+\xi)^\dagger X_m^a
 D_{q(\cdot,\delta y)}Q(y)P,
 \label{eq:v445-middle-local}\\
 Q(x)^\dagger\langle\delta x|E\Gamma
 &=2^{-M/2}
 [Q(x)^\dagger X_m^aQ(x)]
 D_{q(\cdot,\delta(x+\xi))}Q(\xi)P.
 \label{eq:v445-exit-local}
\end{align}
Parent projection is applied on the left of the last expression. The second expression is a sum of parent operators of the same form. Use $Q(y+\xi)=Q(y)Q(\xi)$ and commute the diagonal factors as needed. Conjugation by any $Q(y)$ adds only the other endpoint of a disjoint CZ pair to the phase of a matter $X$.  Those added horizontal variables have columns in $I(S)$, and those added vertical variables have rows in $J(S)$.  Substitution of a fixed green string adds no matter variable. The same statements hold after multiplying by other parent errors supported in $S$: bit translations change coefficients of diagonal polynomials but introduce no new variables.

Choose $i,j\in\{0,\ldots,L-1\}$ outside $I(S),J(S)$, respectively. There is a parent logical $Z$ consisting of a red path from $(0,j)$ to $(i,j)$ and then down to $(i,0)$, followed by the blue path from $(i,0)$ to $(i,L)$.  It avoids every matter bit flip, because all its horizontal edges lie in row $j$ and all its vertical edges in column $i$.  There is a parent logical $X$ with the following support: red horizontal edges in column $i$ at every height; blue horizontal edges in column $i$ at heights $0,\ldots,j$; and blue vertical edges in row $j$ at columns $0,\ldots,i$.  It avoids every phase variable just described.  To verify that this is a logical $X$, use vertex potentials
\begin{equation}
 \phi_r(u,v)=[u\ge i+1],\qquad
 \phi_b(u,v)=[u<i+1]\,[v<j+1].
 \label{eq:v445-folded-logical-x}
\end{equation}
They obey the left and top boundary conditions and have $\phi_r+\phi_b=1$ on $\mathcal L_{rb}$.  The logical $X$ and $Z$ intersect oddly and generate the one-qubit logical matrix algebra. Each resulting parent operator commutes with both logical operators. Its bit flips avoid the logical $Z$, and its diagonal phase avoids the logical $X$.  It therefore acts as a scalar after parent-code projection. Summing all terms proves the assertion.
\end{proof}

The folded parent has code distance $L$: for $|S|<L$, choose $i\notin I_0(S)$ and $j\notin J_0(S)$, so both logical representatives above avoid $S$ and every operator on $S$ acts as a scalar after projection onto the code. Conversely, $\prod_{i=0}^{L-1}Z_{r,e_x(i,L)}$ is a weight-$L$ logical $Z$, as it commutes with all parent checks and intersects the displayed logical $X$ once.

For an explicit conservative circuit constant, let $w$ bound every measured check weight and $q$ bound the number of measured checks containing a code or completion qubit.  Let $w_0$ bound the total number of code and completion qubits in any one extraction gadget, and also bound the arity of the other native locations.  Take these maxima over both the patch and parent measurement families.  Define
\begin{equation}
 K_0=w_0[1+q(w-1)],\qquad \kappa=32K_0.
 \label{eq:v445-conservative-constant}
\end{equation}
These constants are independent of $L$, the number of rounds, and the observed green-star vector.  The factor 32 bounds the support enlargement through at most four two-qubit layers, including the physical $U^\dagger$ and exit CZ correction, followed by the factor-two coordinate enlargement in Eq.~\eqref{eq:v445-coordinate-cover}. The one-qubit $T$ and Pauli layers do not enlarge support.

\begin{theorem}[Linear circuit distance for the \mbox{three-boundary} switch]
\label{thm:v445-three-boundary-linear-distance}
Under the preceding circuit, noise and outer-boundary assumptions, every noise Kraus branch with
\begin{equation}
 f<\min\left\{R_{\rm pre},R_{\rm post},
              \left\lceil L/\kappa\right\rceil\right\}
 \label{eq:v445-three-boundary-fault-bound}
\end{equation}
faulty native locations and a record satisfying the stated tests obeys
\begin{equation}
 P K_{m,z,\mu}P=c_{m,z,\mu}\,\overline T P.
 \label{eq:v445-three-boundary-instrument}
\end{equation}
The scalar may vanish.  In particular, the circuit fault distance is at least the minimum in Eq.~\eqref{eq:v445-three-boundary-fault-bound}; choosing both repeated intervals proportional to $L$ gives a linear lower bound.
\end{theorem}

\begin{proof}
Because $f<R_{\rm pre},R_{\rm post}$, there is a fault-free full round before the gate and another after it. Their projectors are $\widehat P_m$ and $\widehat P$. The propagation argument in Appendix~\ref{sec:locality-sm} confines each fault on a gadget's qubits $B$ to
\[
 N(B)=B\cup\bigcup_{\operatorname{supp}C\cap B\ne\varnothing}
                \operatorname{supp}C,
 \qquad |N(B)|\le K_0.
\]
Checks disjoint from $B$ cancel when commuted past the fault, so this bound is independent of the number of ideal rounds. Use the parent check family in the parent intervals and the patch check family in the patch intervals. Parent errors are placed beside the adjacent switching interface, where Lemma~\ref{lem:v445-folded-coordinate-compression} already allows them. They need not be propagated again through the patch checks. Several faults contribute the union of their individual neighborhoods. The argument applies term by term to arbitrary local Kraus faults, including faults occurring after earlier errors.

The random green-star outcome vector $m$ does not create an extended error support. Exactly,
\begin{equation}
 C_m\widehat P_m=\widehat P C_m,\qquad C_mJ=J.
 \label{eq:v445-sign-removal}
\end{equation}
Move the ideal $C_m$ through all earlier errors and the selected entry projector to the green qubits initialized in $|0\rangle$.  Conjugating an arbitrary operator by this product of single-qubit Paulis preserves its support, irrespective of the length of $\eta_m$.  Faults in its physical implementation remain one-qubit faults.  After this rearrangement, both selected patch projectors are positive.

Consider the interval containing $U^\dagger$.  Its terms have the form
\begin{equation}
 \widehat P E_+\widetilde U^\dagger E_-\widehat P.
 \label{eq:v445-middle-interval}
\end{equation}
Errors from physical correction gates can be included in $E_-$. Expand faults inside $U^\dagger$ at the output of that circuit, and write $\widetilde U^\dagger=E_U U^\dagger$ term by term.  Since $U^\dagger$ preserves the completed code,
\begin{equation}
 \widehat P E_+E_U U^\dagger E_-\widehat P
 =\widehat P F\widehat P U^\dagger,
 \quad F=E_+E_U(U^\dagger E_-U).
 \label{eq:v445-joint-middle-error}
\end{equation}
This combines the errors before, during and after $U^\dagger$. Its fixed number of two-qubit layers enlarges each fault support by only a constant factor, including faults within those layers.

For every term of every interval, let $S$ be the union of the edge supports so obtained, including faults in green readout, the physical exit correction, and any noisy parent rounds.  The conservative count in Eq.~\eqref{eq:v445-conservative-constant} gives $|I(S)|,|J(S)|\le\kappa f<L$.  Completion contractions cannot enlarge edge support.  Lemma~\ref{lem:v445-folded-coordinate-compression} therefore makes the error after switching into the patch, the combined error around the gate $\widehat\Gamma^\dagger F\widehat\Gamma$, and the error after switching back act as scalars on the parent code.  The exit formula includes incorrect green measurement reports as green bit flips.  The exactness test on the reported $z$ forces their surviving total shift to be exact, which is precisely the condition used in that lemma.

The remaining ideal maps are $\widehat\Gamma$, $U^\dagger\widehat\Gamma=\widehat\Gamma\overline T$, and the corrected exit map.  Their product is the intended logical $T$, up to the ideal branch normalization.  Multiplying the scalar factors and summing the expansions proves Eq.~\eqref{eq:v445-three-boundary-instrument} for every Kraus branch, including coherent local faults.
\end{proof}

\section{CSS code instances and native extraction}\label{sec:sm-gross}
This appendix supplies the code-specific checks, gate orders, schedules and fault witnesses for the collective-CNOT resource protocol in Appendix~\ref{sec:prx273-css-source}. We begin with two blocks of the $[[144,12,12]]$ gross bivariate bicycle (BB) code~\cite{Bravyi2024Memory}. Transversal CNOT~\cite{Chang2026Magic} acts as a product of twelve logical CNOTs. Protocol~\ref{alg:commuting-gauging} of the main text measures one sign of this product and returns both blocks to their parent code spaces. The implementation steps are given in Appendix~\ref{sec:sm-implementation}.

\paragraph{Gauging graph and checks.}
Each gross-code block has two sites $L_u,R_u$ per cell $u\in\Lambda=\mathbb{Z}_{12}\times\mathbb{Z}_6$. For shifts $a=((3,0),(0,1),(0,2))$ and $b=((0,3),(1,0),(2,0))$ indexed from zero, the parent checks have supports
\begin{align}
 S_u&=\{L_{u+a_i}\}_{i=0}^2\cup\{R_{u+b_j}\}_{j=0}^2,\nonumber\\
 T_v&=\{L_{v-b_j}\}_{j=0}^2\cup\{R_{v-a_i}\}_{i=0}^2.
\end{align}
Each block measures $X(S_u),Z(T_v)$. Let $H_X$ and $H_Z$ be the binary check matrices with row supports $S_u$ and $T_v$, respectively. Pair the control $a_q$ and target $b_q$ at each vertex $q$. Join the reference vertex $r_u=L_{u+a_0}$ of $S_u$, or $s_v=R_{v-a_0}$ of $T_v$, to its five other vertices. Merging duplicate edges gives a seven-regular graph with 144 vertices and 504 edges. Write $g_{D,q}$ for the gauge qubit on the edge from the reference vertex of $D$ to $q$. For $\mathcal{U}=\prod_q\operatorname{CNOT}(a_q\to b_q)$, the Gauss, modified parent and invariant checks are
\begin{align}
 A_q&=\operatorname{CNOT}(a_q\to b_q)\prod_{e\ni q}X_{g_e},\nonumber\\
 C_u^X&=X_a(S_u)\prod_{q\in S_u\setminus\{r_u\}}
       \operatorname{CNOT}(g_{S_u,q}\to b_q),\nonumber\\
 C_v^Z&=Z_b(T_v)\prod_{q\in T_v\setminus\{s_v\}}
       \operatorname{CZ}(g_{T_v,q},a_q),\nonumber\\
 N_u^X&=X_b(S_u),\qquad N_v^Z=Z_a(T_v).
 \label{eq:sm-gross-raw}
\end{align}
For $v=u+a_i+b_j$, $i=1,2$, let $F_{uij}$ be the product of gauge $Z$ around $(r_u,L_{u+a_i},s_v,R_{u+b_j},r_u)$. It is the group commutator of $C^X_u,C^Z_v$. Their other conflicts are with the Gauss checks at their reference vertices, with labels $N^X_u,N^Z_v$. Assign one completion qubit $t_{uij}$ per square conflict and $t^X_u,t^Z_v$ per conflict at a reference vertex. Abbreviate $X_{uij}=X_{t_{uij}}$, $V_{uij}=V(F_{uij},t_{uij})$, using Equation~\eqref{eq:main-completion} of the main text. The completed checks are
\begin{align}
 \widehat{A}_{r_u}&=A_{r_u}V(N_u^X,t_u^X),\nonumber\\
 \widehat{A}_{s_v}&=A_{s_v}V(N_v^Z,t_v^Z),\nonumber\\
 \widehat{C}^X_u&=C^X_u X_{t_u^X}\prod_{j=0}^2 X_{u1j}V_{u2j},\nonumber\\
 \widehat{C}^Z_v&=C^Z_v X_{t_v^Z}\prod_{j=0}^2 V_{v-a_1-b_j,1,j}X_{v-a_2-b_j,2,j}.
 \label{eq:gross-complete}
\end{align}
Every vertex is either an $r_u$ or an $s_v$, so the first two lines give all 144 Gauss checks. For each square, one modified parent check gains $X_t$ and the other gains $V(F_{uij},t)$ on the same qubit. Their commutator contributes a second $F_{uij}$, canceling the first. The pairs at reference vertices obey the same rule with $F_{uij}$ replaced by $N^X_u$ or $N^Z_v$. In addition to Eq.~\eqref{eq:gross-complete}, measure the invariant checks $N^X_u,N^Z_v$, the flux checks specified below, and one pinning check $T(B,t)=H_t\operatorname{ctrl}_t(-B)H_t$ for each completion qubit $t$ with its corresponding label $B$.

All raw checks in Eq.~\eqref{eq:sm-gross-raw} are Clifford and square to the identity, and all commutator labels are Pauli. The completion and resource arguments in Section~\ref{sec:prx-central} of the main text therefore give commuting Clifford checks and pins, with extraction gates at hierarchy level three or below.

\paragraph{Choice of flux checks.}
Measure the product of gauge $Z$ operators on the 432 squares and on additional cycles chosen so that the full cycle space is generated.

\subsection{Logical input and output}
\label{sec:sm269-toffoli}
Use positive CSS check signs and matched logical Pauli bases in the two blocks, with at least two logical qubits per block. The encoded input, collective observable and branch projection are given in Equations~\eqref{eq:prx273-css-logical}--\eqref{eq:prx273-css-projection}. For either outcome $s=(-1)^{\eta}$, the correction $Z_{B_2}^{\eta}$ gives the Toffoli resource of Equation~\eqref{eq:prx273-css-resource} on $(B_2,A_1,B_1)$. Each branch has probability $1/2$.

For the gross code, the remaining 21 logical qubits are fixed spectators. The preparation needs no logical Hadamard. A Hadamard on the target converts the output to a CCZ state if that encoded operation is available. The extraction below uses physical Toffoli and CCZ gates, and the cost of preparing the encoded stabilizer input is additional. Its assumed input boundary is the one specified in Theorem~\ref{thm:end-distance} of the main text.

\subsection{Explicit native extraction for the gross code}
\label{sec:sm269-gross-native}
Treat CNOT, CZ, Toffoli and CCZ as native gates on their listed participants. A fault can act arbitrarily on all participants of one such location. There is one dedicated reset measurement ancilla per measured check.

\paragraph{Completed Gauss and modified parent checks.}
Prepare the ancilla $m$ in $|+\rangle$, apply the completed check unitary controlled by $m$, and read $m$ in $X$. Use the displayed factor order in Eq.~\eqref{eq:gross-complete}. The support order for $S_u$ is
\begin{equation}
 L_{u+a_0},L_{u+a_1},L_{u+a_2},
 R_{u+b_0},R_{u+b_1},R_{u+b_2},
 \label{eq:sm269-gross-order}
\end{equation}
with the analogous displayed order for $T_v$ and increasing $j$ for the completion factors assigned to square conflicts. Use the controlled-gate rules in Appendix~\ref{sec:prx273-css-source}. Controlling $V(B,t)=\operatorname{ctrl}_t(B)$ gives a product of Toffoli gates for an $X$-type label and CCZ gates for a $Z$-type label.

\paragraph{Joint central-check and pin extraction.}
Each of the 144 invariant checks and 432 square flux checks has one completion qubit. Its central check $B$ and pin are
\begin{align}
 T(B,t)&=P_-(B)+P_+(B)X_t,\label{eq:sm269-pin}\\
 P_{s_B}(B)&=\frac{I+s_B B}{2}.
\end{align}
Initialize $a_B$ in $|1\rangle$ and $a_T$ in $|+\rangle$. For $X$-type $B$, first apply Hadamards to its data support. Accumulate the resulting $Z$ parity into $a_B$ by CNOTs controlled by the checked qubits and targeting $a_B$, apply $\operatorname{Toffoli}(a_B,a_T\to t)$, undo the data Hadamards, and read $a_B$ in $Z$ and $a_T$ in $X$. The central-check sign is $s_B=2q-1$ for readout bit $q$.

After parity accumulation, $a_B=1$ precisely when $B=+1$, so the Toffoli implements the required control of $X_t$ in that sector. The outcome $(s_B,x)$ has Kraus operator
\begin{equation}
 K_{s_B,x}=P_{s_B}(B)\frac{I+xT(B,t)}{2}.
 \label{eq:sm269-joint-instrument}
\end{equation}
Thus, this gadget realizes the two-check projective instrument on arbitrary inputs. The additional flux checks use ordinary $Z$-parity extraction.

\paragraph{Finite resources.}
The 432 square masks have binary rank 352. The graph cycle space has dimension $504-144+1=361$. Supplement the squares by nine cycles of weights $4,10,4,9,7,10,10,7,8$. These cycles are obtained by extending the square span with independent fundamental cycles of a breadth-first spanning tree rooted at $L_{(0,0)}$, with vertices and neighbors in lexicographic order.

There are 288 matter, 504 gauge and 576 completion qubits. Measuring 144 Gauss checks, 144 modified parent checks, 144 invariants, 441 fluxes and 576 pins uses 1449 dedicated measurement ancillas, giving 2817 allocated qubits. One gauged round contains 4677 CNOTs, 1512 Toffolis, 2520 CCZs, 432 CZs and 864 Hadamards, for 10005 unitary locations, plus 1449 preparations and readouts each.

For a reproducible schedule, extract central checks and pins first and completed noncentral checks second. Within each stage, sort gadgets lexicographically and greedily color their overlapping code-qubit supports. Execute colors consecutively and disjoint gadgets in parallel, serializing gates within each gadget. The two stages use ten and twelve colors, of depths 150 and 296. Assigning one tick to every native gate, preparation and readout gives 446 ticks per round. Parent intervals, idles, routing and synthesis into a different hardware gate set are separate costs.

\subsection{A native hook and a flag supplied by completion}
\label{sec:sm269-completion-flag}
Global commutation bounds propagation through repeated rounds, while individual extraction gadgets can still create several data errors. The gross circuit above gives an explicit example.

\paragraph{Eleven faults that pass the detection tests.}
Index parent coordinates by $q=6i+j$ for $L_{(i,j)}$ and $q=72+6i+j$ for $R_{(i,j)}$. During the first central-check stage, insert a $Z$ fault on the central-check readout ancilla $a_B$ immediately after the second data-to-$a_B$ CNOT in the extraction of $N^X_{(3,0)}$. The remaining CNOTs and final Hadamards produce a target-block $X$ hook on $\{20,93,96,102\}$. Insert ten further single-qubit $X_b$ faults after final data Hadamards of central-check gadgets, at
\begin{equation}
 \{17,43,58,85,93,100,101,118,125,141\}.
 \label{eq:sm269-ten-locations}
\end{equation}
The combined target error has support
\begin{equation}
 \begin{split}
 E=\{&17,20,43,58,85,96,100,101,\\
     &102,118,125,141\}.
 \end{split}
 \label{eq:sm269-gross-logical-support}
\end{equation}
It has zero $H_Z$ syndrome and anticommutes with the zero-$H_X$-syndrome $Z$ operator on
\begin{equation}
 \{0,1,4,5,18,20,36,37,40,41,54,56\}.
 \label{eq:sm269-gross-witness}
\end{equation}
The odd overlap certifies that $X_b(E)$ is a nontrivial logical Pauli, whose $+1$ and $-1$ eigenspaces have equal dimensions. Since $\operatorname{Tr}\overline{\mathcal{U}}=2^{12}$, the two measured symmetry eigenspaces have unequal dimensions. Hence, $X_b(E)$ cannot be scalar on both of them. The individual faults leave the central-check and pin readouts unchanged. Their product is established before the first modified parent checks are measured, commutes with every completed check, and commutes with the blockwise CNOT correction at ungauging. It therefore passes the detection tests for any number of subsequent rounds and yields
\begin{equation}
 d_{\mathrm{surg}}\leq11
 \label{eq:sm269-eleven-upper-bound}
\end{equation}
for this native circuit. This witness motivates using the existing completion qubit to detect the hook, as described next.

\paragraph{Using the completion qubit as a flag.}
Add $\operatorname{CNOT}(t\to a_B)$ immediately before and after the CNOTs from the checked qubits to $a_B$, as in Section~\ref{sec:prx-dual-role} of the main text. For an $X$-type label, place both added CNOTs between the data basis changes. Its ideal instrument remains Equation~\eqref{eq:sm269-joint-instrument}. For an input satisfying $B=X_t=+1$ and reported outcomes $B=T=+1$, Lemma~\ref{lem:prx273-completion-flag} of the main text places every single-fault Kraus operator, restricted to this input, in the span of residual operators supported on at most one data qubit and the completion qubit.

\begin{table*}[!tp]
\centering
\begin{tabular}{lrrrrrrrr}
\hline
Parent $(n,k)$ & Gauge & Completion & Readout & Total & Maximum & Colors & Round ticks & Third-level\\
 & $|E|$ & $c$ & $M$ & $N$ & check support & & & gates $G_3$\\
\hline
LCS $(15,3)$ & 27 & 21 & 74 & 152 & 25 & 15 & 221 & 261\\
Mitten $(150,30)$ & 819 & 1032 & 2330 & 4481 & 70 & 51 & 1888 & 10502\\
Lifted product $(560,112)$ & 3053 & 3879 & 8778 & 16830 & 73 & 50 & 1929 & 38984\\
Kasai $(1104,556)$ & 5406 & 4854 & 11382 & 23850 & 80 & 32 & 1253 & 54192\\
\hline
\end{tabular}
\caption{Explicit two-block extraction circuits. $N=2n+|E|+c+M$ includes one dedicated readout ancilla per check. $G_3$ counts native Toffoli and CCZ gates. Disjoint complete gadgets run in parallel, with serial gates within each gadget and one tick per native gate, preparation or readout. All checks and pins enter the support and color counts. These extraction resources exclude routing, gate synthesis and parent-code intervals.}
\label{tab:sm269-native-resources}
\end{table*}

\begin{table}[!htbp]
\centering
\begin{tabular}{lrrrr}
\hline
Parent & $n$ & $k$ & $d$ & $(r_X,r_Z)$\\
\hline
Lift-connected surface & 15 & 3 & 3 & $(6,6)$\\
Mitten & 150 & 30 & 10 & $(60,60)$\\
Abelian lifted product & 560 & 112 & 14 & $(224,224)$\\
Kasai APM & 1104 & 556 & 10 & $(274,274)$\\
\hline
\end{tabular}
\caption{Explicit CSS parents and their parent-code distances $d$. Here, $r_X,r_Z$ denote the binary check-matrix ranks. The fifteen-qubit distance is proved in the text. The other parent distances are those reported or certified in Refs.~\cite{Bhardwaj2026Mitten,Cheng2026DistanceAudit,KasaiCatalogue}.}
\label{tab:sm269-parent-codes}
\end{table}

\subsection{Other CSS parents and their compiled measurements}
\label{sec:sm269-other-css}
The logical preparation in Appendix~\ref{sec:prx273-css-source} requires two matched CSS blocks, each encoding at least two qubits. Table~\ref{tab:sm269-parent-codes} gives four explicit inputs. Let $L_X,L_Z$ have the binary supports of the chosen logical $X,Z$ operators as rows. Their matrices satisfy
\begin{align}
 H_XH_Z^{\mathsf{T}}&=0,\label{eq:sm269-css-commute}\\
 H_ZL_X^{\mathsf{T}}=H_XL_Z^{\mathsf{T}}&=0,\label{eq:sm269-css-kernels}\\
 L_XL_Z^{\mathsf{T}}&=I_k.
 \label{eq:sm269-css-pairing}
\end{align}
These identities specify the matched logical bases and verify Equation~\eqref{eq:prx273-css-logical}.

\paragraph{A fifteen-qubit example with an exact distance proof.}
Use the lift-connected surface construction~\cite{Old2024LCS}. Let $P$ be the $3\times3$ cyclic permutation with $P_{a,b}=1$ for $b=a+1\pmod3$, and $Q=I+P$ over $\mathbb{F}_2$. Its matrices are
\begin{align}
 H_X&=\begin{pmatrix}I&Q&0&0&I\\0&0&I&Q&Q^{\mathsf{T}}\end{pmatrix},
 \label{eq:sm269-lcs-hx}\\
 H_Z&=\begin{pmatrix}I&0&Q&0&I\\0&I&0&Q&Q^{\mathsf{T}}\end{pmatrix}.
 \label{eq:sm269-lcs-hz}
\end{align}
Each block has three columns, numbered consecutively from 0 to 14. The displayed identity blocks give rank six for each matrix, and multiplication gives $H_XH_Z^{\mathsf{T}}=0$. Hence, $k=3$. Each matrix has nonzero, pairwise distinct columns, excluding every weight-one or weight-two zero-syndrome Pauli. Mixed Paulis obey both independent syndrome conditions and are also excluded. The $X$ operator on $\{4,5,11\}$ and $Z$ operator on $\{0,1,4\}$ have zero syndrome and odd overlap. They anticommute, so neither is a stabilizer. Thus, $d=3$. Two copies use 30 matter qubits and implement the preparation of Appendix~\ref{sec:prx273-css-source} on six logical qubits.

\paragraph{Mitten, lifted-product and Kasai instances.}
For the Mitten $[[150,30,10]]$ and abelian lifted-product $[[560,112,14]]$ codes, use the explicit binary matrices and paired logical bases of Ref.~\cite{Bhardwaj2026Mitten}. Their maximum row weight is nine and maximum column weight is six. The larger parent distance 14 has an independent certification~\cite{Cheng2026DistanceAudit}.

For Kasai's affine-permutation construction~\cite{Kasai2026APM}, take the catalogue instance with $P=92$, $J=3$, $L=12$ and seed 92312005~\cite{KasaiCatalogue}. For $x,y\in\mathbb Z_{92}$ and $i=0,\ldots,5$, define
\begin{align}
 (F_i)_{x,y}&=\delta_{y,a_ix+b_i\bmod92},\label{eq:sm269-kasai-f}\\
 (G_i)_{x,y}&=\delta_{y,c_ix+d_i\bmod92},\label{eq:sm269-kasai-g}\\
 a&=(47,1,1,1,1,1),\label{eq:sm269-kasai-a}\\
 b&=(40,73,58,42,48,52),\label{eq:sm269-kasai-b}\\
 c&=(1,1,47,1,1,1),\label{eq:sm269-kasai-c}\\
 d&=(0,48,70,45,16,2).
 \label{eq:sm269-kasai-d}
\end{align}
For $0\leq i<3$ and $0\leq j<6$, its block entries are
\begin{align}
 (H_X)_{i,j}&=F_{j-i},& (H_X)_{i,6+j}&=G_{j-i},\label{eq:sm269-kasai-hx}\\
 (H_Z)_{i,j}&=G_{i-j}^{\mathsf{T}},&
 (H_Z)_{i,6+j}&=F_{i-j}^{\mathsf{T}},
 \label{eq:sm269-kasai-hz}
\end{align}
with subscripts modulo six. The resulting $276\times1104$ matrices have rank 274 each, row weight twelve and column weight three. Their complete paired logical bases supply 556 logical pairs for Equation~\eqref{eq:prx273-css-logical}.

\paragraph{Uniform check and circuit compilation.}
Let $S_u,T_v$ be the supports of rows of $H_X,H_Z$, with reference vertices $r_u,s_v$. Choose graph paths $p_u(q)$ and $p_v(q)$ from each reference vertex to the indicated support vertex. The path to the reference vertex itself is empty. The raw checks are
\begin{align}
 A_q&=\operatorname{CNOT}(a_q\to b_q)\prod_{e\ni q}X_{g_e},
 \label{eq:sm269-generic-gauss}\\
 C_u^X&=X_a(S_u)\prod_{q\in S_u}\prod_{e\in p_u(q)}
 \operatorname{CNOT}(g_e\to b_q),
 \label{eq:sm269-generic-x}\\
 C_v^Z&=Z_b(T_v)\prod_{q\in T_v}\prod_{e\in p_v(q)}
 \operatorname{CZ}(g_e,a_q),
 \label{eq:sm269-generic-z}\\
 N_u^X&=X_b(S_u),\qquad N_v^Z=Z_a(T_v).
 \label{eq:sm269-generic-invariants}
\end{align}
The Gauss conflicts at the reference vertices have labels $N_u^X,N_v^Z$. The remaining conflicts are $Z_g(\eta_{uv})$, where the binary edge chain is
\begin{equation}
 \eta_{uv}=\sum_{q\in S_u\cap T_v}
 \bigl(p_u(q)+p_v(q)\bigr).
 \label{eq:sm269-generic-cycle}
\end{equation}
Its boundary vanishes because CSS orthogonality makes $|S_u\cap T_v|$ even. Thus, these labels are central fluxes. Add one completion qubit for each nonzero conflict, apply Equation~\eqref{eq:main-completion} of the main text, and measure a spanning family of fluxes and every completion pin.

Table~\ref{tab:sm269-native-resources} uses merged stars centered at the reference vertices, one deterministic pass balancing their degrees and one balancing conflict orientations. It retains all nonzero commutator cycles and supplements their span to the full cycle space. Each check is measured by its own ancilla using the controlled-gate rules in Appendix~\ref{sec:prx273-css-source}. For a Pauli label $K=\prod_q P_q$ with overall phase $+1$, the separate pin gadget uses
\begin{equation}
 \operatorname{ctrl}_a[T(K,t)]
 =H_t\operatorname{CZ}_{a,t}
 \left(\prod_q\operatorname{CC}P_{a,t,q}\right)H_t,
 \label{eq:sm269-generic-pin-circuit}
\end{equation}
where $\operatorname{CC}P$ is Toffoli for $P=X$ and CCZ for $P=Z$. This follows by controlling $T=H_t\operatorname{ctrl}_t(-K)H_t$, including its conditional minus sign.

The numbers of measured fluxes are respectively $14,908,3443,4320$, with ranks $13,670,2494,4303$, equal to $|E|-n+1$ for each connected graph. These finite constructions supply the collective observable and native measurement circuits needed for the common magic-state protocol. Completion appends a fixed product state to the raw gauged code space and preserves its distance. The graph conditions of Appendix~\ref{sec:graph-distance} relate the gauged distance to the parent distance, while the specified native gate order determines fault propagation during extraction. The gross-code example makes this circuit dependence explicit and shows how the same completion qubit can flag a harmful hook.

\section{\texorpdfstring{Commuting-check $D(S_3)$}{Commuting-check D(S3)}}
\label{sec:sm-hybrid}

We apply Proposition~\ref{prop:zn-patch-checks} to charge conjugation of the qutrit toric code. The gauging procedure in Appendix~\ref{sec:sm-zn-measurement} measures this $\mathbb{Z}_2$ symmetry by entering the $D(S_3)$ quantum double and ungauging back. Each edge carries a matter qutrit and a gauge qubit. We give the commuting checks and logical measurement protocol, then prove that exact completion with fixed product-state auxiliaries cannot make all checks finite-hierarchy in this encoding.

\subsection{The native hierarchy obstruction for \texorpdfstring{$S_3$}{S3}}
\label{sec:prx273-s3-boundary}
For $S_3=D_3$, completion gives compatible checks and the exact code-space identity on the native system of qubits and qutrits. With $\omega=e^{2\pi i/3}$ and $Z_e$ on an outgoing qutrit,
\begin{equation}
 \widehat{R}_v Z_e\widehat{R}_v^\dagger Z_e^\dagger
 =\omega^{-1}|0\rangle\langle0|_{t_v}
  +\omega|1\rangle\langle1|_{t_v}=D_{t_v}.
 \label{eq:prx-s3-diagnostic}
\end{equation}
The next commutator with $X_{t_v}$ gives $D_{t_v}^2$, and another gives $D_{t_v}$ again. Neither is a Pauli up to phase. A Pauli commutator lowers a finite hierarchy level, so this nonterminating cycle excludes $\widehat{R}_v$ from every finite level of the native hierarchy.

Theorem~\ref{thm:s3-completion-nogo} excludes every finite-hierarchy completion preserving the same native code projector with product-state auxiliaries. Finite-hierarchy commuting checks have a product eigenbasis across the qubit--qutrit split, whereas this code contains no such vector. A padded qubit encoding changes that Pauli structure. The completion theorem establishes compatibility without imposing a finite hierarchy level.

\subsection{Applying the commuting-check construction}

Orient an $L\times L$ square torus ($L\ge4$) rightward and upward. Assign the two outgoing matter qutrits at each vertex $v$ to the symmetry factor $U_v=K_{N(v)}K_{E(v)}$, where $K|a\rangle=|-a\rangle$ for $a\in\mathbb{Z}_3$. Choose the same square lattice as the gauging graph. Write $X_e,Z_e$ for the qutrit Paulis and $\sigma_e^X,\sigma_e^Z$ for the gauge-qubit Paulis. For a vertex or plaquette, $N,E,S,W$ denote its incident or boundary edges. The parent toric-code generators are
\begin{equation}
 \begin{aligned}
 A_v^0&=X_NX_EX_S^{-1}X_W^{-1},\\
 B_p^0&=Z_NZ_E^{-1}Z_S^{-1}Z_W.
 \end{aligned}
 \label{eq:sm-s3-parent-generators}
\end{equation}
Let $J_v^X$ be $+1$ on the $A_v^0=1$ eigenspace and $-1$ on its other two eigenspaces, and define $J_p^Z$ analogously from $B_p^0$. These parent checks commute and have the same code space as the toric-code generators. For each generator, the product of the symmetry factors whose supports meet it sends it to its inverse and therefore preserves the corresponding parent check, as required by Proposition~\ref{prop:zn-patch-checks}.

We now choose the paths used in the construction. For a star at $v$, the matter qutrits belong to the three vertices $\{v,v-\hat{x},v-\hat{y}\}$. Use $v$ as the reference vertex and its west and south edges as paths to the other two vertices. For a plaquette with southwest corner $u$, the corresponding vertices are $\{u,u+\hat{x},u+\hat{y}\}$. Use $u$ as the reference vertex and the south and west boundary edges as paths to the other two vertices. In a gauge configuration $b_e\in\{0,1\}$, with $\sigma_e^Z=(-1)^{b_e}$, the resulting conjugations on the check supports are
\begin{equation}
 Q_v(b)=K_S^{b_S}K_W^{b_W},\qquad
 Q_p(b)=K_N^{b_W}K_E^{b_S}.
 \label{eq:sm-s3-tree-transport}
\end{equation}
Factors acting outside the parent check support cancel in the conjugation. These are precisely the controlled-inversion circuits $Q_v,Q_p$ written below.

Assign each star--plaquette compatibility test to the plaquette check. Two distinct stars or two distinct plaquettes share at most one of the vertices assigned above, so their tests always pass. For a star and a plaquette, the only nonautomatic test occurs at the northeast corner. There the two tree assignments differ by $b_E+b_S$ at the southeast vertex and by $b_N+b_W$ at the northwest vertex, with sums modulo two. They agree up to a constant exactly when
\begin{equation}
 b_E+b_S=b_N+b_W\pmod2,
 \qquad H_p=\prod_{e\in\partial p}\sigma_e^Z=+1.
 \label{eq:sm-s3-compatibility}
\end{equation}
All stars are therefore enabled, and a plaquette is enabled precisely when $H_p=+1$. Proposition~\ref{prop:zn-patch-checks} gives
\begin{equation}
 \begin{aligned}
 E_v(b)&=Q_v(b)J_v^XQ_v(b),\\
 F_p(b)&=\begin{cases}
 Q_p(b)J_p^ZQ_p(b),&H_p=+1,\\
 I,&H_p=-1.
 \end{cases}
 \end{aligned}
 \label{eq:sm-s3-from-parent}
\end{equation}
Together with the Gauss's law operators $G_v=U_v\prod_{e\ni v}\sigma_e^X$ and the flux checks $H_p$, these checks commute on the full Hilbert space. Replacing a disabled check by $I$ leaves the common $+1$ space unchanged because that space already requires $H_p=+1$ on every plaquette. The gauge qubits control whether each check is enabled, without a separate measurement of the compatibility-test outcomes. No completion qubits are needed. The gate products below implement Eq.~\eqref{eq:sm-s3-from-parent}.

For comparison, the qutrit vertex and plaquette checks in Ref.~\cite{LyonsBrown2026S3} can fail to commute when $H_p=-1$. Our construction retains the vertex constraint and makes the plaquette check trivial in that sector.

\subsection{\texorpdfstring{Commuting checks of the $D(S_3)$ code}{Commuting checks of the D(S3) code}}
\label{sec:sm-s3-product}

We use the group coordinates of Refs.~\cite{Lyons2025Protocols,LyonsBrown2026S3,Sajith2025Hybrid}. On each edge $e$, write $|a_e,b_e\rangle$ for the computational basis of the matter qutrit and gauge qubit, with $a_e\in\mathbb{Z}_3$ and $b_e\in\mathbb{Z}_2$. Identify this pair with $g_e=r^{a_e}s^{b_e}\in S_3$, where the rotation $r$ and reflection $s$ obey $r^3=s^2=1$ and $srs=r^{-1}$. Group multiplication in these coordinates is
\begin{equation}
 (a,b)(c,d)=(a+(-1)^bc,b+d).
\end{equation}
Write $X_3|a\rangle=|a+1\rangle$ and $Z_3|a\rangle=\omega^a|a\rangle$, with $\omega=e^{2\pi i/3}$. On edge $e$, $X_e,Z_e,K_e$ act on the qutrit, with $K_e|a\rangle=|-a\rangle$, and $\sigma_e^X,\sigma_e^Z$ are the qubit Paulis. Set $R_3=\operatorname{diag}(1,-1,-1)$ and use the qutrit Fourier transform $\mathsf{F}$ with $\mathsf{F} X_3\mathsf{F}^\dagger=Z_3$. The gate $\mathrm{C}_b(Q)$ applies $Q$ when control qubit $b$ is in $|1\rangle$. Our qutrit controlled shift is
\begin{equation}
 \mathrm{CX}_3(i\to j)|a_i,a_j\rangle
 =|a_i,a_j+a_i\rangle.
\end{equation}
Each check below is a product of local gates. Its measurement controls that product with one readout ancilla, as in Section~\ref{sec:prx-ft}. Appendix~\ref{sec:sm-s3-native-cost} analyzes these checks and their controlled gates in the native qubit--qutrit Pauli hierarchy.

At vertex $v$, $N,E,S,W$ label its four incident edges by compass direction. All gate products act from right to left. For the vertex checks define
\begin{align}
 Q_v&=\mathrm{C}_{b_S}(K_S)\mathrm{C}_{b_W}(K_W),\nonumber\\
 {L_v}&=\mathrm{CX}_3(N\to W)\mathrm{CX}_3(N\to S)
       \mathrm{CX}_3(N\to E)^{-1},\nonumber\\
 V_v&=\mathsf{F}_N {L_v}Q_v.
 \label{eq:sm-s3-Vword}
\end{align}
For plaquette $p$ the compass labels refer to its boundary edges. Define
\begin{align}
 Q_p&=\mathrm{C}_{b_W}(K_N)\mathrm{C}_{b_S}(K_E),\nonumber\\
 {L_p}&=\mathrm{CX}_3(S\to W)^{-1}\mathrm{CX}_3(E\to W)^{-1}
       \mathrm{CX}_3(N\to W),\nonumber\\
 D_p&=\mathrm{CNOT}(b_S\to b_W)\mathrm{CNOT}(b_E\to b_W)
       \nonumber\\
 &\quad\times\mathrm{CNOT}(b_N\to b_W),\nonumber\\
 W_p&=D_pQ_p{L_p}Q_p.
 \label{eq:sm-s3-Wword}
\end{align}
The parent checks used above are $J_v^X=(\mathsf{F}_NL_v)^\dagger R_{3,N}(\mathsf{F}_NL_v)$ and $J_p^Z=L_p^\dagger R_{3,W}L_p$. Call the qutrit and qubit outputs on the west edge after $W_p$ the coordinates $c_p$ and $\tau_p$, respectively. On these matter and gauge qudits let $J_W=\sigma_W^X\mathrm{C}_{b_W}(R_{3,W})\sigma_W^X$. This applies $R_3$ only when the computed $\mathbb{Z}_2$ gauge flux is zero. The sequence $W_p^\dagger J_W W_p$ uses the existing west-edge gauge qubit as the control and then uncomputes the coordinate change. At each vertex measure the $\mathbb{Z}_2$ Gauss's law operator $G_v$ and the rotation-neutrality check $E_v$. At each plaquette measure the $\mathbb{Z}_2$ gauge-flux check $H_p$ and the rotation-flux check $F_p$. The latter tests rotation flux only when $H_p=+1$ and acts as the identity otherwise. They are
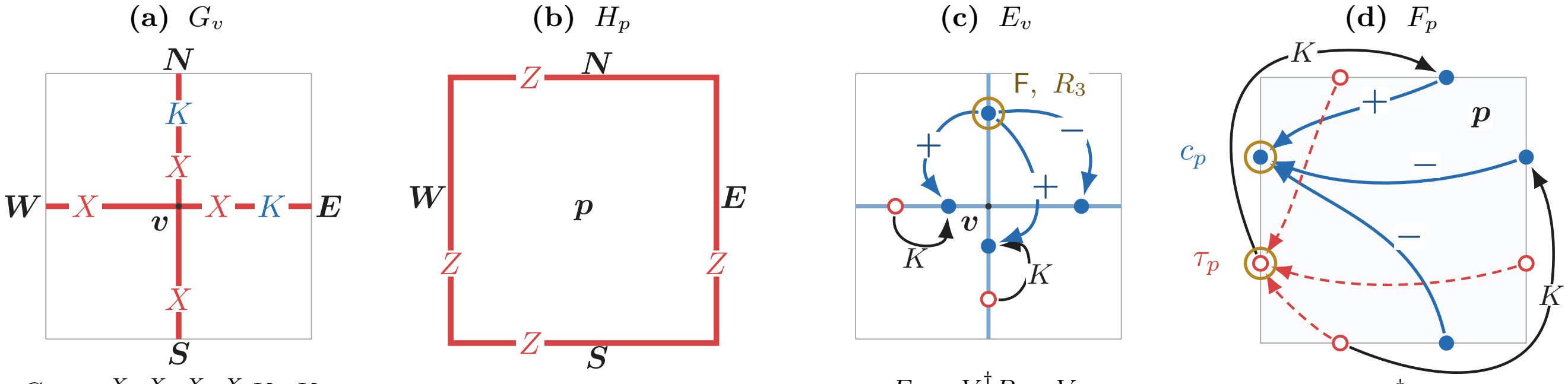
\begin{align}
 G_v&=\sigma_N^X\sigma_E^X\sigma_S^X\sigma_W^XK_NK_E,\nonumber\\
 E_v&=V_v^\dagger R_{3,N}V_v,\nonumber\\
 H_p&=\sigma_N^Z\sigma_E^Z\sigma_S^Z\sigma_W^Z,\nonumber\\
 F_p&=W_p^\dagger J_W W_p.
 \label{eq:sm-s3checks}
\end{align}
Figure~\ref{fig:sm-s3-generators} displays their support and gate factors.

\definecolor{sThreeRed}{HTML}{D84242}
\definecolor{sThreeBlue}{HTML}{276CB0}
\definecolor{sThreeGold}{HTML}{B58720}
\begin{figure*}[t]
\centering
\tikzset{
  hybrid edge label/.style={font=\normalsize\boldmath,text=black!88,inner sep=.9pt},
  hybrid site label/.style={font=\normalsize\boldmath,text=black!88,fill=white,inner sep=.7pt},
  hybrid operator/.style={font=\normalsize,fill=white,inner sep=.7pt},
  hybrid shift/.style={->,sThreeBlue,line width=1pt,shorten >=3.3pt,shorten <=3.2pt},
  hybrid inversion/.style={->,black!88,line width=.85pt,shorten >=3.3pt,shorten <=3.2pt},
  hybrid parity/.style={->,sThreeRed,densely dashed,line width=.8pt,shorten >=3.3pt,shorten <=3.2pt},
  hybrid sign/.style={font=\normalsize\boldmath,text=sThreeBlue!75!black,inner sep=.3pt,fill=white},
  hybrid gate/.style={font=\small,text=black!88,inner sep=.3pt,fill=white}
}
\begin{minipage}[t]{.245\textwidth}\centering
\textbf{(a)}\; $G_v$\par\smallskip
\begin{tikzpicture}[x=1.48cm,y=1.48cm,font=\footnotesize,>=Latex]
\path[use as bounding box] (-1.27,-1.21) rectangle (1.27,1.21);
\draw[step=1,black!35,line width=.35pt] (-1,-1) grid (1,1);
\draw[sThreeRed,line width=1.8pt] (-1,0)--(1,0) (0,-1)--(0,1);
\foreach \x/\y in {.30/0,-.70/0,0/.30,0/-.70}{
  \node[hybrid operator,text=sThreeRed] at (\x,\y) {$X$};}
\node[hybrid operator,text=sThreeBlue] at (.70,0) {$K$};
\node[hybrid operator,text=sThreeBlue] at (0,.70) {$K$};
\fill[black!80] (0,0) circle[radius=1pt];
\node[hybrid site label] at (-.14,-.13) {$v$};
\node[above,hybrid edge label] at (0,1) {$N$};
\node[right,hybrid edge label] at (1,0) {$E$};
\node[below,hybrid edge label] at (0,-1) {$S$};
\node[left,hybrid edge label] at (-1,0) {$W$};
\end{tikzpicture}\par
{\footnotesize $G_v=\sigma_N^X\sigma_E^X\sigma_S^X\sigma_W^X K_NK_E$}
\end{minipage}\hfill
\begin{minipage}[t]{.245\textwidth}\centering
\textbf{(b)}\; $H_p$\par\smallskip
\begin{tikzpicture}[x=2.96cm,y=2.96cm,font=\footnotesize,>=Latex]
\path[use as bounding box] (-.18,-.12) rectangle (1.18,1.12);
\draw[black!35,line width=.35pt] (0,0) rectangle (1,1);
\draw[sThreeRed,line width=1.8pt] (0,0) rectangle (1,1);
\foreach \x/\y in {.30/0,.30/1,0/.30,1/.30}{
  \node[hybrid operator,text=sThreeRed] at (\x,\y) {$Z$};}
\node[hybrid site label] at (.5,.5) {$p$};
\node[above,hybrid edge label] at (.55,1) {$N$};
\node[right,hybrid edge label] at (1,.55) {$E$};
\node[below,hybrid edge label] at (.55,0) {$S$};
\node[left,hybrid edge label] at (0,.55) {$W$};
\end{tikzpicture}\par
{\footnotesize $H_p=\sigma_N^Z\sigma_E^Z\sigma_S^Z\sigma_W^Z$}
\end{minipage}\hfill
\begin{minipage}[t]{.245\textwidth}\centering
\textbf{(c)}\; $E_v$\par\smallskip
\begin{tikzpicture}[x=1.48cm,y=1.48cm,font=\footnotesize,>=Latex]
\path[use as bounding box] (-1.27,-1.21) rectangle (1.27,1.21);
\draw[step=1,black!35,line width=.35pt] (-1,-1) grid (1,1);
\draw[sThreeBlue!60,line width=1.3pt] (-1,0)--(1,0) (0,-1)--(0,1);
\draw[hybrid inversion] (-.70,0) .. controls (-.74,-.34) and (-.33,-.34) .. (-.30,0);
\draw[hybrid inversion] (0,-.70) .. controls (.35,-.72) and (.35,-.28) .. (0,-.30);
\node[hybrid gate] at (-.55,-.39) {$K$};
\node[hybrid gate] at (.39,-.51) {$K$};
\draw[hybrid shift] (0,.70) .. controls (.59,.77) and (.83,.55) .. (.70,0);
\draw[hybrid shift] (0,.70) .. controls (-.39,.74) and (-.55,.31) .. (-.30,0);
\draw[hybrid shift] (0,.70) .. controls (.36,.43) and (.49,-.13) .. (0,-.30);
\node[hybrid sign] at (.63,.57) {$-$};
\node[hybrid sign] at (-.45,.46) {$+$};
\node[hybrid sign] at (.43,.15) {$+$};
\foreach \x/\y in {0/.70,.70/0,0/-.30,-.30/0}{\fill[sThreeBlue] (\x,\y) circle[radius=2.4pt];}
\foreach \x/\y in {0/-.70,-.70/0}{\draw[sThreeRed,fill=white,line width=.9pt] (\x,\y) circle[radius=2.2pt];}
\draw[sThreeGold,line width=1pt] (0,.70) circle[radius=4.8pt];
\node[anchor=west,text=sThreeGold!65!black,font=\small,fill=white,inner sep=.4pt] at (.17,.91) {$\mathsf{F},\ R_3$};
\fill[black!80] (0,0) circle[radius=1pt];
\node[hybrid site label] at (-.14,-.13) {$v$};
\end{tikzpicture}\par
{\footnotesize $E_v=V_v^\dagger R_{3,N}V_v$\par
$V_v=\mathsf{F}_NL_vQ_v$}
\end{minipage}\hfill
\begin{minipage}[t]{.245\textwidth}\centering
\textbf{(d)}\; $F_p$\par\smallskip
\begin{tikzpicture}[x=2.96cm,y=2.96cm,font=\footnotesize,>=Latex]
\path[use as bounding box] (-.18,-.12) rectangle (1.18,1.12);
\fill[sThreeBlue!3] (0,0) rectangle (1,1);
\draw[black!35,line width=.35pt] (0,0) rectangle (1,1);
\draw[hybrid inversion] (0,.30) .. controls (-.18,.72) and (-.15,1.08) .. (.21,1.095)
                                      .. controls (.41,1.12) and (.55,1.08) .. (.70,1);
\draw[hybrid inversion] (.30,0) .. controls (.72,-.18) and (1.08,-.15) .. (1.095,.21)
                                     .. controls (1.12,.41) and (1.08,.55) .. (1,.70);
\node[hybrid gate] at (.16,1.095) {$K$};
\node[hybrid gate] at (1.095,.18) {$K$};
\draw[hybrid shift] (.70,1) .. controls (.46,.89) and (.21,.86) .. (0,.70);
\draw[hybrid shift] (1,.70) .. controls (.65,.58) and (.33,.58) .. (0,.70);
\draw[hybrid shift] (.70,0) .. controls (.61,.39) and (.31,.49) .. (0,.70);
\node[hybrid sign] at (.43,.91) {$+$};
\node[hybrid sign] at (.62,.67) {$-$};
\node[hybrid sign] at (.56,.40) {$-$};
\draw[hybrid parity] (.30,1) .. controls (.19,.77) and (.13,.51) .. (0,.30);
\draw[hybrid parity] (1,.30) .. controls (.68,.20) and (.32,.20) .. (0,.30);
\draw[hybrid parity] (.30,0) .. controls (.19,.06) and (.06,.19) .. (0,.30);
\foreach \x/\y in {.70/0,.70/1,0/.70,1/.70}{\fill[sThreeBlue] (\x,\y) circle[radius=2.4pt];}
\foreach \x/\y in {.30/0,.30/1,0/.30,1/.30}{\draw[sThreeRed,fill=white,line width=.9pt] (\x,\y) circle[radius=2.2pt];}
\draw[sThreeGold,line width=1pt] (0,.70) circle[radius=4.8pt];
\draw[sThreeGold,line width=1pt] (0,.30) circle[radius=4.8pt];
\node[anchor=east,text=sThreeBlue,font=\normalsize,inner sep=.5pt] at (-.19,.70) {$c_p$};
\node[anchor=east,text=sThreeRed,font=\normalsize,inner sep=.5pt] at (-.14,.30) {$\tau_p$};
\node[hybrid site label] at (.83,.85) {$p$};
\end{tikzpicture}\par
{\footnotesize $F_p=W_p^\dagger J_WW_p$\par
$W_p=D_pQ_pL_pQ_p$}
\end{minipage}
\par\medskip
\begin{minipage}[t]{.49\textwidth}\centering\footnotesize
$\begin{aligned}
Q_v&=\mathrm{C}_{b_S}(K_S)\,\mathrm{C}_{b_W}(K_W),\\
L_v&=\mathrm{CX}_3(N\!\to\! W)\,
\mathrm{CX}_3(N\!\to\! S)\,
\mathrm{CX}_3(N\!\to\! E)^{-1}.
\end{aligned}$
\end{minipage}\hfill
\begin{minipage}[t]{.49\textwidth}\centering\footnotesize
$\begin{aligned}
Q_p&=\mathrm{C}_{b_W}(K_N)\,\mathrm{C}_{b_S}(K_E),\\
L_p&=\mathrm{CX}_3(S\!\to\! W)^{-1}\,
\mathrm{CX}_3(E\!\to\! W)^{-1}\,
\mathrm{CX}_3(N\!\to\! W),\\
D_p&=\mathrm{CNOT}(b_S\!\to\! b_W)\,
\mathrm{CNOT}(b_E\!\to\! b_W)\,
\mathrm{CNOT}(b_N\!\to\! b_W).
\end{aligned}$
\end{minipage}
\par\smallskip
{\footnotesize $R_3=\operatorname{diag}(1,-1,-1)$,\qquad $J_W=\sigma^X_W\,\mathrm{C}_{b_W}(R_{3,W})\,\sigma^X_W$.}
\par\smallskip
\begin{tikzpicture}[font=\footnotesize,>=Latex]
\draw[sThreeRed,fill=white,line width=.9pt] (0,0) circle[radius=2.2pt];
\node[anchor=west] at (.12,0) {gauge qubit $b$};
\fill[sThreeBlue] (3.05,0) circle[radius=2.4pt];
\node[anchor=west] at (3.17,0) {matter qutrit $a$};
\draw[->,sThreeBlue,line width=.85pt] (6.35,0)--(6.95,0);
\node[anchor=west] at (7.02,0) {$\mathrm{CX}_3^{\pm1}$};
\draw[->,sThreeRed,densely dashed,line width=.7pt] (9.1,0)--(9.7,0);
\node[anchor=west] at (9.77,0) {$\mathrm{CNOT}$};
\draw[sThreeGold,line width=1pt] (12.0,0) circle[radius=4.2pt];
\node[anchor=west] at (12.18,0) {qudits for $R_3$ or $J_W$};
\end{tikzpicture}
\caption{The commuting checks of the qubit--qutrit $D(S_3)$ code. Open red circles mark gauge qubits and filled blue circles mark matter qutrits. Panels (a,b) give $G_v,H_p$ and label the edge directions used in (c,d). Panels (c,d) show the forward coordinate changes for $E_v,F_p$, followed by the diagonal gate and the inverse coordinate changes. Gold rings mark the qudits acted on by $R_{3,N}$ in (c) and $J_W$ in (d). In (d), $c_p$ is the qutrit target and $\tau_p$ is the gauge-qubit control. The gate $J_W$ applies $R_3$ to $c_p$ only when $\tau_p=0$. Blue arrows denote $\mathrm{CX}_3^{\pm1}$, red dashed arrows CNOTs, and black $K$ arrows controlled inversion. In (c) $Q_v$ precedes the shifts and Fourier transform. In (d) $Q_pL_pQ_p$ precedes the parity circuit $D_p$.}
\label{fig:sm-s3-generators}
\end{figure*}

\begin{proposition}[Commuting checks and code space]
The checks in Eq.~\eqref{eq:sm-s3checks} are mutually commuting Hermitian unitaries on the full physical Hilbert space. Their common $+1$ space is the untwisted Kitaev $D(S_3)$ code space.
\end{proposition}
\begin{proof}
For $h\in S_3$, let $A_v(h)$ left-multiply outgoing group labels by $h$ and right-multiply incoming labels by $h^{-1}$. Write $X_e^{-\sigma_e^Z}$ for the qutrit shift by $-(-1)^{b_e}$ on edge $e$. The rotation generator is
\begin{equation}
 A_v(r)=X_NX_EX_S^{-\sigma_S^Z}X_W^{-\sigma_W^Z}.
\end{equation}
The circuit $Q_v$ removes the controlled signs, ${L_v}$ concentrates the rotation on the north qutrit, and $\mathsf{F}_N$ turns it into $Z_N$. Hence, $E_v$ has eigenvalue $+1$ precisely on rotation charge zero and $-1$ on the other two charges. Since $G_v=A_v(s)$ conjugates $A_v(r)$ to its inverse, it preserves these two eigenspaces and commutes with $E_v$. Actions at different vertices commute by left--right multiplication on their shared edge.

The circuit $W_p$ stores, on the existing west-edge matter and gauge qudits, the coordinates
\begin{align}
 c_p&=a_W+(-1)^{b_W}a_N-(-1)^{b_S}a_E-a_S,\nonumber\\
 \tau_p&=b_N\oplus b_E\oplus b_S\oplus b_W.
 \label{eq:sm-s3-plaquette-coordinates}
\end{align}
Here, $\oplus$ is addition modulo two. For $h_p=g_Sg_Eg_N^{-1}g_W^{-1}$, the rotation coordinate is $-c_p$ when $\tau_p=0$. $H_p$ has eigenvalue $(-1)^{\tau_p}$, and $F_p$ is negative exactly when $\tau_p=0$, $c_p\ne0$. Their signs distinguish identity, nontrivial rotation and reflection holonomy. These diagonal checks commute with each other. Their values are unchanged under $h_p\mapsto gh_pg^{-1}$, so they also commute with every vertex action.

The projector onto the $+1$ eigenspace of the rotation-neutrality check is $(I+E_v)/2=(I+A_v(r)+A_v(r)^2)/3$. Writing $\delta_{h_p,1}$ for the diagonal projector onto identity holonomy, the joint $+1$ projectors obey
\begin{align}
 \frac{I+E_v}{2}\frac{I+G_v}{2}
 &=\frac16\sum_{h\in S_3}A_v(h),\nonumber\\
 \frac{I+H_p}{2}\frac{I+F_p}{2}&=\delta_{h_p,1}.
 \label{eq:sm-s3vacuum}
\end{align}
Together these checks impose gauge invariance at vertices and identity group product around each plaquette, as required for the $D(S_3)$ code space.
\end{proof}

Each gauge qubit is flipped twice and each matter qutrit conjugated once, so in every flux sector
\begin{equation}
 \prod_vG_v=\prod_eK_e=:\mathcal{U}
 \label{eq:sm-s3-global}
\end{equation}
$E_v=+1$ enforces invariance under the order-three rotation, and $G_v=+1$ enforces invariance under the reflection. Together they impose full $S_3$ invariance at the vertex. For a plaquette, $H_p=-1$ means that its edge-group product is a reflection, and $F_p=+1$ is then automatic. If $H_p=+1$, $F_p$ distinguishes the identity from a nonidentity rotation. Their common positive eigenspace therefore imposes all the $D(S_3)$ code conditions. The checks commute even on inputs that violate these conditions. The outcome $E_v=-1$ groups the two nonzero rotation charges together. We do not measure them separately because $G_v$ exchanges them. The controlled-$E_v$ circuit preserves their coherence.

\subsection{Logical measurement by gauging and ungauging}
\label{sec:sm-s3-instrument}
Let $P$ be the parent qutrit toric-code projector and let $J$ append $|0\rangle$ to every gauge qubit. Write $V$ for the vertex set and $s(e),t(e)$ for the source and target of edge $e$. Choose a reference vertex $v_0$. For a binary vertex string $y=(y_v)_{v\in V}$ define
\begin{equation}
 U(y)=\prod_eK_e^{y_{s(e)}},\qquad
 (\delta y)_e=y_{s(e)}\oplus y_{t(e)}.
\end{equation}
The symmetry $\mathcal{U}$ preserves $P$. For encoded matter $|\psi\rangle$, direct multiplication gives
\begin{equation}
 \prod_vG_v^{y_v}J|\psi\rangle
 =U(y)|\psi\rangle\otimes|\delta y\rangle_b.
 \label{eq:sm-s3gaugingaction}
\end{equation}
On the initialized code the commuting $E_v,H_p,F_p$ have value $+1$. In the procedure of Appendix~\ref{sec:sm-zn-measurement}, take $U_v=K_{N(v)}K_{E(v)}$. Let $g_v$ be the last outcome of $G_v$ and $s=\prod_vg_v$. For an exact gauge record $b=\delta y$ with $y_{v_0}=0$, the corrected Kraus operator is
\begin{align}
 K_{g,b}&=U(y)\,{}_b\!\langle b|
          \prod_v\frac{I+g_vG_v}{2}\,J\nonumber\\
 &=2^{1-|V|}\left(\prod_vg_v^{y_v}\right)\Pi_s(\mathcal{U}),\nonumber\\
 \Pi_s(\mathcal{U})&=\frac{I+s\mathcal{U}}{2}.
 \label{eq:sm-s3instrument}
\end{align}
By Eq.~\eqref{eq:em-zn-instrument}, summing these branches at fixed $s$ gives $\mathcal{M}_s(\rho)=\Pi_s(\mathcal{U})\rho\Pi_s(\mathcal{U})$, including for inputs entangled with a reference.

Let $|j,k\rangle_L$ denote a logical basis state of the two encoded torus qutrits, with $j,k\in\mathbb{Z}_3$. The induced logical operator $\overline{\mathcal{U}}=\overline{K}_1\overline{K}_2$ conjugates each label, sending $|j,k\rangle_L$ to $|-j,-k\rangle_L$. The positive and negative eigenspaces have dimensions five and four. For example,
\begin{equation}
 |1,0\rangle_L\longmapsto
 \frac{|1,0\rangle_L+s|2,0\rangle_L}{\sqrt2}
\end{equation}
with probability $1/2$ for either sign. Both outputs are nonstabilizer qutrit states. For both logical outcomes, the gauge record has the form $b=\delta y$. The negative outcome is carried by the product of the Gauss-law outcomes, $\prod_vg_v=-1$.

\paragraph{Controlled conjugation on independent inputs.}
Let $c$ be an encoded control qubit and $\mathcal{U}$ the target inversion. On the torus, fixing the second logical qutrit in $|0\rangle_L$ gives $\mathcal{U}=\bar{K}_1$ on the first. Prepare a mediator qubit $a$ in $|+\rangle_L$. Measure $\overline{Z}_c\overline{Z}_a$, $\overline{X}_a \mathcal{U}$ and $\overline{Z}_a$ in order, with sign bits $p,q,r$ (bit zero means $+1$). Gauge the middle product on a single connected graph. On a control basis state $|i\rangle_L$, the first sign sets the mediator to $i\oplus p$. The final sign selects $\mathcal{U}^{i\oplus p\oplus r}$. The branch on the two inputs is therefore
\[
 2^{-3/2}(-1)^{q(p\oplus r)}\overline{Z}_c^q \mathcal{U}^{p\oplus r}\,\mathrm{ctrl}_c(\mathcal{U}).
\]
Correct the two known byproducts. All eight records implement the same coherent controlled conjugation, also for reference-entangled inputs. For a surface-code interface based on non-Abelian code switching see Ref.~\cite{Sajith2025Hybrid}.

For fault tolerance, stack the qubit and qutrit toric-code layers over the same $L\times L$ torus and assign outgoing matter qubits and qutrits to their source vertex. The three observables factor over these bounded sets. Use the square gauging graph and the local commuting checks of Proposition~\ref{prop:zn-patch-checks}. The reduction in Theorem~\ref{thm:zn-accepted-distance} applies to each measurement, using the existence of a row and a column outside the residual error support as in the proof of Eq.~\eqref{eq:em-zn-distance}. Take $R$ gauged rounds per measurement and $R$ rounds of parent-code syndrome extraction between measurements, requiring $+1$ outcomes of the parent checks in the detection tests. Fewer than $R$ faults leave a fault-free round in each intervening parent-code interval, so the three measurements can be composed using their parent-code projections. With ideal encoded input, ideal final projection and the known byproducts tracked in the logical frame, the resulting gate has circuit fault distance at least $\min\{R,\lceil L/\kappa\rceil\}$ for a circuit-dependent constant $\kappa$.

\subsection{Fault tolerance from the general analysis}
\label{sec:sm-s3-ft-general}
The checks in Eq.~\eqref{eq:sm-s3checks} specialize the $D(\mathbb{Z}_m\rtimes\mathbb{Z}_N)$ construction in Eq.~\eqref{eq:em-zn-generators} to $(m,N,q)=(3,2,2)$. Measure each check by controlling its displayed gate product with one measurement ancilla. Each circuit stays within one star or plaquette, and overlapping circuits are serialized. The general bound in Eq.~\eqref{eq:em-zn-distance} therefore gives
\begin{equation}
 d_{\rm surg}\ge\min\{R,\lceil L/4\rceil\},
 \label{eq:sm-s3-general-distance}
\end{equation}
Taking $R=\Theta(L)$ gives circuit fault distance $\Omega(L)$ in $O(L)$ rounds. This uses the detection tests, ideal encoded input, and ideal final parent-code projection of Theorem~\ref{thm:zn-accepted-distance}.

\subsection{Hierarchy of the native gates and checks}
\label{sec:sm-s3-native-cost}

The hierarchy must use the physical Pauli operators of both species. Write $\mathcal{H}_2=(\mathbb{C}^2)^{\otimes n}$ and $\mathcal{H}_3=(\mathbb{C}^3)^{\otimes m}$, grouping all physical and completion qudits by their prime dimensions. Let $\mathcal{P}_{n,m}$ consist, up to scalar phase, of tensor products of qubit Paulis $X_2,Z_2$ and qutrit Paulis $X_3,Z_3$. Define
\begin{equation}
 \begin{aligned}
 \mathcal{C}_1&=\mathcal{P}_{n,m},\\
 \mathcal{C}_{k+1}&=\{V:VPV^\dagger\in\mathcal{C}_k
       \text{ for all }P\in\mathcal{P}_{n,m}\},
 \end{aligned}
 \label{eq:sm-native-hierarchy}
\end{equation}
where $V$ is unitary. Scalar phases do not affect hierarchy membership. This definition uses the qutrit hierarchy on the matter qudits, not a qubit encoding of their three basis states.

\paragraph{Charge conjugation and its controlled gate.}
Charge conjugation is qutrit Clifford because
\begin{equation}
 KX_3K^\dagger=X_3^{-1},\qquad
 KZ_3K^\dagger=Z_3^{-1}.
\end{equation}
Its qubit-controlled version $V=\operatorname{ctrl}_b(K)$, used in $Q_v$ and $Q_p$, has a different cost. With $[A,B]_{\rm grp}=ABA^\dagger B^\dagger$, direct calculation gives
\begin{align}
 [V,I\otimes X_3]_{\rm grp}&=\operatorname{diag}(I,X_3)=W,\nonumber\\
 [W,X_2\otimes I]_{\rm grp}&=\operatorname{diag}(X_3^{-1},X_3)=D,\nonumber\\
 [D,X_2\otimes I]_{\rm grp}&=D^{-1},\qquad
 [D^{-1},X_2\otimes I]_{\rm grp}=D.
 \label{eq:sm-s3-controlled-inversion-cycle}
\end{align}
Neither $D$ nor $D^{-1}$ is a product Pauli. A commutator with a Pauli lowers a finite hierarchy level until a Pauli is reached. This nonterminating cycle therefore places $V$ outside every finite level of the native hierarchy.

\paragraph{The displayed commuting checks.}
The checks $G_v$ are Clifford and $H_p$ are Pauli. The level-three bound on individual gates for measuring qubit Clifford checks does not apply to $G_v$, which acts on both qubits and qutrits. At gauge bits $b=0$, the checks $E_v$ and $F_p$ are qutrit Clifford conjugates of $R_3$ on one matter qutrit. For a diagonal unitary $D$, write $\Delta(D)=[D,X_3]_{\rm grp}$. Then
\begin{equation}
 \begin{aligned}
 R_3&=\operatorname{diag}(1,-1,-1),\\
 \Delta(R_3)&=\operatorname{diag}(-1,-1,1),\\
 \Delta^2(R_3)&=\operatorname{diag}(-1,1,-1),\\
 \Delta^3(R_3)&=R_3.
 \end{aligned}
\end{equation}
None of these sign patterns is a qutrit Pauli up to phase. Thus, $R_3$ lies outside every finite qutrit hierarchy level~\cite{Cui2017Diagonal,Glaudell2022QutritMetaplectic}. Restricting a finite-level unitary that preserves each computational gauge sector to any such sector gives a finite-level qutrit unitary. This follows inductively from Eq.~\eqref{eq:sm-native-hierarchy} by conjugating matter Paulis. Consequently, the full operators $E_v$ and $F_p$ also lie outside every finite level.

The hierarchy levels of these check operators and the gates used to implement them are distinct. Up to a Clifford permutation and global phase, $R_3$ is the qutrit metaplectic reflection, which admits an exact two-qutrit Clifford+$T$ circuit~\cite{Glaudell2022QutritMetaplectic}. Such a decomposition is possible because higher hierarchy levels are not closed under composition. The reflection argument applies to the displayed operators, while Theorem~\ref{thm:s3-completion-nogo} below covers every globally commuting check family with the same native code space and finite product-state auxiliaries.

\subsection{Obstruction to finite-hierarchy completion}
\label{sec:sm-s3-hierarchy}

We now prove an obstruction that applies to every exact commuting-check completion in the native qubit--qutrit encoding.

\begin{lemma}[Factorization between qubits and qutrits]
\label{lem:sm-prime-hierarchy-factorization}
Every $V\in\mathcal{C}_k$ at a finite level $k$ factors as $V=V_2\otimes V_3$ on $\mathcal{H}_2\otimes\mathcal{H}_3$.
\end{lemma}
\begin{proof}
The claim is immediate if there are no qubits or no qutrits. Assume $n,m\ge1$ and put $d_2=2^n$, $d_3=3^m$. We induct on $k$. The claim holds for $\mathcal{C}_1$. At the next level, the induction hypothesis makes the image of every Pauli generator under conjugation by $V$ a product of two unitaries.

We first record the constraints on such products. If $A\otimes B$ and $A'\otimes B'$ commute up to a scalar $q$, equality of nonzero elementary tensors gives scalars $\alpha,\beta$ such that
\begin{equation}
 \begin{aligned}
 AA'&=\alpha A'A,& BB'&=\beta B'B,\\
 \alpha\beta&=q,& \alpha^{d_2}&=\beta^{d_3}=1.
 \end{aligned}
 \label{eq:sm-prime-commutation-scalars}
\end{equation}
The last identities follow by taking determinants. If $q=\pm1$, then $q^{d_2}=1$, so $\beta^{d_2}=1$. Coprimality of $d_2$ and $d_3$ implies $\beta=1$ and $\alpha=q$. In particular, when $q=1$, both local commutation scalars are one.

Write the conjugated qubit generators as
\begin{equation}
 \begin{aligned}
 V(X_{2,i}\otimes I)V^\dagger&=A_i\otimes B_i,\\
 V(Z_{2,i}\otimes I)V^\dagger&=C_i\otimes D_i.
 \end{aligned}
\end{equation}
Equation~\eqref{eq:sm-prime-commutation-scalars} implies that $A_i,C_i$ satisfy the qubit Pauli commutation relations. Each of their squares is scalar because the corresponding product generator squares to $I$. Rescale these local factors by phases so that $A_i^2=C_i^2=I$. These relations then give a unital representation of the full matrix algebra $M_{2^n}(\mathbb{C})$ on $\mathcal{H}_2$. Its dimension is $2^n$, so this representation is irreducible.

Every conjugated qutrit generator commutes with every conjugated qubit generator. Applying Eq.~\eqref{eq:sm-prime-commutation-scalars} with $q=1$ shows that its qubit factor commutes with all $A_i,C_i$, and is therefore scalar. Thus,
\begin{equation}
 V\bigl(I\otimes M_{3^m}(\mathbb{C})\bigr)V^\dagger
   =I\otimes M_{3^m}(\mathbb{C}),
\end{equation}
where equality follows from the equal dimensions. Every automorphism of this full matrix algebra is inner. Choose $V_3$ implementing this automorphism. Then $(I\otimes V_3^\dagger)V$ commutes with $I\otimes M_{3^m}(\mathbb{C})$ and hence equals $V_2\otimes I$. This completes the induction.
\end{proof}

\begin{corollary}[Product basis for commuting checks]
\label{cor:sm-prime-check-product-basis}
Every common eigenspace of a globally commuting family of finite-level native-hierarchy unitaries has an orthonormal basis of product vectors across $\mathcal{H}_2\mid\mathcal{H}_3$.
\end{corollary}
\begin{proof}
By Lemma~\ref{lem:sm-prime-hierarchy-factorization}, write each check as $T_a=A_a\otimes B_a$. Commutativity of $T_a,T_b$ and Eq.~\eqref{eq:sm-prime-commutation-scalars} with $q=1$ imply that $A_a,A_b$ commute and $B_a,B_b$ commute. Both local families therefore admit common orthonormal eigenbases. Their tensor products diagonalize all checks, and the product basis vectors with the prescribed eigenvalues span each common eigenspace.
\end{proof}

\begin{theorem}[Obstruction to exact native completion]
\label{thm:s3-completion-nogo}
Let $P_{S_3}$ be the common $+1$ projector of the native $D(S_3)$ checks in Eq.~\eqref{eq:sm-s3checks} on an $L\times L$ square torus with $L\ge4$. Adjoin any finite number of completion qubits and qutrits in a fixed product state $|\mathrm{aux}\rangle$. There is no globally commuting family of unitary checks, each belonging to a finite level of the native hierarchy in Eq.~\eqref{eq:sm-native-hierarchy}, whose common $+1$ projector is
\begin{equation}
 \widehat{P}=P_{S_3}\otimes|\mathrm{aux}\rangle
                  \langle\mathrm{aux}|.
 \label{eq:sm-s3-completion-nogo}
\end{equation}
No locality bound or common upper bound on the hierarchy levels is assumed.
\end{theorem}
\begin{proof}
We first show that the range of $P_{S_3}$ contains no nonzero product vector across the gauge qubits and matter qutrits. Suppose that $|\psi\rangle\otimes|\phi\rangle$ were such a vector, with normalized factors. Write
\begin{equation}
 S_v^X=\prod_{e\ni v}\sigma_e^X,
 \qquad K_v=K_NK_E,
 \qquad G_v=S_v^X\otimes K_v.
\end{equation}
The condition $G_v(|\psi\rangle\otimes|\phi\rangle) =|\psi\rangle\otimes|\phi\rangle$ implies $S_v^X|\psi\rangle=\epsilon_v|\psi\rangle$ and $K_v|\phi\rangle=\epsilon_v|\phi\rangle$ for some $\epsilon_v\in\{+1,-1\}$. Write $|\psi\rangle=\sum_b\psi(b)|b\rangle$, where $b=(b_e)_{e\in E}$ lists the computational-basis bits of all gauge qubits. Let $\delta_v$ be one on the four edges incident on $v$ and zero elsewhere. The gauge-qubit amplitudes obey
\begin{equation}
 \psi(b+\delta_v)=\epsilon_v\psi(b).
\end{equation}
Thus, if the bit string $b$ has nonzero amplitude, so does $b+\delta_v$, obtained by flipping the four gauge qubits at $v$. Repeating these flips preserves nonzero amplitude.

For every supported $b$, the condition $E_v(b)|\phi\rangle=|\phi\rangle$ from Eq.~\eqref{eq:sm-s3-from-parent} is equivalent to
\begin{equation}
 \begin{aligned}
 A_v(b)|\phi\rangle&=|\phi\rangle,\\
 A_v(b)&=X_NX_E
       X_S^{-(-1)^{b_S}}X_W^{-(-1)^{b_W}}.
 \end{aligned}
 \label{eq:sm-s3-product-rotation}
\end{equation}
This uses the fact that $J_v^X$ is the reflection whose $+1$ eigenspace is the $A_v^0=1$ eigenspace. The gauge move at $v-\hat{y}$ flips $b_S$ but not $b_W$. Both $b$ and $b+\delta_{v-\hat{y}}$ are supported. Dividing their two commuting rotation constraints gives
\begin{equation}
 X_S^{2(-1)^{b_S}}|\phi\rangle=|\phi\rangle,
\end{equation}
and hence $X_S|\phi\rangle=|\phi\rangle$, since $X_S^3=I$. The gauge move at $v-\hat{x}$ similarly gives $X_W|\phi\rangle=|\phi\rangle$. These comparisons are valid on the stated lattice because the two incoming edges are distinct and each neighbor move flips only the indicated one. Applying them at every vertex covers every edge, so
\begin{equation}
 |\phi\rangle=e^{i\theta}|+_3\rangle^{\otimes |E|}.
 \label{eq:sm-s3-product-uniform}
\end{equation}

The constraints $H_p=+1$ require every supported $b$ to have zero $\mathbb{Z}_2$ gauge flux on every plaquette. Thus, $F_p(b)=+1$ requires
\begin{equation}
 \begin{aligned}
 B_p(b)|\phi\rangle&=|\phi\rangle,\\
 B_p(b)&=Q_p(b)B_p^0Q_p(b)\\
 &=Z_N^{(-1)^{b_W}}Z_E^{-(-1)^{b_S}}Z_S^{-1}Z_W.
 \end{aligned}
 \label{eq:sm-s3-product-flux}
\end{equation}
This nontrivial $Z$-type Pauli sends the uniform state in Eq.~\eqref{eq:sm-s3-product-uniform} to an orthogonal state. This is a contradiction, so the code space contains no nonzero product vector between the gauge qubits and matter qutrits.

The completion state factors as $|\mathrm{aux}_2\rangle\otimes|\mathrm{aux}_3\rangle$ across the same sectors. If the range of $\widehat{P}$ contained a nonzero product vector, contracting its completion qudits with $\langle\mathrm{aux}_2|\otimes\langle\mathrm{aux}_3|$ would give a nonzero product vector in the range of $P_{S_3}$. Thus, the range of $\widehat{P}$ also contains no such vector. Corollary~\ref{cor:sm-prime-check-product-basis} excludes it as the nonzero common $+1$ space of commuting finite-hierarchy checks.
\end{proof}

Theorem~\ref{thm:s3-completion-nogo} excludes alternative commuting generators and finite product-state completions with exactly the same native code projector. The theorem also fixes the native tensor-product Pauli structure. For example, the encoding
\begin{equation}
 |0\rangle\mapsto|00\rangle,\qquad
 |1\rangle\mapsto|01\rangle,\qquad
 |2\rangle\mapsto|10\rangle
\end{equation}
extends charge conjugation to a two-qubit SWAP. Its qubit-controlled version extends to a Fredkin gate, which is at level three of the qubit hierarchy. This changes the physical encoding and Pauli structure, so it lies outside the theorem.

\section{Dihedral circuits and distance}
\label{sec:sm-prx-semidirect}
We give the ordered measurement gates, the square-lattice $D_8$ example, gate counts and the faults detected by completion qubits. We then prove the circuit-distance bound.

\subsection{Edge operations for the measurement circuits}
\label{sec:sm-prx-dihedral-circuits}
Use the group law, edge orientation and vertex operators $R_v,S_v$ of Section~\ref{sec:prx273-dihedral-checks}. For inversion, $D=2\mathbb{Z}_m$. The projector selecting rotation charges unchanged by reflection is
\begin{equation}
 E_v=\frac{1}{n_m}\sum_{j=0}^{n_m-1}R_v^{2j}
     =\Pi(R_v^2=1),\qquad n_m=\frac{m}{\gcd(m,2)}.
 \label{eq:sm-prx-enable}
\end{equation}
We use the completed checks and plaquette checks in Eqs.~\eqref{eq:prx-dihedral-completion} and~\eqref{eq:prx273-dihedral-flux}.

The outcome operators below verify that no additional observables are measured. Circuits sharing matter, gauge or completion qudits run sequentially.

\paragraph{Rotation charge and pin.}
Let $\omega_m=e^{2\pi i/m}$ and $\Pi_k=\Pi(\widehat{R}_v=\omega_m^k)$. Prepare an $m$-level readout qudit $y$ in its uniform state and perform
\begin{equation}
 U_{\mathrm{phase}}=\sum_{y=0}^{m-1}|y\rangle\langle y|
                      \otimes\widehat{R}_v^y.
 \label{eq:sm-prx-phase-query}
\end{equation}
In the order $N,E,S,W$, its outgoing and incoming edge updates are
\begin{align}
 a_e&\longmapsto a_e+(-1)^t y,&&e\text{ outgoing},\\
 a_e&\longmapsto a_e-(-1)^{b_e+t}y,&&e\text{ incoming}.
 \label{eq:sm-prx-edge-query}
\end{align}
All control labels remain unchanged. Here $t\in\{0,1\}$ is the computational-basis label of the completion qubit $t_v$. The operator $X_t$ below acts on that qubit. Apply $\mathsf{F}_m^\dagger$ to $y$, where $\mathsf{F}_m|k\rangle=m^{-1/2}\sum_y\omega_m^{ky}|y\rangle$. The readout qudit now holds $k$. Prepare a pin readout $u$ in $|+\rangle$, and apply $X_t$ controlled by $u=1$ and $2k=0\pmod m$. Measure $y$ in its computational basis and $u$ in $X$. This fixed circuit has branch operator
\begin{align}
 K_{k,x}
 &=\Pi_k\left[
  \mathbf{1}_{2k=0}\frac{I+xX_t}{2}
  +\mathbf{1}_{2k\ne0}\delta_{x,+1}I\right]\nonumber\\
 &=\Pi_k\frac{I+xT_v}{2}.
 \label{eq:sm-prx-joint-instrument}
\end{align}
Here, the conditions on $2k$ are modulo $m$. The identity holds because $\widehat{R}_v^2=1$ selects exactly $E_v$, and $X_t$ commutes with $\widehat{R}_v$ on that subspace. The circuit measures the pin itself. It does not measure the condition $R_v^2=1$ as an additional observable.

For $m=2^q$, the $q$ phase qubits resolve the binary spectral checks
\begin{equation}
 C_{v,j}=\sum_{k=0}^{m-1}(-1)^{\operatorname{bit}_j(k)}\Pi_k,
 \qquad 0\leq j<q.
 \label{eq:sm-prx-charge-bits}
\end{equation}
There is one readout qubit per binary charge check and one for the pin. The pin condition is that the lowest $q-1$ charge bits vanish. For general $m$, the circuit uses an $m$-level readout qudit and its exact Fourier transform.

\paragraph{Completed reflection.}
Prepare a readout $a$ in $|+\rangle$. Apply controlled $K_mX_{b_e}$ on each outgoing edge and controlled $X_{b_e}$ on each incoming edge, where $K_m|j\rangle=|-j\rangle$ and $a$ is the control. Finally, apply $\operatorname{CNOT}(a\to t_v)$ and measure $a$ in $X$. This extracts $S_vX_{t_v}$. The optional flag uses a CZ gate before and after the controlled edge actions, with the final CNOT after both CZ gates.

\paragraph{Flatness.}
Initialize temporary qudits $w=(c,d)$ to the identity group label. For $h_p=g_Sg_Eg_N^{-1}g_W^{-1}$, compute the four factors in this order. Right multiplication by $g_e=(a,b)$ uses
\begin{equation}
 c\longmapsto c+(-1)^d a,\qquad d\longmapsto d+b.
 \label{eq:sm-prx-holonomy-update}
\end{equation}
For an inverse factor, update $d\mapsto d+b$ first and then subtract $(-1)^d a$ from $c$. Test only whether $w$ is the identity into one readout qubit, then reverse the entire computation before reading that qubit. The temporary qudits return to the identity, and the two data Kraus operators are $\Pi(h_p=1)$ and $I-\Pi(h_p=1)$. Measuring the entire group product would reveal extra information and is not used. To extract $B_p$ separately, prepare $a_B$ in $|0\rangle$, apply four CNOTs from the edge gauge qubits to $a_B$, and measure $a_B$ in $Z$.

\subsection{Square-lattice \texorpdfstring{$D_8$}{D8} checks and qubit circuits}
\label{sec:prx278-d8}
We use the binary qubit realization of the dihedral edge operators in Section~IV of Ref.~\cite{WarmanSchaferNameki2026}, and apply the completion construction to the measured family specified here. Our convention is that $D_8$ has order $16$, with an order-eight rotation. Each oriented edge has four physical qubits,
\begin{equation}
 |r^{a_e}s^{b_e}\rangle
 =|a_{e,0},a_{e,1},a_{e,2},b_e\rangle,\qquad
 a_e=a_{e,0}+2a_{e,1}+4a_{e,2}.
 \label{eq:prx279-d8-bits}
\end{equation}
Bit zero is least significant. Define the following three-qubit operators, with the rightmost gate applied first:
\begin{align}
 \mathsf{X}_{8,e}
 &=X_{e,0}\operatorname{CNOT}_{e,0\to e,1}
       \operatorname{Toffoli}_{e,0;e,1\to e,2},
 \label{eq:prx279-d8-increment}\\
 J_e&=X_{e,0}X_{e,1}X_{e,2},\qquad
 K_{8,e}=\mathsf{X}_{8,e}J_e,
 \label{eq:prx279-d8-inversion}\\
 \mathsf{Z}_{8,e}&=T_{e,0}S_{e,1}Z_{e,2}.
 \label{eq:prx279-d8-clock}
\end{align}
They act as $a_e\mapsto a_e+1$, $a_e\mapsto7-a_e$, $a_e\mapsto-a_e$, and multiplication by $e^{2\pi i a_e/8}$, respectively. Here, $T=\operatorname{diag}(1,e^{i\pi/4})$ and $S=T^2$. This fixes the edge-level qubit operators before completion. The raw vertex commutator $S_vR_vS_v^{-1}R_v^{-1}=R_v^{-2}$ is noncentral, so this example uses Theorem~\ref{thm:prx-semidirect}.

Orient edges north and east and put one completion qubit $t_v$ at each vertex. Specializing Eq.~\eqref{eq:prx-dihedral-completion} to $m=8$ gives the pin condition
\begin{equation}
 E_v=\frac{I+R_v^2+R_v^4+R_v^6}{4}.
 \label{eq:prx278-d8-pin}
\end{equation}
The completed rotation and reflection have the qubit gate products
\begin{align}
 \mathcal{D}_v
 &=\prod_{e\ni v}J_e^{t_v}
    \prod_{e=S,W}J_e^{1+b_e},
 \label{eq:prx279-d8-frame}\\
 \widehat{R}_v
 &=\mathcal{D}_v\left(\prod_{e\ni v}\mathsf{X}_{8,e}\right)
       \mathcal{D}_v,
 \label{eq:prx279-d8-qubit-rotation}\\
 \widehat{S}_v
 &=X_{t_v}\prod_{e\ni v}X_{b_e}\prod_{e=N,E}K_{8,e}.
 \label{eq:prx279-d8-qubit-reflection}
\end{align}
The exponent in $J_e^{1+b_e}$ is binary. Each $J_e^{t_v}$ is three CNOTs from $t_v$ to the three rotation bits. The incoming-edge factor $J_e^{1+b_e}$ adds three CNOTs from $b_e$ and an $X$ on each rotation bit. These gates commute, and $\mathcal{D}_v^2=I$. Thus, the completed rotation has an explicit three-step implementation. Apply $\mathcal{D}_v$, the four three-qubit increments, then $\mathcal{D}_v$ again. On an outgoing edge its shift is $(-1)^{t_v}$, and on an incoming edge it is $-(-1)^{b_e+t_v}$.

\begin{figure*}[t]
\centering
\begingroup
\definecolor{dEightHigh}{HTML}{B87900}
\tikzset{
 dEgrid/.style={draw=black!28,line width=.35pt},
 dEbox/.style={draw=black!55,fill=black!3,rounded corners=1.5pt,line width=.55pt},
 dEdiamond/.style={diamond,aspect=1.05,draw=black!75,fill=white,
   minimum size=4.5mm,inner sep=.5pt,font=\scriptsize},
 dEop/.style={fill=white,inner sep=.6pt,font=\small},
 dEwire/.style={draw=black!65,line width=.45pt}}
\newcommand{\dEface}{\foreach \j in {0,1}{
  \draw[dEgrid,-{Latex[length=1.4mm]}] (0,\j)--(1.04,\j);
  \draw[dEgrid,-{Latex[length=1.4mm]}] (\j,0)--(\j,1.04);
  \foreach \p/\c in {.18/dFourGreen,.39/dFourBlue,.60/dEightHigh,.81/dFourRed}{
   \fill[\c] (\p,\j) circle(1.8pt);
   \fill[\c] (\j,\p) circle(1.8pt);}}}
\newcommand{\dEstar}{\draw[dEgrid,step=1] (-1,-1) grid (1,1);
 \draw[black!45,-{Latex[length=1.5mm]},line width=.5pt](-1,0)--(1.06,0);
 \draw[black!45,-{Latex[length=1.5mm]},line width=.5pt](0,-1)--(0,1.06);
 \foreach \p/\c in {.34/dFourGreen,.52/dFourBlue,.70/dEightHigh,.88/dFourRed}{
  \fill[\c](\p,0)circle(1.8pt);\fill[\c](0,\p)circle(1.8pt);
  \fill[\c]({\p-1.22},0)circle(1.8pt);\fill[\c](0,{\p-1.22})circle(1.8pt);}}
\newcommand{\dETarget}[2]{\draw[fill=white,line width=.6pt](#1,#2)circle(2.9pt);
 \draw[line width=.6pt]([xshift=-2.9pt]#1,#2)--([xshift=2.9pt]#1,#2);
 \draw[line width=.6pt]([yshift=-2.9pt]#1,#2)--([yshift=2.9pt]#1,#2);}
\newcommand{\dECX}[3]{\draw[line width=.55pt](#1,#2)--(#1,#3);\fill(#1,#2)circle(1.7pt);\dETarget{#1}{#3}}
\newcommand{\dECCX}[4]{\draw[line width=.55pt](#1,#2)--(#1,#3)--(#1,#4);\fill(#1,#2)circle(1.7pt);\fill(#1,#3)circle(1.7pt);\dETarget{#1}{#4}}

\begin{minipage}[t]{.32\textwidth}\centering
\textbf{(a) Qubit layout}\par\smallskip
\begin{tikzpicture}[x=2.75cm,y=2.75cm,font=\scriptsize]
 \path[use as bounding box](-.34,-.21)rectangle(1.34,1.22);
 \dEface
 \foreach \x/\y in {0/0,1/0,0/1,1/1}
  \node[dEdiamond]at(\x,\y){$t$};
 \foreach \p/\lab/\c in {.18/a_0/dFourGreen,.39/a_1/dFourBlue,.60/a_2/dEightHigh,.81/b/dFourRed}
  \node[anchor=south,text=\c!85!black,inner sep=1pt]at(\p,1.09){$\lab$};
 \node[font=\small]at(.5,.5){$p$};
\end{tikzpicture}\par
{\footnotesize Four qubits per edge.\par One completion qubit at each vertex.}
\end{minipage}\hfill
\begin{minipage}[t]{.32\textwidth}\centering
\textbf{(b) Rotation core}\par\smallskip
\begin{tikzpicture}[x=1.62cm,y=1.62cm,font=\scriptsize]
 \path[use as bounding box](-1.38,-1.21)rectangle(1.38,1.22);
 \draw[dEbox](.26,-.10)rectangle(.78,.10);
 \draw[dEbox](-.96,-.10)rectangle(-.44,.10);
 \draw[dEbox](-.10,.26)rectangle(.10,.78);
 \draw[dEbox](-.10,-.96)rectangle(.10,-.44);
 \dEstar
 \node[dEdiamond]at(0,0){$t_v$};
 \node[fill=white,inner sep=1pt]at(.52,-.25){$\mathsf X_8$};
 \node[fill=white,inner sep=1pt]at(-.70,.25){$\mathsf X_8$};
 \node[fill=white,inner sep=1pt]at(.32,.52){$\mathsf X_8$};
 \node[fill=white,inner sep=1pt]at(-.32,-.70){$\mathsf X_8$};
 \node[anchor=south,inner sep=1pt]at(0,1.09){$N$};
 \node[anchor=west,inner sep=1pt]at(1.09,0){$E$};
 \node[anchor=north,inner sep=1pt]at(0,-1.09){$S$};
 \node[anchor=east,inner sep=1pt]at(-1.09,0){$W$};
\end{tikzpicture}\par
{\footnotesize Increment the four sets of rotation bits.\par
Apply $\mathcal D_v$ from (c) before and after.}
\end{minipage}\hfill
\begin{minipage}[t]{.32\textwidth}\centering
\textbf{(c) $X$ and CNOT gates}\par\smallskip
\begin{tikzpicture}[x=.51cm,y=.53cm,font=\scriptsize]
 \path[use as bounding box](-1.05,-1.05)rectangle(8.95,6.37);
 \fill[black!3,rounded corners=2pt](3.85,-.35)rectangle(8.05,3.35);
 \foreach \y/\lab/\c in {4/t_v/black,3/b_e/dFourRed,2/a_{e,0}/dFourGreen,1/a_{e,1}/dFourBlue,0/a_{e,2}/dEightHigh}{
  \draw[dEwire](0,\y)--(8.55,\y);
  \node[anchor=east,text=\c!85!black,inner sep=2pt]at(0,\y){$\lab$};}
 \foreach \x/\y in {.75/2,1.75/1,2.75/0}{\dECX{\x}{4}{\y}}
 \foreach \x/\y in {4.45/2,5.45/1,6.45/0}{\dECX{\x}{3}{\y}}
 \foreach \y in {0,1,2}{\node[draw,fill=white,minimum size=3.3mm,inner sep=.5pt]at(7.45,\y){$X$};}
 \draw[decorate,decoration={brace,amplitude=3pt}](.42,4.45)--(3.08,4.45);
 \node[anchor=south,inner sep=2pt]at(1.75,4.62){all edges};
 \node[align=center]at(5.95,5.25){extra gates on\\incoming edges $S,W$};
 \node[anchor=north,inner sep=2pt]at(5.95,-.50){$J_e^{b_e}J_e$};
\end{tikzpicture}\par
{\footnotesize $J_e=X_{e,0}X_{e,1}X_{e,2}$.\par
Every edge receives $J_e^{t_v}$.}
\end{minipage}

\par\medskip
\begin{minipage}[t]{.32\textwidth}\centering
\textbf{(d) Completed reflection}\par\smallskip
\begin{tikzpicture}[x=1.62cm,y=1.62cm,font=\scriptsize]
 \path[use as bounding box](-1.38,-1.21)rectangle(1.38,1.22);
 \draw[dEbox](.26,-.10)rectangle(.78,.10);
 \draw[dEbox](-.10,.26)rectangle(.10,.78);
 \dEstar
 \node[dEdiamond]at(0,0){$X$};
 \foreach \x/\y in {.88/0,0/.88,-.34/0,0/-.34}
  \node[dEop,text=dFourRed!85!black]at(\x,\y){$X$};
 \node[fill=white,inner sep=1pt]at(.52,-.25){$K_8$};
 \node[fill=white,inner sep=1pt]at(.32,.52){$K_8$};
 \node[anchor=south,inner sep=1pt]at(0,1.09){$N$};
 \node[anchor=west,inner sep=1pt]at(1.09,0){$E$};
 \node[anchor=north,inner sep=1pt]at(0,-1.09){$S$};
 \node[anchor=east,inner sep=1pt]at(-1.09,0){$W$};
\end{tikzpicture}\par
{\footnotesize $K_8$ acts on north and east only.\par
Flip all four red qubits and $t_v$.}
\end{minipage}\hfill
\begin{minipage}[t]{.32\textwidth}\centering
\textbf{(e) $\mathbb{Z}_2$ gauge flux}\par\smallskip
\begin{tikzpicture}[x=2.75cm,y=2.75cm,font=\scriptsize]
 \path[use as bounding box](-.34,-.21)rectangle(1.34,1.22);
 \draw[dFourRed!60,line width=1.1pt](0,0)rectangle(1,1);
 \dEface
 \foreach \x/\y in {.81/0,.81/1,0/.81,1/.81}
  \node[dEop,text=dFourRed!85!black]at(\x,\y){$Z$};
 \foreach \x/\y in {0/0,0/1,1/0,1/1}\fill[black!55](\x,\y)circle(.8pt);
 \node[font=\small]at(.5,.5){$p$};
\end{tikzpicture}\par
{\footnotesize $B_p=\prod_{e\in\partial p}Z_{b_e}$.\par
Only the red qubits carry $Z$.}
\end{minipage}\hfill
\begin{minipage}[t]{.32\textwidth}\centering
\textbf{(f) Full flatness}\par\smallskip
\begin{tikzpicture}[x=2.75cm,y=2.75cm,font=\scriptsize]
 \path[use as bounding box](-.34,-.21)rectangle(1.34,1.22);
 \fill[black!3](0,0)rectangle(1,1);
 \dEface
 \foreach \x/\y in {0/0,0/1,1/0,1/1}\fill[black!55](\x,\y)circle(.8pt);
 \node[anchor=north,inner sep=1pt]at(.5,-.08){$g_S$};
 \node[anchor=south,inner sep=1pt]at(.5,1.08){$g_N^{-1}$};
 \node[anchor=west,inner sep=1pt]at(1.08,.5){$g_E$};
 \node[anchor=east,inner sep=1pt]at(-.08,.5){$g_W^{-1}$};
 \node[align=center,fill=white,inner sep=3pt]at(.5,.5){$h_p=1$\\[-1pt]$\alpha_p=\beta_p=0$};
\end{tikzpicture}\par
{\footnotesize $h_p=g_Sg_Eg_N^{-1}g_W^{-1}$.\par
Measure only whether $h_p=1$.}
\end{minipage}

\par\medskip
\noindent\rule{.96\textwidth}{.3pt}\par\smallskip
{\footnotesize\textbf{Gates on the three rotation bits}\quad Time runs left to right. Bit 0 is least significant.}\par\smallskip
\begin{minipage}[t]{.46\textwidth}\centering
\begin{tikzpicture}[x=.80cm,y=.48cm,font=\scriptsize]
 \path[use as bounding box](-1.55,-.4)rectangle(5.4,2.5);
 \node[anchor=east,font=\small]at(-.85,1){$\mathsf X_8:$};
 \foreach \y/\lab/\c in {2/a_0/dFourGreen,1/a_1/dFourBlue,0/a_2/dEightHigh}{
  \draw[dEwire](0,\y)--(4.6,\y);\node[anchor=east,text=\c!85!black,inner sep=2pt]at(0,\y){$\lab$};}
 \dECCX{1}{2}{1}{0}\dECX{2.3}{2}{1}
 \node[draw,fill=white,minimum size=3.5mm,inner sep=.5pt]at(3.6,2){$X$};
\end{tikzpicture}
\end{minipage}\hfill
\begin{minipage}[t]{.50\textwidth}\centering
\begin{tikzpicture}[x=.80cm,y=.48cm,font=\scriptsize]
 \path[use as bounding box](-1.55,-.4)rectangle(6.05,2.5);
 \node[anchor=east,font=\small]at(-.85,1){$K_8:$};
 \foreach \y/\lab/\c in {2/a_0/dFourGreen,1/a_1/dFourBlue,0/a_2/dEightHigh}{
  \draw[dEwire](0,\y)--(5.55,\y);\node[anchor=east,text=\c!85!black,inner sep=2pt]at(0,\y){$\lab$};}
 \dECCX{.85}{2}{1}{0}\dECX{2.05}{1}{0}\dECX{3.25}{2}{0}\dECX{4.45}{2}{1}
\end{tikzpicture}
\end{minipage}
\endgroup
\caption{Completed $D(D_8)$ checks on the square lattice. Green, blue and ochre dots are the rotation bits $a_0,a_1,a_2$. Red dots are gauge qubits $b$. Diamonds are completion qubits. Edges point north and east. (b) Rounded outlines group the three rotation qubits acted on by each increment, whose exact gates are shown below the panels. Together with (c), the completed rotation is $\widehat R_v=\mathcal D_v(\prod_{e\ni v}\mathsf X_{8,e})\mathcal D_v$. Apply $\mathcal D_v$, the four increments, then $\mathcal D_v$ again. (c) All displayed $X$ and CNOT gates commute. The shaded extra gates act only on incoming edges. (d) The outlined outgoing rotation qubits receive $K_8$ in the displayed gate order. Each red qubit and $t_v$ receives $X$. (e) The $\mathbb{Z}_2$ gauge-flux check acts on the four red qubits. (f) Full flatness checks the identity holonomy, using the coordinates in Eq.~\eqref{eq:prx279-d8-holonomy-bits}. Its readout circuit is shown in Fig.~\ref{fig:prx278-d8-circuits}(c). These are check operators. Their measurement uses that figure and the circuits in the text.}
\label{fig:prx278-d8-checks}
\end{figure*}
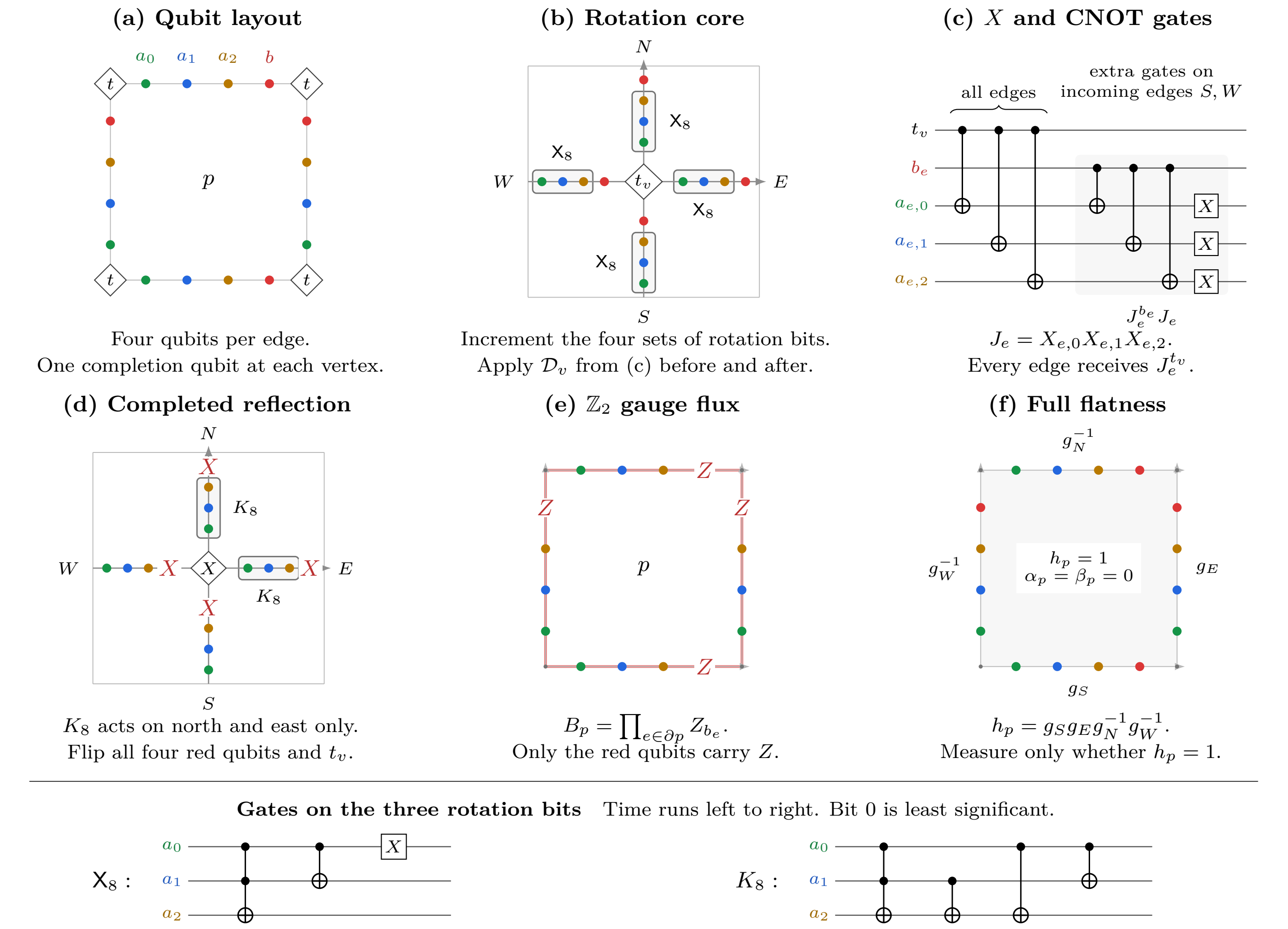

Figure~\ref{fig:prx278-d8-checks} shows the physical qubits and these gate products. For its reflection panel we use the equivalent form
\begin{equation}
 \begin{aligned}
 K_{8,e}={}&\operatorname{CNOT}_{e,0\to e,1}
 \operatorname{CNOT}_{e,0\to e,2}\\
 &{}\times\operatorname{CNOT}_{e,1\to e,2}
 \operatorname{Toffoli}_{e,0;e,1\to e,2}.
 \end{aligned}
 \label{eq:prx279-d8-inversion-word}
\end{equation}
The three CNOTs and the Toffoli in this expression are ordered factors. The native measurement circuit used for the resource and distance bounds is the controlled-addition measurement circuit specified below.

For the plaquette checks in Eq.~\eqref{eq:prx273-dihedral-flux}, the product $h_p=g_Sg_Eg_N^{-1}g_W^{-1}$ has $\mathbb{Z}_2$ gauge coordinate $\beta_p$ and rotation coordinate $\alpha_p$ given by
\begin{align}
 \beta_p&=b_S+b_E+b_N+b_W\pmod2,\\
 \alpha_p&=a_S+(-1)^{b_S}a_E
 -(-1)^{b_S+b_E+b_N}a_N\nonumber\\
 &\quad{}-(-1)^{\beta_p}a_W\pmod8.
 \label{eq:prx279-d8-holonomy-bits}
\end{align}
The flatness readout records only whether $\beta_p=\alpha_p=0$ and returns the four bits storing the group product to zero by reversing the computation. It does not measure the four bits separately. The completed common positive space is the $D(D_8)$ code with product-state completion qubits.

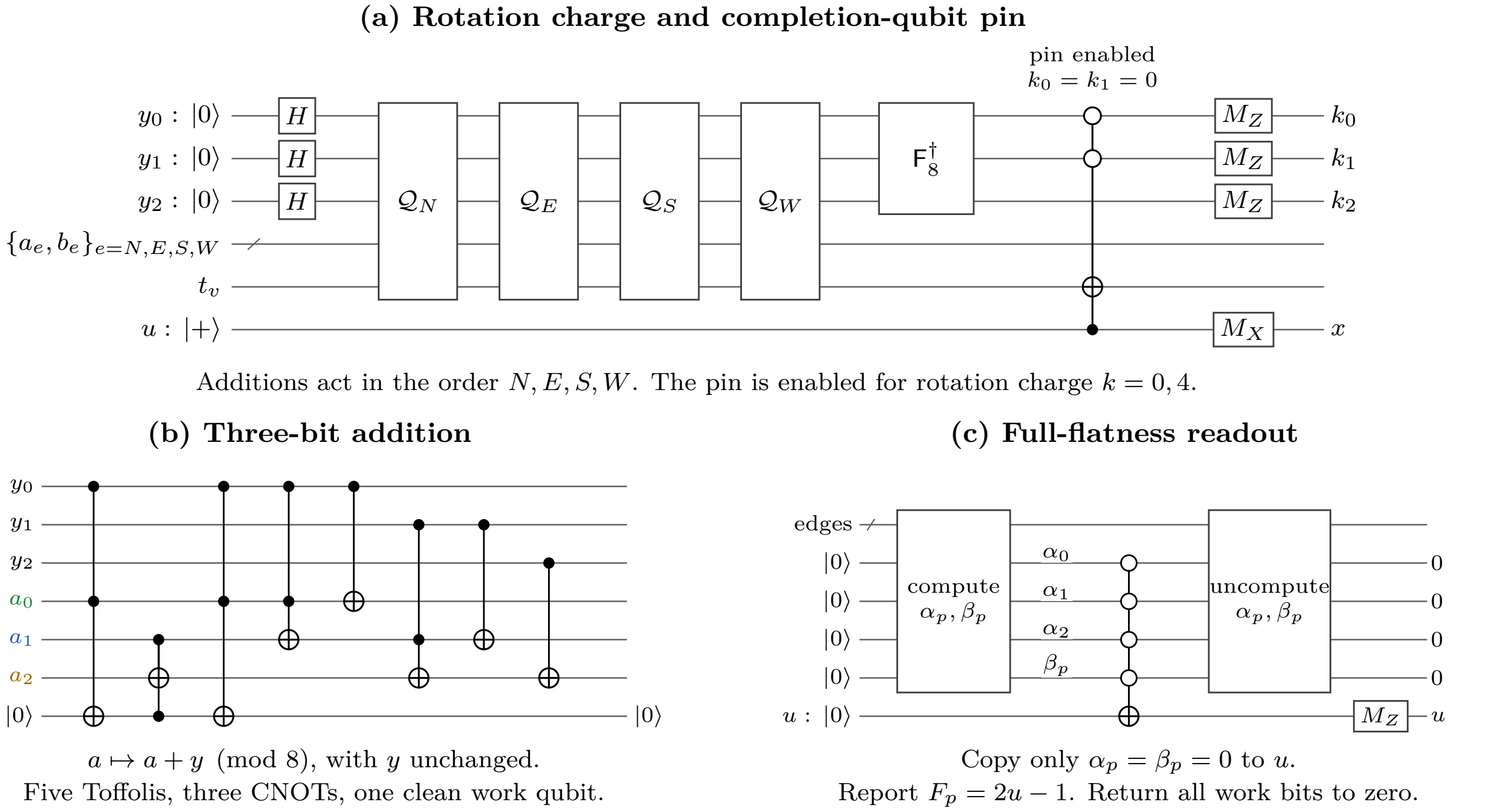
\begin{figure*}[t]
\centering
\begingroup
\definecolor{dEightHigh}{HTML}{B87900}
\tikzset{dEwire/.style={draw=black!65,line width=.45pt},
 dEgate/.style={draw=black!75,fill=white,line width=.55pt,inner sep=2pt}}
\newcommand{\dETarget}[2]{\draw[fill=white,line width=.6pt](#1,#2)circle(3.1pt);
 \draw[line width=.6pt]([xshift=-3.1pt]#1,#2)--([xshift=3.1pt]#1,#2);
 \draw[line width=.6pt]([yshift=-3.1pt]#1,#2)--([yshift=3.1pt]#1,#2);}
\newcommand{\dECX}[3]{\draw[line width=.55pt](#1,#2)--(#1,#3);\fill(#1,#2)circle(1.8pt);\dETarget{#1}{#3}}
\newcommand{\dECCX}[4]{\draw[line width=.55pt](#1,#2)--(#1,#3)--(#1,#4);\fill(#1,#2)circle(1.8pt);\fill(#1,#3)circle(1.8pt);\dETarget{#1}{#4}}

\textbf{(a) Rotation charge and completion-qubit pin}\par\smallskip
\begin{tikzpicture}[x=1.13cm,y=.48cm,font=\footnotesize]
 \path[use as bounding box](-2.25,-.65)rectangle(11.45,6.65);
 \foreach \y/\lab in {5/{y_0:\,|0\rangle},4/{y_1:\,|0\rangle},3/{y_2:\,|0\rangle},2/{\{a_e,b_e\}_{e=N,E,S,W}},1/t_v,0/{u:\,|+\rangle}}{
  \draw[dEwire](0,\y)--(10.85,\y);
  \node[anchor=east,inner sep=3pt]at(0,\y){$\lab$};}
 \draw[black!65](.16,1.87)--(.28,2.13);
 \foreach \y in {3,4,5}\node[dEgate,minimum size=4mm]at(.65,\y){$H$};
 \foreach \x/\lab in {1.85/N,3.05/E,4.25/S,5.45/W}{
  \draw[dEgate](\x-.39,.70)rectangle(\x+.39,5.30);
  \node[fill=white,inner sep=1pt]at(\x,3){$\mathcal Q_{\lab}$};}
 \draw[dEgate](6.43,2.70)rectangle(7.37,5.30);
 \node[fill=white,inner sep=1pt]at(6.90,4){$\mathsf F_8^\dagger$};
 \draw[line width=.6pt](8.55,0)--(8.55,5);
 \foreach \y in {4,5}\draw[fill=white,line width=.6pt](8.55,\y)circle(2.7pt);
 \fill(8.55,0)circle(1.8pt);\dETarget{8.55}{1}
 \node[anchor=south,align=center,font=\scriptsize,inner sep=2pt]at(8.55,5.45){pin enabled\\$k_0=k_1=0$};
 \foreach \y/\lab in {5/k_0,4/k_1,3/k_2}{
  \node[dEgate,minimum width=5mm]at(10.05,\y){$M_Z$};
  \node[anchor=west,inner sep=2pt]at(10.85,\y){$\lab$};}
 \node[dEgate,minimum width=5mm]at(10.05,0){$M_X$};
 \node[anchor=west,inner sep=2pt]at(10.85,0){$x$};
\end{tikzpicture}\par
{\footnotesize Additions act in the order $N,E,S,W$. The pin is enabled for rotation charge $k=0,4$.}\par\medskip

\begin{minipage}[t]{.52\textwidth}\centering
\textbf{(b) Three-bit addition}\par\smallskip
\begin{tikzpicture}[x=.73cm,y=.43cm,font=\scriptsize]
 \path[use as bounding box](-1.25,-.55)rectangle(9.55,6.65);
 \foreach \y/\lab/\c in {6/y_0/black,5/y_1/black,4/y_2/black,3/a_0/dFourGreen,2/a_1/dFourBlue,1/a_2/dEightHigh,0/{w:\,|0\rangle}/black}{
  \draw[dEwire](0,\y)--(9.00,\y);
  \node[anchor=east,text=\c!85!black,inner sep=2pt]at(0,\y){$\lab$};}
 \dECCX{.8}{6}{3}{0}
 \dECCX{1.8}{0}{2}{1}
 \dECCX{2.8}{6}{3}{0}
 \dECCX{3.8}{6}{3}{2}
 \dECX{4.8}{6}{3}
 \dECCX{5.8}{5}{2}{1}
 \dECX{6.8}{5}{2}
 \dECX{7.8}{4}{1}
 \node[anchor=west,inner sep=2pt]at(9,0){$|0\rangle$};
\end{tikzpicture}\par
{\footnotesize $a\mapsto a+y\pmod8$, with $y$ unchanged.\par
Five Toffolis, three CNOTs, one clean work qubit.}
\end{minipage}\hfill
\begin{minipage}[t]{.46\textwidth}\centering
\textbf{(c) Full-flatness readout}\par\smallskip
\begin{tikzpicture}[x=.65cm,y=.43cm,font=\scriptsize]
 \path[use as bounding box](-1.15,-.55)rectangle(10.35,6.65);
 \foreach \y/\lab in {5/\mathrm{edges},4/{|0\rangle},3/{|0\rangle},2/{|0\rangle},1/{|0\rangle},0/{u:\,|0\rangle}}{
  \draw[dEwire](0,\y)--(9.8,\y);
  \node[anchor=east,inner sep=2pt]at(0,\y){$\lab$};}
 \draw[black!65](.13,4.85)--(.26,5.15);
 \draw[dEgate](.65,.62)rectangle(2.60,5.38);
 \node[align=center]at(1.625,3){compute\\$\alpha_p,\beta_p$};
 \foreach \y/\lab in {4/\alpha_0,3/\alpha_1,2/\alpha_2,1/\beta_p}
  \node[anchor=south,fill=white,inner sep=.5pt]at(3.40,\y){$\lab$};
 \draw[line width=.55pt](4.65,0)--(4.65,4);
 \foreach \y in {1,2,3,4}\draw[fill=white,line width=.55pt](4.65,\y)circle(2.5pt);
 \dETarget{4.65}{0}
 \draw[dEgate](6.03,.62)rectangle(8.13,5.38);
 \node[align=center]at(7.08,3){uncompute\\$\alpha_p,\beta_p$};
 \foreach \y in {1,2,3,4}\node[anchor=west,inner sep=1pt]at(9.8,\y){$0$};
 \node[dEgate,minimum width=4.5mm]at(9.0,0){$M_Z$};
 \node[anchor=west,inner sep=1pt]at(9.8,0){$u$};
\end{tikzpicture}\par
{\footnotesize Copy only $\alpha_p=\beta_p=0$ to $u$.\par
Report $F_p=2u-1$. Return all work bits to zero.}
\end{minipage}
\endgroup
\caption{Native extraction for the $D_8$ example. Time runs left to right. Filled and open controls test bits 1 and 0, respectively. Circles with a plus sign are $X$ targets. (a) The three phase qubits and pin readout are separate. Each controlled addition touches only its indicated edge, the readout qubits and $t_v$. Surrounding the adder in (b) by the controlled complements in Eqs.~\eqref{eq:prx278-d8-out-compiler} and \eqref{eq:prx278-d8-in-compiler} implements the signed addition. After the inverse Fourier transform, $k=k_0+2k_1+4k_2$ is read in $Z$. The pin's two open controls are used coherently. Together with the readout control, they compile into three Toffolis and four $X$ gates using the adder's work qubit. (b) The first three gates compute, use and uncompute a carry. The circuit adds modulo eight on all 64 input pairs with $w=0$. Reversing the gate order subtracts. (c) The readout records only whether the product of group labels around the plaquette is the identity, without measuring its four bits separately. The four-open-control gate compiles into five Toffolis and eight $X$ gates with two clean work qubits (Appendix~\ref{sec:sm-prx-dihedral-cost}). The inverse three-qubit Fourier circuit in (a) uses three Hadamards, three controlled phases and one SWAP.}
\label{fig:prx278-d8-circuits}
\end{figure*}

\paragraph{Explicit extraction.}
The rotation and pin circuit in Fig.~\ref{fig:prx278-d8-circuits}(a) uses three readout qubits $y$ and a pin readout $u$. Its four controlled additions act in the order $N,E,S,W$. They implement
\begin{align}
 \mathcal{Q}_e:\quad a_e&\longmapsto a_e+(-1)^{t_v}y,
       &&e=N,E,
 \label{eq:prx278-d8-query-out}\\
 a_e&\longmapsto a_e-(-1)^{b_e+t_v}y,
       &&e=S,W.
 \label{eq:prx278-d8-query-in}
\end{align}
After $\mathsf{F}_8^\dagger$, the readout stores $k=k_0+2k_1+4k_2$. The pin is enabled only at $k_0=k_1=0$, or $k=0,4$. With $\Pi_k=\Pi(\widehat{R}_v=e^{2\pi i k/8})$, the joint branch is $\Pi_k(I+xT_v)/2$.

These additions have explicit qubit circuits. Let $J|a\rangle=|7-a\rangle$ and let $\operatorname{ADD}_y$ add $y$ modulo eight. In operator order,
\begin{align}
 \mathcal{Q}_{N,E}&=J^{t_v}\operatorname{ADD}_yJ^{t_v},
 \label{eq:prx278-d8-out-compiler}\\
 \mathcal{Q}_{S,W}&=J^{t_v}J^{b_e}
   \operatorname{ADD}_y^{-1}J^{b_e}J^{t_v}.
 \label{eq:prx278-d8-in-compiler}
\end{align}
Each controlled $J$ is three CNOTs. Figure~\ref{fig:prx278-d8-circuits}(b) gives the complete adder with five Toffolis, three CNOTs and one clean work qubit. Reversing its gate order gives subtraction. The pin's control on two zero bits uses three Toffolis and the same work qubit, with four $X$ gates implementing its zero controls.

Here $a_S$ labels the reflection readout qubit. Reflection extraction applies the native controlled-$K_8$ on outgoing edges and controlled-$X_b$ on all four edges in $N,E,S,W$ order, then applies $\operatorname{CNOT}(a_S\to t_v)$ and reads $X_{a_S}$. The uncontrolled edge operators are shown in Fig.~\ref{fig:prx278-d8-checks}(d). Four CNOTs extract the $\mathbb{Z}_2$ gauge-flux check. The optional flag gates use the same completion qubit and satisfy Lemma~\ref{lem:sm-prx-flags}.

\paragraph{Resources and the surgery guarantee.}
Table~\ref{tab:prx278-d8-cost} gives the per-cell, per-round native counts, including both pairs of flag gates and basis-change Hadamards for computational-basis preparation and readout. Every native unitary gate in these circuits has qubit Clifford-hierarchy level at most four.

\begin{table*}[t]
\centering\small
\begin{tabular}{lrrrrrrr}
\toprule
$D_8$ circuit & Toffoli & CNOT & CZ & $H$ & Controlled phase & SWAP & $X$\\
\midrule
Rotation and pin &23&48&0&8&3&1&4\\
Reflection with flag &8&13&2&2&0&0&0\\
Gauge flux with flag &0&6&0&0&0&0&0\\
Full flatness &45&80&0&0&0&0&8\\
\midrule
Total &76&147&2&10&3&1&12\\
\bottomrule
\end{tabular}
\caption{Native unitary gate counts for the explicit $D_8$ circuits. Preparations, measurements, work-qubit resets, idles, routing and synthesis of the listed gates are excluded. Both pairs of completion-qubit flag gates are included.}
\label{tab:prx278-d8-cost}
\end{table*}

An $L\times L$ torus has $9L^2$ persistent data qubits, comprising $8L^2$ edge qubits and $L^2$ completion qubits. Seven dedicated readout qubits per cell resolve three rotation bits, the pin, reflection and the two flux checks. Rotation and pin extraction additionally use one clean work qubit per active circuit. Full flatness uses four qubits for the group product and at most two work qubits. The number of temporary qubits depends on the number of concurrent disjoint circuits.

Gauging and ungauging the $\mathbb{Z}_8$ toric-code parent measures charge conjugation through these completed $D(D_8)$ checks. Theorem~\ref{thm:prx273-dihedral-distance} of the main text supplies the circuit-distance guarantee.

\subsection{Qubit gate counts and the \texorpdfstring{$D_{16}$}{D16} example}
\label{sec:sm-prx-dihedral-cost}
For $m=2^q$, $q\geq2$, a conservative exact adder uses
\begin{equation}
 A_q=\frac{q(q-1)(2q-1)}{6}
 \label{eq:sm-prx-adder-cost}
\end{equation}
Toffoli gates, $q$ CNOTs, and $q-2$ clean work qubits. One construction adds each control bit by a controlled increment of the corresponding target suffix. Within a suffix, flip high bits before low bits. A NOT with $k$ controls uses $2k-3$ Toffolis and $k-2$ clean work qubits when $k\geq2$. Summing these counts gives Equation~\eqref{eq:sm-prx-adder-cost}. Work qubits are reset between edge blocks.

Let $J|x\rangle=|m-1-x\rangle$. Since $J\operatorname{ADD}_yJ=\operatorname{ADD}_{-y}$, outgoing signed addition is $J^t\operatorname{ADD}_yJ^t$. Incoming signed subtraction is $J^tJ^{b_e}\operatorname{ADD}_y^{-1}J^{b_e}J^t$. The four rotation interactions use $4A_q$ Toffolis and $16q$ CNOTs. Uniform phase preparation and inverse Fourier transform add $2q$ Hadamards, $q(q-1)/2$ controlled phases, and $\lfloor q/2\rfloor$ swaps. The pin uses $2q-3$ Toffolis, $2(q-1)$ $X$ gates, and two readout Hadamards. A controlled modular negation uses $(q-1)^2$ Toffolis and $q+1$ CNOTs. Two such negations, the four controlled gauge-qubit flips, the completion flip, and the flag gates give the reflection count below. The flatness computation and its inverse use eight adders and $24q+8$ CNOTs. Its identity test adds $2q-1$ Toffolis and $2(q+1)$ $X$ gates.

\begin{table*}[t]
\centering\small
\begin{tabular}{lrrrrrrr}
\toprule
$D_{16}$ circuit & Toffoli & CNOT & CZ & $H$ & Controlled phase & SWAP & $X$\\
\midrule
Rotation and pin &61&64&0&10&6&2&6\\
Reflection with flag &18&15&2&2&0&0&0\\
Gauge flux with flag &0&6&0&0&0&0&0\\
Full flatness &119&104&0&0&0&0&10\\
\midrule
Total &198&189&2&12&6&2&16\\
\bottomrule
\end{tabular}
\caption{Unitary gate counts by circuit and per vertex--face cell for $D_{16}$. Preparations, measurements, work-qubit resets, idles, routing, and synthesis of the listed native gates are excluded. Both pairs of completion-qubit flag gates are included. Rotation and pin use five readout qubits and two work qubits. Flatness uses five temporary group-product qubits and at most three work qubits, in addition to its readout.}
\label{tab:sm-prx-d16-cost}
\end{table*}

For $D_{16}$, $q=4$ and the group order is $32$. An $L\times L$ torus has $11L^2$ persistent data qubits, comprising $10L^2$ edge qubits and $L^2$ completion qubits, and $8L^2$ dedicated readout qubits. Additional temporary qubits depend on how many disjoint circuits run concurrently. No separate flag qubit is used. For general $q$, the cell count is $2(q-1)(2q^2+1)$ Toffolis, $42q+21$ CNOTs, two CZs, $2q+4$ Hadamards, $q(q-1)/2$ controlled phases, $\lfloor q/2\rfloor$ swaps, and $4q$ $X$ gates. Thus, the unitary gate count is $O(q^3)$ and active workspace is $O(q)$.

For fixed $m$, bounded overlap of the star and face measurement circuits gives a constant number of sequential layers per round by graph coloring. The round duration may depend on $m$. The native gate library has qubit Clifford-hierarchy level at most $\max\{3,q+1\}$, set by Toffoli and Fourier controlled-phase gates. This gives level five for $D_{16}$.

\subsection{Faults detected by the completion qubit}
\label{sec:sm-prx-dual-role}
The completion qubit makes rotation and reflection measurements compatible. It can also detect particular faults that spread between edges, without an additional flag qubit. The ideal final tests used below require neutral rotation charge and a positive pin outcome. Their joint projector is
\begin{equation}
 \Pi(\widehat{R}_v=1)\Pi(T_v=+1)
 =\Pi(R_v=1)\otimes|+\rangle\langle+|_{t_v}.
 \label{eq:sm-prx-flag-acceptance}
\end{equation}
Both tests matter because $T_v=I$ when the pin is disabled.

To measure the $\mathbb{Z}_2$ gauge-flux check $B_p$, prepare the readout qubit $a_B$ in $|0\rangle$ and report its $Z$-readout sign as the outcome of $B_p$. Apply the following operations in chronological order.
\begin{equation}
 \begin{aligned}
 &\operatorname{CNOT}(t_v\to a_B),\quad
   \{\operatorname{CNOT}(b_e\to a_B)\}_{e\in\partial p},\\
 &\operatorname{CNOT}(t_v\to a_B),\quad Z_{a_B}\text{ readout}.
 \end{aligned}
 \label{eq:sm-prx-b-wrapper}
\end{equation}
For reflection, prepare the readout qubit $a_S$ in $|+\rangle$ and report its $X$-readout sign as the outcome of $\widehat S_v$. Write $S_v=\prod_{e\ni v}S_{v,e}$, where $S_{v,e}=K_{m,e}X_{b_e}$ on outgoing edges and $S_{v,e}=X_{b_e}$ on incoming edges. Use
\begin{equation}
 \begin{aligned}
 &\operatorname{CZ}(a_S,t_v),\quad
   \{\operatorname{ctrl}_{a_S}(S_{v,e})\}_{e\ni v},\\
 &\operatorname{CZ}(a_S,t_v),\quad
   \operatorname{CNOT}(a_S\to t_v),\quad X_{a_S}\text{ readout}.
 \end{aligned}
 \label{eq:sm-prx-s-wrapper}
\end{equation}
Assign a face to its southwest vertex for the first sequence. The two additional flag gates---CNOTs in the gauge-flux circuit and CZs in the reflection circuit---commute through the intervening ideal edge operations and cancel. The measurement operators are therefore unchanged on every input sector.

\begin{lemma}[Detection of a single extraction fault]
\label{lem:sm-prx-flags}
Assume a neutral-charge, positive-pin input. For the gauge-flux circuit assume an input with $B_p=+1$, and for the reflection circuit assume a known input reflection eigenvalue. With one arbitrary native fault in either circuit, every branch passing the expected measurement outcome and ideal final tests in Equation~\eqref{eq:sm-prx-flag-acceptance} has an equivalent residual operator supported on at most one data edge. Edge arithmetic must preserve the readout's computational control and use separate work qudits for each edge block or reset them between blocks. Faults on the completion qubit are included.
\end{lemma}
\begin{proof}
Let $a$ denote the readout of the circuit under discussion, $A_e$ an operator on the current data edge, and $u,w\in\{0,1\}$ Pauli exponents. For the parity circuit, complete the current edge block and expand each resulting error operator as $A_eX_a^uZ_a^w$. The subsequent $\operatorname{CNOT}(b_e\to a_B)$ gates and the final flag gate give
\begin{equation}
 A_e\left(\prod_{j>e}Z_{b_j}\right)^wX_a^uZ_a^wZ_t^w.
 \label{eq:sm-prx-b-hook}
\end{equation}
The expected readout rejects an inappropriate bit-flip component, and the final charge and pin projection rejects $w=1$. For reflection, an interior $X_a$ component similarly applies the remaining $S_{v,e}$ factors and creates a $Z_t$ at the final CZ. The final controlled $X_t$ retains that $Z_t$ parity, so the same final tests reject this component. The surviving terms act on the current edge only.

A fault at the first flag gate can propagate a full $B_p$ or $S_v$, which is scalar on the specified input. A fault at the final flag gate or later has no subsequent edge interactions. A completion-qubit idle fault alone does not spread to the data edges. The same expansion covers preparation, measurement, data idle, and arithmetic faults. Resetting the work qudits prevents a fault in one edge block from contaminating later edge blocks. Write $R=R_v=\prod_{e\ni v}R_e$, where $R_e$ is the raw rotation action on edge $e$, and let $P_R=\Pi(R_v=1)$. Finally, if $J_0$ embeds a neutral input, the final charge projection satisfies
\begin{align}
 P_R A_eJ_0
 &=\frac{1}{m}\sum_{j=0}^{m-1}R^j A_eR^{-j}J_0\nonumber\\
 &=\frac{1}{m}\sum_{j=0}^{m-1}R_e^j A_eR_e^{-j}J_0,
 \label{eq:sm-prx-one-edge-twirl}
\end{align}
which remains on that edge. Linearity gives the result for arbitrary local Kraus operators.
\end{proof}

\subsection{Proof of the dihedral circuit-distance bound}
\label{sec:sm-prx-dihedral-distance}
We use the protocol and fault model of Section~\ref{sec:prx273-dihedral-distance}. Each extraction circuit is supported on one star or one face, possibly with its assigned completion qubit. Work qudits are reset before reuse, and circuits sharing code qudits are performed one after another. The encoded input and final code-space projection are ideal. The same bound holds with or without the optional completion-qubit flag gates.

\begin{proof}[Proof of Theorem~\ref{thm:prx273-dihedral-distance} of the main text]
Fix the measurement record and all noise Kraus labels. Contract each faulty circuit's preparations, work qudits and readouts to an operator on the code qudits it touches. Expand the intervening ideal projectors into powers of their commuting checks. Moving these check products to an ideal boundary or a fault-free round conjugates each faulty operator. A check disjoint from its original support $S$ commutes past the other check factors and cancels. The resulting effective error is supported in
\begin{equation}
 \mathcal{N}(S)=S\cup
 \bigcup_{C:\operatorname{supp}(C)\cap S\ne\varnothing}
 \operatorname{supp}(C).
 \label{eq:sm-prx-neighborhood}
\end{equation}
This enlargement depends on the original support, not on the enlarged support after an earlier round. With $f$ faults there are at most $f$ such neighborhoods, even if several faults occur in one check-measurement circuit.

Index edges by their source coordinate. A star neighborhood occupies at most four columns $i-2,\ldots,i+1$ and four rows $j-2,\ldots,j+1$. A face neighborhood occupies at most four columns $i-1,\ldots,i+2$ and four rows $j-1,\ldots,j+2$. The southwest completion qubit adds no new vertex neighborhood because that vertex already meets the face. Thus, if $4f<L$, some source row and column avoid every effective error.

For memory, we show directly that errors on the remaining edges act as scalars after code projection, consistent with the correctability results for quantum doubles~\cite{Cui2020QuantumDoubleQEC}. Enlarge the support to all edges $A$ whose sources avoid this row and column, and let $C$ contain the other edges. The graph $C$ is connected, contains both winding cycles, and contains every boundary vertex of $A$. Choose a flat edge configuration $h_{x,y}$ with every $A$-edge label at the identity and its two seam labels $x,y\in G=D_m$ on $C$. The products $x,y$ around the two winding loops commute. Gauge transformations identify $(x,y)$ with $(kxk^{-1},kyk^{-1})$ for $k\in G$. Let $Z_G(x,y)$ be the subgroup commuting with both $x$ and $y$. Averaging over vertex gauge transformations $g\in G^V$ gives the normalized ground state
\begin{equation}
 |\psi_{x,y}\rangle=
 \frac{\sum_{g\in G^V}|g\cdot h_{x,y}\rangle}
 {\sqrt{|G|^{|V|}|Z_G(x,y)|}}.
 \label{eq:sm-prx-orbit-state}
\end{equation}
In a partial trace over $C$, equal labels on $C$ force the relative gauge transformation to be one common element of $Z_G(x,y)$ on $V(C)$. Its values on interior vertices $I=V\setminus V(C)$ are free. The common element cancels from every $A$-edge label. Summing over it cancels the factor $|Z_G(x,y)|$ in the normalization. Extending $\ell\in G^I$ by identity on $V(C)$ gives
\begin{equation}
 \rho_A=|G|^{-|V|}\sum_{g\in G^V}\sum_{\ell\in G^I}
 |g_A\rangle\langle(g\ell)_A|,
 \label{eq:sm-prx-reduced-state}
\end{equation}
where $(g_A)_e=g_{s(e)}g_{t(e)}^{-1}$. This is independent of the loop products $x,y$. Cross terms between pairs not related by simultaneous conjugation vanish because the two loops in $C$ distinguish them. Every operator on $A$ is consequently scalar after code projection. Completion operators contract against the fixed plus states, proving the memory bound. For correction of all branches with at most $t$ faults, apply this argument to a product of two error operators. Their combined support contains at most $2t$ neighborhoods, so the Knill--Laflamme conditions hold when $8t<L$.

For surgery, $f<R$ leaves a fault-free gauged round. Theorem~\ref{thm:prx-semidirect} identifies its projector, with the recorded reflection outcomes, as the corresponding raw gauging projector tensored with the completion plus state. For a chosen reference vertex $v_0$, the gauging isometry is
\begin{equation}
 \Gamma_{\boldsymbol{s}}|\psi\rangle
 =\frac{1}{\sqrt{2^{|V|-1}}}
 \sum_{x:x_{v_0}=0}\left(\prod_v s_v^{x_v}\right)
 U(x)|\psi\rangle\otimes|\delta x\rangle.
 \label{eq:sm-prx-rooted-map}
\end{equation}
Different $x$ with the reference value fixed have distinct link records, proving isometry. The Gauss condition at the reference vertex imposes the remaining global eigenvalue $s$. For an exact gauge record $b=\delta x$, put $\widetilde a_e=(-1)^{x_{s(e)}}a_e$. Conjugating the rotation star in this fixed gauge configuration by $U(x)^\dagger$ gives the oriented parent rotation star raised to $(-1)^{x_v}$. Its outgoing shift coefficient is $(-1)^{x_v}$. The incoming coefficient is $-(-1)^{b_e+x_{s(e)}}=-(-1)^{x_v}$. The plaquette holonomy has $\mathbb{Z}_2$ gauge coordinate zero and rotation coordinate $(-1)^{x_{\mathrm{SW}(p)}} (\widetilde a_S+\widetilde a_E-\widetilde a_N-\widetilde a_W)$. Thus neutral rotations and full flatness impose exactly the parent star and plaquette constraints. Let $Q_{\boldsymbol{s}}$ be the joint completed-check projector for neutral rotation charge, positive flux and pin outcomes, and reflection outcomes $\boldsymbol{s}$. In the exact-link sector,
\begin{equation}
 Q_{\boldsymbol{s}}P_{\mathrm{ex}}
 =\Gamma_{\boldsymbol{s}}\frac{I+s\mathcal{U}}{2}P
 \Gamma_{\boldsymbol{s}}^\dagger
 \otimes|+\rangle\langle+|^{\otimes|V|}.
 \label{eq:sm-prx-rooted-sector}
\end{equation}
Expand each effective matter and link error in the Pauli basis. The gauge-flux tests $B_p=+1$ retain only incoming gauge-bit flip patterns with even parity around every face. Avoiding the unused row and column also gives zero parity around both noncontractible cycles. Each surviving pattern therefore has the exact form $\eta_{uv}=y_u+y_v\pmod2$ for vertex bits $y_v$. The exact final link record forces the outgoing pattern to have the same vertex-bit form. Both patterns vanish on the connected graph $C$, so their vertex potentials are constant on $V(C)$. Subtract these constants to make both potentials zero there. This only contributes the measured global-symmetry eigenvalue. Since each $U_v$ acts only on outgoing matter edges, the symmetry corrections and direct matter errors avoid the supports of the two logical $X$ operators and two logical $Z$ operators of the parent code on that row and column. Their products span all operators on the encoded states of the $\mathbb{Z}_m$ toric code, including for composite $m$. Every parent error is therefore scalar after projection onto the code. The input and output contractions are Eqs.~\eqref{eq:sm-zn-input} and~\eqref{eq:sm-zn-output}, with gauge order $N=2$ and completion state $|+\rangle^{\otimes|V|}$. The unused row and column bound the matter support in those contractions. Applying this statement before and after the fault-free round gives Equation~\eqref{eq:prx273-dihedral-branch} of the main text for each Pauli term and hence for their sum. The scalar identity also preserves coherence with an external reference.
\end{proof}

\section{Additional logical protocols}
\label{sec:v278-additional-protocols}
We give CSS, Steane-code and qutrit preparations and the Toffoli consumption circuit. Appendix~\ref{sec:sm-gross} supplies the compiled CSS instances. Table~\ref{tab:prx271-protocols} collects the outputs and their circuit guarantees.

\subsection{CSS codes: a collective CNOT measurement}
\label{sec:prx273-css-source}
Take two identical CSS blocks $A,B$, each encoding $k\geq2$ logical qubits. Use positive parent-check signs and matched logical Pauli bases. Transversal physical CNOT then implements
\begin{equation}
 \overline{\mathcal{U}}=\prod_{j=1}^{k}
 \overline{\operatorname{CNOT}}_{A_j\to B_j}.
 \label{eq:prx273-css-logical}
\end{equation}

Prepare the encoded stabilizer input
\begin{equation}
 |\phi\rangle=|+0\rangle_{A_1B_1}|10\rangle_{A_2B_2}
 \bigotimes_{j=3}^{k}|00\rangle_{A_jB_j}.
 \label{eq:prx273-css-input}
\end{equation}
Run the completed gauging and ungauging measurement of Eq.~\eqref{eq:prx273-css-logical}. The CNOTs on $j\geq3$ fix their inputs, while the state $|1\rangle_{A_2}$ makes the second CNOT act as $X_{B_2}$. On the active order $(B_2,A_1,B_1)$ the input is therefore $|0+0\rangle$ and the measured observable is $X_1\operatorname{CNOT}_{2\to3}$. For the reported sign $s=(-1)^{\eta}$,
\begin{equation}
 \frac{I+sX_1\operatorname{CNOT}_{2\to3}}{2}|0+0\rangle
 =\frac{1}{\sqrt{2}}Z_1^{\eta}|\mathrm{Tof}\rangle,
 \label{eq:prx273-css-projection}
\end{equation}
where
\begin{equation}
 |\mathrm{Tof}\rangle
 =\frac{|000\rangle+|010\rangle+|100\rangle+|111\rangle}{2}.
 \label{eq:prx273-css-resource}
\end{equation}
The correction $Z_{B_2}^{\eta}$ removes the relative sign between the first-bit sectors. Each branch has probability $1/2$ and produces one three-qubit resource from two logical pairs. The other logical qubits retain their specified stabilizer states.

\paragraph{A compiled gross-code instance.}
For the gross $[[144,12,12]]$ code, three of the 24 parent logical qubits carry the resource and the remaining 21 are fixed spectators. The completed measurement in Appendix~\ref{sec:sm-gross} implements the logical measurement above. Its completed Gauss and modified-parent checks use CNOT, CZ and Pauli factors. With one readout ancilla $a$ per check, the controlled gates are obtained factor by factor from
\begin{align}
 \operatorname{ctrl}_a(X_q)&=\operatorname{CNOT}_{a\to q},\\
 \operatorname{ctrl}_a(Z_q)&=\operatorname{CZ}_{aq},\\
 \operatorname{ctrl}_a(\operatorname{CNOT}_{c\to q})
 &=\operatorname{Toffoli}_{a,c\to q},\\
 \operatorname{ctrl}_a(\operatorname{CZ}_{cq})
 &=\operatorname{CCZ}_{acq}.
 \label{eq:prx273-css-native}
\end{align}
The readout is in $X$ after preparation in $|+\rangle$. Central checks and their pins use the joint parity--pin gadget of Section~\ref{sec:prx-dual-role} of the main text. These rules give an explicit circuit from the check factor lists, with arbitrary local faults allowed on all participants of each displayed gate.

Appendix~\ref{sec:sm-gross} gives the complete qubit and gate counts, the fixed schedule and the fault analysis. An explicit eleven-fault pattern establishes $d_{\mathrm{surg}}\le11$ for the unflagged gross-code circuit. Using the retained completion qubits as flags rejects this pattern.

The same encoded preparation applies to the Mitten, lifted-product and Kasai examples in Appendix~\ref{sec:sm269-other-css}. Their check matrices, logical bases and native circuit counts provide additional finite instances of this protocol.

\subsection{Steane code: explicit checks for a single-qubit magic state}
\label{sec:prx271-steane}
This example specifies the parent checks, gauging graph, completion and logical output. It uses one Steane $[[7,1,3]]$ block, whose transversal Hadamard implements logical Hadamard~\cite{Gottesman1997Steane}. Label the matter qubits $0,\ldots,6$ and take the three supports
\begin{align}
 D_1&=\{3,4,5,6\},\\
 D_2&=\{1,2,5,6\},\\
 D_3&=\{0,2,4,6\}.
 \label{eq:prx271-steane-supports}
\end{align}
The parent generators are $X(D_i),Z(D_i)$, and we choose the logical representatives $X_L=\prod_{q=0}^{6}X_q$ and $Z_L=\prod_{q=0}^{6}Z_q$. The common reference vertex $6$ gives a star gauging graph, with gauge qubit $g_q$ on edge $(6,q)$ for $q=0,\ldots,5$. This graph is a tree, so it has no independent flux checks. Define
\begin{equation}
 V_q=|0\rangle\!\langle0|_{g_q}X_q+
     |1\rangle\!\langle1|_{g_q}Z_q.
 \label{eq:prx271-steane-local-transport}
\end{equation}
An explicit set of raw checks is
\begin{align}
 A_q&=H_qX_{g_q},\qquad 0\le q<6,\\
 A_6&=H_6\prod_{q=0}^{5}X_{g_q},\\
 C_i&=X_6\prod_{q\in D_i\setminus\{6\}}V_q,\\
 N_i&=Y(D_i)\prod_{q\in D_i\setminus\{6\}}Z_{g_q}.
 \label{eq:prx271-steane-raw}
\end{align}
Here, $C_i$ is the modified parent $X$ check and $N_iC_i$ is the modified parent $Z$ check. All $N_i$ are central Pauli involutions. All pairs among the $A_q$ commute, all pairs among the $C_i$ commute, and the only noncommuting pairs are
\begin{equation}
 A_6 C_i=N_i C_i A_6.
 \label{eq:prx271-steane-conflicts}
\end{equation}
To verify these relations, conjugating $V_q$ by $H_qX_{g_q}$ leaves it unchanged. Conjugating $C_i$ by $A_6$ exchanges every $X$ with $Z$ in that modified parent check. Their product is $N_i$, since $|D_i|=4$. Two distinct $D_i$ intersect on two matter qubits, so the corresponding Pauli signs cancel. In $[N_i,A_6]$, the one sign from $H_6Y_6H_6=-Y_6$ cancels the three signs from the incident gauge $Z$ factors. These observations also prove centrality in the entire raw algebra.

Add three completion qubits $t_i$ and use the functions $V(B,t)$ and $T(B,t)$ of Equations~\eqref{eq:main-completion} and~\eqref{eq:main-pinning} of the main text. The measurement checks are
\begin{align}
 \widehat{A}_q&=A_q,\qquad 0\le q<6,\\
 \widehat{A}_6&=A_6\prod_{i=1}^{3}X_{t_i},\\
 \widehat{C}_i&=C_i V(N_i,t_i),\\
 N_i&\quad\text{and}\quad T_i=T(N_i,t_i).
 \label{eq:prx271-steane-complete}
\end{align}
The identity $X_tV(N,t)=NV(N,t)X_t$ cancels each of the three raw commutators. The pinning check commutes with $X_t$ and with $V(N,t)$, as follows by resolving the two eigenspaces of $N$. Consequently, all 16 checks in Equation~\eqref{eq:prx271-steane-complete} commute globally. In the sector $N_i=+1$, each pin becomes $X_{t_i}$ and each controlled factor $V(N_i,t_i)$ becomes the identity. Thus, the common positive projector is exactly the raw positive projector tensored with $|+\rangle\!\langle+|^{\otimes3}$. The three completion qubits remove the three raw conflicts without changing the selected encoded subspace.

\paragraph{Extraction and qubit count.}
There are seven matter, six gauge and three completion qubits, for a total of 16 qubits retained between rounds. One readout ancilla per listed check gives 32 allocated qubits. Every check is a Hermitian Clifford operator. For example,
\begin{equation}
 V_q=X_q S_{g_q}^{\dagger}\mathrm{CY}_{g_q\to q},
 \label{eq:prx271-steane-V-circuit}
\end{equation}
with the rightmost gate applied first. Preparing a readout ancilla in $|+\rangle$, applying the check unitary controlled by that ancilla and measuring the ancilla in $X$ therefore gives an explicit native extraction circuit. The factors in Equation~\eqref{eq:prx271-steane-V-circuit} use CNOT, controlled $S^{\dagger}$ and doubly controlled $Y$ after adding the readout control. The Gauss checks use controlled $H$ and CNOT. The completion and pin factors use the same controlled-Pauli construction as before. Each elementary gate has hierarchy level at most three. For a fixed schedule, extract $N_i,T_i$ first, then $\widehat{A}_q$, then $\widehat{C}_i$, in increasing index order. Complete gadgets run sequentially, resetting each readout ancilla before its gadget. The star graph is used here for this seven-qubit instance.

\paragraph{Output magic state.}
Initialize the parent block in $|0\rangle_L$, perform the completed measurement and ungauge. For outcome $(-1)^{\eta}$, the normalized output is the corresponding logical Hadamard eigenstate,
\begin{align}
 |H_+\rangle&=\cos(\pi/8)|0\rangle+\sin(\pi/8)|1\rangle,\\
 |H_-\rangle&=\sin(\pi/8)|0\rangle-\cos(\pi/8)|1\rangle,
 \label{eq:prx271-H-states}\\
 p_\eta&=\frac{1+(-1)^{\eta}/\sqrt{2}}{2}.
 \label{eq:prx271-H-probability}
\end{align}
Indeed, $(I+(-1)^{\eta}H)|0\rangle/2$ has squared norm $p_\eta$ and the displayed normalized vector. Since $Y|H_-\rangle=i|H_+\rangle$, a logical $Y$ correction for $\eta=1$ gives the same magic state in both branches. This correction can be tracked as a Pauli frame. The state is Clifford equivalent to the standard $T$ resource because
\begin{equation}
 H S^{\dagger}H|H_+\rangle
 =e^{-i\pi/8}T|+\rangle,
 \qquad T=\operatorname{diag}(1,e^{i\pi/4}).
 \label{eq:prx271-H-to-T}
\end{equation}

\subsection{Consuming the logical Toffoli resource}
\label{sec:prx271-consumption}
\begin{figure*}[t]
\centering
\begin{tikzpicture}[x=1.35cm,y=.42cm,line width=.55pt,font=\small]
\foreach \y/\name in {5/q_1,4/q_2,3/q_3}{
 \draw (0,\y)--(7,\y);
 \node[left] at (0,\y) {$\name$};
}
\foreach \y/\name in {2/r_1,1/r_2,0/r_3}{
 \draw (0,\y)--(4.55,\y);
 \node[left] at (0,\y) {$\name$};
}
\draw[decorate,decoration={brace,amplitude=4pt}] (-.65,2.65)--(-.65,5.35)
 node[midway,left=6pt] {$|\psi\rangle$};
\draw[decorate,decoration={brace,amplitude=4pt}] (-.65,-.35)--(-.65,2.35)
 node[midway,left=6pt] {$|\mathrm{Tof}\rangle$};
\foreach \x/\c/\t in {1.6/5/2,2.6/4/1,3.6/0/3}{
 \draw (\x,\c)--(\x,\t);
 \fill (\x,\c) circle(2pt);
 \draw[fill=white] (\x,\t) circle(3.2pt);
 \draw (\x,\t-.22)--(\x,\t+.22);
 \draw (\x-.085,\t)--(\x+.085,\t);
}
\foreach \y/\basis/\bit in {2/Z/m_1,1/Z/m_2,0/X/m_3}{
 \node[draw,fill=white,minimum width=.7cm,minimum height=.34cm] at (4.7,\y) {$M_{\basis}$};
 \node[right] at (5.05,\y) {$\bit$};
}
\node[draw,fill=white,minimum width=.9cm,minimum height=1.23cm] at (6,4) {$C_m$};
\draw[decorate,decoration={brace,amplitude=4pt}] (7.2,5.35)--(7.2,2.65)
 node[midway,right=7pt] {$\mathrm{CCX}|\psi\rangle$};
\end{tikzpicture}
\caption{Ideal encoded circuit for consuming a logical Toffoli resource. Three logical CNOTs and the indicated Pauli measurements implement a Toffoli gate on arbitrary data after the outcome-dependent Clifford correction $C_m$ in Equation~\eqref{eq:prx271-consumption-correction}. The identity holds for data entangled with a reference. The resource can come from either the bilayer toric or CSS preparation, with its known source Pauli frame incorporated into $C_m$.}
\label{fig:prx271-consumption}
\end{figure*}
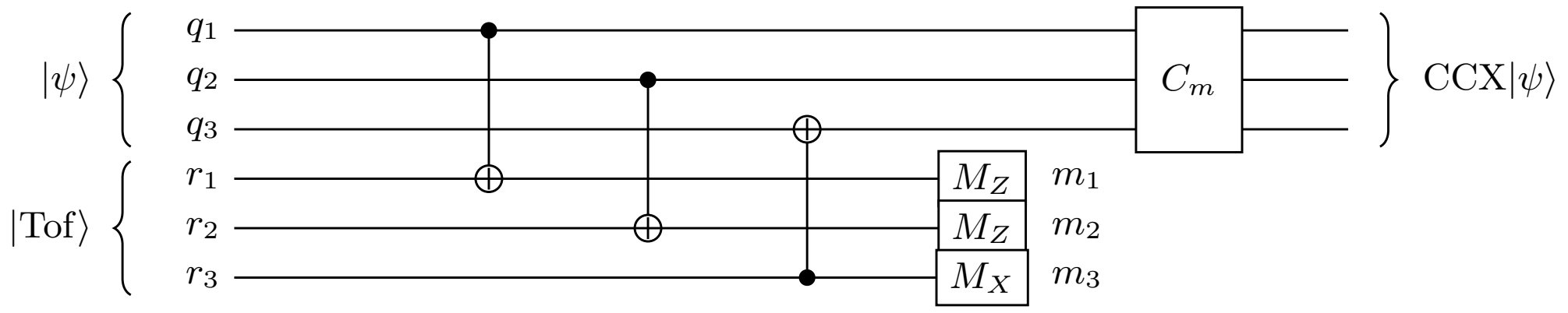

The toric protocol in Equation~\eqref{eq:main-d4-toffoli} of the main text and the CSS protocol in Eq.~\eqref{eq:prx273-css-projection} prepare the same resource after their recorded Pauli corrections. Its consumption uses gate teleportation~\cite{GottesmanChuang1999}, with the encoded circuit shown in Fig.~\ref{fig:prx271-consumption}. We give all branches explicitly to connect the prepared resource to a logical gate. Label the resource's logical qubits $r_1,r_2,r_3$, with $r_3$ the Toffoli target. The logical data qubits $q_1,q_2,q_3$ may be in any state, including one entangled with a reference. Apply
\begin{equation}
 \mathrm{CNOT}_{q_1\to r_1},\qquad
 \mathrm{CNOT}_{q_2\to r_2},\qquad
 \mathrm{CNOT}_{r_3\to q_3},
 \label{eq:prx271-consumption-cnot}
\end{equation}
and read $Z_{r_1},Z_{r_2},X_{r_3}$, with outcomes $(-1)^{m_1}$, $(-1)^{m_2}$ and $(-1)^{m_3}$. Define the conditional Clifford
\begin{align}
 C_m={}&X_{q_3}^{m_1m_2}
 Z_{q_1}^{m_2m_3}Z_{q_2}^{m_1m_3}\nonumber\\
 &\times\mathrm{CNOT}_{q_1\to q_3}^{m_2}
 \mathrm{CNOT}_{q_2\to q_3}^{m_1}
 \mathrm{CZ}_{q_1q_2}^{m_3}.
 \label{eq:prx271-consumption-correction}
\end{align}
The factors commute. Applying $C_m$ gives logical Toffoli on the data in every branch. Specifically, the uncorrected branch operator obeys
\begin{equation}
 C_m K_m=\frac{(-1)^{m_1m_2m_3}}{\sqrt{8}}
 \mathrm{CCX}_{q_1,q_2\to q_3}.
 \label{eq:prx271-consumption-branch}
\end{equation}
Thus, each of the eight outcomes has probability $1/8$, independent of the data state.

\begin{proof}
The two $Z$ readouts set the first two resource bits to $x\oplus m_1$ and $y\oplus m_2$ on a data basis state $|x,y,z\rangle$. The resource constraint fixes its third bit to their product. Its $X$ readout gives
\begin{equation}
 \begin{aligned}
 K_m|x,y,z\rangle={}&
 \frac{(-1)^{m_3(x\oplus m_1)(y\oplus m_2)}}{\sqrt{8}}\\
 &\times|x,y,z\oplus(x\oplus m_1)(y\oplus m_2)\rangle.
 \end{aligned}
 \label{eq:prx271-consumption-basis}
\end{equation}
Expanding the Boolean product as $xy\oplus m_2x\oplus m_1y\oplus m_1m_2$ shows that the target $X$ and two CNOT corrections remove the extra flips, while the CZ and two $Z$ corrections remove all data-dependent phases. The remaining phase is $(-1)^{m_1m_2m_3}$. Linearity proves the operator identity for arbitrary inputs and reference systems.
\end{proof}

The source Pauli correction need not be physically applied first. A known $X_{r_3}^{\eta}$ source byproduct, as in the toric protocol, replaces $C_m$ by $X_{q_3}^{\eta}C_m$. A known $Z_{r_1}^{\eta}$ source byproduct, as in the CSS protocol, replaces it by $Z_{q_1}^{\eta}C_m$. These statements follow by propagating the source Pauli through the three CNOTs and their readouts. The CNOT and CZ corrections in Equation~\eqref{eq:prx271-consumption-correction} must be performed or tracked in a Clifford frame.

For codes without a convenient encoded CNOT or CZ, the required controlled-Pauli gates have a measurement implementation. In the following gadget, $q$ and $t$ label the logical control and target qubits. For $P=X$ or $Z$, prepare a helper $h$ in $|+\rangle$ and measure $Z_qZ_h$, $X_hP_t$, then $Z_h$, with outcome bits $u,v,w$. The branch operator is
\begin{equation}
 2^{-3/2}(-1)^{v(u\oplus w)}
 Z_q^v P_t^{u\oplus w}\operatorname{ctrl}_q(P_t).
 \label{eq:prx271-CP-teleportation}
\end{equation}
This implements the controlled-Pauli gate using Pauli measurements and known Pauli corrections. Three mandatory and at most three conditional controlled-Pauli operations give at most twelve joint Pauli measurements per resource consumption, or nine on average. These counts exclude helper preparation and readout, the resource readouts, source preparation and source reset.

\subsection{Qutrit code: two distinct nonstabilizer resources}
\label{sec:prx271-qutrit-resources}
The $S_3$ construction gives a compatible charge-conjugation measurement on a qutrit toric parent, with logical action $\overline{K}_1\overline{K}_2$, where $K|j\rangle=|-j\rangle$ over $\mathbb{Z}_3$. Initialize the two logical qutrits in $|1,0\rangle$. Since $K|0\rangle=|0\rangle$,
\begin{align}
 \frac{I+(-1)^{\eta}\overline{K}_1\overline{K}_2}{2}|1,0\rangle
 &=\frac{1}{\sqrt{2}}|\psi_\eta\rangle|0\rangle,\\
 |\psi_\eta\rangle&=\frac{|1\rangle+(-1)^{\eta}|2\rangle}{\sqrt{2}}.
 \label{eq:prx271-qutrit-branches}
\end{align}
Each sign occurs with probability $1/2$. Both states are nonstabilizer, because a pure single-qutrit stabilizer state has either one or three nonzero computational-basis amplitudes. They are distinct resources, not branches related by a qutrit Pauli correction. A nontrivial power of $X$ changes the support $\{1,2\}$, and powers of $Z$ can change the relative phase only by a third root of unity, never by $-1$. If a subsequent protocol requires one chosen sign, postselection succeeds with probability $1/2$. Using both signs requires a consumption protocol appropriate to each resource. A compatible realization uses the controlled-check circuit on qubits and qutrits in Appendix~\ref{sec:sm-s3-ft-general}, whose bound is $d_{\mathrm{surg}}\ge\min\{R,\lceil L/4\rceil\}$ under its stated ideal outer boundaries and detection tests. Theorem~\ref{thm:prx-semidirect} supplies another compatible set of checks, with the native hierarchy obstruction in Eq.~\eqref{eq:prx-s3-diagnostic}.

\begin{table*}[t]
\caption{Explicit logical resources from compatible measurements. The outcome is $(-1)^{\eta}$ and $|\mathrm{Tof}\rangle=\mathrm{CCX}|++0\rangle$. All qubit preparations succeed for either ideal outcome after the indicated Pauli correction. The two qutrit branches are retained as distinct resources. The fault-distance claims concern the specified measurement circuits and their stated ideal outer boundaries.}
\label{tab:prx271-protocols}
\setlength{\tabcolsep}{4pt}
\renewcommand{\arraystretch}{1.25}
\begin{tabular}{@{}>{\raggedright\arraybackslash}p{0.19\textwidth}>{\raggedright\arraybackslash}p{0.25\textwidth}>{\raggedright\arraybackslash}p{0.30\textwidth}>{\raggedright\arraybackslash}p{0.20\textwidth}@{}}
\toprule
Parent and measurement & Encoded input & Resource and correction & Circuit result\\
\midrule
Two toric codes\newline $\overline{\mathrm{CZ}}_{12}\overline{\mathrm{CZ}}_{34}$
& $|++1+\rangle$
& $|\mathrm{Tof}\rangle_{1,2,4}$ after $\overline{X}_4^{\eta}$\newline $p_\eta=1/2$
& $d_{\mathrm{surg}}=\min(L,R)$ for the specified even-$L$ circuit.\\[4pt]
Two identical CSS blocks\newline $\prod_j\overline{\mathrm{CNOT}}_{A_j\to B_j}$
& $|+0\rangle_{A_1B_1}|10\rangle_{A_2B_2}$\newline Other pairs: $|00\rangle$
& $|\mathrm{Tof}\rangle_{B_2,A_1,B_1}$ after $Z_{B_2}^{\eta}$\newline $p_\eta=1/2$
& Explicit native circuits and resource counts.\\[4pt]
Steane code\newline $\overline{H}$
& $|0\rangle_L$
& $|H_+\rangle$ after $Y_L^{\eta}$\newline $p_\eta=(1+(-1)^{\eta}/\sqrt{2})/2$
& Explicit native circuits using 32 qubits including readout.\\[4pt]
Qutrit toric code\newline $\overline{K}_1\overline{K}_2$
& $|1,0\rangle_L$\newline $K|j\rangle=|-j\rangle$
& $(|1\rangle+(-1)^{\eta}|2\rangle)/\sqrt{2}$\newline $p_\eta=1/2$. Distinct qutrit resources.
& Linear distance for the specified circuits. Uses native qubit--qutrit gates outside the finite hierarchy.\\
\bottomrule
\end{tabular}
\end{table*}

\section{Hadamard-gauged color codes and a logical \texorpdfstring{$T$}{T} source}
\label{sec:v278-t-details}
\label{sec:sm278-t-source}
This appendix gives an alternative $T$ source by measuring logical Hadamard on a color-code parent. We choose the triangular $4.8.8$ family because each patch encodes one qubit and supports the transversal Clifford gates used below. We give explicit native circuits and prove a linear circuit-distance bound for the regular patches specified in Appendix~\ref{sec:prx278-t-geometric-proof}. Let a patch have distance $d$, odd $n$ matter vertices and face supports of weight $4$ or $8$. The parent stabilizer generators are $X(f),Z(f)$, and choose $X_L=X^{\otimes n}$ and $Z_L=Z^{\otimes n}$. The gauge graph is its connected edge graph $G=(V,E)$. Its bounded faces form the face set $\mathcal{P}$, and their boundaries span the binary cycle space of $G$. Thus, positive face fluxes select exact gauge strings on this planar patch. No outer-face measurement of growing weight is required.

\subsection{Logical preparation by Hadamard measurement}
\label{sec:v385-color-preparation}
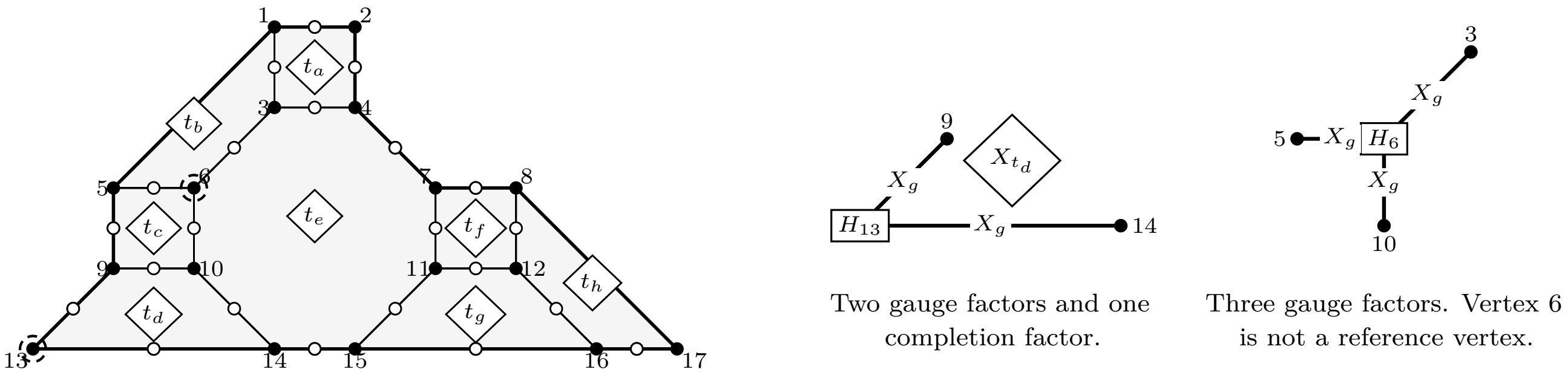
\begin{figure*}[t]
\centering
\begin{minipage}[t]{.52\textwidth}\centering
\textbf{(a)} A finite square--octagon patch\par\smallskip
\begin{tikzpicture}[x=.88cm,y=.88cm,font=\scriptsize,line width=.55pt,
 every node/.style={text=black},every path/.style={draw=black}]
\path[use as bounding box,draw=none](-.45,-.4)rectangle(8.45,4.5);
\foreach \i/\x/\y in {1/3/4,2/4/4,3/3/3,4/4/3,5/1/2,6/2/2,7/5/2,8/6/2,9/1/1,10/2/1,11/5/1,12/6/1,13/0/0,14/3/0,15/4/0,16/7/0,17/8/0}
 {\coordinate(v\i)at(\x,\y);}
\foreach \p in {(v1)--(v3)--(v4)--(v2)--cycle,
 (v1)--(v3)--(v6)--(v5)--cycle,
 (v5)--(v6)--(v10)--(v9)--cycle,
 (v13)--(v9)--(v10)--(v14)--cycle,
 (v14)--(v10)--(v6)--(v3)--(v4)--(v7)--(v11)--(v15)--cycle,
 (v8)--(v12)--(v11)--(v7)--cycle,
 (v15)--(v11)--(v12)--(v16)--cycle,
 (v17)--(v16)--(v12)--(v8)--cycle}
 {\path[fill=black!4] \p;}
\foreach \i/\j in {1/2,1/3,1/5,2/4,3/4,3/6,4/7,5/6,5/9,6/10,7/8,7/11,8/12,8/17,9/10,9/13,10/14,11/12,11/15,12/16,13/14,14/15,15/16,16/17}{
 \draw(v\i)--(v\j);
 \path(v\i)--node[circle,draw,fill=white,inner sep=0pt,minimum size=3.7pt]{}(v\j);}
\foreach \i/\j in {1/2,1/5,2/4,4/7,5/9,7/8,8/17,9/13,13/14,14/15,15/16,16/17}
 {\draw[line width=1.05pt](v\i)--(v\j);}
\foreach \i/\j in {1/2,1/5,2/4,4/7,5/9,7/8,8/17,9/13,13/14,14/15,15/16,16/17}
 {\path(v\i)--node[circle,draw,fill=white,inner sep=0pt,minimum size=3.7pt]{}(v\j);}
\foreach \i in {1,...,17}{\fill[black](v\i)circle(1.8pt);}
\foreach \i/\pos in {1/above left,2/above right,3/left,4/right,5/left,6/above right,7/above left,8/above right,9/left,10/right,11/left,12/right,13/below left,14/below,15/below,16/below,17/below right}
 {\node[\pos,inner sep=1.1pt]at(v\i){$\i$};}
\foreach \f/\x/\y in {a/3.5/3.5,b/2.0/2.8,c/1.5/1.5,d/1.5/.43,e/3.5/1.65,f/5.5/1.5,g/5.5/.43,h/6.95/.82}
 {\node[diamond,aspect=1.05,draw,fill=white,inner sep=1.2pt]at(\x,\y){$t_{\f}$};}
\draw[densely dashed,line width=.8pt](v6)circle(.16);
\draw[densely dashed,line width=.8pt](v13)circle(.16);
\end{tikzpicture}\par
{\footnotesize Filled dots: matter. Open circles: gauge.\par
Diamonds: completion. The heavy perimeter is physical.}
\end{minipage}\hfill
\begin{minipage}[t]{.225\textwidth}\centering
\textbf{(b)} Corner Gauss check\par\smallskip
\begin{tikzpicture}[x=.95cm,y=.95cm,font=\scriptsize,line width=.6pt,
 every node/.style={text=black},every path/.style={draw=black}]
\path[use as bounding box,draw=none](-.3,-.6)rectangle(3.3,2.75);
\coordinate(c)at(0,0);\coordinate(n)at(1,1);\coordinate(e)at(3,0);
\draw[line width=1.2pt](c)--(n) (c)--(e);
\node[draw,fill=white,inner sep=2pt]at(c){$H_{13}$};
\fill[black](n)circle(1.8pt);\node[above]at(n){$9$};
\fill[black](e)circle(1.8pt);\node[right]at(e){$14$};
\node[fill=white,inner sep=1pt]at(.5,.5){$X_g$};
\node[fill=white,inner sep=1pt]at(1.5,0){$X_g$};
\node[diamond,aspect=1.05,draw,fill=white,inner sep=2pt]at(1.75,.75){$X_{t_d}$};
\end{tikzpicture}\par
{\footnotesize Two gauge factors and one completion factor.}
\end{minipage}\hfill
\begin{minipage}[t]{.225\textwidth}\centering
\textbf{(c)} Bulk Gauss check\par\smallskip
\begin{tikzpicture}[x=.95cm,y=.95cm,font=\scriptsize,line width=.6pt,
 every node/.style={text=black},every path/.style={draw=black}]
\path[use as bounding box,draw=none](-1.55,-1.6)rectangle(1.55,1.75);
\draw[line width=1.2pt](-1,0)--(0,0)--(1,1) (0,0)--(0,-1);
\foreach \x/\y/\q/\pos in {-1/0/5/left,1/1/3/above,0/-1/10/below}
 {\fill[black](\x,\y)circle(1.8pt);\node[\pos]at(\x,\y){$\q$};}
\node[draw,fill=white,inner sep=2pt]at(0,0){$H_6$};
\foreach \x/\y in {-.5/0,.5/.5,0/-.5}
 {\node[fill=white,inner sep=1pt]at(\x,\y){$X_g$};}
\end{tikzpicture}\par
{\footnotesize Three gauge factors. Vertex $6$ is not a reference vertex.}
\end{minipage}
\caption{Physical boundaries for the Hadamard-gauged color-code source. (a) A $17$-vertex, $24$-edge, eight-face patch, drawn with the incidence of the standard triangular $4.8.8$ code. The octagon is face $e$ and the other faces are quadrilaterals. The eight diamonds are sufficient for the reference vertices specified in Appendix~\ref{sec:v279-color-boundary}. (b),(c) The complete operators $\widehat{A}_{13}$ and $\widehat{A}_6$ are supported only on the displayed matter, gauge and completion qubits. The bulk and corner checks belong to one globally commuting family. The square-lattice bilayer boundary is a separate realization.}
\label{fig:v279-color-boundary}
\end{figure*}

The transversal Hadamard implements logical Hadamard. Gauging it gives $D(D_4)$ bulk order because it exchanges the two layers in the toric-code correspondence~\cite{PremWilliamson2019Gauging}. Figure~\ref{fig:v279-color-boundary} uses a $[[17,1,5]]$ member of this family to display the physical boundary checks and the placement of completion qubits. Corner Gauss checks use two incident gauge qubits. All boundary-face checks and completion factors are retained and commute with the bulk checks (Appendix~\ref{sec:v279-color-boundary}). The general completion construction does not require a triangular shape.

Huang and Chen have proposed $D(D_4)$-assisted $T$-state preparation starting from a $\mathbb{Z}_4$ surface code and using charge-conjugation gauging followed by anyon condensation~\cite{HuangChen2026Magic}. Here, the logical resource comes from a completed Hadamard measurement on a qubit color code. This differs from converting the preceding bilayer Toffoli resource. No finite number of Toffoli or CCZ states can yield an exact $T$ state using only stabilizer operations, even with postselection, by Theorem~2.8 of Ref.~\cite{Beverland2020LowerBounds}.

\begin{figure*}[t]
\centering
\begin{minipage}[t]{.47\textwidth}\centering
\textbf{(a)} Gauss-check extraction\par\smallskip
\begin{tikzpicture}[x=.68cm,y=.48cm,line width=.55pt,font=\scriptsize]
\foreach \y/\name in {3/{a:\ |+\rangle},2/{v},1/{g_e},0/{t_f}}{
 \draw(0,\y)--(7.1,\y);
 \node[left]at(0,\y){$\name$};}
\draw(1.1,3)--(1.1,2);\fill(1.1,3)circle(1.8pt);
\node[draw,fill=white,minimum width=.38cm,minimum height=.28cm]at(1.1,2){$H$};
\foreach \x/\y in {2.8/1,4.7/0}{
 \draw(\x,3)--(\x,\y);\fill(\x,3)circle(1.8pt);
 \draw[fill=white](\x,\y)circle(2.6pt);
 \draw(\x-.09,\y)--(\x+.09,\y);
 \draw(\x,\y-.14)--(\x,\y+.14);}
\node[below]at(2.8,.75){each $e\ni v$};
\node[below]at(4.7,-.25){each $r_f=v$};
\node[draw,fill=white,minimum width=.46cm,minimum height=.28cm]at(6.4,3){$M_X$};
\end{tikzpicture}
\end{minipage}\hfill
\begin{minipage}[t]{.50\textwidth}\centering
\textbf{(b)} One factor of a modified parent check\par\smallskip
\begin{tikzpicture}[x=.70cm,y=.48cm,line width=.55pt,font=\scriptsize]
\foreach \y/\name in {2/{a},1/{p},0/{v_j}}{
 \draw(0,\y)--(7.1,\y);\node[left]at(0,\y){$\name$};}
\draw(1.3,2)--(1.3,0);\fill(1.3,2)circle(1.8pt);\fill(1.3,1)circle(1.8pt);
\node[draw,fill=white,minimum width=.38cm,minimum height=.28cm]at(1.3,0){$Y$};
\draw(3.3,2)--(3.3,1);\fill(3.3,2)circle(1.8pt);
\node[draw,fill=white,minimum width=.48cm,minimum height=.28cm]at(3.3,1){$S^{\dagger}$};
\draw(5.3,2)--(5.3,0);\fill(5.3,2)circle(1.8pt);
\draw[fill=white](5.3,0)circle(2.6pt);
\draw(5.21,0)--(5.39,0);\draw(5.3,-.14)--(5.3,.14);
\node[below,align=center]at(3.5,-.42){$p=\bigoplus_{e\in p_f(v_j)}z_e$\quad (coherent work value)};
\end{tikzpicture}
\end{minipage}
\caption{Native circuits for the Hadamard-gauged color-code source. (a) The completion orientation gives $\widehat{A}_v=H_v\prod_{e\ni v}X_{g_e}\prod_{f:r_f=v}X_{t_f}$. Each displayed repeated CNOT has the same readout control. (b) Gates run from left to right. The circuit implements $\operatorname{ctrl}_a(V_{p,v_j})$, where $V_{p,v}=|0\rangle\langle0|_pX_v+|1\rangle\langle1|_pZ_v$. The doubly controlled $Y$ is $S_v\operatorname{Toffoli}S_v^{\dagger}$. Appendix~\ref{sec:prx279-general-color-checks} gives the parity accumulation, uncomputation and every completion and pin factor.}
\label{fig:prx278-t-circuits}
\end{figure*}
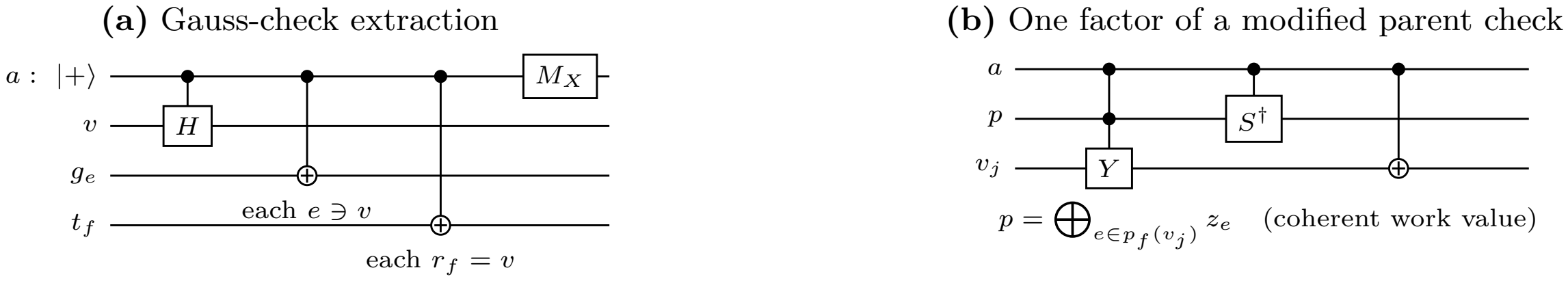

Use the color-code edge graph for the gauge qubits. On each face, number the vertices cyclically as $v_0,\ldots,v_{w-1}$, take $r_f=v_0$, and let $p_f(v_j)$ follow the first $j$ boundary edges. For a computational-basis gauge string $z=(z_e)_{e\in E}$, put $a_{f,j}=\sum_{e\in p_f(v_j)}z_e\bmod2$. The raw checks are
\begin{align}
 A_v&=H_v\prod_{e\ni v}X_{g_e},
 \label{eq:prx278-color-gauss}\\
 C_f&=\sum_z|z\rangle\langle z|_g
       \prod_{j=0}^{w-1}H_{v_j}^{a_{f,j}}X_{v_j}H_{v_j}^{a_{f,j}}.
 \label{eq:prx278-color-raw}
\end{align}
The central family consists of face fluxes $B_f=Z_g(\partial f)$ and the parent $Y(f)$ checks dressed by gauge $Z$ operators
\begin{equation}
 N_f=Y(f)Z_g(\xi_f),\qquad
 \xi_f=e_0+e_2+\cdots+e_{w-2}.
 \label{eq:prx278-color-invariant}
\end{equation}
Here, $e_j=(v_j,v_{j+1})$, and edge sums are binary. The conflicts between Gauss and face checks are labeled by $N_f$. Conflicts between neighboring face checks are labeled by products of face fluxes. All these labels are central Pauli checks, so the completion construction applies. Appendix~\ref{sec:prx279-general-color-checks} derives their exact commutators and gives the completed checks.

Each Gauss readout uses one controlled-Hadamard and CNOTs to its incident gauge and completion qubits. A face circuit accumulates the successive path parities in the clean work qubit $p$ using CNOTs from the path gauge qubits to $p$ and applies the three-gate controlled factor in Fig.~\ref{fig:prx278-t-circuits}. It then uncomputes the parity and applies the completion factors. Central and pinning extraction use the controlled-Pauli circuits in Appendix~\ref{sec:prx279-general-color-checks}. The native library consists of Clifford gates, controlled $H$, controlled $S^{\dagger}$, Toffoli and CCZ, all of hierarchy level at most three. Complete overlapping gadgets are serialized, and disjoint gadgets run in parallel. There are $n$ controlled-Hadamard gates per gauged round. All qubit and gate counts are linear in the patch size.

Proposition~\ref{prop:prx278-t-distance} proves a circuit-distance lower bound proportional to $d$ when $R=\Theta(d)$, under its stated ideal boundary conditions and record tests.

Initialize the parent in $|0\rangle_L$ and follow Protocol~\ref{alg:commuting-gauging}, using the face-measurement circuits above with their additional work qubits for storing path parities. The product of the Gauss outcomes is $(-1)^{\eta}$, giving
\begin{align}
 |H_+\rangle_L&=\cos(\pi/8)|0\rangle_L+\sin(\pi/8)|1\rangle_L,
 \label{eq:prx278-color-Hplus}\\
 |H_-\rangle_L&=\sin(\pi/8)|0\rangle_L-\cos(\pi/8)|1\rangle_L,
 \label{eq:prx278-color-Hminus}\\
 p_\eta&=\frac{1+(-1)^{\eta}/\sqrt{2}}{2}.
 \label{eq:prx278-color-probability}
\end{align}
Track $Y_L^{\eta}$ in the Pauli frame. Applying the logical Clifford $H_LS_L^{\dagger}H_L$ then yields
\begin{equation}
 H_LS_L^{\dagger}H_L|H_+\rangle_L
 =e^{-i\pi/8}\frac{|0\rangle_L+e^{i\pi/4}|1\rangle_L}{\sqrt{2}}.
 \label{eq:prx278-color-T}
\end{equation}
The tracked correction becomes $Z_L^{\eta}$. The final Clifford has a depth-three transversal implementation on this family and precedes the ideal final parent projection. Both branches therefore prepare the same $T$ resource from a stabilizer input, using non-Clifford controlled-Hadamard extraction gates.

\subsection{An explicit color-code boundary}
\label{sec:v279-color-boundary}
The Hadamard-gauged $T$ source uses the square--octagon color-code cellulation, rather than the three-edge-qubit square-lattice realization in Figure~\ref{fig:d4-completed-checks} of the main text. Figure~\ref{fig:v279-color-boundary} shows a finite $[[17,1,5]]$ patch of the triangular $4.8.8$ family~\cite{LandahlRyanAnderson2014}. Matter qubits occupy its vertices and gauge qubits occupy its actual edges. The four-vertex boundary faces are retained as complete face checks. At a corner the Gauss check contains the two incident gauge qubits. No check contains an edge outside the patch.

One explicit choice of cyclic face orders is
\begin{equation}
\begin{aligned}
 a&=(1,3,4,2),& b&=(1,3,6,5),\\
 c&=(5,6,10,9),& d&=(13,9,10,14),\\
 e&=(14,10,6,3,4,7,11,15),\\
 f&=(8,12,11,7),&g&=(15,11,12,16),\\
 h&=(17,16,12,8).
\end{aligned}
\label{eq:v279-color-face-cycles}
\end{equation}
The first vertex is the reference vertex and the last-to-first edge is the omitted closing edge. In this small patch every closing edge can be placed on the physical perimeter. Consequently, the face--face commutators vanish and only one completion qubit $t_f$ per face is needed. For these reference vertices,
\begin{align}
 \widehat{A}_{13}&=H_{13}X_{g_{13,9}}X_{g_{13,14}}X_{t_d},
 \label{eq:v279-color-corner-A}\\
 \widehat{A}_6&=H_6X_{g_{6,3}}X_{g_{6,5}}X_{g_{6,10}},
 \label{eq:v279-color-bulk-A}\\
 \widehat{C}_d&=C_dV(N_d,t_d),
 \label{eq:v279-color-boundary-C}\\
 N_d&=Y_{13}Y_9Y_{10}Y_{14}Z_{g_{13,9}}Z_{g_{10,14}}.
 \label{eq:v279-color-boundary-N}
\end{align}
Here, $g_{u,v}$ denotes the gauge qubit on edge $(u,v)$. The raw $C_d$ uses the three successive path parities along $13\to9\to10\to14$. Its completion factor is explicitly
\begin{equation}
 V(N_d,t_d)=\operatorname{CZ}_{t_d,g_{13,9}}
 \operatorname{CZ}_{t_d,g_{10,14}}
 \prod_{v\in\{13,9,10,14\}}\operatorname{CY}_{t_d,v}.
 \label{eq:v279-color-boundary-V}
\end{equation}
Also measure $B_d=Z_{g_{13,9}}Z_{g_{9,10}}Z_{g_{10,14}}Z_{g_{14,13}}$, $N_d$ and the pin $T_d=P_-(N_d)+P_+(N_d)X_{t_d}$, and the corresponding operators on every other face.

The boundary changes neither the commutation proof nor its hypotheses. Each retained face is a closed even cycle, so its gauge flux commutes with every Gauss check, including a corner check. Its alternating edge mask meets each face vertex once, so the gauge sign cancels the sign from $H_vY_vH_v=-Y_v$ in $N_f$. Distinct retained faces overlap on zero or two matter vertices. Thus, the $N_f$ remain central. The relation at the reference vertex is still $C_fA_{r_f}=N_fA_{r_f}C_f$, canceled by $X_{t_f}V(N_f,t_f)=N_fV(N_f,t_f)X_{t_f}$. These identities prove that every displayed boundary check commutes with every other boundary or bulk check, on the full Hilbert space. On larger patches an omitted closing edge may be shared by two faces, whose checks can then fail to commute. Add the completion qubits of Equations~\eqref{eq:sm278-color-face-conflict}--\eqref{eq:sm278-color-secondary-pin} for these face--face conflicts. Their supports remain bounded near the two faces.

\subsection{General checks and extraction}
\label{sec:prx279-general-color-checks}
\paragraph{Raw checks and all commutators.}
For each face $f$, choose a cyclic order $v_0,\ldots,v_{w_f-1}$ with reference vertex $r_f=v_0$. Write $e_j=(v_j,v_{j+1})$, with indices modulo $w_f$. The paths are
\begin{equation}
 p_f(v_j)=e_0+\cdots+e_{j-1},\qquad p_f(v_0)=0.
 \label{eq:sm278-color-paths}
\end{equation}
Only the closing edge $e_{w_f-1}$ is omitted. For a gauge basis string $z$, let $a_{f,j}=p_f(v_j)\cdot z\bmod2$, and define
\begin{align}
 A_v&=H_v\prod_{e\ni v}X_{g_e},
 \label{eq:sm278-color-A}\\
 C_f&=\sum_z|z\rangle\langle z|_g
        \prod_{j=0}^{w_f-1}H_{v_j}^{a_{f,j}}X_{v_j}H_{v_j}^{a_{f,j}},
 \label{eq:sm278-color-C}\\
 B_f&=Z_g(\partial f),
 \label{eq:sm278-color-B}\\
 N_f&=Y(f)Z_g(\xi_f),\qquad
 \xi_f=\sum_jp_f(v_j)=\sum_{j\,\mathrm{even}}e_j.
 \label{eq:sm278-color-N}
\end{align}
The modified parent $Z(f)$ check is $N_fC_f$, so retaining $C_f,N_f$ retains both parent constraints. This uses $w_f=0\bmod4$.

The Gauss checks commute with one another. For $v\ne r_f$, their flips change just the path parity at $v$ when $v\in f$. Conjugation by $H_v$ cancels that change. At the reference vertex, all other face path parities flip and the matter factors exchange $X$ and $Z$, giving
\begin{equation}
 C_fA_v=N_f^{\delta_{v,r_f}}A_vC_f.
 \label{eq:sm278-color-root-conflict}
\end{equation}
For two faces, the local $X$ or $Z$ factors anticommute precisely where their path parities differ. Consequently,
\begin{align}
 C_fC_{f'}&=K_{ff'}C_{f'}C_f,
 \label{eq:sm278-color-face-conflict}\\
 K_{ff'}&=Z_g(\gamma_{ff'}),\qquad
 \gamma_{ff'}=\sum_{v\in f\cap f'}[p_f(v)+p_{f'}(v)].
 \label{eq:sm278-color-gamma}
\end{align}
Distinct faces either are disjoint or share one edge. For a shared edge $e=(u,v)$, put $\epsilon_f(e)=1$ if $e$ is the omitted closing edge of $f$, and zero otherwise. The identity $p_f(u)+p_f(v)=e+\epsilon_f(e)\partial f$ yields
\begin{equation}
 K_{ff'}=B_f^{\epsilon_f(e)}B_{f'}^{\epsilon_{f'}(e)}.
 \label{eq:sm278-color-local-K}
\end{equation}
Every $B_f$ commutes with every raw check. Each $N_f$ commutes with each $C_{f'}$ because their matter overlap has even size. Its sign under the matter Hadamard in $A_v$ is canceled by the incident gauge-$Z$ sign. The alternating edges in $\xi_f$ meet each vertex of $f$ exactly once. Thus, all $N_f$ are central as well. Equations~\eqref{eq:sm278-color-root-conflict} and~\eqref{eq:sm278-color-local-K} exhaust the raw conflicts.

\paragraph{Completed checks and qubit placement.}
Give each conflict at a reference vertex a completion qubit $t_f$ placed near $r_f$ and orient it from $A_{r_f}$ to $C_f$. Let $\mathcal{E}_{\rm s}$ be the set of neighboring-face pairs with $K_{ff'}\ne I$. Order the faces and orient each pair from the earlier face to the later face. Put one qubit $t_{ff'}$ near their shared edge. For an oriented face--face conflict $e:f\to f'$, write $K_e=K_{ff'}$ and $t_e=t_{ff'}$. With $V(K,t)=|0\rangle\langle0|_t+|1\rangle\langle1|_tK$, the checks are
\begin{align}
 \widehat{A}_v&=A_v\prod_{f:r_f=v}X_{t_f},
 \label{eq:sm278-color-Ahat}\\
 \widehat{C}_f&=C_fV(N_f,t_f)
   \prod_{e:f\to f'}X_{t_e}
   \prod_{e:f'\to f}V(K_e,t_e),
 \label{eq:sm278-color-Chat}\\
 T_f&=\frac{I-N_f}{2}+\frac{I+N_f}{2}X_{t_f},
 \label{eq:sm278-color-primary-pin}\\
 T_e&=\frac{I-K_e}{2}+\frac{I+K_e}{2}X_{t_e}
       \quad(e\in\mathcal{E}_{\rm s}).
 \label{eq:sm278-color-secondary-pin}
\end{align}
Measure these checks, every $N_f$, and every $B_f$. The central completion theorem proves global commutation. It identifies the common positive code space with the raw code space and the completion qubits fixed in $|+\rangle$ states. It also preserves every encoded measurement branch and the static distance of the raw gauged code.

\paragraph{Native extraction of a face.}
Use one readout $a$ initialized in $|+\rangle$ and one private work qubit $p$ reset to $|0\rangle$. The work qubit is separate from the readout and completion qubits. First apply $\operatorname{CNOT}_{a\to v_0}$. For $j=1,\ldots,w_f-1$, apply $\operatorname{CNOT}_{g_{e_{j-1}}\to p}$ and then the gates, in time order,
\begin{equation}
 \operatorname{CC}Y_{a,p\to v_j},\qquad
 \operatorname{CS}^{\dagger}_{a,p},\qquad
 \operatorname{CNOT}_{a\to v_j}.
 \label{eq:sm278-color-transport-circuit}
\end{equation}
Here, $S=\operatorname{diag}(1,i)$ and $\operatorname{CC}Y=S_{v_j}\operatorname{Toffoli}_{a,p\to v_j}S_{v_j}^{\dagger}$, with the rightmost factor applied first. The work value is the prefix parity $a_{f,j}$ when this factor is applied. The identity
\begin{equation}
 |0\rangle\langle0|_pX_v+|1\rangle\langle1|_pZ_v
 =X_vS_p^{\dagger}\operatorname{CY}_{p\to v}
 \label{eq:sm278-color-V-factor}
\end{equation}
proves the controlled factor in Equation~\eqref{eq:sm278-color-transport-circuit}. Finally, undo the CNOTs $\operatorname{CNOT}_{g_{e_{j-1}}\to p}$ in reverse edge order. This returns $p$ to $|0\rangle$ for every input on the matter, gauge and readout qubits. No parity is measured. The resulting circuit is exactly $\operatorname{ctrl}_a(C_f)$. It uses $3w_f-2$ CNOTs, $w_f-1$ Toffolis, $w_f-1$ controlled $S^{\dagger}$ gates and $2(w_f-1)$ single-qubit $S^{\pm1}$ gates.

Append the controlled completion factors before measuring $a$ in $X$. An outgoing $X_t$ factor uses one CNOT. For a central Pauli product $K=\prod_{q\in\operatorname{supp}K}P_q$, an incoming factor uses
\begin{equation}
 \operatorname{ctrl}_a(V(K,t))
   =\prod_q\operatorname{CC}P_{a,t;q}.
 \label{eq:sm278-color-incoming-circuit}
\end{equation}
All Pauli products $K$ used here have overall phase $+1$. Their factors are $Y$ or $Z$, so the right side uses $\operatorname{CC}Y$ or CCZ. The Gauss circuit has one controlled $H_v$, one CNOT to each incident gauge qubit and one CNOT to each completion qubit $t_f$ with reference vertex $r_f=v$. These factors commute within the ideal Gauss check.

\paragraph{Central checks, pins and scheduling.}
A readout in $|+\rangle$ controls each Pauli factor of a central check and is measured in $X$. Controlled $Y$ is implemented as $S_q\operatorname{CNOT}_{a\to q}S_q^{\dagger}$, and controlled $Z$ is CZ. For a pin $T(K,t)$, its readout controls the following exact factorization:
\begin{equation}
 \operatorname{ctrl}_a(T(K,t))
 =H_t\operatorname{CZ}_{a,t}
    \left(\prod_q\operatorname{CC}P_{a,t;q}\right)H_t.
 \label{eq:sm278-color-pin-circuit}
\end{equation}
The unconditional Hadamards on $t$ cancel in the $a=0$ branch. In the $a=1$ branch they conjugate $\operatorname{ctrl}_t(-K)$ to the required pin. The elementary operations are therefore explicit and have hierarchy level at most three. In particular, controlled $H$ is an additional native resource compared with the bilayer-CZ circuit.

Extract central checks and pins first, then Gauss checks, and then completed face checks. Within each stage, partition the complete gadgets into groups that do not share code qubits. Run one group at a time, with its gadgets in parallel. Finish every gadget in its stated gate order before starting another gadget on any of the same code qubits. Face size, vertex degree and the number of neighboring faces are uniformly bounded, so this gives a constant-depth gauged round for the local native library. Reset readouts and work qubits between gadgets. Completion qubits remain part of the code through the entire gauged interval. Faults may act on any participants of any native gate, including these work qubits.

\paragraph{Qubit and gate counts.}
Write $F=|\mathcal{P}|$, $c_{\rm s}=|\mathcal{E}_{\rm s}|$, $c=F+c_{\rm s}$, $W=\sum_fw_f$, and $K_{\Sigma}=\sum_{e\in\mathcal{E}_{\rm s}}\operatorname{wt}(K_e)$. Each face omits one edge, so $c_{\rm s}\le F$. A dedicated readout for every measured check and one work qubit for every face give
\begin{align}
 n_{\rm persistent}&=n+|E|+F+c_{\rm s},
 \label{eq:sm278-color-persistent}\\
 n_{\rm readout}&=n+4F+c_{\rm s},
 \label{eq:sm278-color-readouts}\\
 n_{\rm work}&=F,
 \label{eq:sm278-color-work}\\
 n_{\rm total}&=2n+|E|+6F+2c_{\rm s}.
 \label{eq:sm278-color-total}
\end{align}
Work qubits can be reused when gadgets are serialized. For the displayed circuits, one gauged round uses
\begin{align}
 N_{\rm CH}&=n,
 \label{eq:sm278-color-count-CH}\\
 N_{\rm CNOT}&=2|E|+4W-F+c_{\rm s},
 \label{eq:sm278-color-count-CNOT}\\
 N_{\rm Toffoli}&=3W-F,
 \label{eq:sm278-color-count-Tof}\\
 N_{\rm CS^{\dagger}}&=W-F,
 \label{eq:sm278-color-count-CS}\\
 N_{\rm CCZ}&=W+2K_{\Sigma},
 \label{eq:sm278-color-count-CCZ}\\
 N_{\rm CZ}&=3W/2+F+c_{\rm s},
 \label{eq:sm278-color-count-CZ}\\
 N_{S^{\pm1}}&=8W-2F,
 \label{eq:sm278-color-count-S}\\
 N_H&=2(F+c_{\rm s}).
 \label{eq:sm278-color-count-H}
\end{align}
There are $n_{\rm readout}$ readout preparations and measurements, and $F$ work resets per round. The counts treat $X$-basis preparation and readout as native, expand each controlled $Y$ into the stated Clifford gates and each doubly controlled $Y$ into Toffoli and phase gates, and exclude idles, routing and synthesis of CH or controlled $S^{\dagger}$ into another hardware gate set. They also exclude initial parent-state preparation and the final transversal Clifford.

\paragraph{The output branch and its final Clifford.}
For ideal completed gauging and ungauging on an encoded parent input, the branch with logical measurement outcome $(-1)^{\eta}$ is
\begin{equation}
 \rho\longmapsto\Pi_\eta\rho\Pi_\eta,\qquad
 \Pi_\eta=\frac{I+(-1)^{\eta}H_L}{2}.
 \label{eq:sm278-color-instrument}
\end{equation}
Equations~\eqref{eq:prx278-color-Hplus}--\eqref{eq:prx278-color-T} follow by applying this projector to $|0\rangle_L$. The $\eta=1$ correction is $Y_L$, since $Y_L|H_-\rangle_L=i|H_+\rangle_L$. For the chosen logical Paulis, define $\epsilon=(-1)^{(n-1)/2}$. Because every face weight is divisible by four, $S^{\otimes n}$ preserves the code and acts as $S_L^{\epsilon}$ up to global phase. The final Clifford therefore has the transversal circuit
\begin{equation}
 H_LS_L^{\dagger}H_L
 \ \widehat{=}\
 H^{\otimes n}(S^{-\epsilon})^{\otimes n}H^{\otimes n}.
 \label{eq:sm278-color-output-Clifford}
\end{equation}
Here, $\widehat{=}$ denotes equality on the encoded subspace up to global phase. It uses $2n$ single-qubit Hadamards and $n$ single-qubit phase gates, with no enlargement of a fault's matter support. The Pauli correction can remain in the frame, which becomes $Z_L^{\eta}$ after the conversion because $H_LS_L^{\dagger}H_LY_LH_LS_LH_L=-Z_L$. Apply the transversal conversion before the final ideal parent-code projection, so its native faults are included in the preparation's output-boundary analysis.

\subsection{A geometric bound for the planar surgery}
\label{sec:prx278-t-geometric-proof}
\label{sec:sm278-t-distance}
The Hadamard construction admits a geometric bound without an expander graph. Use regular triangular patches with uniformly bounded face diameter and vertex density. Each patch domain is obtained from a fixed triangle of linear size $L$ by a deformation that changes distances by at most a fixed factor in either direction, independent of $L$. This excludes holes and narrow necks and includes the usual undistorted family. Measure every bounded face flux. Parent generators and extraction gadgets have bounded diameter independent of $d$, and $d=\Theta(L)$.

We first record the geometric fact used at the two surgery interfaces. A collar is a fixed-width strip around a region. Fix a collar width larger than both the mesh diameter and the parent interaction range. There is a constant $c_{\mathrm{box}}$ such that the $f$ bounded-diameter fault neighborhoods obtained after commuting-check propagation can be enclosed by regions $W_1,\ldots,W_m$ with the following properties whenever $c_{\mathrm{box}}f$ is smaller than the patch linear size:
\begin{enumerate}
\item Each $W_j$ lies in a square of side at most $c_{\mathrm{box}}f$.
\item No parent stabilizer generator meets two distinct $W_j$.
\item The vertex set outside $W=\bigcup_jW_j$ is nonempty, and every vertex in it can be joined to every other such vertex by a graph path that avoids the original fault neighborhoods, including the endpoints of the affected gauge edges.
\end{enumerate}
The path in the third property may enter the collars within $W$. Only avoidance of these fault neighborhoods is needed.

Here, we give a construction and its size bound. Begin with constant-size rectangles around the fault neighborhoods in fixed coarse coordinates. Merge rectangles whenever their three-collar enlargements meet. The sum of their widths and heights increases by only a constant per merge, so its total is $O(f)$. When a rectangle approaches two sides of the triangle and cuts off their corner, enlarge it to include that corner and merge again. Fixed boundary angles bound this enlargement by a constant factor. Once a rectangle contains a corner, every later rectangle formed by merging it with others still contains that corner. Each initial rectangle's size can therefore acquire at most three such factors, one for each corner. Including the constant cost of each merge still gives total width and height $O(f)$, even if many rectangles successively reach the same corner. Regions that come within one collar of the boundary are enlarged to meet it.

Let $R_j$ be the resulting inner rectangles and let $W_j$ include their two-collar enlargements. Three-collar separation ensures the second property. For a sufficiently small constant in $f=O(d)$, no rectangle spans the patch. The complement of the one-collar enlargements is connected. It is a triangle with separated interior holes and boundary caps, with any isolated corner caps already filled. A polygonal path in that complement can be moved onto boundaries of the mesh faces it crosses. The collar exceeds the face diameter, so the resulting graph path avoids the inner rectangles. Every vertex outside $W$ can use such a path. This proves the third property without assuming that the graph complement of $W$ itself is connected. The allowed deformation only changes the constants.

For a two-dimensional local stabilizer code, the correctability argument based on boundary size gives a constant $c_{\mathrm{cor}}>0$ such that every square of side less than $c_{\mathrm{cor}}d$ is a correctable erasure region~\cite{BravyiPoulinTerhal2010}. This statement also applies at the physical boundary by padding the outside with fixed product-state qubits. One can obtain it by growing a small correctable square by a full collar containing its inner and outer interaction-range boundaries. This collar contains only a number of qubits proportional to the current side length. While that number is below $d$, the correctable-region expansion lemma allows the square to grow by another interaction-range layer.

Consequently, every $W_j$ above is correctable for $f<cd$, for a constant $c>0$ fixed by the geometry and the gadgets. Their union is correctable as well. To see this directly for a stabilizer parent, expand an operator on $W$ in Paulis. If a Pauli term has a nonzero syndrome then its parent-code compression vanishes. Otherwise its restriction to each $W_j$ separately has zero syndrome, since no stabilizer generator meets two of the regions. Each restriction acts as a scalar on the code by correctability of $W_j$. Their product therefore also acts as a scalar.

\begin{proposition}[Planar Hadamard-surgery distance]
\label{prop:prx278-t-distance}
Consider the completed Hadamard gauging on the regular simply connected triangular patches specified above, using bounded-diameter native gadgets, the extraction schedule in Appendix~\ref{sec:prx279-general-color-checks}, reset readout and work qubits, and $R$ gauged rounds. Assume an ideally encoded input and an ideal final parent-code projection. The detection tests require positive outcomes for the parent, central, flux, modified-parent and pin checks, the same Gauss record in every gauged round, and an exact final gauge record. There is a constant $c>0$, independent of $d$ and $R$, such that
\begin{equation}
 d_{\mathrm{surg}}\geq\min\{R,\lceil cd\rceil\}.
 \label{eq:prx278-t-distance}
\end{equation}
The bound allows arbitrary Kraus faults at native gates, preparations, readouts and idles, including faults on completion and work qubits. It also permits the two added CNOTs that use a completion qubit as a flag, provided they are included in the bounded gadget neighborhoods.
\end{proposition}
\begin{proof}
Fix a measurement record satisfying these detection tests and a corresponding noise Kraus branch containing $f<\min\{R,cd\}$ faults. Some complete gauged round is ideal. The commuting-check propagation reduction places the residual operators on the two sides of this round in the union of $f$ bounded-diameter neighborhoods. Expanding these residual operators in Paulis gives matter factors $M$ and gauge flips $\eta$, as in the boundary transfer identities of the general surgery theorem.

The incoming flip pattern passes every face-flux check. Since the gauging cellulation is a disk, it is exact, $\eta=\delta y$. The exact final gauge-record test gives the corresponding outgoing exactness. Apply the geometric cover above to the matter supports and the endpoints of the flipped gauge edges. The potential $y$ is constant along every graph path avoiding those edge endpoints. It therefore has one common value on all vertices outside $W$. Replacing $y$ by $y+\mathbf{1}$ if necessary sets that value to zero. This replacement multiplies the boundary operator by the measured global Hadamard symmetry, which is the scalar reported sign on the selected branch.

Write $H(u)=\prod_{v\in V}H_v^{u_v}$ for a binary vertex string $u$, and let $x$ be the vertex potential of the final exact gauge record $z=\delta x$. For these representatives, the input boundary operator $H(y)M$ is supported within $W$. The output boundary operator has the form $H(x)^{\dagger}M H(x)H(y)$ and has the same support bound because $H(x)$ acts separately on the matter qubits. The correctability argument above makes both parent-code compressions scalar. Substitution into the two transfer identities proves
\begin{equation}
 P K_{F,m,\alpha}P
 =c_{F,m,\alpha}\frac{I+s\mathcal{U}}{2}P
 \label{eq:prx278-t-branch-distance}
\end{equation}
for every residual Pauli term and hence for their coherent sum. This is the required logical-measurement condition.
\end{proof}

The final logical Clifford converting the Hadamard eigenstate to the standard $T$ state can be implemented transversally on the doubly-even color-code parent and is applied before the final ideal parent-code projection. For the circuit including this conversion, the intended branch operator before the reported Pauli-frame correction is $H_LS_L^{\dagger}H_L\Pi_sP$. The same fault-distance lower bound holds relative to this converted instrument: its one-qubit native faults only change the constant-size neighborhoods. The reported branch correction can be retained in a Pauli frame.

\section{Recovery of the logical measurement across a faulty temporal boundary}
\label{sec:sm297-boundary-recovery}
Completion also supports correction of detected boundary errors. We construct a coherent repair for green-$X$ errors at the input to gauging, with all later operations ideal, and compare it with a repair using controlled checks and additional matter measurements. We use the bilayer symmetry $\mathcal{U}$ of Eq.~\eqref{eq:smd4-U} and write $\Pi_s(\mathcal{U})=(I+s\mathcal{U})/2$. In this appendix, $P$ abbreviates the parent-code projector $P_{\rm par}$.

\subsection{Coherent recovery using the completion qubits}
\label{sec:sm297-coherent-recovery}
Use the even-$L$ bilayer torus and completed checks of Section~\ref{sec:prx-d4}. A negative green flux releases its completion qubit from the pinning constraint, allowing it to retain the coherent gauge label used by the local Clifford repair. Keeping this register through the repair preserves coherence between the stored labels in Eq.~\eqref{eq:prx412-recovery-state}. The repair acts before ungauging.

\paragraph{Single-edge identity used in the main text.}
Section~\ref{sec:prx-d4-recovery} gives the single-edge state and repair in Eqs.~\eqref{eq:prx412-recovery-state} and~\eqref{eq:prx412-recovery-repair}. We retain its notation: $D=X_{g,e}$, $G_a$ is the completed red-star projector, and $M_q$ measures the two active completion qubits in $Z$ and resets them to $|+\rangle$. For each term $y$, the CNOT supplies $Z_{b(e)}^{y_v}$ and $X_{g,e}$ then removes $D$. Measuring an active completion qubit gives
\begin{equation}
 \langle z_f|y_f\rangle_X=2^{-1/2}q_f^{y_f},
 \qquad q_f=(-1)^{z_f}.
 \label{eq:sm412-recovery-contraction}
\end{equation}
The two contractions multiply $a^y$ by $q_v^{y_v}q_w^{y_w}/2=(a')^y/(2a^y)$, proving Eq.~\eqref{eq:prx412-recovery-identity}. The result holds on arbitrary encoded input and preserves coherence within each logical measurement outcome. We next treat errors on several green-layer qubits and then account for faults in the repair.

\paragraph{Green-$X$ error chains.}
Let $E$ be the set of edges whose green-layer qubits have Pauli $X$ (bit-flip) errors, and write $D_E=\prod_{e\in E}X_{g,e}$. Each error flips the green flux on the two adjacent faces. Thus, the faces incident to an odd number of edges in $E$ have negative green flux. They form the boundary $\partial E$ of the error chain on the dual lattice. In the expansion of Eq.~\eqref{eq:prx412-recovery-state}, the completion qubits carry $|y_f\rangle_X$ precisely on these faces. Choose a correction chain $C$ pairing them on the dual lattice. To move a negative-flux excitation from $f$ to a neighboring positive-flux face $h$, apply
\[
 \operatorname{SWAP}(t_{g,f},t_{g,h}),\qquad
 \operatorname{CZ}(r_{fh},t_{g,h}).
\]
The red link stores $y_f\mathbin{\oplus}y_h$, so these operations transform $|y_f\rangle_X|+\rangle$ into $|+\rangle|y_h\rangle_X$. Apply the dressed edge correction $R_e$ before this transfer if $v_e=f$, or afterward if $v_e=h$. When two negative-flux excitations meet, apply $R_e$ and measure and reset both completion qubits as above. Each step preserves the same expansion, with the updated error chain, completion states and Gauss signs. After all defects are paired, the result is ideal gauging of $D_{E\triangle C}|\psi\rangle$.

For minimum-weight flux decoding, $|C|\leq|E|$. If $|E|<L/2$, the closed residual has weight below $L$ and cannot contain a noncontractible cycle. It is a product of parent green $X$ stars and acts trivially on encoded input. Thus, this recovery implements the ideal logical measurement for every such boundary pattern, assuming all subsequent operations are ideal. No negative-flux pattern is discarded.

\paragraph{Local faults in a repair.}
The single-edge repair is Clifford. Move a Pauli fault to just before the two completion measurements and write it as $E_mX_{t_v}^{p_v}Z_{t_v}^{r_v}X_{t_w}^{p_w}Z_{t_w}^{r_w}$, where $t_f=t_{g,f}$ and $E_m$ acts on matter. For actual computational-basis outcomes $z_v,z_w$ and ideal resets, the resulting branch is
\[
 \frac{e^{i\phi}}2E_mG_{a'}JP_{\rm par},\qquad
 a'_f=a_f(-1)^{z_f\oplus p_f}\quad(f\in\{v,w\}).
\]
The other Gauss signs are unchanged. This follows from $\langle z|X^pZ^r\propto\langle z\oplus p|$ and Eq.~\eqref{eq:prx412-recovery-identity}. A fault in the added CNOT has matter support only on $b(e)$, and a fault in the $X_{g,e}$ correction only on $g_e$. Measurement faults after the CNOT do not spread back to matter, while reset faults leave a local completion error. Expanding an arbitrary local Kraus operator in Paulis gives the corresponding linear span of these branches. The repair therefore converts local faults into local data errors and changes in the Gauss signs for subsequent error correction.

\paragraph{Recovery paths and circuit guarantees.}
We bound noisy recovery by counting the cells visited by the full recovery network. A $Z_t$ fault can reach the blue control of each later CNOT along its path, while flux-readout faults can alter the chosen paths. The cell count accounts for both effects.

Each recovery step is contained in a cell $\{g_e,b(e),r_{fh},t_{g,f},t_{g,h}\}$. For $e=e_x(i,j)$, the blue and red edges are $b_{e_y(i+1,j)}$ and $r_{e_y(i,j-1)}$. For $e=e_y(i,j)$, they are $b_{e_x(i,j+1)}$ and $r_{e_x(i-1,j)}$. Each cell occupies at most one coordinate in each coordinate set of Eq.~\eqref{eq:smd4-cuts}. For a fixed record, let $\ell$ count the distinct visited recovery cells. Let $F$ be the set of faulty native locations, including those in recovery, and write $f=|F|$. After recovery, perform $R$ verification rounds. Require positive non-red check outcomes and the same red-star vector $a$ in every verification round. Retain the exact gauge-readout consistency test and ideal final parent projection, while allowing earlier records to contain detected errors. For every measurement record satisfying these final verification conditions and every corresponding noise Kraus branch,
\[
 \begin{gathered}
 P_{\rm par}K_{F,m,\alpha}P_{\rm par}
 \in\mathbb{C}\,\Pi_{\prod_v a_v}P_{\rm par}
 \\
 \text{if}\quad f<R,\qquad f+\ell<L.
 \end{gathered}
\]
To see this, expand each resolved Kraus operator of a recovery step in Paulis on its cell. At least one verification round is fault-free and supplies $\widetilde Q_a$. The propagation bounds for the completed checks place the residual coordinate sets within the union of the native-fault coordinates and the coordinates of visited recovery cells. Use the input transfer identity of Lemma~\ref{lem:smd4-transfer} first, then its output identity. Both effective errors have fewer than $L$ occupied coordinates, giving the stated scalar multiple of the intended logical projector. Repeated use of a cell adds no new coordinates, while multiple recovery passes require the union over all passes.

If a decoder additionally guarantees $\ell\leq cf$, a resolved branch whose parent-code compression is not proportional to the intended logical projector, with a record passing the final verification tests, requires at least $\min\{R,\lceil L/(1+c)\rceil\}$ faults. The decoder-dependent constant $c$ accounts for propagation along the chosen recovery network.

\subsection{Boundary recovery without completion}
\label{sec:sm297-no-completion-recovery}

For comparison with the coherent repair, we construct a recovery using the controlled checks of Equation~\eqref{eq:conditional-central-completion} and additional measurements on the matter qubits. Throughout this subsection, faults occur only on the green matter at the input boundary. All subsequent operations, including the recovery, are ideal.

Keep every flux and red Gauss check. Assign the red--green and red--blue conflicts to their respective matter checks, and the green--blue conflict to green. With $P_+(B)=(I+B)/2$ and $\Delta=(1,1)$, the controlled matter checks use
\begin{align}
 E_{g,v}&=P_+(B_{b,v})P_+(B_{r,v-\Delta}),\nonumber\\
 E_{b,v}&=P_+(B_{g,v-\Delta}),\nonumber\\
 A^{\mathrm{ctrl}}_{c,v}&=I-E_{c,v}+E_{c,v}A_{c,v},
 \qquad c=g,b.
 \label{eq:sm297-conditional-assignment}
\end{align}
These checks commute globally. For a boundary error $D_E=\prod_{e\in E}X_{g,e}$, all controlled matter stars initially have eigenvalue $+1$. The red edges are zero and $D_E$ commutes with the parent $X$ stars. They retain this eigenvalue through the Gauss measurements. The green flux defects disable precisely the blue stars at vertices $\partial E+\Delta$.

Let $n=L^2$, $a$ be the red-check record, $s=\prod_v a_v$, and $z=\delta y$ the final red-edge record. After the usual $U(y)^\dagger$ correction, the matter branch is
\begin{equation}
 K_{a,z}P=2^{1-n}a^y\Pi_s(\mathcal{U})D_EP,
 \qquad \Pi_s(\mathcal{U})=\frac{I+s\mathcal{U}}{2}.
 \label{eq:sm297-conditional-boundary-branch}
\end{equation}
This expression already includes all ideal controlled matter-check measurements. In particular, those measurements have not resolved the coherence needed for the following recovery.

For one erroneous green edge, write $D=X_{g,e}$ and $C=Z_{b(e)}$, where $b(e)$ is its blue CZ partner. Then $\mathcal{U} D\mathcal{U}^\dagger=DC$, and $C$ commutes with $D$ and $\mathcal{U}$. The endpoints of $b(e)$ are exactly the two disabled blue-star vertices. For example, $e_x(i,j)$ is paired with $e_y(i+1,j)$, whose endpoints are the adjacent green faces $(i,j),(i,j-1)$ shifted by $\Delta$. Thus, the active controlled checks do not measure away this degree of freedom.

Measure the single-qubit Pauli $C$ with outcome $c$, apply $D$, and then extract the ordinary parent syndrome. If $Q_0$ and $Q_1$ project onto the trivial blue syndrome and the syndrome of $C$, respectively, apply $C^r$ for outcome $r=0,1$. The physical branch obeys
\begin{equation}
 C^rQ_rD\Pi_c(C)K_{a,z}P
 =2^{-n}a^yc^r\Pi_{sc}(\mathcal{U})P.
 \label{eq:sm297-conditional-single-repair}
\end{equation}
Indeed, after the $D$ correction the state is proportional to $\Pi_c(C)\Pi_{sc}(\mathcal{U})|\psi\rangle$. Its two terms have distinct parent syndromes and become the same encoded state after recovery. Reporting $\lambda=sc$ therefore implements the full ideal projective measurement, including its conditional output state. The extra $C$ measurement must precede ordinary parent extraction, which would otherwise resolve these two terms. It may follow the full controlled gauged-check sweep, or be performed while gauged, since $C$ commutes with $U(y)$.

\begin{proposition}[Recovery with controlled checks]
\label{prop:sm297-conditional-boundary-recovery}
Use Equation~\eqref{eq:sm297-conditional-assignment} on the even $L\times L$ torus. Suppose the input error is a green-$X$ pattern $D_E$ with $|E|<L/2$, and all subsequent operations are ideal. Minimum-weight green-flux decoding, the additional Pauli measurements described below, and parent-code erasure recovery implement the ideal logical measurement without rejecting any negative-flux syndrome and without completion qubits.
\end{proposition}
\begin{proof}
Choose a minimum-weight correction chain $F$ with $\partial F=\partial E$. Then $|F|\leq|E|$ and $E\mathbin{\triangle}F$ is homologically trivial, so $D_EP=D_FP$. Let $\pi(F)$ be the paired blue-edge set and $C_F=\prod_{e\in F}Z_{b(e)}$. The disjoint CZ pairing gives $\mathcal{U} D_F\mathcal{U}^\dagger=D_FC_F$.

After ungauging and before ordinary parent extraction, measure each blue $Z$ on $\pi(F)$. Write $M_m$ for their joint projector and $c$ for the product of their outcomes. Apply $D_F$. Equation~\eqref{eq:sm297-conditional-boundary-branch} then becomes
\begin{equation}
 2^{1-n}a^y M_m\Pi_{sc}(\mathcal{U})P.
 \label{eq:sm297-conditional-erasure-state}
\end{equation}
The known support $\pi(F)$ has size $|F|<L$ and is a correctable erasure for the distance-$L$ parent code. Explicitly, expand $M_m$ in $Z$ Paulis on this support. Two terms with the same parent syndrome differ by a stabilizer, since their product has weight less than $L$. For each parent syndrome, a Pauli correction on $\pi(F)$ therefore maps $M_mP$ to a scalar multiple of $P$. Every recovered branch of Equation~\eqref{eq:sm297-conditional-erasure-state} is a scalar multiple of $\Pi_{sc}(\mathcal{U})P$. Completeness of the physical recovery gives the ideal instrument when the reported result is $\lambda=sc$.
\end{proof}

The same argument applies to Kraus operators in the linear span of these correctable green-$X$ patterns. Within a fixed flux syndrome they have the same action as the selected $D_F$ on the code. The recovery therefore also handles coherent errors in this correctable boundary span.

All added measurements and corrections above are Pauli or Clifford operations. This recovery uses extra measurements of the matter qubits, which the coherent repair in Appendix~\ref{sec:sm297-coherent-recovery} avoids. Section~\ref{sec:prx277-local-comparison} compares the added qubits, check operators and recovery operations in these two procedures.

\paragraph{Pauli logical measurements.}
For a Pauli target $\mathcal{U}$ and a fixed Pauli error $D$, $\mathcal{U} D=\eta D\mathcal{U}$ with $\eta=\pm1$, so
\begin{equation}
 \Pi_s(\mathcal{U})D=D\Pi_{\eta s}(\mathcal{U}).
 \label{eq:sm297-pauli-boundary-sign}
\end{equation}
A known Pauli error therefore preserves or reverses a definite logical sign. The randomization discussed above occurs for a Clifford target whose commutator with $D$ is a nontrivial observable on the data.

\Needspace{8\baselineskip}
\section{Details of the comparison with completion}
\label{sec:v460-comparison-details}
The local examples below show how completion retains joint check outcomes and keeps the measured-check hierarchy bounded as the number of assigned conflicts grows. We give explicit extraction circuits and then derive the hierarchy levels of the square-lattice $D(D_4)$ stars. The bilayer recovery advantage is demonstrated in Section~\ref{sec:prx277-local-comparison}, using the protocols of Appendix~\ref{sec:sm297-boundary-recovery}.

\subsection{Local checks}
We compare the raw, controlled and completed constructions for
\begin{equation}
 C_0=X_q,\qquad C_j=\operatorname{CZ}_{c_jq},\qquad
 \mathsf{K}_j=Z_{c_j},\quad j=1,2.
 \label{eq:prx277-comparison-raw}
\end{equation}
All use the same original qubits $c_1,c_2,q$ and retain the central checks $\mathsf{K}_j=Z_{c_j}$. The projector onto the common $+1$ space of the raw and central checks is
\begin{equation}
 P_0=|00\rangle\langle00|_{c_1c_2}
       \otimes|+\rangle\langle+|_q.
 \label{eq:prx277-local-projector}
\end{equation}
Assigning both conflicts to $C_0$ gives the controlled checks
\begin{equation}
 \begin{aligned}
 C_0^{\mathrm{ctrl}}&=I-E+EX_q,\\
 E&=\frac{(I+Z_{c_1})(I+Z_{c_2})}{4},\\
 C_j^{\mathrm{ctrl}}&=C_j.
 \end{aligned}
 \label{eq:prx277-local-conditional}
\end{equation}
For completion, orient both conflicts from $C_0$ to $C_j$ and use
\begin{align}
 \widehat{C}_0&=X_qX_{t_1}X_{t_2},\nonumber\\
 \widehat{C}_j&=\operatorname{CZ}_{c_jq}
                     \operatorname{CZ}_{c_jt_j},\nonumber\\
 T_j&=\frac{I-Z_{c_j}}{2}
                  +\frac{I+Z_{c_j}}{2}X_{t_j},\qquad j=1,2.
 \label{eq:prx277-local-completed}
\end{align}
Including the retained central checks, the controlled family's common $+1$ projector is $P_0$. Including the central checks and pins, the completed family's projector is $P_0\otimes|++\rangle\langle++|_{t_1t_2}$. Both controlled and completed families commute globally, while the raw family does not.

\subsection{Hierarchy levels of the local checks}
The check $C_0^{\mathrm{ctrl}}$ is a Toffoli activated when both control bits are zero. Conjugating its target by $H_q$ gives the phase $(-1)^{q(1+c_1)(1+c_2)}$, whose Boolean polynomial has degree three. In these phase polynomials, $q,c_j$ and the readout label $r$ denote computational-basis bits, and arithmetic is modulo two. Adding the separate readout control $r$ gives degree four. Successive Pauli-$X$ commutators take finite differences of these polynomials, proving exact hierarchy levels three and four. Each completed check and pin is Clifford, and its readout-controlled unitary has level at most three. More generally, for $C_0=X_q$ and $C_j=\operatorname{CZ}_{c_jq}$, $1\leq j\leq m$, assigning all conflicts to $C_0$ gives the degree-$(m+1)$ phase $q\prod_j(1+c_j)$ and degree $m+2$ after readout control. Separate completion qubits keep the measured checks Clifford. These levels use the stated conflict assignment. Assigning each conflict to $C_j$ makes $C_j^{\mathrm{ctrl}}=I$ in this example, preserving $P_0$ while discarding that check's outcome outside the code space.

\subsection{Extraction circuits and gate counts}
\begin{table}[tbp]
\caption{Qubit and gate counts for one extraction round of the local example. Readout qubits are dedicated one per check. The work qubit for the controlled construction is separate from the readouts and returned to zero ideally.}
\label{tab:prx277-local-counts}
\centering
\setlength{\tabcolsep}{4pt}
\begin{tabular}{lrrr}
\toprule
Resource & Raw & Controlled & Completed\\
\midrule
Completion qubits &0&0&2\\
Work qubits &0&1&0\\
Readout qubits &5&5&7\\
Work resets &0&1&0\\
CNOT &1&0&3\\
CZ &2&2&2\\
Toffoli &0&3&2\\
CCZ &2&2&4\\
Pauli $X$ &0&4&4\\
\bottomrule
\end{tabular}
\end{table}

Use one readout qubit per check, initialized in $|+\rangle$ and measured in $X$, and regard CNOT, CZ, Toffoli, CCZ and one-qubit Pauli gates as the elementary gates. Extracting the raw check $C_0$ uses one CNOT, extracting each raw check $C_j$ uses one CCZ, and extracting each central check uses one CZ. The same central extractions are used in the other two constructions.

For the controlled check, reset a private work qubit $w$ to $|0\rangle$. Apply $X_{c_1}X_{c_2}$, then in time order
\begin{equation}
 \operatorname{Toffoli}_{c_1c_2\to w},\quad
 \operatorname{Toffoli}_{rw\to q},\quad
 \operatorname{Toffoli}_{c_1c_2\to w},
 \label{eq:prx277-conditional-native}
\end{equation}
and finally $X_{c_1}X_{c_2}$. This implements $\operatorname{ctrl}_r(C_0^{\mathrm{ctrl}})$ and returns $w$ to zero for every input on the original and readout qubits. The control condition and work qubit are not measured, so no additional information is revealed. The decomposition uses level-three elementary gates for this level-four readout-controlled operation.

Extracting the completed check $\widehat{C}_0$ uses three CNOTs, and extracting each $\widehat{C}_j$ uses two CCZs. The readout-controlled pin circuit applies $X_{c_j}$, $\operatorname{Toffoli}_{r c_j\to t_j}$, and $X_{c_j}$ in that time order. Table~\ref{tab:prx277-local-counts} includes every central check and pin. The work qubit is reset before each controlled $C_0$ extraction. Completion qubits are initialized once and retained between rounds. Counts exclude idles, routing, and initial preparation of the original and completion qubits. Each check readout also requires one $X$-basis preparation. The table records gate counts for these explicit circuits.

\subsection{Measurements outside the code space}
\label{sec:v460-comparison-outside}
Fix $c_1=1,c_2=0$ and put the second completion qubit in $|+\rangle_{t_2}$. On $q,t_1$, the completed pair is $X_qX_{t_1},Z_qZ_{t_1}$, with joint eigenstates given by the Bell basis. The two states $|\Phi_\pm\rangle=(|00\rangle\pm|11\rangle)/\sqrt2$ have opposite $X_qX_{t_1}$ outcomes and the same $Z_qZ_{t_1}=+1$ outcome. Controlled checks, extended by identity on qubit $t_1$, measure only $Z_q$ in this sector. Its two outcomes have probability $1/2$ in either Bell state. Completion thus distinguishes these Bell states through $X_qX_{t_1}$, a joint observable on the data and completion qubit. In the sector $c_1=c_2=0$ with both pins positive, the completion qubits are fixed to $|+\rangle$ and the completed checks reproduce the original compatible measurement branches.

\subsection{Hierarchy levels of the square-lattice stars}
\label{sec:v460-comparison-d4}
For the bulk $D(D_4)$ stars, we derive the hierarchy levels underlying the comparison in Section~\ref{sec:prx277-local-comparison}. For the assignment in Eq.~\eqref{eq:sm297-conditional-assignment}, the red stars remain Clifford, while the controlled blue and green stars have exact hierarchy levels three and four, respectively. Their readout-controlled unitaries have levels three, four and five for red, blue and green. To verify these levels, apply Hadamards to the four qubits flipped by each star. The operator becomes diagonal, with a sign determined by the original linear and quadratic terms multiplied by its flux conditions. The one blue-star condition gives degree three, and the two green-star conditions give degree four. Direct expansion using Eq.~\eqref{eq:prx273-d4-quadratics} gives nonzero highest-degree terms. A readout control raises the degree by one. All completed stars are Clifford, and their readout-controlled unitaries have hierarchy level three. These are the hierarchy levels of the full star checks and their readout-controlled unitaries.

\section{SPT Hamiltonians and gauging maps}
\label{sec:prx314-spt-details}
\label{sec:prx-physical-principle}
This appendix gives the physical gauging map, ideal preparation protocol and associated square-lattice SPT Hamiltonians. We display the cluster resource and its realization with completion qubits, including their lattice terms and protecting symmetries. Local reference qubits, initially paired with matter qubits in Bell states, let the resource encode the coherent circuit's action on arbitrary matter inputs. The final subsection gives the type-III SPT ancestor of the bilayer protocol.

\subsection{Physical gauging of local interactions}
\label{sec:prx306-duality}
Gauging replaces a local symmetric interaction by an interaction that also acts on the gauge qubits and commutes with the local gauge transformations~\cite{VijayHaahFu2016,Shirley2019Gauging}. We verify this correspondence for the circuit used here. Let $h_i$ be a local interaction supported on a union $D_i$ of symmetry blocks and invariant under $\prod_{v\in D_i}U_v$. Choose a root $r_i\in D_i$ and connecting paths for these blocks, and define $W_i(z)$ by Eq.~\eqref{eq:general-raw-gauging} in Appendix~\ref{sec:sm-clifford-derivation}, with its product over the blocks in $D_i$. The paths lie in a bounded routing neighborhood when a local gauging map is required. The gauged interaction is
\begin{equation}
 h_i^{\mathrm g}=\sum_z|z\rangle\langle z|_g\otimes h_i(z),
 \qquad h_i(z)=W_i(z)h_iW_i(z)^\dagger.
 \label{eq:prx308-local-gauging}
\end{equation}
A vertex transformation $z\mapsto z+\delta u$ changes each path potential by $u_v+u_{r_i}$. The common root factor cancels by the symmetry of $h_i$, giving
\begin{equation}
 h_i(z+\delta u)=U(u)h_i(z)U(u)^\dagger,
 \qquad U(u)=\prod_vU_v^{u_v}.
 \label{eq:prx308-gauge-covariance}
\end{equation}
Thus, $h_i^{\mathrm g}$ commutes with every Gauss operator $A_v$ on the full Hilbert space. Closed paths give the flux observables $B_p$ in Eq.~\eqref{eq:prx273-gauss-flux}. Imposing $B_p=+1$ for the chosen cycle relations makes the corresponding path potentials consistent.

For configurations $z=\delta x$, all interactions obey $h_i(\delta x)=U(x)h_iU(x)^\dagger$. Their products and relations are therefore preserved in this sector. The gauging isometry of Eq.~\eqref{eq:prx273-rooted-isometry} consequently satisfies the concrete interaction correspondence
\begin{equation}
 (h_i^{\mathrm g}\otimes I_t)\Gamma_a=\Gamma_a h_i.
 \label{eq:prx306-duality-intertwining}
\end{equation}
Each bracketed sum in the parent Hamiltonian in Eq.~\eqref{eq:prx284-symmetric-parent} is symmetric and obeys this correspondence. The completed checks commute on the full Hilbert space, including after errors.

This local interaction map underlies the state correspondence. Since $\prod_vA_v=\mathcal{U}$, the parent symmetry sign fixes the product of the Gauss signs. On the symmetry-even parent ground space, $\Gamma_{\boldsymbol{+}}$ is an isometry onto the corresponding gauged ground-state sector, with inverse $\Gamma_{\boldsymbol{+}}^\dagger$ on its image. The link initialization selects configurations $z=\delta x$, with zero parity around every closed path, including paths that wind around the torus. The correspondence is restricted to these configurations. After correction of the Gauss signs, the gauging protocol with completed checks implements this isometry on these ground states, as established in Section~\ref{sec:prx273-physical} and made explicit below.

Gauging can change intrinsic topological order, as in the transition from two toric codes to $D(D_4)$. The constructions with and without completion realize the same ideal ground-state map after omitting the fixed completion qubits. Their measured checks and boundary-recovery procedures differ, as compared in Section~\ref{sec:prx277-local-comparison}.

\subsection{Preparation of the gauged state}
Keeping the gauge qubits after gauging prepares a state of the gauged code. Here we show that, for a symmetry-even parent-code input, every allowed measurement outcome can be corrected to the same gauged output. This gives an ideal preparation without postselection and identifies the coherent gauging stage represented by the SPT resource in Appendix~\ref{sec:prx314-spt-example}. Choose the symmetric parent Hamiltonian
\begin{equation}
 H_{\mathrm{par}}=\sum_i\left[
 \frac{I-S_i}{2}+\frac{I-\mathcal{U}S_i\mathcal{U}^{\dagger}}{2}\right].
 \label{eq:prx284-symmetric-parent}
\end{equation}
Both check sets belong to the parent stabilizer group, so the terms commute and have ground projector $P$. Conjugation by $\mathcal{U}$ exchanges the paired terms. A local parent and bounded local symmetry blocks therefore give a local symmetric Hamiltonian.

Start with a symmetry-even parent ground state $\rho$, satisfying $P\rho P=\rho$ and $\mathcal{U}\rho=\rho$. On a connected gauging graph with the locality bounds of Section~\ref{sec:prx-central}, proceed as follows.
\begin{enumerate}[label=\arabic*.,leftmargin=1.6em,labelsep=.4em, topsep=3pt,itemsep=3pt,parsep=0pt]
\item \emph{Enrich the symmetry.} Prepare vertex ancillas $\mu_v$ in $|+\rangle$ and apply $\operatorname{ctrl}_{\mu_v}(U_v)$. The enriched subspace obeys $X_{\mu_v}U_v=1$, so the product of ancilla $X$ operators represents the matter symmetry \cite{Verresen2021Enrichment,Tantivasadakarn2024LongRange}.
\item \emph{Introduce gauge links.} Prepare each $g_e$ in $|0\rangle$ and apply CNOTs from its endpoint ancillas. The links store $z=\delta x$. Dress the parent checks along the chosen paths and retain every flux and invariant constraint.
\item \emph{Complete the constraints.} Add $t_e$ in $|+\rangle$ for each conflict, dress the endpoint checks and include its pin. Apply $\bm{E}_{\mathrm{comp}}$ before vertex readout, as in Equation~\eqref{eq:prx273-physical-factorization}.
\item \emph{Measure and correct.} Read the vertex ancillas in $X$ and measure the remaining completed, central and pinning checks. Correct the Gauss signs as below. Retain the gauge and completion qubits. The measured vertex ancillas can be reset and reused.
\end{enumerate}

This ideal preparation needs no postselection. Write $a_v=(-1)^{b_v}$. The even input implies $\prod_va_v=+1$, so connectedness gives an edge vector $c$ with incidence boundary $\partial_1c=b$. Applying $Z_g(c)=\prod_eZ_{g_e}^{c_e}$ gives
\begin{align}
 Z_g(c)\Gamma_a&=\Gamma_{\boldsymbol{+}},
 \label{eq:prx284-gauss-correction}\\
 \Phi_{\mathrm{gauge}}(\rho)
 &=\Gamma_{\boldsymbol{+}}\rho\Gamma_{\boldsymbol{+}}^\dagger.
 \label{eq:prx284-preparation-channel}
\end{align}
Here, $\Gamma_a$ is the gauging isometry of Equation~\eqref{eq:prx273-rooted-isometry} and $\boldsymbol{+}$ is the all-positive record. On $|\delta x\rangle_g$, the correction contributes $(-1)^{c\cdot\delta x}=(-1)^{b\cdot x}=a^x$, canceling the branch phase. Each allowed record has probability $2^{1-|V|}$, giving the normalized channel above. The correction flips the Gauss signs and commutes with the gauge-$Z$-diagonal modified-parent, central and pinning checks. Link initialization selects configurations $z=\delta x$, including zero gauge parity along both directions around the torus.

For a general parent ground state, $\prod_va_v$ measures the symmetry sector. Retaining the gauge system gives that gauged branch. Reading out the links and applying the matter correction gives the outcome-$s$ branch $\rho\mapsto\Pi_s\rho\Pi_s$ of the logical measurement. Thus, physical gauging and logical surgery use the same operations, with different choices of which qubits to retain. Their circuit-distance guarantees use the specified native schedules and boundary tests.

\subsection{The cluster resource of the gauging map}
\label{sec:prx313-resource}
Place a qubit $\mu_v$ at each vertex and a qubit $r_e$ on each edge of a closed square lattice. The cluster Hamiltonian is~\cite{Yoshida2016Generalized,Tantivasadakarn2024LongRange}
\begin{equation}
 \begin{aligned}
 H_{\rm cl}&=-\sum_vK_v^{\rm cl}-\sum_{e=uw}K_e^{\rm cl},\\
 K_v^{\rm cl}&=X_{\mu_v}\prod_{e\ni v}Z_{r_e},\qquad
 K_{uw}^{\rm cl}=X_{r_{uw}}Z_{\mu_u}Z_{\mu_w}.
 \end{aligned}
 \label{eq:v468-cluster-hamiltonian}
\end{equation}
Figure~\ref{fig:v468-spt-terms}(a) shows these terms. They are obtained from independent $X$ operators by applying CZ between every edge qubit and its endpoint qubits. This is the gauge-link preparation circuit after Hadamards on the links.

The protecting symmetries are $S_0=\prod_vX_{\mu_v}$ and $S_1(\gamma)=\prod_{e\in\gamma}X_{r_e}$ for every closed loop $\gamma$. These are an ordinary $\mathbb Z_2$ symmetry and a one-form $\mathbb Z_2$ symmetry. On the ground state, opening a loop from $u$ to $w$ gives endpoint operators $Z_{\mu_u}Z_{\mu_w}$, each charged under $S_0$. This endpoint charge identifies the mixed $\mathbb Z_2^{(0)}\times\mathbb Z_2^{(1)}$ cluster SPT~\cite{Yoshida2016Generalized,Tantivasadakarn2024LongRange}.

To represent a circuit as a state, prepare a Bell pair between each input qubit and a reference qubit, then apply the circuit to the input qubits. Correlations with the references then encode the circuit's action on arbitrary inputs. This construction relates Kramers--Wannier duality to a cluster resource~\cite{GorantlaShaoTantivasadakarn2025}. For the bilayer protocol, the cluster resource and its CCZ enrichment are the established construction of Refs.~\cite{Tantivasadakarn2023Shortest,Christos2026NonAbelian}. We now include the matter and reference qubits explicitly before incorporating completion.

Use the physical CZ pairs of Eq.~\eqref{eq:main-d4-physical}, grouped into disjoint local blocks $v$. Let $\mathcal P_v$ denote the two pairs in block $v$, so $U_v=\prod_{(q,p)\in\mathcal P_v}\mathrm{CZ}_{q,p}$. Write $r_e$ for the new gauge links, in their Hadamard-transformed basis. Pair each matter qubit $q$ with a local reference $q'$ in $|\Phi^+\rangle=(|00\rangle+|11\rangle)/\sqrt2$. Define
\begin{equation}
 \begin{aligned}
 V_{\mathrm{cl}}&=\prod_{e=vw}
 \mathrm{CZ}_{\mu_v,r_e}\mathrm{CZ}_{\mu_w,r_e},\\
 V_{\mathrm{enr}}&=\prod_v\prod_{(q,p)\in\mathcal P_v}
 \mathrm{CCZ}_{\mu_v,q,p},\\
 |\mathcal R_{\mathcal U}\rangle
 &=V_{\mathrm{cl}}V_{\mathrm{enr}}
 \left(|+\rangle_\mu|+\rangle_r
 \bigotimes_q|\Phi^+\rangle_{q'q}\right).
 \end{aligned}
 \label{eq:prx313-resource-state}
\end{equation}
The products of $|+\rangle$ states run over all vertices and edges. These bounded-depth local circuits give a short-range-entangled state. If $p(q)$ is the CZ partner of $q$ in block $v_q$, its complete commuting checks are
\begin{equation}
 \begin{aligned}
 K_{\mu_v}&=X_{\mu_v}U_v\prod_{e\ni v}Z_{r_e},\\
 K_{r_{vw}}&=X_{r_{vw}}Z_{\mu_v}Z_{\mu_w},\\
 K_q^X&=X_{q'}X_q\,\mathrm{CZ}_{\mu_{v_q},p(q)},\\
 K_q^Z&=Z_{q'}Z_q.
 \end{aligned}
 \label{eq:prx313-resource-checks}
\end{equation}
They are obtained by conjugating the independent Pauli operators $X_{\mu_v}$, $X_{r_e}$, $X_{q'}X_q$ and $Z_{q'}Z_q$ by $V_{\mathrm{cl}}V_{\mathrm{enr}}$. Hence, $H_{\mathcal R}=\sum_K(I-K)/2$ has a unique ground state and unit gap.

On a closed lattice the protecting symmetries are
\begin{equation}
 S_0=\prod_vX_{\mu_v}U_v,\qquad
 S_1(\gamma)=\prod_{e\in\gamma}X_{r_e},\quad\partial\gamma=0.
 \label{eq:prx313-resource-symmetry}
\end{equation}
The first is onsite on the enlarged local blocks. The second is supported on closed loops. For an open path $\gamma$ from $v$ to $w$, multiplying the edge checks gives
\begin{equation}
 X_r(\gamma)|\mathcal R_{\mathcal U}\rangle
 =Z_{\mu_v}Z_{\mu_w}|\mathcal R_{\mathcal U}\rangle.
 \label{eq:prx313-resource-endpoints}
\end{equation}
Each endpoint carries an $S_0$ charge. This symmetry relation gives the SPT content beyond the mutual commutation of the checks. The basis change $V_{\mathrm{enr}}^\dagger$, acting independently within each local block, maps the state and its protecting symmetry to the ordinary cluster SPT with local Bell pairs.

In higher-form notation, the protecting symmetry is $\mathbb Z_2^{(0)}\times\mathbb Z_2^{(1)}$. This is the split 2-group with trivial action and vanishing Postnikov class~\cite{BeniniCordovaHsin2019}. Its SPT response on a closed three-dimensional spacetime $M_3$ is
\begin{equation}
 Z_{\mathrm{SPT}}[A,B]=(-1)^{\int_{M_3}A\cup B},\qquad
 \delta A=0,\quad\delta B=0,
 \label{eq:prx313-resource-response}
\end{equation}
where $A$ and $B$ are background $\mathbb Z_2$ cochains of degrees one and two~\cite{Tantivasadakarn2024LongRange}. The vanishing Postnikov class means that $B$ remains closed independently of $A$. A nonzero class would instead couple their flatness conditions through $\delta B=\beta(A)$. The response in Eq.~\eqref{eq:prx313-resource-response} describes the SPT state without changing these symmetry relations.

Define the unnormalized raw gauging branch by
\begin{equation}
 K_a|\psi\rangle=2^{-|V|}\sum_x a^x
 |\delta x\rangle_r U(x)|\psi\rangle.
 \label{eq:prx313-resource-branch}
\end{equation}
Its restriction to parent-code inputs agrees with Eq.~\eqref{eq:prx273-gauging-sum} after appending the fixed completion qubits. Let $d_m$ be the matter Hilbert-space dimension and $\mathcal H_r=\bigotimes_e H_{r_e}$. Write $m'$ for all reference qubits $q'$ together. Then
\begin{equation}
 \left({}_\mu\langle a^X|\otimes{}_{m'}\langle\psi^*|\right)
 |\mathcal R_{\mathcal U}\rangle
 =d_m^{-1/2}\mathcal H_r K_a|\psi\rangle.
 \label{eq:prx313-resource-contraction}
\end{equation}
The conjugate is taken in the computational basis. This identity holds for every input and record and shows how the resource represents the gauging map. The SPT description applies with the reference qubits retained. Using an actual toric-code bilayer as input instead gives intrinsic topological order before readout. Completion makes the subsequent check measurements on the gauged output compatible.

The completed circuit also gives an explicit commuting parent Hamiltonian for the state. Set $W_{\mathrm{comp}}=\mathcal H_r\bm{E}_{\mathrm{comp}} \mathcal H_r$. Then
\begin{equation}
 \begin{aligned}
 |\widehat{\mathcal R}_{\mathcal U}\rangle
 &=W_{\mathrm{comp}}
 (|\mathcal R_{\mathcal U}\rangle\otimes|+\rangle_t),\\
 H_{\widehat{\mathcal R}}
 &=W_{\mathrm{comp}}\left[H_{\mathcal R}
 +\sum_t\frac{I-X_t}{2}\right]W_{\mathrm{comp}}^\dagger.
 \end{aligned}
 \label{eq:prx313-completed-resource}
\end{equation}
Its terms commute, and it has a unique ground state and unit gap. The bounded-depth completion circuit preserves locality. Conjugating the symmetry by the same circuit preserves its relations while changing its local action. Appendix~\ref{sec:prx314-onsite-resource} gives a Hamiltonian for the same resource with an ordinary symmetry acting separately on local blocks. The two Hamiltonians and their symmetry actions are compared there.

\subsection{Explicit Hamiltonians for the completed gauging resource}
\label{sec:prx314-completed-resource}
We now expand the completed resource of Eq.~\eqref{eq:prx313-completed-resource} for the square-lattice bilayer construction.

Let $f_v$ be the face with southwest vertex $v$, and write $B_{g,v}=\prod_{e\in\partial f_v}Z_{g_e}$. The local symmetry factor in Eq.~\eqref{eq:main-d4-physical} is
\begin{equation}
 U_v=\operatorname{CZ}_{g_{e_x(v)},b_{e_y(v+\hat x)}}
     \operatorname{CZ}_{g_{e_y(v)},b_{e_x(v+\hat y)}}.
 \label{eq:prx314-resource-local-symmetry}
\end{equation}
Recall $V(B,t)=(I+B)/2+(I-B)Z_t/2$. The red Gauss completion factor is $D_v=X_{t_{b,v}}V(B_{g,v},t_{g,v})$. It contains no red gauge links, so the circuit in Eq.~\eqref{eq:prx313-completed-resource} becomes
\begin{equation}
 W_{\rm comp}
 =\prod_v\operatorname{CNOT}_{\mu_v\to t_{b,v}}
   \prod_v\prod_{e\in\partial f_v}
       \operatorname{CCZ}_{\mu_v,t_{g,v},g_e}.
 \label{eq:prx314-resource-completion-circuit}
\end{equation}
These gates commute with one another.

Writing the Hamiltonian directly as a sum of unitary checks gives
\begin{equation}
 \mathsf H_{\mathcal R}
 =-\sum_vK_{\mu_v}-\sum_eK_{r_e}
  -\sum_q(K_q^X+K_q^Z),
 \label{eq:prx314-unitary-resource-hamiltonian}
\end{equation}
where the generators are those of Eq.~\eqref{eq:prx313-resource-checks}, and $q$ runs over physical green and blue matter qubits. This normalization differs from $H_{\mathcal R}$ only by an additive constant and an overall factor. For a green matter qubit $q$, $\mathcal I_g(q)=\{v:q\in\partial f_v\}$ lists its two adjacent face labels. It is empty for blue $q$. Conjugating every generator, including the completion-qubit generators, gives
\begin{align}
 \widehat K_{\mu_v}
 &=X_{\mu_v}U_v\prod_{e\ni v}Z_{r_e}\,
     X_{t_{b,v}}V(B_{g,v},t_{g,v}),\nonumber\\
 \widehat K_{r_{vw}}&=X_{r_{vw}}Z_{\mu_v}Z_{\mu_w},\nonumber\\
 \widehat K_q^X
 &=X_{q'}X_q\operatorname{CZ}_{\mu_{v_q},p(q)}
   \prod_{v\in\mathcal I_g(q)}
       \operatorname{CZ}_{\mu_v,t_{g,v}},\nonumber\\
 \widehat K_q^Z&=Z_{q'}Z_q,\nonumber\\
 L_{g,v}&=X_{t_{g,v}}V(B_{g,v},\mu_v),\nonumber\\
 L_{b,v}&=X_{t_{b,v}},\qquad L_{r,v}=X_{t_{r,v}}.
 \label{eq:prx314-natural-resource-generators}
\end{align}
Here, $v_q$ and $p(q)$ are the symmetry block and CZ partner of $q$ defined above Eq.~\eqref{eq:prx313-resource-checks}. The Hamiltonian directly obtained from the completion circuit is
\begin{align}
 \mathsf H_{\rm nat}
 &=-\sum_v\widehat K_{\mu_v}-\sum_e\widehat K_{r_e}\nonumber\\
 &\quad-\sum_q(\widehat K_q^X+\widehat K_q^Z)
   -\sum_{c,v}L_{c,v}\nonumber\\
 &=W_{\rm comp}\left(
       \mathsf H_{\mathcal R}-\sum_{c,v}X_{t_{c,v}}
       \right)W_{\rm comp}^{\dagger}.
 \label{eq:prx314-natural-resource-hamiltonian}
\end{align}
All displayed generators are local Clifford operators that square to identity. They commute and are independent because they are unitary conjugates of the mutually commuting, independent Pauli operators $X_{\mu_v}$, $X_{r_e}$, $X_{q'}X_q$, $Z_{q'}Z_q$ and $X_{t_{c,v}}$. Their common $+1$ eigenstate is $|\widehat{\mathcal R}_{\mathcal U}\rangle$, and the gap of $\mathsf H_{\rm nat}$ is two. The operators $L_{c,v}$ are resource generators. They are distinct from the conditional output pins $T(B,t)$ in Eq.~\eqref{eq:main-pinning}.

The green completion qubits are entangled with this resource. For a square face with distinct edges, its single-qubit reduced state is
\begin{equation}
 \rho_{t_{g,v}}=\frac12\left(I+\frac12X_{t_{g,v}}\right).
 \label{eq:prx314-completion-reduced-state}
\end{equation}
Indeed, the computational-basis distribution of $|\mathcal R_{\mathcal U}\rangle$ makes $\mu_v$ and the green flux parity independent and uniform. The corresponding completion phase applies $Z_{t_{g,v}}$ precisely when $\mu_v=1$ and $B_{g,v}=-1$. This event has probability $1/4$, giving eigenvalues $3/4$ and $1/4$ in Eq.~\eqref{eq:prx314-completion-reduced-state}. The blue completion qubits remain in $|+\rangle$ because they occur only as CNOT targets initialized in that state. The red completion qubits are untouched by this coherent gauging circuit. Thus, the green completion qubits are entangled with the resource, while the blue and red completion qubits remain in $|+\rangle$.

The symmetry inherited directly by $\mathsf H_{\rm nat}$ is
\begin{equation}
 \begin{aligned}
 \widehat S_0
 &=S_0\prod_vX_{t_{b,v}}V(B_{g,v},t_{g,v}),\\
 \widehat S_1(\gamma)&=S_1(\gamma).
 \end{aligned}
 \label{eq:prx314-natural-resource-symmetry}
\end{equation}
It is obtained by conjugating Eq.~\eqref{eq:prx313-resource-symmetry} with $W_{\rm comp}$. Its zero-form generator is generally not onsite on the original symmetry blocks. An onsite realization of the same resource state is given next.

\subsubsection{An onsite-symmetric parent of the same resource}
\label{sec:prx314-onsite-resource}
\label{sec:prx314-spt-example}
The reference qubits allow a strictly local block assignment. With $f_v$ the face whose southwest corner is $v$, define
\begin{equation}
 F_v=Z_{g_{e_x(v)}}Z_{g_{e_y(v)}}
     Z_{g'_{e_y(v+\hat x)}}Z_{g'_{e_x(v+\hat y)}}.
 \label{eq:prx314-reference-flux}
\end{equation}
This uses physical green qubits on the bottom and left edges of $f_v$ and their references on the top and right edges. Block $v$ contains $\mu_v$, its three completion qubits, the two outgoing physical green qubits, the two physical blue qubits in $U_v$, the two green reference qubits appearing in $F_v$, and the references of those physical blue qubits. These blocks are disjoint and have bounded diameter. Define a circuit acting independently within these blocks
\begin{equation}
 W_{\rm loc}=\prod_v\operatorname{ctrl}_{\mu_v}
       \left[X_{t_{b,v}}V(F_v,t_{g,v})\right].
 \label{eq:prx314-onsite-completion-circuit}
\end{equation}
The $Z_qZ_{q'}=1$ constraints are preserved by the diagonal resource preparation. Consequently, $F_v$ and $B_{g,v}$ agree on that subspace, including throughout the completion circuit, and
\begin{equation}
 W_{\rm loc}\bigl(|\mathcal R_{\mathcal U}\rangle|+\rangle_t\bigr)
 =W_{\rm comp}\bigl(|\mathcal R_{\mathcal U}\rangle|+\rangle_t\bigr).
 \label{eq:prx314-resource-state-equality}
\end{equation}
This equality is an equality of states, not of the two circuits on the entire Hilbert space.

An onsite-symmetric Hamiltonian for this state is
\begin{equation}
 \mathsf H_{\rm loc}
 =W_{\rm loc}\left(
       \mathsf H_{\mathcal R}-\sum_{c,v}X_{t_{c,v}}
       \right)W_{\rm loc}^{\dagger}.
 \label{eq:prx314-onsite-resource-hamiltonian}
\end{equation}
Its vertex and green-completion generators are
\begin{align}
 A_v^{\rm loc}
 &=X_{\mu_v}U_v\prod_{e\ni v}Z_{r_e}\,
       X_{t_{b,v}}V(F_v,t_{g,v}),\nonumber\\
 L_{g,v}^{\rm loc}&=X_{t_{g,v}}V(F_v,\mu_v).
 \label{eq:prx314-onsite-resource-generators}
\end{align}
The other generators are $\widehat K_{r_e}$, $\widehat K_q^X$, $\widehat K_q^Z$, $L_{b,v}$ and $L_{r,v}$ in Eq.~\eqref{eq:prx314-natural-resource-generators}. In particular, the $X_{q'}X_q$ factor flips $q$ and $q'$ together, so moving a green $Z$ factor to its reference leaves its conjugation factor unchanged. The sum of these generators with minus signs, each included once, is exactly $\mathsf H_{\rm loc}$.

On a closed lattice its protecting symmetry is
\begin{equation}
 \begin{aligned}
 S_0^{\rm loc}
 &=\prod_vX_{\mu_v}U_vX_{t_{b,v}}V(F_v,t_{g,v}),\\
 S_1(\gamma)&=\prod_{e\in\gamma}X_{r_e},\qquad\partial\gamma=0.
 \end{aligned}
 \label{eq:prx314-onsite-resource-symmetry}
\end{equation}
The first line is a product of disjoint block operators, each squaring to identity. It also equals $\prod_vA_v^{\rm loc}$. Both $W_{\rm loc}$ and $V_{\rm enr}$ act independently within these blocks and leave $S_1$ unchanged. Their inverse maps the state and its symmetry to the ordinary cluster resource with fixed Bell pairs and completion states. The Bell pairs have bounded range and are neutral under the remaining cluster symmetries. This local change of basis gives another realization of the split $\mathbb Z_2^{(0)}\times\mathbb Z_2^{(1)}$ SPT discussed in Eq.~\eqref{eq:prx313-resource-response}, with vanishing Postnikov class. In particular, multiplying the unchanged link generators along an open path still gives endpoint operators $Z_{\mu_v}Z_{\mu_w}$, each charged under $S_0^{\rm loc}$. The mixed response remains $(-1)^{\int A\cup B}$. Completion therefore gives an alternative commuting Hamiltonian for the same SPT phase.

\begin{figure*}[tbp]
\centering
\begin{minipage}[t]{.30\textwidth}
\centering
\textbf{(a)} Cluster terms\par\smallskip
\begin{tikzpicture}[x=.80cm,y=.80cm,font=\footnotesize,text=black,
 lattice/.style={draw=black!23,line width=.45pt},
 vertex/.style={circle,draw=black,fill=white,inner sep=1pt,minimum size=12pt},
 link/.style={rectangle,draw=black,fill=white,inner sep=1pt,minimum size=12pt}]
\path[use as bounding box](-2.05,-2.65)rectangle(2.05,3.30);
\draw[lattice](-1.55,1.55)--(1.55,1.55);
\draw[lattice](0,0)--(0,3.10);
\foreach \x/\y in {-1.55/1.55,1.55/1.55,0/0,0/3.10}
  {\fill[black!25](\x,\y)circle(1.5pt);}
\node[vertex]at(0,1.55){$X$};
\node[above left,inner sep=7pt]at(0,1.55){$\mu_v$};
\foreach \x/\y in {-.78/1.55,.78/1.55,0/.77,0/2.33}
  {\node[link]at(\x,\y){$Z$};}
\node at(0,-.42){$K_v^{\rm cl}=X_{\mu_v}\prod_{e\ni v}Z_{r_e}$};
\draw[lattice](-1.30,-1.35)--(1.30,-1.35);
\node[vertex]at(-1.30,-1.35){$Z$};
\node[link]at(0,-1.35){$X$};
\node[vertex]at(1.30,-1.35){$Z$};
\node[below,inner sep=9pt]at(-1.30,-1.35){$\mu_u$};
\node[below,inner sep=9pt]at(0,-1.35){$r_e$};
\node[below,inner sep=9pt]at(1.30,-1.35){$\mu_w$};
\node at(0,-2.30){$K_e^{\rm cl}=X_{r_e}Z_{\mu_u}Z_{\mu_w}$};
\end{tikzpicture}
\end{minipage}\hfill
\begin{minipage}[t]{.37\textwidth}
\centering
\textbf{(b)} Completed vertex term $A_v^{\rm loc}$\par\smallskip
\begin{tikzpicture}[x=.80cm,y=.80cm,font=\footnotesize,text=black,
 lattice/.style={draw=black!23,line width=.45pt},
 phase/.style={draw=black,line width=.80pt},
 vertex/.style={circle,draw=black,fill=white,inner sep=1pt,minimum size=12pt},
 link/.style={rectangle,draw=black,fill=white,inner sep=1pt,minimum size=12pt},
 completion/.style={diamond,draw=black,fill=white,inner sep=1pt,minimum size=12pt},
 green/.style={circle,draw=green!45!black,fill=green!45!black,inner sep=0pt,minimum size=5pt},
 greenref/.style={circle,draw=green!45!black,fill=white,inner sep=0pt,minimum size=5pt},
 blue/.style={circle,draw=blue!65!black,fill=blue!65!black,inner sep=0pt,minimum size=5pt}]
\path[use as bounding box](-1.55,-1.65)rectangle(3.70,4.30);
\draw[lattice](0,0)rectangle(2.80,2.80);
\draw[lattice](-1.40,0)--(0,0)--(0,-1.40);
\foreach \x/\y in {2.80/0,0/2.80,2.80/2.80,-1.40/0,0/-1.40}
  {\fill[black!25](\x,\y)circle(1.5pt);}
\coordinate(gE)at(1.40,-.06);
\coordinate(gN)at(-.06,1.40);
\coordinate(gRp)at(2.62,1.40);
\coordinate(gTp)at(1.40,2.62);
\coordinate(bR)at(2.98,1.40);
\coordinate(bT)at(1.40,2.98);
\coordinate(tg)at(1.00,1.00);
\draw[phase](gE)to[bend right=9](bR);
\draw[phase](gN)to[bend left=9](bT);
\foreach \s in {gE,gN,gRp,gTp}
  {\draw[phase](tg)--(\s);}
\node[vertex]at(0,0){$X$};
\node[above left,inner sep=7pt]at(0,0){$\mu_v$};
\foreach \x/\y in {-.72/0,0/-.72,1.40/-.48,-.48/1.40}
  {\node[link]at(\x,\y){$Z$};}
\node[completion]at(.70,-.78){$X$};
\node[below right,inner sep=8pt]at(.70,-.78){$t_{b,v}$};
\node[completion,minimum size=7pt]at(tg){};
\node[below left,inner sep=6pt]at(tg){$t_{g,v}$};
\node[green]at(gE){};\node[green]at(gN){};
\node[greenref]at(gRp){};\node[greenref]at(gTp){};
\node[blue]at(bR){};\node[blue]at(bT){};
\node[above left,inner sep=3pt]at(gE){$g$};
\node[above right,inner sep=3pt]at(gN){$g$};
\node[below left,inner sep=3pt]at(gRp){$g'$};
\node[below right,inner sep=3pt]at(gTp){$g'$};
\node[right,inner sep=3pt]at(bR){$b$};
\node[above,inner sep=3pt]at(bT){$b$};
\node[align=center,text width=4.2cm]at(1.05,3.76)
 {$X_{\mu_v}$, $X_{t_{b,v}}$, four link $Z$ factors,\\and six CZ factors};
\end{tikzpicture}
\end{minipage}\hfill
\begin{minipage}[t]{.30\textwidth}
\centering
\textbf{(c)} Completion term $L_{g,v}^{\rm loc}$\par\smallskip
\begin{tikzpicture}[x=.80cm,y=.80cm,font=\footnotesize,text=black,
 lattice/.style={draw=black!23,line width=.45pt},
 phase/.style={draw=black,line width=.80pt},
 vertex/.style={circle,draw=black,fill=white,inner sep=1pt,minimum size=7pt},
 completion/.style={diamond,draw=black,fill=white,inner sep=1pt,minimum size=12pt},
 green/.style={circle,draw=green!45!black,fill=green!45!black,inner sep=0pt,minimum size=5pt},
 greenref/.style={circle,draw=green!45!black,fill=white,inner sep=0pt,minimum size=5pt}]
\path[use as bounding box](-.75,-1.65)rectangle(3.55,4.30);
\draw[lattice](0,0)rectangle(2.80,2.80);
\foreach \x/\y in {2.80/0,0/2.80,2.80/2.80}
  {\fill[black!25](\x,\y)circle(1.5pt);}
\coordinate(gE)at(1.40,-.06);
\coordinate(gN)at(-.06,1.40);
\coordinate(gRp)at(2.62,1.40);
\coordinate(gTp)at(1.40,2.62);
\draw[phase](0,0)to[bend right=14](gE);
\draw[phase](0,0)to[bend left=14](gN);
\draw[phase](0,0)--(gRp);
\draw[phase](0,0)--(gTp);
\node[vertex]at(0,0){};
\node[below left,inner sep=3pt]at(0,0){$\mu_v$};
\node[completion]at(1.06,1.06){$X$};
\node[above right,inner sep=8pt]at(1.06,1.06){$t_{g,v}$};
\node[green]at(gE){};\node[green]at(gN){};
\node[greenref]at(gRp){};\node[greenref]at(gTp){};
\node[below,inner sep=3pt]at(gE){$g$};
\node[left,inner sep=3pt]at(gN){$g$};
\node[right,inner sep=3pt]at(gRp){$g'$};
\node[above,inner sep=3pt]at(gTp){$g'$};
\node[align=center,text width=3.5cm]at(1.40,3.76)
 {$X_{t_{g,v}}$ and four CZ factors\\from $\mu_v$ to the green sites};
\end{tikzpicture}
\end{minipage}
\caption{Square-lattice terms of the gauging-resource Hamiltonians. (a) The cluster Hamiltonian has one vertex term and one edge term per vertex and edge. Squares are gauge-link qubits and circles are vertex qubits. (b),(c) Representative terms of the completed bilayer resource. The drawn face has southwest vertex $v$. Filled green and blue dots are physical matter qubits. Hollow green dots labeled $g'$ are the reference qubits on the top and right edges. Diamonds are completion qubits. A thick bond denotes a CZ factor, and a displayed $X$ or $Z$ is a Pauli factor. Thin gray lines show the lattice only. Qubit species on the same edge are separated slightly for clarity. Panel (b) includes the two CZ factors of $U_v$ and the four CZ factors of $V(F_v,t_{g,v})$. Panel (c) includes the four CZ factors of $V(F_v,\mu_v)$. The dressed $X_{q'}X_q$ terms, $Z_{q'}Z_q$ terms and single-qubit completion $X$ terms are specified in the Hamiltonian.}
\label{fig:v468-spt-terms}
\end{figure*}
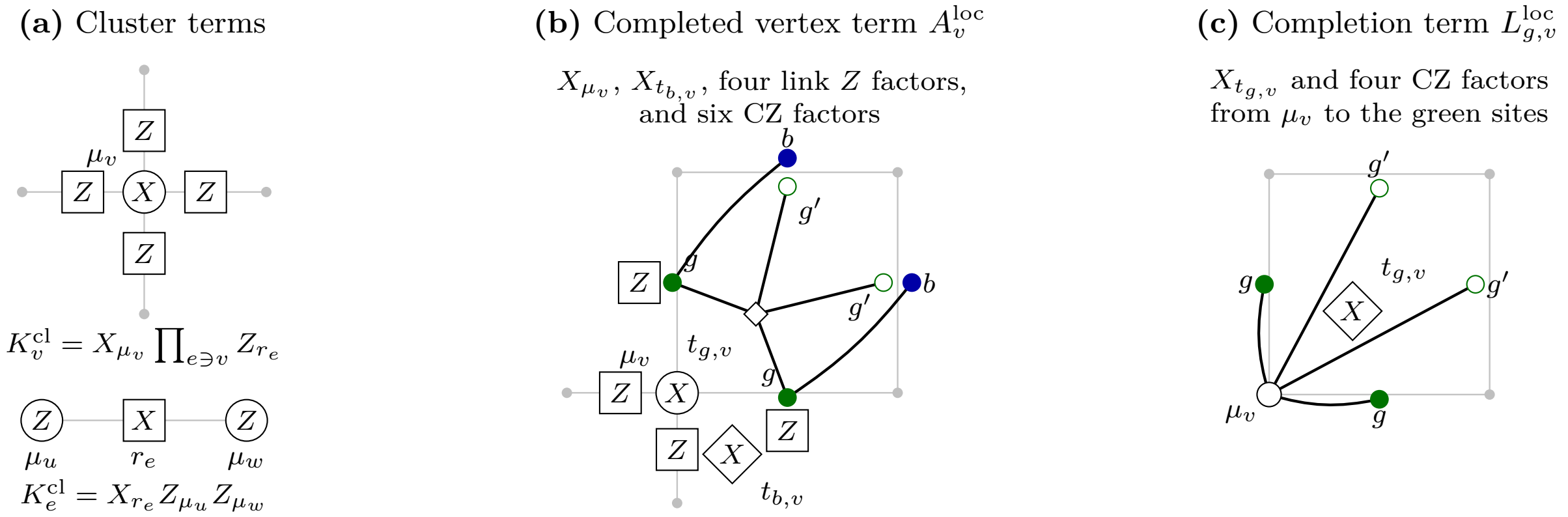

Combining the generator families gives the complete Hamiltonian $H_{\rm comp}=\mathsf H_{\rm loc}$ in explicit form,
\begin{equation}
 \begin{aligned}
 H_{\rm comp}={}&-\sum_v(A_v^{\rm loc}+L_{g,v}^{\rm loc}
                  +X_{t_{b,v}}+X_{t_{r,v}})\\
 &-\sum_eK_e^{\rm cl}-\sum_q(\widehat K_q^X+K_q^Z),\\
 A_v^{\rm loc}={}&X_{\mu_v}U_v\prod_{e\ni v}Z_{r_e}\,
                 X_{t_{b,v}}V(F_v,t_{g,v}),\\
 L_{g,v}^{\rm loc}={}&X_{t_{g,v}}V(F_v,\mu_v),\\
 \widehat K_q^X={}&X_{q'}X_q\operatorname{CZ}_{\mu_{v_q},p(q)}
       \prod_{v\in\mathcal I_g(q)}\operatorname{CZ}_{\mu_v,t_{g,v}},\\
 K_q^Z={}&Z_{q'}Z_q.
 \end{aligned}
 \label{eq:v468-completed-spt-hamiltonian}
\end{equation}
where $K_e^{\rm cl}=\widehat K_{r_e}$ is the unchanged link generator in Eq.~\eqref{eq:v468-cluster-hamiltonian}. The generators commute, square to identity and have a unique common $+1$ state. Figure~\ref{fig:v468-spt-terms}(b,c) shows the vertex and green-completion terms. The other terms are the unchanged link generators, $X_{q'}X_q$ generators dressed by CZ factors, $Z_{q'}Z_q$ generators and single-qubit completion $X$ operators.

The Hamiltonians $\mathsf H_{\rm nat}$ and $\mathsf H_{\rm loc}$ have the same unique ground state, but different terms and different displayed symmetry actions. For example,
\begin{equation}
 S_0^{\rm loc}L_{g,v}S_0^{{\rm loc}\dagger}
 =B_{g,v}F_vL_{g,v},
 \label{eq:prx314-resource-symmetry-distinction}
\end{equation}
so $S_0^{\rm loc}$ need not preserve the natural resource generator outside the subspace satisfying $Z_qZ_{q'}=1$ for every matter qubit $q$ and its reference $q'$. Equation~\eqref{eq:prx314-resource-state-equality} does not identify the two Hamiltonians under one fixed symmetry action.

\subsubsection{Restriction to the ideal input and the measurement record}
For an ideal toric-code bilayer input, the green fluxes already have $B_{g,v}=+1$. Enrichment and gauge-link preparation preserve these fluxes. Equation~\eqref{eq:prx314-resource-completion-circuit} then acts as the identity on the initialized completion qubits. Accordingly, the physical ideal gauged state has completion qubits in $|+\rangle_t$, as required by Eq.~\eqref{eq:main-code-distance}. By contrast, the resource with the reference qubits retained represents arbitrary matter inputs. It includes flux sectors in which the completion circuit couples to $t_g$, giving the reduced state in Eq.~\eqref{eq:prx314-completion-reduced-state}.

The Hamiltonians above encode the coherent gauging stage of Eq.~\eqref{eq:prx273-physical-factorization}, before vertex readout. The surgery protocol then measures the completed parent checks, central checks and conditional pins using the extraction circuits whose fault-distance and recovery guarantees are established in the main text.

\subsection{Relation to earlier SPT and gauging constructions}
\label{sec:prx314-spt-comparison}
The cluster resource used here builds on established gauging constructions~\cite{Tantivasadakarn2024LongRange,GorantlaShaoTantivasadakarn2025}. The bilayer construction of Christos \emph{et al.} uses the same zero-form/one-form cluster structure and CCZ symmetry enrichment before vertex readout~\cite{Christos2026NonAbelian}. Completion makes the gauged output checks commute on the full Hilbert space, including on error states, while preserving the ideal ground-state gauging map on the specified input sector. Equations~\eqref{eq:prx314-natural-resource-generators} and~\eqref{eq:prx314-onsite-resource-generators} give explicit parent Hamiltonians containing the completion-dependent couplings. The main text establishes the extraction circuits, fault-distance bounds and recovery procedures for quantum code surgery.

Explicit commuting higher-group SPT Hamiltonians also precede this work. Chen, Ellison and Tantivasadakarn construct a three-spatial-dimensional 2-group SPT Hamiltonian by an explicit finite-depth circuit conjugation, with explicit symmetry operators and local couplings~\cite{ChenEllisonTantivasadakarn2021}. Its dimension and degrees of freedom differ from the two-dimensional resource here. Our same-phase statement follows from the local change of basis above. On a general interaction graph, a spatial SPT interpretation additionally requires geometric locality.

\subsection{A two-dimensional SPT ancestor of the bilayer protocol}
\label{sec:prx313-typeiii}
The bilayer construction has an explicit short-range-entangled ancestor. It is a square-lattice representative of the two-dimensional type-III $\mathbb{Z}_2^3$ symmetry-protected topological (SPT) phase underlying measurement preparations of $D(D_4)$ order \cite{Tantivasadakarn2023Shortest,Tantivasadakarn2024LongRange,Christos2026NonAbelian}. Put three vertex qubits $\phi_{c,v}$, with $c=r,g,b$, on a closed square lattice. Write $\phi_c$ for their computational-basis bits and $a_c=\delta\phi_c$ for the edge differences. With $\hat x=(1,0)$ and $\hat y=(0,1)$, define
\begin{equation}
 \begin{aligned}
 F(\phi)&=\sum_v\phi_{r,v}(a_g\cup a_b)_v,\\
 (a_g\cup a_b)_v
 &=a_{g,e_x(v)}a_{b,e_y(v+\hat x)}
   +a_{g,e_y(v)}a_{b,e_x(v+\hat y)}.
 \end{aligned}
 \label{eq:prx313-typeiii-phase}
\end{equation}
All sums are modulo two. The diagonal circuit $W_{\mathrm{III}}|\phi\rangle=(-1)^{F(\phi)}|\phi\rangle$ is a bounded-depth product of local CCZ gates. Its state and Hamiltonian are
\begin{equation}
 \begin{gathered}
 |\mathrm{SPT}\rangle=W_{\mathrm{III}}|+\rangle^{\otimes3|V|},\qquad
 H_{\mathrm{SPT}}=\sum_{c,v}\frac{I-K_{c,v}}2,\\
 K_{c,v}=W_{\mathrm{III}}X_{\phi_{c,v}}W_{\mathrm{III}}^\dagger
       =X_{\phi_{c,v}}D_{q_{c,v}(\delta\phi)}.
 \end{gathered}
 \label{eq:prx313-typeiii-hamiltonian}
\end{equation}
Here, $D_q|\phi\rangle=(-1)^{q(\phi)}|\phi\rangle$. The bit string $e_{c,v}$ equals one at $(c,v)$ and zero elsewhere. The finite difference $F(\phi+e_{c,v})+F(\phi)$ equals the corresponding quadratic $q_{c,v}(\delta\phi)$ of Eq.~\eqref{eq:prx273-d4-quadratics} for all three colors. Thus, these are the raw star phases expressed in the vertex variables $\phi$. The $K_{c,v}$ are mutually commuting local Clifford operators that square to identity, and their common $+1$ eigenstate is unique.

The protecting symmetry is $S_c=\prod_vX_{\phi_{c,v}}$ for each color. Green and blue flips leave their differences unchanged. A red flip changes $F$ by $\int\delta\phi_g\cup\delta\phi_b=0$ on the closed lattice. More locally, let $R$ be a union of square faces and flip the red vertex at the southwest corner of each face in $R$. The phase changes by
\begin{equation}
 (-1)^{\int_R\delta\phi_g\cup\delta\phi_b}
 =(-1)^{\int_{\partial R}\phi_g\cup\delta\phi_b}.
 \label{eq:prx313-typeiii-domainwall}
\end{equation}
The red domain wall therefore supports a one-dimensional $\mathbb{Z}_2^g\times\mathbb{Z}_2^b$ SPT state. This identifies the type-III class, represented by the cocycle $\omega(a,b,c)=(-1)^{a_r b_g c_b}$. The fields $A_r,A_g,A_b$ are degree-one $\mathbb Z_2$ background cochains for the red, green and blue symmetries, respectively. Its background response $(-1)^{\int A_r\cup A_g\cup A_b}$ is the pullback of Eq.~\eqref{eq:prx313-resource-response} under $A=A_r$ and $B=A_g\cup A_b$. When $\delta A_g=\delta A_b=0$, this substitution also gives $\delta B=0$.

Now gauge the green and blue symmetries, replacing vertex differences by edge variables $g,b$. For each color, the normalized map $\mathcal G_c$ sends $(|\phi_c\rangle+|\phi_c+\mathbf1\rangle)/\sqrt2$ to $|\delta\phi_c\rangle$, with one representative per global-flip orbit. It is an isometry on the $S_c=+1$ sector. The output is
\begin{equation}
 \begin{aligned}
 &\mathcal{G}_g\mathcal{G}_b|\mathrm{SPT}\rangle\\
 &\quad=\bm{E}_{\mathrm{enr}}
 \bigl(|+\rangle_\mu^{\otimes|V|}
       |\mathrm{TC}_0\rangle_g|\mathrm{TC}_0\rangle_b\bigr),\\
 \bm{E}_{\mathrm{enr}}&=\prod_v
 \mathrm{CCZ}_{\mu_v,g_{e_x(v)},b_{e_y(v+\hat x)}}
 \mathrm{CCZ}_{\mu_v,g_{e_y(v)},b_{e_x(v+\hat y)}}.
 \end{aligned}
 \label{eq:prx313-typeiii-partial-gauging}
\end{equation}
Here, $|\mathrm{TC}_0\rangle$ is the uniform superposition of exact edge configurations, and $\mu_v$ is the remaining red vertex qubit. This bilayer input is $\mathcal U$-even because $\int g\cup b=0$ for exact $g,b$ on the closed lattice. The residual red symmetry acts on the bilayer through exactly $\mathcal U$ in Eq.~\eqref{eq:main-d4-physical}, by Eq.~\eqref{eq:prx273-enrichment-intertwining}. Gauging red with $\bm{E}_{\mathrm{link}}$, $X$ readout and the recorded sign correction then gives the $D(D_4)$ ground-state representative. Completion appends its fixed product-state qubits on this branch and supplies compatible output checks on excitations. The ancestor is an SPT, while its partially gauged state already has bilayer topological order. These equalities select zero parity around every closed path. General logical inputs are described by the branch maps of Section~\ref{sec:main-surgery}.

\bibliography{references}
\end{document}